\documentclass[10pt]{iopart}
\usepackage[T1]{fontenc}  
\usepackage{iopams}  
\expandafter\let\csname equation*\endcsname\relax
\expandafter\let\csname endequation*\endcsname\relax
\usepackage{amsmath}
\usepackage{mathtools}
\usepackage{bbold}
\usepackage{graphicx,subcaption}
\usepackage{tabularx}
\usepackage{bbm}
\usepackage[dvipsnames]{xcolor}
\usepackage{makecell}

\usepackage{multirow}
\usepackage[menucolor=false,linktoc=none]{hyperref}
\usepackage{braket}
\usepackage{orcidlink}
\usepackage{natbib}
\newcommand{\newblock}{}
\setcitestyle{square,numbers}
\usepackage{doi}

\begin{document}

\review{Quantum algorithms for scientific computing} 

\author{R. Au-Yeung$^1$, B. Camino$^2$, O. Rathore$^3$ and V. Kendon$^1$}
\address{$^1$ Department of Physics, University of Strathclyde, Glasgow G4 0NG, United Kingdom}

\address{$^2$ Department of Chemistry, UCL, London WC1E 6BT, United Kingdom}

\address{$^3$ Department of Physics, Durham University, Durham DH1 3LE, United Kingdom}

\eads{\mailto{rhonda.au-yeung@strath.ac.uk}, \mailto{viv.kendon@strath.ac.uk}}
\vspace{10pt}
\begin{indented}
\item[]\today
\end{indented}

\begin{abstract}
Quantum computing promises to provide the next step up in computational power for diverse application areas.  In this review, we examine the science behind the quantum hype, and the breakthroughs required to achieve true quantum advantage in real world applications. Areas that are likely to have the greatest impact on high performance computing (HPC) include simulation of quantum systems, optimization, and machine learning. We draw our examples from electronic structure calculations and computational fluid dynamics which account for a large fraction of current scientific and engineering use of HPC. Potential challenges include encoding and decoding classical data for quantum devices, and mismatched clock speeds between classical and quantum processors. Even a modest quantum enhancement to current classical techniques would have far-reaching impacts in areas such as weather forecasting, aerospace engineering, and the design of ``green'' materials for sustainable development. This requires significant effort from the computational science, engineering and quantum computing communities working together.
\end{abstract}

\vspace{2pc}

\submitto{\RPP}
\maketitle
\ioptwocol

\tableofcontents 
\markboth{Quantum algorithms for scientific applications}{Quantum algorithms for scientific applications}


\section{Introduction}\label{sec:introduction}

The largest quantum computers are now approaching a size and capability where it is no longer possible to efficiently simulate them classically, even using the largest available near-exascale classical computing capabilities. For most scientific and engineering applications, ``exascale'' means the system's aggregate performance crosses a threshold of $10^{18}$ IEEE 754 double precision (64-bit) operations (multiplications and/or additions) per second (exaFLOPS). To put this in context, the Sony PlayStation 5 digital edition is listed as having a peak of $\sim$10 teraFLOPS ($\sim10^{12}$ FLOPS), a million times smaller. Classical machines just below exascale have been used to verify quantum supremacy claims. For example, random circuit sampling and Gaussian boson sampling methods respectively used the Summit ($\sim$150 petaFLOPs) \cite{Arute2019} and Sunway TaihuLight ($\sim$93 petaFLOPs) \cite{Li2022g} supercomputers.  These supremacy tasks are not yet known to have any industrial or other useful applications, and the classical algorithms have since been improved (see section \ref{sssec:quantifyperformance}). However, these supremacy headlines do indicate that we are nearing the time when a quantum co-processor running a well-chosen sub-task could provide a significant boost to large scale computational capabilities.

\begin{figure*}[ht!]
\centering
\includegraphics[width=\linewidth]{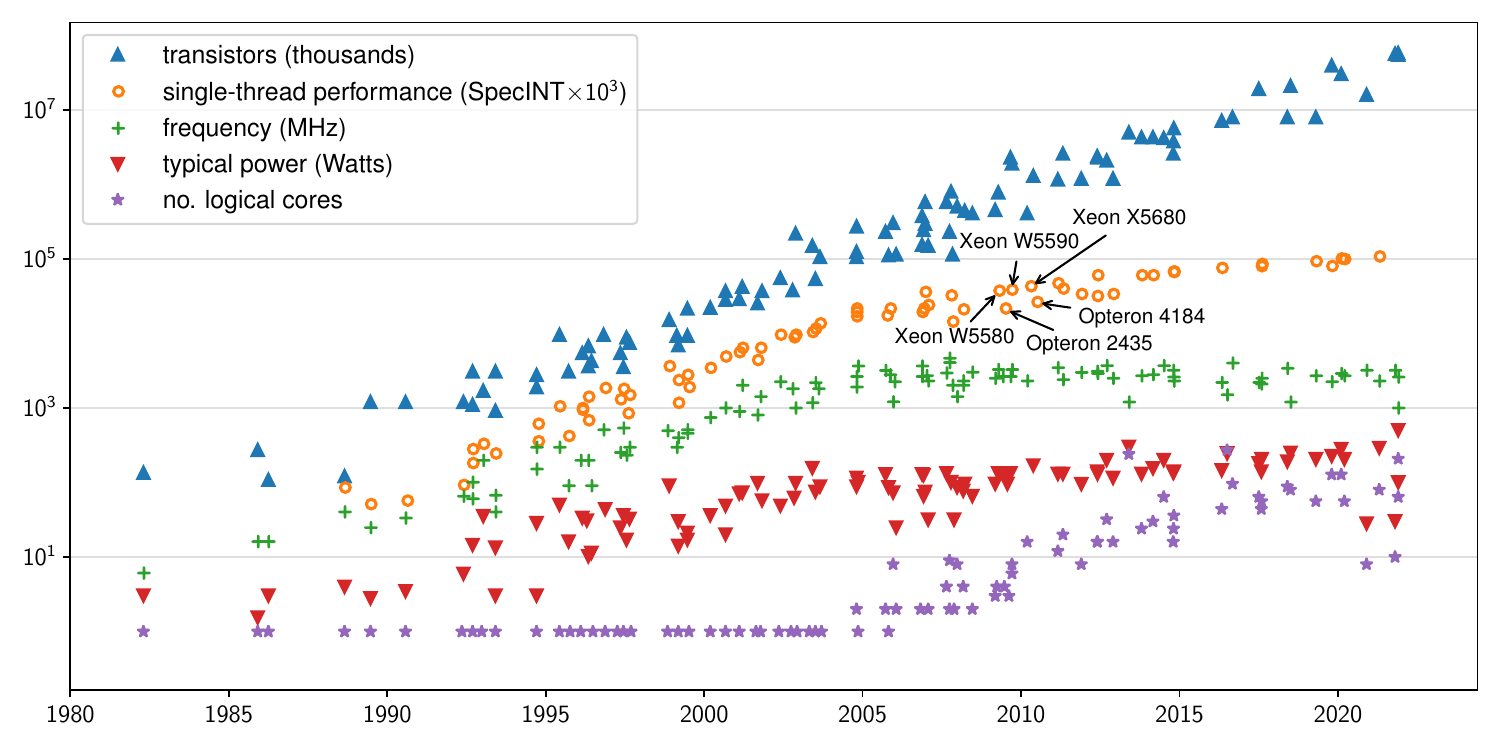}
\caption{Microprocessor trend data, 1980-2021. Data up to year 2010 collected by M. Horowitz, F. Labonte, O. Shacham, K. Olukotun, L. Hammond, and C. Batten, data for 2010-2021 by K. Rupp. Both datasets taken from \cite{Rupp2022}.}
\label{fig:microprocessor_trends}
\end{figure*}

Reaching for exascale computing calls for a paradigm shift from traditional computing advances \cite{Mann2020,Betcke2022}. Historical advances in computing were achieved via increasing clock speed and memory performance, adding more cores and parallelization. Moore's law \cite{Moore1965} predicts that the exponential growth of silicon-based transistors on computer chips is about to hit physical limits \cite{Khan2018,Leiserson2020}. The growth predicted was borne out by major chip and computer system vendors until about 2005, when CPU clock speeds reached around 4 GHz (Figure \ref{fig:microprocessor_trends}). Significantly increasing the clock speed beyond this would require enormous effort to cool the processor to prevent malfunction or permanent hardware damage from overheating.

Launched in 1993, the TOP500 project assembles and maintains a list of the world's 500 most powerful computer systems \cite{top500}. As of June 2024, Oak Ridge National Laboratory's Frontier system and Argonne National Laboratory's Aurora system are the only true exascale machines worldwide. In recent years, GPUs have enabled a step change in processing power -- all of the GREEN500 \cite{green500} use them. Only Fugaku in the top ten of the current TOP500 list \cite{top500} does not use GPUs. Physical constraints and power requirements are forcing computer vendors to develop new strategies to achieve more compute power. Novel hardware systems, including quantum computing, will be required to significantly extend global computational capacity beyond current capabilities.

The idea of using quantum systems to process information more efficiently than classical von Neumann computers was introduced some 40 years ago \cite{Feynman1982,Deutsch1985}. Since then there has been steady progress in quantum computing, building on the initial achievements of Shor's algorithm for factorization \cite{Shor1994} that could break current methods of encryption, and Grover's algorithm \cite{Grover1996,Grover1997} for searching unsorted data. Quantum simulation is another promising field of quantum information processing \cite{Georgescu2014,Altman2021}. Many near-term applications of quantum computers fall under this umbrella. Quantum simulation involves modelling the quantum properties of systems that are directly relevant to understanding modern materials science, condensed matter physics, high-energy physics, and quantum chemistry. 

As the practical relevance of quantum computing becomes clearer, interest has grown beyond the confines of academia and many countries now have national strategies to develop quantum computing and quantum technology more generally, e.g., \cite{DSIT2023,Acin2018,Monroe2019,Australia2023}. In June 2024, the United Nations proclaimed 2025 as the International Year of Quantum Science and Technology \cite{UN_IYQ}. The primary goal is to recognize the importance of quantum science and increase public awareness of its future impact. 

The consensus is that quantum devices will not simply replace HPC. It makes little sense (in terms of engineering and economics) to use quantum methods for problems that classical computers already do well.  Hence, a hybrid solution is the most efficient, cost-effective and productive way to approach quantum computation. Instead of simply sending jobs to one or more CPUs, hybrid architectures delegate different parts of the problem to different types of processors, the most common currently being GPUs. Conceptually, quantum processor units (QPUs) are a natural extension to enhance the processing power of HPC. However, increasing the number and diversity of processors compounds the challenges for efficient programming and scheduling. Quantum computing hardware is not yet mature enough to seamlessly integrate with HPC (see section \ref{sec:hpc} for discussion of some of the engineering hurdles).

\begin{figure*}[ht!]
\centering
\includegraphics[width=\linewidth,trim={0.5cm 6cm 0.5cm 6cm},clip]{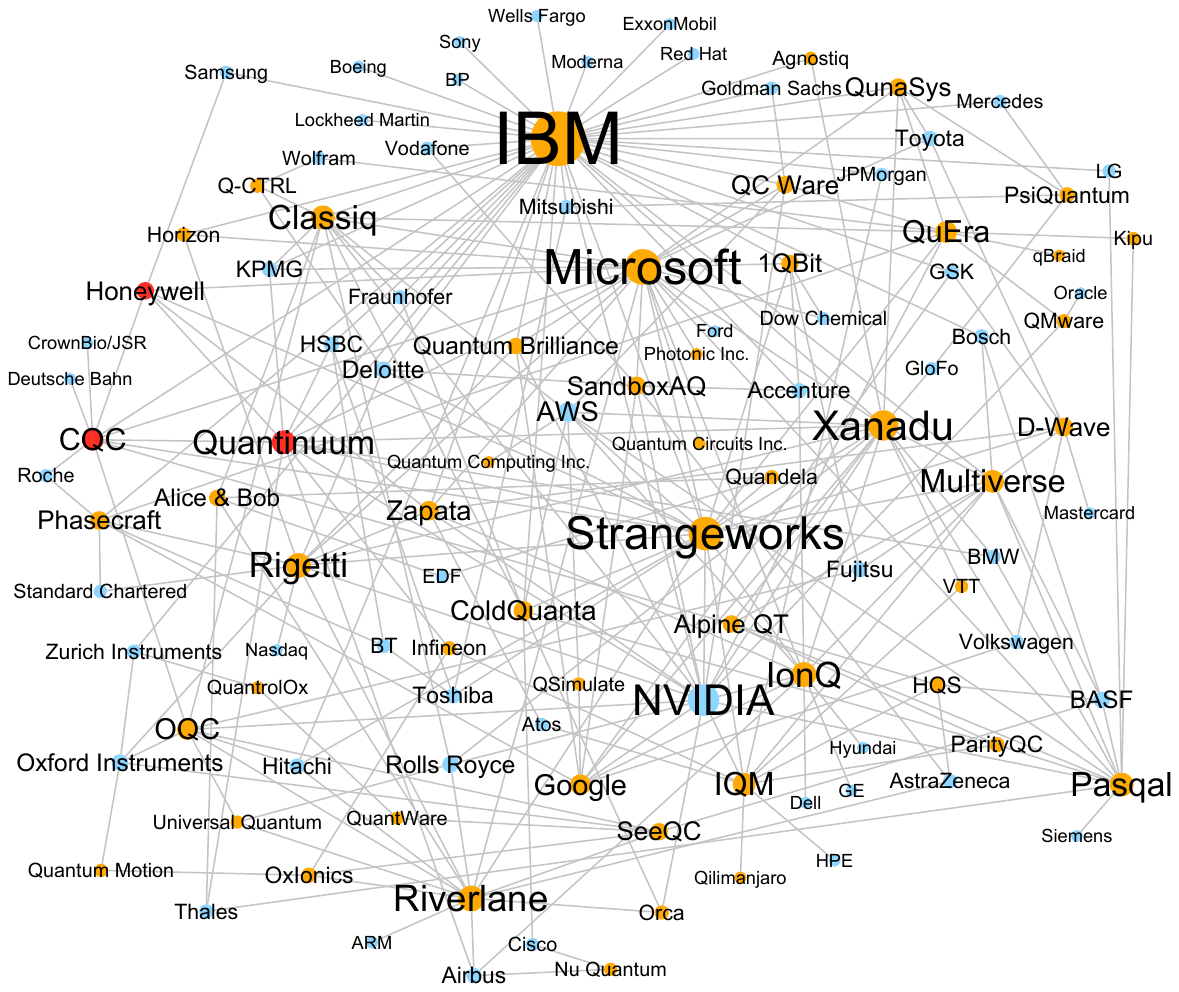}
\caption{Quantum computing commercial landscape. Some examples of companies (blue) partnered with quantum tech companies (orange) to advance QC research. Cambridge Quantum Computing (CQC) and Honeywell now merged as Quantinuum (red). We include CQC, Honeywell and Quantinuum separately as some collaborations were formed before the merger in Dec 2021. Universities, research institutes and government organizations not shown. Compiled by R. Au-Yeung from press releases and publicly disclosed interactions with end users.
}
\label{fig:commercial_landscape}
\end{figure*}

After several decades of largely academic-based theoretical and experimental development, growing commercial interest has resulted in venture capital \cite{MacQuarrie2020,Gibney2019,Bova2021} providing funding for many existing and start-up companies focused on aspects of quantum computing. The current landscape shows growing collaborations between quantum start-ups, established big players, and with well-established financial, pharmaceutical and automotive companies who wish to integrate quantum technologies into their development and production processes (Figure \ref{fig:commercial_landscape}). Notable players include Google, IBM and Rigetti, who are developing superconducting qubit systems, and Intel who are building systems based on quantum dots, with chips that can be made in its existing foundries. Cloud providers such as Amazon Web Services (AWS), are now selling time on different quantum systems, including QuEra, which uses neutral atoms in tweezer arrays, IonQ ion trap quantum processors, and superconducting qubit systems from Rigetti. Current systems are small and imperfect, known as noisy, intermediate scale quantum computers (NISQ) \cite{Preskill2018,Bharti2022}. They are not yet powerful enough to solve useful problems, but do allow researchers to test and develop quantum algorithms and applications to prepare for more advanced quantum hardware.

Several commercial partnerships are focused on integrating quantum processors with HPC. Finnish quantum start-up IQM is mapping quantum applications and algorithms directly to quantum processors to develop application-specific superconducting computers \cite{IQMtech}. The result is a quantum system optimized to run particular applications such as HPC workloads. French company Atos and partners Pasqal and IQM are involved in two major quantum hybridization projects in France and Germany \cite{Atos2020,Atos2021}: the European HPCQS (Quantum Simulation) project \cite{HPCQS2021} aims to build the first European hybrid HPC with an integrated quantum accelerator, and the German Government's Q-EXA project to integrate the first German quantum computer into HPC \cite{IQM2024}. NVIDIA's cuQuantum software development kit (SDK) provides the tools needed to integrate and run quantum simulations in HPC environments using GPUs \cite{NVIDIAcuQuantum2022}, with the goal of adding QPUs to the SDK. Quantum software vendor Zapata anticipates the convergence of quantum computing, HPC, and machine learning, and has created the Orquestra platform to develop and deploy generative AI applications \cite{ZapataOrquestra}.

Given the growing government and commercial investment in the development of quantum computing, it is timely to consider how it might accelerate computational science and engineering. Now considered the third pillar of scientific research alongside theoretical and experimental science, scientific computing has become ubiquitous in scientific and engineering applications. It leverages the exponential growth in computing power of the last decades, and receives even more government investment than quantum technology, given its essential role in science, engineering, and innovation. 

In the UK, HPC investments by the Engineering and Physical Sciences Research Council (EPSRC) totaled \pounds466 million over 10 years up to 2019 \cite{LondonEconomics2019,DSIT2022}. This is anticipated to add between \pounds3 billion and \pounds9.1 billion to the UK economy. The European High-Performance Computing Joint Undertaking (EuroHPC JU) is a \texteuro7 billion joint initiative between the EU, European countries and private partners to advance the HPC ecosystem in Europe \cite{EuroHPC}. These examples show that governments are willing to invest significant resources into developing computational technologies that can provide numerous societal and economic benefits.  Future UK plans include a significant increase in digital research infrastructure across the whole UKRI remit \cite{UKRI2024} to enable all disciplines to take advantage of computational and data based methods in their research, including quantum computing, when it can offer an advantage.

\subsection{Scope of this review}

Scientific computing has a broad reach across applications, so this review necessarily has to select which areas to focus on. We choose two important applications: electronic structure calculations as used in computational chemistry, materials, and life sciences; and fluid dynamics simulations, important across all length scales from cells to cosmology. In computational chemistry, the hard problems are fully quantum and therefore natural for quantum computers \cite{Feynman1982}. In fluid dynamics, the problem is purely classical, the hardness stems from the need to solve nonlinear differential equations. It is less clear that there will be substantial quantum advantages here, although there are multiple interesting proposals for possible quantum algorithms. Our chosen applications make up a significant fraction of current HPC use. Between them, they are diverse enough to offer insights for other areas of computational science and engineering simulations.  Although we discuss quantum machine learning techniques relevant for chemistry and fluid simulations, the wider field of ``big data'' and artificial intelligence (AI) is outside the scope of this review. There are indeed potential quantum enhancements in AI, but they merit their own dedicated reviews (see, e.g., \cite{Dunjko2018}).

In this review, we begin by describing the concepts behind quantum computers and giving examples of important foundational quantum algorithms in section \ref{sec:design}. For conciseness, we minimize the amount of mathematical detail where possible. Interested readers are encouraged to consult references such as \cite{Nielsen2010,Abhijith2022} for details, and refer to the original papers for proofs. There are already many excellent reviews that cover different classes of quantum algorithms (see e.g., \cite{Montanaro2016,Albash2018,Childs2010r}), including a ``Quantum Algorithm Zoo'' website \cite{QuantumAlgorithmZoo} which cites over 400 papers. In this review, we aim to emphasize the physical mechanisms and qualitative insights into quantum algorithms and their real-world applications. We will not discuss abstract topics such as complexity analysis \cite{Bernstein1993,Vazirani2002} in depth. We also cover some of the alternative quantum computing methods, quantum annealing (section \ref{ssec:qopt}) for chemistry and materials science simulations (section \ref{sec:quantumoptimizationchemistry}) and variational quantum algorithms (section \ref{ssec:VQAs}). We outline some of the many proposed performance benchmarks for quantum computers in section \ref{ssec:benchmark}.

Next, we focus on our two example fields: quantum simulation of quantum chemical systems (section \ref{sec:qsim}), and quantum algorithms for classical fluids simulation (section \ref{sec:qfluid}). We summarize the current (classical) methods and shortcomings, then explore how quantum computers may improve performance. With the theoretical potential established, we turn to the practical challenges of how to combine quantum components into HPC architectures in section \ref{sec:hpc}. This includes the clock speed mismatches. and how to encode classical data into quantum states. Finally, we summarize and set out future research directions in section \ref{sec:outlook}.

There are other important topics on the road to practical quantum computing that already have their own reviews. These include: quantum error correction \cite{Terhal2015,Roffe2019}; verification and testing \cite{Eisert2020,Carrasco2021} (we only cover benchmarks in section \ref{ssec:benchmark}); and comparisons between different hardware and software platforms available on the market \cite{MacQuarrie2020}.  Quantum annealing has been well-reviewed recently: \citeauthor{Hauke2020} \cite{Hauke2020} cover methods and implementations while \citeauthor{Yarkoni2022} \cite{Yarkoni2022} focus on industrial applications of optimization problems.

Despite numerous publications hyping the quantum revolution (see, e.g., \cite{Arute2019,Wu2021,Kim2023a}), it pays to read the fine print \cite{Leymann2020,Coveney2020,Aaronson2015}. The methodology in \cite{Chancellor2020} is an instructive example of how to fairly and objectively evaluate different use cases. This helps to avoid evaluations that focus on the positives and ignore weaknesses. 
A key message from the IQM-Atos 2021 white paper \cite{IQM2021} is that HPC centres must create a long-term strategic plan to successfully integrate quantum computing into their workflow. In particular, they recommend establishing a three-step roadmap:
\begin{enumerate}
\item Gap analysis and quantum solution identification (now). Where are the bottlenecks in classical HPC? How can we improve the solution accuracy?
\item Quantum solution design and integration (mid-term). What quantum methods can we use to resolve the issues identified in gap analysis? How do we integrate this into HPC architectures?
\item Quantum computing use-case development and implementation (long-term). Which applications and problems can get the most benefit from quantum speedup?
\end{enumerate}
Our review provides an introduction to the tools and topics needed to make such plans.


\section{Designing quantum algorithms}\label{sec:design}

Developing useful quantum algorithms is extremely challenging, especially before quantum computing hardware is widely available. It is like building a car without having a road or fuel to take it for a test drive. Since quantum computing has a different logical basis from classical computing, it is not as straightforward as taking a successful classical algorithm and mapping it to the quantum domain. For example, a classical CNOT gate makes a copy of a classical bit. In quantum theory, the no-cloning theorem \cite{Wootters1982} states that it is impossible to make copies of unknown quantum states. 

\begin{table}[t!]
\centering
\caption{Notation in our paper.}
\bgroup
\def\arraystretch{1.5}
\begin{tabular}{p{.6in}p{2.25in}}
\textbf{Notation} & \textbf{Description} \\
\hline
$a$, $A$ & scalar, can be real or complex; greek letters often used for complex scalars \\
$a(t)$, $A(t)$ & scalar function defined in terms of variable $t$ \\
$\vec{a}$ & vector, contains elements $\vec{a}_0$, $\vec{a}_1$, ... \\
$\mathbf{A}$ & matrix \\
$\ket{a}$, $\ket{A}$ & quantum state vector (with scalar label is usually a basis state)\\
$\ket{\psi}$ & general quantum state vector (with greek letter label) \\
$\hat{A}$ & quantum operator \\
$j$, $k$ & index in vector or sum\\
$i$ & imaginary unit $\sqrt{-1}$\\
\hline
\end{tabular}
\egroup
\label{tab:notation}
\end{table}

In this section, we outline the basic concepts underpinning quantum computing. Then, we describe some existing quantum algorithms that can potentially be adapted for practical scientific applications. To avoid confusion, we use the notation in Table \ref{tab:notation} to differentiate between different quantities.  These include operators (hats), matrices (bold font), vectors (arrows), scalars and functions (italics), and bra-ket notation for quantum states.

\subsection{Universal quantum computers}\label{ssec:universal}

The most widely used model for quantum computing is the gate or circuit model. The basic unit of quantum information is the quantum bit (qubit), a two-state system. Qubits are the quantum analogue of classical bits. They are two-level quantum systems that represent linear combinations of two basis states 
\begin{equation}
\ket{\psi} = \alpha\ket{0} + \beta\ket{1}
\end{equation} 
where $\alpha,\beta \in \mathbb{C}$ and $\vert\alpha\vert^2 + \vert\beta\vert^2 = 1$. 
In vector notation, we usually choose 
\begin{equation}\label{eq:compbasis}
\ket{0} = \begin{bmatrix} 1 \\ 0 \end{bmatrix}, \quad \ket{1} = \begin{bmatrix} 0 \\ 1 \end{bmatrix}.
\end{equation}

Multiple qubits can be entangled, forming systems with an exponentially large state space of size $N=2^n$ for $n$ qubits.  This corresponds to the tensor product of matrices, for example, 
\begin{equation}
\ket{0}\otimes\ket{1} = \begin{bmatrix} 0 \\ 1 \\ 0 \\ 0 \end{bmatrix}.
\end{equation} 
The superposition property means that an $N$-qubit system can represent $2^N$ states whereas classical bits represent only one state at a time (a single $N$-bit string).  
\begin{figure}[t!]
\centering
\includegraphics[width=\linewidth]{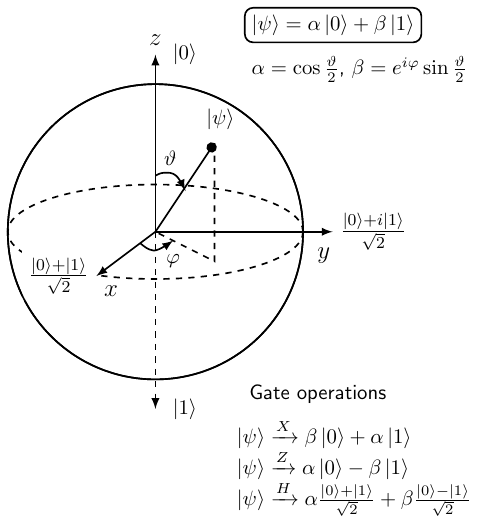}
\caption{Bloch sphere representation of qubit state $\ket{\psi}$ with examples of gate operations.}
\label{fig:bloch-sphere}
\end{figure}
We visualize single qubits using a Bloch sphere (Figure \ref{fig:bloch-sphere}). To perform calculations, we manipulate the qubit state using (reversible) gate operations. These are analogous to the classical logic gates, such as AND, OR and NOT.

\begin{table}[t!]
\centering
\caption{Examples of commonly used quantum gates. Middle column uses circuit notation where horizontal lines represent qubits. Boxes, crosses and dots indicate the gate applied. Time flows from left to right. Right-hand column shows the matrix representation for acting on computational basis states, \eqref{eq:compbasis}.}
\bgroup
\def\arraystretch{1}
\begin{tabular}{p{1in}p{.8in}p{.8in}}
\makecell{\textbf{One-qubit} \\ \textbf{gate}} & 
\thead{\textbf{Circuit} \\ \textbf{symbol}} & 
\thead{\textbf{Matrix}} \\[4pt]
\hline \\
\makecell{Pauli-X or NOT \\ {\small (bit flip)}} &
\makecell{\includegraphics[width=.8in]{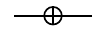}} &
$\begin{bmatrix} 0 & 1 \\ 1 & 0\end{bmatrix}$
\\
\makecell{Pauli-Z \\ {\small (phase flip)}} &
\makecell{\includegraphics[width=.8in]{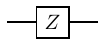}} &
$\begin{bmatrix} 1 & 0 \\ 0 & -1\end{bmatrix}$
\\
\makecell{Hadamard \\ {\small (create} \\ {\small superposition)}} &
\makecell{\includegraphics[width=.8in]{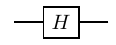}} &
$\dfrac{1}{\sqrt{2}}\begin{bmatrix} 1 & 1 \\ 1 & -1\end{bmatrix}$
\\
T gate &
\makecell{\includegraphics[width=.8in]{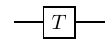}} &
$\begin{bmatrix} 1 & 0 \\ 0 & e^{i\pi/4} \end{bmatrix}$
\\
Phase shift &
\makecell{\includegraphics[width=.8in]{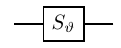}} &
$\begin{bmatrix} 1 & 0 \\ 0 & e^{i\vartheta} \end{bmatrix}$
\\
Z-rotation &
\makecell{\includegraphics[width=.8in]{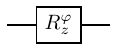}} &
$\begin{bmatrix} e^{-i\varphi/2} & 0 \\ 0 & e^{i\varphi/2} \end{bmatrix}$
\\
\hline\hline \\
\makecell{\textbf{Two-qubit} \\ \textbf{gates}} & 
\thead{\textbf{Circuit} \\ \textbf{symbol}} & 
\thead{\textbf{Matrix}} \\[4pt]
\hline \\
\makecell{CNOT \\ {\small (flip target qubit if} \\ {\small control qubit is $\ket{1}$)}} &
\makecell{\includegraphics[width=.8in]{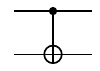}} &
$\begin{bmatrix} 1 & 0 & 0 & 0 \\ 0 & 1 & 0 & 0 \\ 0 & 0 & 0 & 1 \\ 0 & 0 & 1 & 0 \end{bmatrix}$
\\
\makecell{SWAP \\ {\small (swap two qubit} \\ {\small states)}} &
\makecell{\includegraphics[width=.8in]{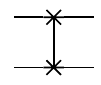}} &
$\begin{bmatrix} 1 & 0 & 0 & 0 \\ 0 & 0 & 1 & 0 \\ 0 & 1 & 0 & 0 \\ 0 & 0 & 0 & 1 \end{bmatrix}$
\\
\hline
\hline
\end{tabular}
\egroup
\label{tab:basic-gates}
\end{table}

Gate operations are unitary operators applied to the qubits. Applying such an operator $\hat{U}$ to a qubit state, then its conjugate transpose $\hat{U}^\dagger$ brings the qubit back to its original state ($\hat{U}\hat{U}^\dagger = \mathbb{1}$). Gate operations can be written as matrices which act on the state vectors (see examples in Table \ref{tab:basic-gates}). The building blocks of the transformations are $2 \times 2$ and $4 \times 4$ unitary matrices on single qubits and two qubits respectively. As in classical computing, any computation can be built up from a small set of universal gates.  A universal gate set includes at least one entangling gate acting on two or more qubits, and one or more single qubit rotations.  The subset of gates that are efficiently classically simulable and thus not sufficient for universal quantum computing are known as Clifford gates, and can be generated by the Hadamard, CNOT and a $\pi/2$ phase gate $S=\sqrt{Z}$.  The single qubit Pauli gates ($X$, $Y$, and $Z$) are included in the set of Clifford gates.  The addition of a $T$ gate or a $Z$-rotation with a well-chosen angle is sufficient to generate universal quantum computing.  Symbols for diagramatic representation of quantum circuits are also shown in table \ref{tab:basic-gates}.

\subsection{Quantum computing stack}

\begin{figure}[t!]
\centering
\includegraphics[width=\linewidth]{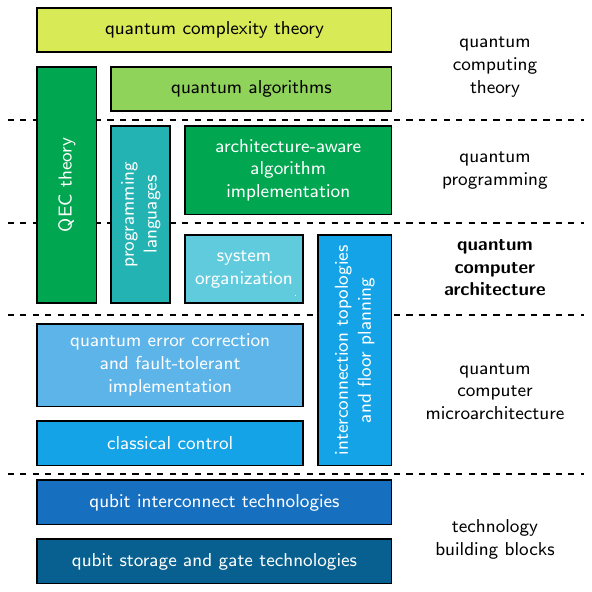}
\caption{Quantum computer architecture among some sub-fields of quantum computation. Reproduced from \cite{vanMeter2013}}
\label{fig:stack2013}
\end{figure}

As with classical computers, we can decompose the quantum computer architecture in terms of its various levels of abstraction, called the ``stack''. In gate-based models, this extends from the user interface and compiler down to the low-level gate operations on the physical hardware itself (Figure \ref{fig:stack2013}). 
At the highest levels, quantum algorithms give instructions for solving the numerical problem. Compiling and translating quantum gates in the middle levels (software stack) compress the algorithms to accelerate performance, while error-correction techniques mitigate quantum hardware errors. The lower levels comprise the quantum computer's building blocks (hardware) for manipulating and storing qubits \cite{vanMeter2013}.

Inside a quantum computer, physical qubits are physical devices that behave as two-level quantum systems.  They are usually not perfect, and error correction \cite{Terhal2015} is designed to combine groups of physical qubits into one or a few logical qubits with lower error rates.  This is implemented at a low level in the stack, so that users can work with near perfect logical qubits in a quantum algorithm.  Many physical qubits are needed for quantum error correction procedures \cite{Roffe2019} to produce one logical qubit to perform useful computations. Engineering qubits in real physical hardware is highly challenging.  The requirements for scalable quantum computing laid out by \citeauthor{DiVincenzo2000} \cite{DiVincenzo2000} are still relevant today: 
scalable, well-characterized qubits;
the ability to initialize the qubits;
long decoherence times;
a ``universal'' set of quantum gates;
and the ability to measure the qubits.

For example, \citeauthor{Reiher2017} \cite{Reiher2017} examine the cost of performing quantum simulations to find the potential energy landscape of a quantum chemical structure (FeMoco, more details in section \ref{sec:fertilizers}). They estimated 313 physical qubits are needed for each logical qubit, if there is one error for every $10^9$ quantum gates. By extension, we would likely need thousands of physical qubits per logical qubit, or much better error rates, for complicated real-world quantum chemistry simulations. In the rest of our review, we will discuss in terms of logical qubits unless otherwise stated.

\subsection{Foundational quantum algorithms}

In this section, we introduce some foundational quantum algorithms which can be adapted or used as components in quantum algorithms for a wide variety of applications. They also illustrate how quantum algorithms can provide powerful speed ups over classical algorithms for the same problems.  For readers seeking a comprehensive graduate-level introduction, we recommend \cite{Abhijith2022,Portugal2022,Nielsen2010} plus IBM's Qiskit software documentation \cite{Qiskit}.

\subsubsection{Quantum Fourier transform}\label{sssec:QFT}\hfill

The quantum Fourier transform (QFT) is a key subroutine in many quantum algorithms, most prominently Shor's factoring algorithm \cite{Shor1994}. The QFT operation changes the quantum state from the computational (Z) basis to the Fourier basis. It is the quantum analogue of the discrete Fourier transform and has exponential speed-up (complexity $O((\log N)^2)$) compared to the fast Fourier transform's $O(N \log N)$ complexity for problem size $N$. The QFT performs a discrete Fourier transform on a list of complex numbers encoded as the amplitudes of a quantum state vector. It stores the result ``in place'' as amplitudes of the updated quantum state vector.  Measurements performed on the quantum state identify individual Fourier components -- the QFT is not directly useful for determining the Fourier-transformed coefficients of the original list of numbers, since these are stored as amplitudes (see section \ref{ssec:encoding} for more details on quantum encoding).

When applying the QFT to an arbitrary multi-qubit input state, we express the operation as
\begin{equation}
\hat{U}_\textrm{QFT} = \frac{1}{\sqrt{N}} \sum_{j,k=0}^{N-1} e^{2\pi i jk/N} \ket{j}\!\bra{k}
\end{equation}
which acts on a quantum state in $N$-dimensional Hilbert space $\ket{x} = \sum_{k=0}^{N-1} x_k \ket{k}$ and maps it to $\ket{y} = \sum_{j=0}^{N-1} y_j \ket{j}$. 
This is a unitary operation on $n$ qubits that can also be expressed as the complex $N \times N$ matrix
\begin{align}
&\mathbf{F}_N = \nonumber\\
&\frac{1}{\sqrt{N}} \begin{bmatrix}
1 & 1 & 1 & 1 & ... & 1 \\
1 & \omega & \omega^2 & \omega^3 & ... & \omega^{N-1} \\
1 & \omega^2 & \omega^4 & \omega^6 & ... & \omega^{2(N-1)} \\
1 & \omega^3 & \omega^6 & \omega^9 & ... & \omega^{3(N-1)} \\
\vdots & \vdots & \vdots & \vdots & \ddots & \vdots \\
1 & \omega^{N-1} & \omega^{2(N-1)} & \omega^{3(N-1)} & ... & \omega^{(N-1)^2}
\end{bmatrix} \label{eq:qft-matrix}
\end{align}
where $\omega=e^{2\pi i/N}$ and $N=2^n$.
Note that the Hadamard gate (Table \ref{tab:basic-gates}) acts as the single-qubit QFT. It transforms between the Z-basis states $\ket{0}$ and $\ket{1}$ to the X-basis states 
\begin{equation}
\ket{\pm} = \frac{1}{\sqrt{2}} (\ket{0} \pm \ket{1}).
\end{equation}

\begin{figure*}[ht!]
\centering
\includegraphics[width=\linewidth]{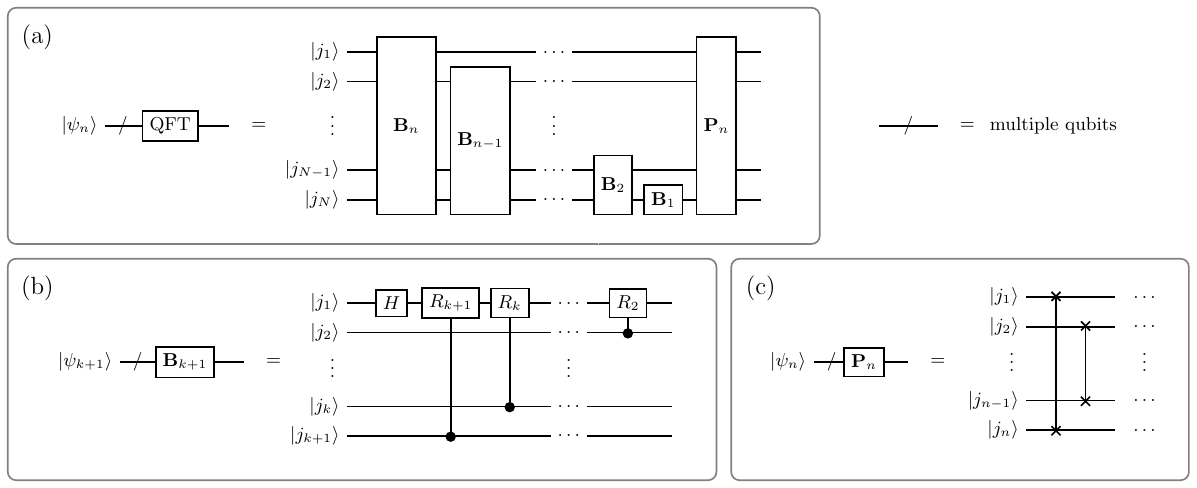}
\caption{Quantum Fourier transform operation on $n$ qubits. QFT = $\mathbf{F}_N$, \eqref{eq:qft-matrix}. Adapted from \cite{Camps2020}.}
\label{fig:qft}
\end{figure*}

Figure \ref{fig:qft}(a) shows the quantum circuit for the QFT operation. Each multi-qubit gate block can be decomposed into a sequence of single or two-qubit gates. For example, the $\mathbf{B}_{k+1}$ block is written in terms of controlled operations involving two qubits (Figure \ref{fig:qft}(b)). The $\mathbf{B}_{k+1}$ gate is applied to the last $k+1$ qubits. All controlled-$R$ gates are diagonal matrices and commute,
\begin{equation}
\mathbf{R}_N = \begin{bmatrix} 1 & 0 \\ 0 & e^{-2\pi i/N} \end{bmatrix}
\end{equation}
so the order in which they are applied does not change the outcome. The bit-reversal permutation matrix $\mathbf{P}_n$ can first be written in terms of $\lfloor n/2 \rfloor$ swap gates (Figure \ref{fig:qft}(c)). Since each swap can be implemented with three CNOT gates, $\mathbf{P}_n$ requires $\lfloor 3n/2 \rfloor$ CNOT gates. Hence the QFT requires $O(n^2)$ elementary gates in total.

The semi-classical nature of the QFT circuit means it can be efficiently simulated on classical computers \cite{Griffiths1996,Browne2007} when used at either the start or end of a quantum computation. The idea is to measure a qubit, then use the result to produce a classical signal which controls a one-qubit transformation on the next qubit before it is measured, and so forth.

\subsubsection{Quantum phase estimation}\label{sssec:QPE}\hfill

Kitaev's quantum phase estimation (QPE) algorithm \cite{Kitaev1996} is a core component of many quantum algorithms and an important technique in algorithm design. While the standard procedure is based on the QFT circuit, various improvements can reduce the dependence on QFT while retaining its accuracy by using, for example, approximations to relax the QFT constraints \cite{Ahmadi2012} or a modified algorithm from quantum metrology \cite{Ni2023}. We outline the QPE algorithm below.

The basic phase estimation problem is, given a unitary operator
\begin{equation}
\hat{U} \ket{\psi} = \lambda \ket{\psi},
\end{equation}
we want to find an eigenvalue $\lambda$ of the eigenvector $\ket{\psi}$. Because $\hat{U}$ is unitary, the eigenvalue can be expressed as $\lambda = e^{2\pi i \theta}$ and the real aim is to estimate the eigenvalue phase $\theta$.

\begin{figure}[ht!]
\centering
\includegraphics[width=\linewidth]{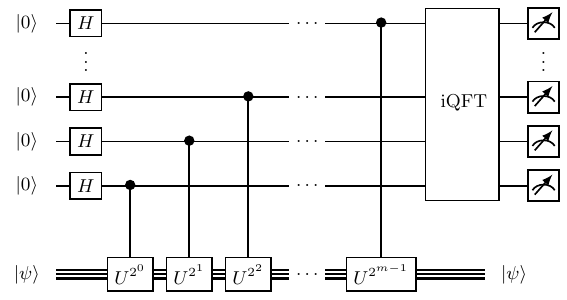}
\caption{Structure of quantum phase estimation circuit, where iQFT is the inverse of the QFT operation.}
\label{fig:qpe}
\end{figure}

The phase estimation algorithm contains two main steps (Figure \ref{fig:qpe}). In the circuit, the top register contains $n$ ``counting'' qubits in state $\vert 0 \rangle$. The second register contains as many qubits as needed to store $\ket{\psi}$, the eigenstate of $\hat{U}$. We apply a set of Hadamard gates to the first register (this is equivalent to a QFT on the all zeros state) and controlled-$\hat{U}$ operations on the second register with $\hat{U}$ raised to successive powers of two. The final state of the first register is 
\begin{align}\label{eq:qpe1}
\frac{1}{\sqrt{2^n}} \sum_{k=0}^{2^n-1}e^{2k\pi i\theta} \vert k \rangle.
\end{align}
Applying the inverse QFT (iQFT) gives
\begin{equation}\label{eq:qpe2}
\frac{1}{\sqrt{2^n}} \sum_{k=0}^{2^n-1}e^{2k\pi i\theta} \vert k \rangle \vert \psi \rangle \to \vert\tilde{\theta}\rangle \vert \psi \rangle
\end{equation}
where $\tilde{\theta}$ is an estimate for $\theta$. The final stage is to read out the state of the first register by measuring in the computational basis. Hence, we estimate $\theta$ to $n$ bits of precision.

Quantum phase estimation is equivalent to simulating quantum system dynamics. Given an eigenstate $\ket{\psi}$, we can estimate the eigenvalue for Hamiltonian $\hat{H}$ by applying $\hat{H}$ via the time-evolution operator $\hat{U}=e^{-i\hat{H}t}$. When the algorithm implements $\hat{U}^k$ for increasing powers $k$, this corresponds to evolving the system for increasing times $kt$ \cite{AspuruGuzik2005}. The phases $\theta$ are eigenvalues of time evolution operator $\hat{U}$ which we can map back on to the eigenenergies $E$ of $\hat{H}$ \cite{AspuruGuzik2005},
\begin{gather}
\hat{U}\ket{\psi} = e^{i\hat{H}t}\ket{\psi} = e^{2\pi i \theta}\ket{\psi}, \\
E = 2\pi\theta/t.
\end{gather}
Computational chemistry often requires finding the eigenenergies, as they can help deduce many chemical properties like ionization potential and equilibrium constants. We discuss some of these applications in section \ref{sec:qsim}.

The phase estimation circuit outputs an eigenphase as the measurement outcomes of the top register. It also prepares the corresponding eigenstate $\ket{\psi}$ in the lower register. This is a useful starting point for calculating other observables besides energy. Once we have the eigenstates and eigenenergies of interest, we can understand how a (closed) system evolves in time by decomposing the initial state into a sum of eigenstates and evolving each eigenstate according to the phase found in QPE. The final state is an interferometric sum of all components.

There are two important performance metrics for the QPE: maximum run time, and total run time over all repetitions from each circuit in the algorithm. Maximum and total run time approximately measure the circuit depth and total cost of the algorithm respectively. The QPE algorithm in Figure \ref{fig:qpe} is unsuitable for early fault-tolerant quantum computers since these devices can only implement low circuit depths and limited numbers of logical qubits \cite{Ni2023}.

\subsubsection{Shor's algorithm}\hfill

While Feynman's 1982 arguments \cite{Feynman1982} formalized the idea of using quantum computers to solve quantum problems, what really caught people's attention was Shor's prime factoring algorithm in 1994 \cite{Shor1994,Ekert1996}. It provided a way for large enough (future) quantum computers to break RSA encryption \cite{Rivest1978}, then considered unbreakable due to the exponential computational cost of factoring large numbers. RSA encryption works by multiplying two large prime numbers $p$ and $q$. The product $C=pq$ is so large that the computational cost of factorizing -- and hence recovering the encrypted message -- is impractical. Shor's algorithm showed that factoring could be solved by quantum computers in polynomial time. 

Shor's factoring algorithm reduces the factorization problem into an order-finding problem. Given the modular exponential function 
\begin{equation}\label{eq:shor}
f(x)=a^x\text{mod}\,C, \; 0<a<C
\end{equation}
where $a$ is a random number coprime to $C$, the order-finding problem involves finding the period $r$ of function $f(x)$. For an even $r$ and if $a^{r/2} \neq -1 \text{mod}C$, at least one prime factor of $C$ is given by $\text{gcd}(a^{r/2} \pm 1, C)$.

\begin{figure}[ht!]
\centering
\includegraphics[width=\linewidth]{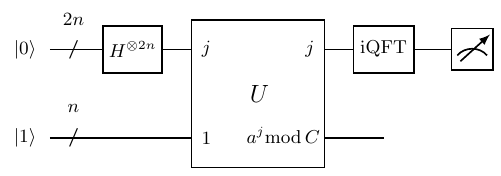}
\caption{Quantum circuit schematics for Shor's algorithm.}
\label{fig:shor-circuit-eq}
\end{figure}

There are two main components of Shor's algorithm: modular exponentiation (calculating $a^x\text{mod}C$) and inverse quantum Fourier transform (iQFT). Figure \ref{fig:shor-circuit-eq} shows a high-level circuit diagram of the algorithm. First we initialize the registers: we use $n=\lceil\log_2(C+1)\rceil$ qubits to store the number $C$. To extract the period $r$ of $f(x)$, we need one register initialized as $\ket{0}$ and containing $2n$ qubits to store values $x$, and another register initialized as $\ket{1}$ with $n$ qubits to store $f(x)$.

Applying the Hadamard gate on the qubits in $\ket{0}$ produces a superposition state of all integers, such that $\ket{0} \to \sum_{j=0}^{2^n-1} \ket{j}/\sqrt{2^n}$. Then the unitary $\hat{U}$ associates each input $j$ with the value $a^j\text{mod}\,C$ \eqref{eq:shor}, for some random $a$ which has no common factors with $C$, to get
\begin{equation}
\frac{1}{\sqrt{2^n}} \sum_{j=1}^{2^n-1} \ket{j}\ket{a^j\text{mod}C}.
\end{equation}
The first register is in a superposition of $2^n$ terms $\ket{j}$ and the circuit calculates the modular exponentiation for $2^n$ values of $j$ in parallel. Next, inverse QFT on the first register gives
\begin{equation}
\frac{1}{2^n} \sum_{k=0}^{2^n-1} \sum_{j=0}^{2^n-1} e^{2\pi i jk/2^n} \ket{k}\ket{a^j\text{mod}C},
\end{equation}
where interference causes only $k$ terms with integer multiples of $2^n/r$ to have a substantial amplitude. This is analogous to the quantum phase estimation in \eqref{eq:qpe1} and \eqref{eq:qpe2}. Finally, measuring the first register gives the period $r$. Ideally, the measurement outcome is an integer multiple of $2^n/r$ with high probability. We can then deduce $r$ classically, with continued fractions.

Note that Kitaev's QPE algorithm \cite{Kitaev1996} gives an alternate derivation of Shor's algorithm. It reduces the order-finding problem to one involving phase estimation by essentially replacing the unitary $\hat{U}$ (Figure \ref{fig:shor-circuit-eq}) with a QPE subroutine.

Classical computers have to date factored semi-primes up to 829 bits long \cite{classicalfactoring}. Each extra bit doubles the classical computational cost. However, quantum computers large enough to beat this are still some way into the future. This is due to the error correction overheads required to carry out the quantum computation accurately enough \cite{vanMeter2016}.

\subsubsection{Grover's algorithm}\label{sssec:search}\hfill

Grover's algorithm \cite{Grover1996,Grover1997} is a technique for searching unstructured data. The canonical example is finding the name corresponding to a given phone number.  It can be used as a building block in other applications, such as database search and constraint satisfaction problems \cite{Ambainis2006}.
\begin{figure*}[ht!]
\centering
\includegraphics[width=\linewidth]{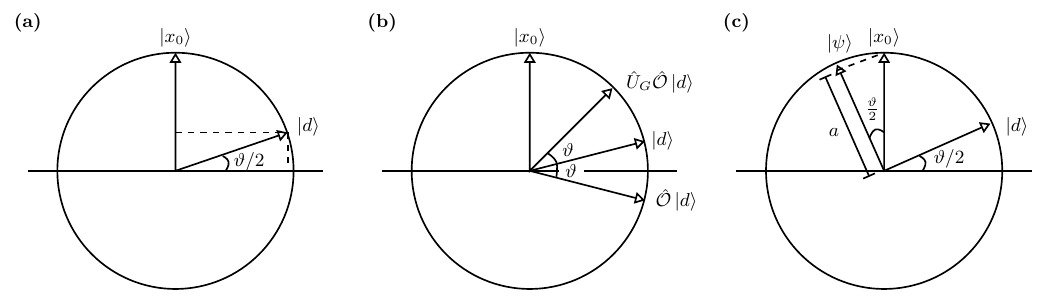}
\caption{Grover's algorithm. \textbf{(a)} Initial state vector $\ket{d}$ with angle $\vartheta/2$. \textbf{(b)} Vector $\ket{d}$ after inversion and reflection. \textbf{(c)} Final state vector $\ket{\psi}$.}
\label{fig:grover}
\end{figure*}
Suppose there are $N$ items labelled by bit strings $x=0$ to $N-1$, and there is a unique element $x_0$ such that $f(x)=1$ if and only if $x=x_0$. The quantum algorithm uses $n$-qubit basis states $\ket{x}$ corresponding to the labels of the items, with $n=\lceil\log_2 N\rceil$.

Grover's algorithm uses an oracle, which in computer science is a mathematical device for providing part of the computation that is not included in the analysis.  Consequently, Grover's algorithm is only of practical use in situations where we can perform the oracle function efficiently, i.e., $f(x)$ can be evaluated for a superposition of all possible inputs $\ket{x}$.  This can happen when the algorithm is used as a subroutine, but is not generally possible when the data are classical.
The oracle $\hat{\mathcal{O}} = (-1)^{f(x)}$ has the effect
\begin{equation}\label{eq:searchoracle}
\hat{\mathcal{O}} \ket{x} =
\begin{cases}
-\ket{x_0}, & x=x_0 \\
\ket{x}, & \textrm{otherwise.} \\
\end{cases}
\end{equation}

Figure \ref{fig:grover} illustrates in two dimensions the amplitude amplification part of Grover's algorithm, through inversion about the mean.  The initial state 
\begin{equation}
\ket{d} = \frac{1}{\sqrt{N}} \sum_x \ket{x}
\end{equation}
is an equal superposition of all $\ket{x}$. For large $N$, note that $\ket{d}$ is almost orthogonal to the marked state $\ket{x_0}$. In Figure \ref{fig:grover}(a), the small angle $\vartheta/2\simeq 1/\sqrt{N}$.  The oracle $\hat{\mathcal{O}}$ acts on only $\ket{x_0}$ by inverting the sign of its amplitude. Geometrically, $\hat{\mathcal{O}}$ reflects vector $\ket{d}$ about the horizontal axis. Then applying the Grover operator
\begin{equation}\label{eq:functionoracle}
\hat{U}_G = 2 \ket{d} \bra{d} - \mathbbm{1}
\end{equation}
reflects $\hat{\mathcal{O}}\ket{d}$ about the axis defined by $\ket{d}$. Figure \ref{fig:grover}(b) shows the action of $\hat{\mathcal{O}}$ and $\hat{U}_G$.  Since $\vartheta$ is a small angle, this achievement is modest. Solving for $r\vartheta = \pi/2$ gives an approximate number of required iterations,
\begin{equation}\label{eq:Grovrep}
r = 
\Big\lfloor \frac{\pi}{2\vartheta} \Big\rfloor 
=
\Big\lfloor \frac{\pi}{4} \sqrt{N} \Big\rfloor
\end{equation}
to reach the final state within $\vartheta/2$ of $\ket{x_0}$. The probability of measuring $\ket{x_0}$ is then better than $1-1/N$. Since the number of iterations $r$ in \eqref{eq:Grovrep} scales as $\sqrt{N}$, Grover's algorithm generally provides a square root speed up over a classical search, which would need to check on average $N/2$ items to find $x_0$. There are alternative quantum algorithms that solve the search problem, such as quantum walks \cite{Shenvi2002} (section \ref{sssec:qw}). These are easily generalized to find multiple items \cite{Childs2003} and are intuitive for graph-based search problems.


\subsubsection{Quantum amplitude amplification}\label{sssec:qaa}\hfill

Grover's algorithm was subsequently generalized to the framework of quantum amplitude amplification (QAA) \cite{Brassard2002}. We can understand QAA as an iterative process that starts with a uniform superposition of all states. Then the algorithm uses a generalization of the Grover operator to increase the probability amplitude of some target state while reducing all other probability amplitudes at each iteration. Amplitude amplification is a valuable subroutine that appears in many quantum algorithms. In the Harrow-Hassidim-Lloyd (HHL) algorithm for solving linear systems (section \ref{sssec:HHL}), the variable time QAA algorithm reduces the number of repetitions needed to obtain the correct solution, which reduces the algorithm run-time \cite{Ambainis2012}. QAA also appears in applications that involve optimization \cite{Hogg2000,Durr1996}, determining graph connectivity \cite{Durr2004}, pattern matching \cite{Ramesh2003}, quantum counting \cite{Brassard1998}, and crypto-key search \cite{Davenport2020}.


\subsubsection{Quantum walks}\label{sssec:qw}\hfill

In the last few decades, classical random walks and Markov chains have generated powerful new algorithms in computer science and mathematics \cite{Motwani1995}. They were thus a natural place to look for quantum equivalents. Early in the development of quantum algorithms both discrete-time \cite{Aharonov2000,Ambainis2001} and continuous-time quantum walks \cite{Farhi1998} were introduced with algorithmic applications in mind.  The first proven speed ups for quantum walk algorithms came soon after, with a search algorithm \cite{Shenvi2002} equivalent to Grover's algorithm \cite{Grover1996} providing a quadratic speed up, and an algorithm for transport across a particular type of disordered graph \cite{Childs2003b} that provides an exponential speed up with respect to an oracle.  There are many comprehensive reviews of quantum walks and their applications, including \cite{VenegasAndraca2012} and \cite{Kadian2021}.

Quantum walks are widely studied for their mathematical properties, and for providing simple models of physical phenomena.  In an algorithmic context, they provide quantum speed up for sampling problems. The quantum walk dynamics provide both faster spreading \cite{Ambainis2001} and faster mixing \cite{Aharonov2000}, but also localization. These are exploited in the search algorithm \cite{Shenvi2002}.  The fast mixing dynamics can also be exploited in a quantum annealing (section \ref{ssec:qopt}) context, see, for example, \cite{Callison2019}.  

Quantum walks also provide models of universal quantum computing \cite{Childs2009, Lovett2010}, but they are not suitable for physical implementation. Instead these are mostly useful for complexity proofs. Quantum walks with multiple non-interacting walkers and quantum particle statistics include the boson sampling model \cite{Aaronson2011}, a computational model intermediate between classical and quantum computing.
Quantum walks with multiple interacting walkers \cite{Childs2013} are a full universal model of quantum computing.  They are a special case of quantum cellular automata (see, for example, \cite{Schumacher2004}) and can provide an alternative physical architecture for quantum computing, most suitable for neutral atoms in optical lattices \cite{Karski2009}.


\subsubsection{Harrow-Hassidim-Lloyd algorithm}\label{sssec:HHL}\hfill

In 2009, Harrow, Hassadim and Lloyd (HHL) \cite{Harrow2009} presented their algorithm to solve one of the most basic classical problems in scientific computing. The HHL algorithm efficiently calculates the solution $\vec{x} = \mathbf{A}^{-1}\vec{b}$ of the linear system $\mathbf{A}\vec{x} = \vec{b}$ for a sparse, regular $N \times N$ Hermitian matrix $\mathbf{A}$. The hard part of this problem is to invert $\mathbf{A}$ and obtain $\mathbf{A}^{-1}$.  To see how a quantum method might help, we present a high level description of the HHL algorithm. For further details and analysis, see \cite{Dervovic2018,Abhijith2022}.

\begin{figure*}[ht!]
\centering
\includegraphics[width=\linewidth]{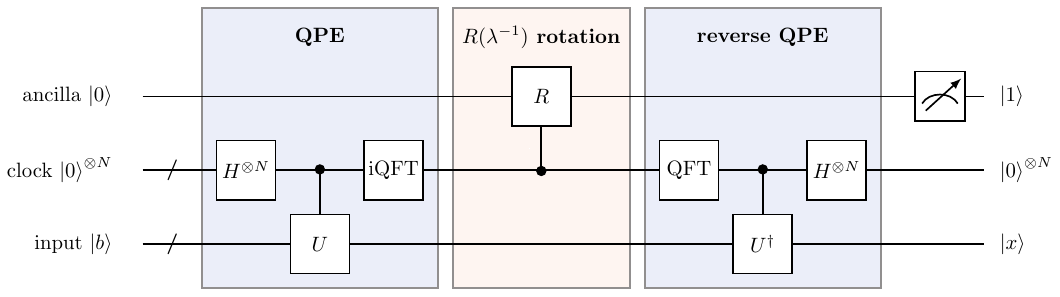}
\caption{High level HHL circuit. Adapted from \cite{Dervovic2018}.}
\label{fig:hhl_circuit}
\end{figure*}

We start with the ancilla, clock, and input registers (Figure \ref{fig:hhl_circuit}). The HHL algorithm represents $\vec{b}$ as a quantum state in the computational basis
\begin{equation}
\ket{\vec{b}} = \sum_j b_j\ket{j}.
\end{equation}
and stores it in the input register with the amplitude encoding method (section \ref{ssec:encoding}). Then using Hamiltonian simulation, we apply 
\begin{equation}\label{eq:hhl_qpe}
\hat{U} = e^{i\hat{A}t}
\end{equation}
to $\ket{b}$ for a superposition of times $t$ determined by the clock register. This step corresponds to quantum phase estimation (QPE) which expresses $\ket{b}$ in terms of the eigenbasis $\ket{u_j}$ of $\mathbf{A}$ and finds the corresponding eigenvalues $\lambda_j$ which are stored in the clock register. Now we can express the system state as
\begin{equation}
\sum_j \beta_j \ket{u_j} \ket{\lambda_j}
\end{equation}
with $\beta_j = \langle u_j \vert \vec{b} \rangle$ and
\begin{equation}
\ket{\vec{b}} = \sum_j \beta_j\ket{u_j}.
\end{equation}
Note that we do not know what the eigenstates $\ket{u_j}$ are.

In the next step, we perform a controlled rotation on the ancilla register with the clock register as the control. The procedure extracts the eigenvalues of $\mathbf{A}^{-1}$ by performing a $\sigma_y$ rotation, conditioned on $\lambda_j$. It is equivalent to the linear map, $\ket{\lambda_j} \to \lambda_j^{-1} \ket{\lambda_j}$. This transforms the system to
\begin{equation}
\sum_j \beta_j \ket{u_j} \ket{\lambda_j} 
\left( 
\sqrt{1-\frac{c^2}{\lambda_j^2}\ket{0}} + \frac{c}{\lambda_j}\ket{1}
\right)
\end{equation}
with normalization constant $c$.

The final step uses inverse QPE to uncompute the $\ket{\lambda_j}$ contained in the clock register. This leaves the remaining state as
\begin{equation}
\sum_j \beta_j \ket{u_j} \ket{0}
\left( 
\sqrt{1-\frac{c^2}{\lambda_j^2}}\ket{0} + \frac{c}{\lambda_j}\ket{1}
\right).
\end{equation}
Measuring and post-selecting on the ancilla qubit being $\ket{1}$ outputs the state
\begin{equation}
\ket{x} = A^{-1} \ket{b} \approx c \sum_j\frac{\beta_j}{\lambda_j}\ket{u_j},
\end{equation}
so we have the solution amplitude encoded in the input register as a quantum state.
If the post-selection is unsuccessful and the ancilla measurement outputs $\ket{0}$, we have to rerun the algorithm.

There are many limitations of the HHL algorithm \cite{Aaronson2015,Dervovic2018}. Key examples include the requirement to be able to efficiently prepare the state $\ket{b}$ by somehow loading the vector components into the quantum computer. This may be easy if the state is an output from another quantum subroutine but in general this is not trivial.  Additionally it must be possible to efficiently implement unitaries of the form $e^{i\hat{A}t}$ to carry out QPE. Moreover, the matrix $\mathbf{A}$ must be Hermitian with unit determinant, although this can be somewhat relaxed by virtue of the fact that any non-Hermitian matrix can be padded into Hermitian form and the determinant can be re-scaled. However, even if these requirements are met, it is not possible to efficiently recover the full solution $\ket{x}$. Instead, we may carry out further processing to compute the expectation value of some operator $\hat{M}$, such as $\bra{x}\hat{M}\ket{x}$. 

The importance of HHL for practical applications has led to many improvements on the original algorithm, such as using time variable amplitude amplification to increase the success probability during ancilla measurement \cite{Ambainis2012}. This is a more general case of amplitude amplification (section \ref{sssec:qaa}), better suited to quantum algorithms where individual branches finish at different times. Individual components, such as Hamiltonian simulation in the QPE step \eqref{eq:hhl_qpe}, have also been improved. QPE remains a bottleneck by requiring $O(1/\epsilon)$ applications of a unitary operation to estimate its eigenvalues to precision $\epsilon$. Hence, even if we use the fastest Hamiltonian simulation methods, the overall improvement would likely be modest with total complexity still being polynomial in $(1/ \epsilon)$. One remedy involves directly applying the matrix inverse to bypass the phase estimation subroutine. This can be achieved by decomposing the inverse operator into a linear combination of unitaries that are easier to apply, using Fourier or Chebyshev expansions \cite{Childs2017}. This approach exponentially improves dependence on the precision parameter and has subsequently found good use in applications ranging from accelerating finite element methods (FEMs) \cite{Montanaro2016a} to approximating the hitting times of Markov chains \cite{Chowdhury2017}. 

Alternatively, HHL can be ``hybridized'' into a quantum-classical method. Given the resource limitations of NISQ-era hardware, hybrid approaches will likely pave the way for meaningful applications in the near future. For example, an iterative-HHL approach \cite{Saito2021} contains a classical iterative process to improve accuracy beyond the limit imposed by number of qubits in QPE. However, the convergence rate for more general cases remains uncertain. Other approaches include streamlining the algorithm, such as using a classical information feed forward step after the initial QPE to reduce the circuit depth for subsequent steps in HHL \cite{lee2019hybrid}. Of course, if the reduced circuit is not applicable, it would then require repeating QPE to determine an appropriate reduction. A hybrid approach can achieve comparable precision to conventional HHL with fewer qubits and multiple phase estimation modules \cite{Gao2023h}. It can also improve phase estimation itself to mitigate the long coherence times \cite{angara2020hybrid}. 

Solving dense matrices, particularly relevant in machine learning and kernel methods \cite{Wilson2015}, has been tackled by using quantum singular value estimation (QSVE) to gain a polynomial advantage over traditional HHL approaches \cite{Wossnig2018,BravoPrieto2020}. There are many quantum machine learning algorithms where HHL is an important subroutine, such as data classification using quantum support vector machines \cite{Rebentrost2014}, quantum principal component analysis \cite{Lloyd2014}, or quantum linear discriminant analysis \cite{Cong2016}; quantum ordinary linear regression (QOLR) \cite{Wiebe2012,Wang2017a} and QOLR for prediction \cite{Schuld2016}; quantum ridge regression (QRR) \cite{Yu2021a}; quantum recommendation systems \cite{Kerenidis2017}; quantum singular value thresholding \cite{Duan2018}; and quantum Hopfield neural networks \cite{Rebentrost2018}.  See section \ref{sssec:qnn} for more on quantum neural networks.  

In a classical context, it is usually the complete solution state that is the desired output from solving a system of linear equations. This could be the velocity at each point in a fluid simulation for example. Thus, it is not immediately clear how recovering the expectation value of some operator such as $\bra{x}\hat{M}\ket{x}$ could be meaningful. Moreover, extracting the individual components of $\ket{x}$ is a costly process that can severely mitigate the inherent speedup offered by the HHL algorithm \cite{Aaronson2015}. Thus, while HHL has proven useful as a subroutine for other quantum algorithms, direct application to classical problems requires detailed consideration of the bottlenecks associated with data transfer between classical and quantum realms \cite{rathore2024integrating} (see section \ref{sec:hpc}). Although for some configurations it is possible to efficiently prepare the quantum state by exploiting an underlying functional form \cite{bharadwaj2023hybrid}, this is not always possible and usually requires some knowledge of the problem \textit{a priori}. 


\subsection{Quantum optimization}\label{ssec:qopt}

\begin{figure*}[ht!]
\centering
\includegraphics[width=\linewidth]{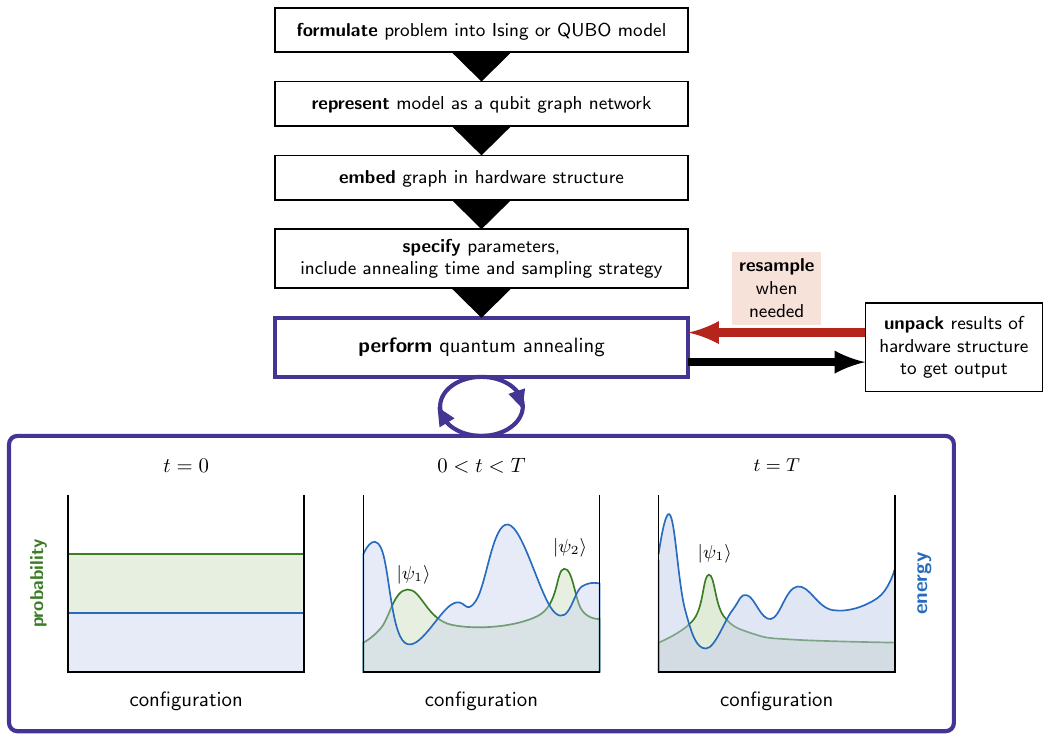}
\caption{The full quantum annealing process from problem formulation to solution.  The graphs illustrate how the initially uniform distribution evolves under the driver Hamiltonian to concentrate in the global minimum representing the solution. Adapted from \cite{AuYeung2023}.}
\label{fig:annealing}
\end{figure*}

Combinatorial optimization problems are widely encountered across industry \cite{Yarkoni2022}, academia \cite{Hauke2020}, and medical science \cite{Emani2021}.  Typically, the problem is to find the best of an exponentially large set of solutions. Mathematically, a basic optimization problem consists of the objective (cost) function which is the output we want to maximize or minimize, the input variables, and optional constraints.  
Many optimization problems can be expressed as quadratic unconstrained binary optimization (QUBO) problems.
These are encoded using an upper-triangular $N \times N$ real matrix $\mathbf{Q}$ and vector of binary variables $\vec{x}$. The aim is to minimize the cost function
\begin{equation}\label{eq:qubo}
f(\mathbf{x}) = \sum_j \mathbf{Q}_{jj} \vec{x}_j + \sum_{j<k} \mathbf{Q}_{jk} \vec{x}_j \vec{x}_k.
\end{equation}
Many cost functions contain a large number of false local minima, making it difficult for classical algorithms to find the true global minimum.

Classical simulated annealing \cite{Kirkpatrick1983} is based on the idea that thermal fluctuations move the system out of local minima and toward the lowest potential energy state as temperature decreases. The simulated annealing algorithm simulates a random walker that travels through the search space or optimization landscape.  The rate at which the temperature is reduced determines how fast the system evolves and how likely it is to avoid becoming stuck in local minima. Quantum annealing \cite{Finnila1994,Kadowaki1998} was conceived as an alternative that adds quantum tunnelling to the system evolution, allowing it to escape from local minima more easily.  Figure \ref{fig:annealing} illustrates the quantum annealing process. 

Conveniently, classical optimization problems can be efficiently mapped to finding the ground state of a classical Ising Hamiltonian \cite{Lucas2014,Choi2011}.  This leads naturally to a quantum version using the transverse field Ising Hamiltonian,
\begin{align}\label{eq:TI}
\hat{H}_\text{Ising} = A(t) \underbrace{\sum_j \hat{\sigma}^x_j}_{\hat{H}_0} + 
B(t) \underbrace{\sum_{j,k} 
\left( 
h_j \hat{\sigma}^z_j + 
J_{jk} \hat{\sigma}^z_j \hat{\sigma}^z_k 
\right)}_{\hat{H}_p}
\end{align}
with the Pauli $x$- and $z$-matrices $\hat{\sigma}^{x,z}$, symmetric interaction strength $J_{jk}=J_{kj}$ of qubit spins $q_j$ and $q_k$, and on-site energy $h_j$. Note that the qubit spin basis states are $\ket{\pm 1}$. There is a straightforward translation from QUBO $\vec{x}$ to Ising $\vec{q}$, using $\vec{x}_j = (\vec{q}_j+1)/2$. The resulting values of the $\{h_j, J_{jk}\}$ variables then encode the problem into the Ising Hamiltonian $\hat{H}_p$.

The transverse field in the $x$-direction $\hat{H}_0$ provides the dynamics that rotate the qubits from their initial state to the final target state, ideally the ground state of problem Hamiltonian $\hat{H}_p$. The control functions $A(t)$ and $B(t)$ must be specified. Usually $A(t)$ varies from 1 to 0, and $B(t)$ from 0 to 1. The choice of how they vary determines the method used to find the ground state and hence the problem solution.

Adiabatic quantum computing (AQC), introduced by \citeauthor{Farhi2001} \cite{Farhi2001} in \citeyear{Farhi2001}, uses the quantum adiabatic theorem to guarantee that the system remains in the ground state. It assumes that the control functions $A(t)$ and $B(t)$ are varied slowly enough to keep the system in the instantaneous ground state. A useful property of AQC is that it is inherently robust against noise \cite{Childs2001}. When used with more general Hamiltonians than in \eqref{eq:TI}, it is computationally equivalent to gate-based quantum computers \cite{Aharonov2004}, making it a universal quantum computing paradigm \cite{Aharonov2008}. 

While AQC provides a sound theoretical underpinning for quantum annealing, in reality, the adiabatic conditions imposed by AQC are rarely met. Meeting them would require long run times that are inefficient and impractical. At the other extreme, continuous-time quantum walk algorithms \cite{Farhi1998} (section \ref{sssec:qw}) can locate ground states for certain problems \cite{Callison2019} where it is viable to make many short repeats. Further diabatic methods, which use other mechanisms with shorter run times, are reviewed in \cite{Crosson2020}. Many of these, including the quantum approximate optimization algorithm (QAOA) or quantum alternating operator ansatz \cite{Farhi2014,Hadfield2019} use classical optimization in the controls to produce an efficient quantum process. We describe QAOA in subsection \ref{sssec:QAOA} after introducing variational quantum algorithms more generally. Exploiting the non-adiabatic and open system effects in quantum annealing is the topic of many recent efforts to derive efficient controls \cite{Banks2024,Schulz2024}.
For more details, \citeauthor{Albash2018} \cite{Albash2018} provide a comprehensive introduction to adiabatic quantum computing and quantum annealing. Hardware currently available for quantum annealing includes superconducting systems, trapped ions, and Rydberg atoms (see \cite{Hauke2020} for an overview).

\subsection{Variational quantum algorithms}\label{ssec:VQAs}

\begin{figure}[t!]
\centering
\includegraphics[width=\linewidth]{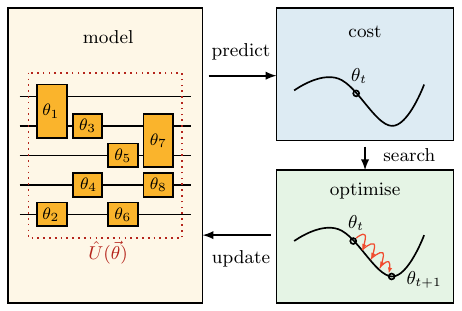}
\caption{High level diagram of hybrid quantum-classical training algorithm for variational circuit. Quantum device (yellow) calculates terms of a cost function, then classical device calculates better circuit parameters $\vec{\theta}$ (blue, green). Repeat cycle until desired accuracy is achieved.}
\label{fig:variationalcircuit}
\end{figure}

Variational quantum algorithms (VQAs) distribute the job of solving a problem between a parameterized quantum circuit and a conventional classical optimizer. Figure \ref{fig:variationalcircuit} shows schematically how it works; for a detailed recent review, see \citeauthor{Cerezo2021} \cite{Cerezo2021}.  The quantum circuit is defined by a set of parameters $\vec{\theta}$ that determine the quantum gates. We evaluate a classical cost function $C(\vec{\theta})$ that can be calculated from the measured outputs  of the quantum circuit.  Classical optimization then updates the parameters $\vec{\theta}$, guided by the cost function. 
Mathematically, the aim is to minimize a linear cost function which typically takes the form
\begin{equation}\label{eq:vqa-cost}
C(\vec{\theta}) = tr \left( \hat{U}(\vec{\theta}) \rho \hat{U}^\dagger(\vec{\theta}) \hat{H} \right)
\end{equation}
with initial state $\rho$, trainable parameterized quantum circuit $\hat{U}(\vec{\theta})$, and Hermitian operator $\hat{H}$.
This general framework can be applied to a wide range of linear optimization problems and extended to solve nonlinear variational problems \cite{Lubasch2020,Kyriienko2021}. Applications include our chosen example of electronic structure for quantum chemistry and materials science \cite{Kokail2019,Kandala2017}, for which we present further details in section \ref{sec:qsim}). 

The past years have seen rapid developments in the variational approach, and it is clear there are beneficial properties for NISQ-era devices \cite{Cerezo2021}. It is possible to use many separate repeats of a shallow (short) quantum circuit. This avoids the need for long coherence times.  Moreover, the variational parameter tuning can compensate for several types of errors. Suitable error mitigation strategies have been developed for NISQ devices \cite{Botelho2022,Cai2023} that can be integrated into VQA circuits. 

In this section, we discuss the most relevant VQAs for computational science: the quantum approximate optimization algorithm (section \ref{sssec:QAOA}); variational quantum eigensolvers (section \ref{sssec:VQE}); and simulation of quantum systems (section \ref{sssec:VQS}).  We also briefly cover quantum neural networks in section \ref{sssec:qnn}. Then we outline some known limitations common to all VQAs, due to noise and the barren plateau problem (section \ref{sec:vqa-limitations}), along with pointers to current research to overcome these and extend VQAs in multiple directions.

\subsubsection{Quantum approximate optimization algorithm}\label{sssec:QAOA}\hfill

The quantum approximate optimization algorithm (QAOA) was originally introduced to solve combinatorial optimization problems \cite{Farhi2014} on gate-based architectures. This was generalized as the quantum alternating operator ansatz \cite{Hadfield2019}.  Instead of applying the full Hamiltonian $\hat{H} = \hat{H}_M + \hat{H}_P$ continuously in time, the aim is to map some input state to the ground state of a problem Hamiltonian $\hat{H}_P$ by sequentially applying the ansatz
\begin{equation}
\hat{U}(\gamma,\beta) = \prod_k e^{-i\beta_k\hat{H}_M} e^{-i\gamma_k\hat{H}_P},
\end{equation}
with mixer Hamiltonian $\hat{H}_M$. The ansatz is applied in short discrete steps, that approximate the continuous time evolution, with parameters $\vec{\theta}=(\vec{\gamma},\vec{\beta})$ to be optimized.
The full Hamiltonian is often the transverse Ising Hamiltonian \eqref{eq:TI}, i.e., $\hat{H}_M=\hat{H}_0$, and $\hat{H}_P=\hat{H}_p$, but the method can be applied more generally to any Hamiltonian with non-commuting parts.

In-depth studies of QAOA performance \cite{Zhou2020,Wang2020} developed an efficient parameter-optimization procedure to find high-quality parameters.  This becomes increasingly important as the depth and complexity of the QAOA circuit increase. The original QAOA proposal suggests a random selection of initial parameters within a range believed to be close to the optimal parameters. However, this method can often hinder the algorithm performance when the cost function landscape is rugged and contains numerous local minima or barren plateaus (section \ref{sec:vqa-limitations}). \citeauthor{Blekos2024}'s review \cite{Blekos2024} outlines various methods that address QAOA optimization. For example, we can provide a good initial guess \cite{Zhou2020} by warm-starting initial parameters \cite{Jain2022} or through parameters transfer \cite{Galda2021}, or by developing a more efficient optimization subroutine \cite{Moussa2022,Cheng2024,Alam2020,Dong2020}. Other approaches involve exploiting the problem structure and symmetries to enhance the optimization process \cite{Shaydulin2021,Shaydulin2021s}. 

Note that many proposed techniques rely on heuristics and empirical observations rather than theoretical guarantees.
However, \citeauthor{Brady2021} \cite{Brady2021} (building on the work of \citeauthor{Yang2017} \cite{Yang2017}) put the method on a sound theoretical base by showing that QAOA converges to an optimal quantum annealing schedule in the limit of many stages.  \citeauthor{Gerblich2024} \cite{Gerblich2024} used the same methods to show that a sequence of quantum walks also approximates optimal quantum annealing in the limit of many stages, and shows better convergence for the same number of stages.  Each stage requires its own parameters to be optimized, so it is helpful to minimize the number of stages to keep the parameter tuning costs down.  Most studies focus on the few-stage regime where the theoretical results don't apply and are thus based on numerical simulations.  It is important to count the optimization costs correctly to assess the full quantum cost, since each round of classical optimization requires at least one quantum run to evaluate the cost function. In a slightly different setting (quantum annealing approximated by B\'ezier curves), \citeauthor{Schulz2024} \cite{Schulz2024} show how ignoring optimization costs can distort the apparent scaling in numerical results, and they also show how to account for the optimization costs in the full analysis, recovering the expected scaling.

\subsubsection{Variational quantum eigensolvers}\label{sssec:VQE}\hfill

Variational quantum eigensolvers (VQEs) were originally developed to find the ground state energy of molecules \cite{McClean2016,Kandala2017,Tilly2022}. 
They have since been adapted to solve a wider range of problems, including optimization \cite{Cerezo2021,Amaro2022}. 
The cost function is defined as the expectation value of the quantum state energy, $\bra{\psi(\theta)}\hat{H}\ket{\psi(\theta)}$. The aim is to minimize the expectation value of Hamiltonian $\hat{H}$ over a trial state $\ket{\psi(\theta)} = \hat{U}(\theta) \ket{\psi_0}$ for ansatz $\hat{U}(\vec{\theta})$ and initial state $\ket{\psi_0}$. 

The Rayleigh-Ritz principle states that the lowest eigenvalue of a Hermitian operator $\hat{H}$ is upper-bounded by the minimum expectation value found by varying the state $\ket{\psi}$. In other words, the statement 
\begin{equation}
\braket{\hat{H}}(\vec{\theta}) \equiv \bra{\psi(\theta)}\hat{H}\ket{\psi(\theta)} \geq E_0
\end{equation}
is valid with the equality holding if $\ket{\psi(\theta)}$ is the ground state $\ket{E_0}$ of $\hat{H}$. Hence, the optimized vector $\vec{\theta}$ that approximates the ground state (or eigenvector corresponding to the lowest eigenvalue) is the choice that minimizes $\braket{\hat{H}}$. Using this principle, we can break down the VQE methodology into three distinct steps, which are then repeated:
\begin{enumerate}
\item Prepare a parameterized initial state as the starting ansatz $\hat{U}(\vec{\theta})\ket{\psi_0}$ using initial values of $\vec{\theta}$.
\item Estimate expectation value $\braket{\hat{H}}(\vec{\theta})$ in the quantum processor (this may require multiple runs using quantum phase estimation).
\item Use a classical optimizer to find a new set of $\vec{\theta}$ values that decreases $\braket{\hat{H}}(\vec{\theta})$.
\item Repeat the above procedure to achieve convergence in $\langle\hat{H}\rangle(\vec{\theta})$. The final values of the parameters $\vec{\theta}$ at convergence define the desired state.
\end{enumerate}
An application in quantum chemistry uses VQEs to generate parameterized quantum circuits (PQCs) which can be used to produce ans{\"a}tze states to approximate molecular ground states \cite{Kandala2017}.  These are an improvement over unitary coupled-cluster (UCC) circuits \cite{Lee2019} based on the coupled-cluster method in computational chemistry, which are too deep to be easily implemented on NISQ quantum computers. See section \ref{sssec:PostHFQC} for more details and context. PQCs are composed of gates that are natural operations on the quantum hardware and hence more efficient than standard quantum circuits.

\subsubsection{Variational quantum simulation}\label{sssec:VQS}\hfill

Simulating the dynamics of quantum systems is expected to be one of the first useful applications of quantum computers.  Variational approaches make best use of limited quantum hardware by optimizing the parameters classically. There are two basic algorithms for variational quantum simulation, specifically for solving real- and imaginary-time evolution of quantum systems. These can be described respectively by the time-dependent Schr{\"o}dinger equation
\begin{equation}
\frac{d\ket{\psi(t)}}{dt} = -i\hat{H} \ket{\psi(t)}
\end{equation}
with Hamiltonian $\hat{H}$ and state $\ket{\psi(t)}$, and normalized Wick-rotated Schr{\"o}dinger equation
\begin{equation}
\frac{d\ket{\psi(\tau)}}{d\tau} = -(\hat{H} - \braket{\hat{H}}) \ket{\psi(\tau)}, 
\end{equation}
where $\tau=-it$, and $\braket{\hat{H}} = \bra{\psi(\tau)} \hat{H} \ket{\psi(\tau)}$ preserves the norm of state $\ket{\psi(\tau)}$.

The direct approach to quantum simulation is to apply a unitary circuit $e^{-i\hat{H}t}$ to the initial state, using numerical integration. However, this increases the circuit depth polynomially with respect to the evolution time $t$ \cite{low2019hamiltonian} which is difficult to achieve in hardware with short coherence times. On the other hand, variational quantum simulations use an ansatz quantum circuit
\begin{equation}
\ket{\phi(\vec{\theta}(t))} = \hat{U}(\vec{\theta}(t)) \ket{\phi_0}
\end{equation}
applied to an initial state $\ket{\phi_0}$ to represent $\ket{\psi(t)}$. 
Therefore, we must choose the variational parameters to map the time evolution of the Schr{\"o}dinger equation for $\ket{\psi(t)}$ on to the evolution of parameters $\vec{\theta}(t)$. The aim is to obtain a shorter circuit than would be needed for direct simulation. How the parameters evolve and simulate the time evolution depends on the variational principle used \cite{Yuan2019}. 

The variational method for generalized time evolution can also be used to perform matrix multiplication and to solve linear systems of equations \cite{Endo2020}. It is an alternative method to those presented in \cite{BravoPrieto2020,Xu2021v}, or the HHL algorithm (section \ref{sssec:HHL}).


\subsubsection{Quantum neural networks}\label{sssec:qnn}\hfill

Artificial neural networks (NNs) \cite{LeCun2015} allow us to classify and cluster large datasets by training the parameters associated with the neural connections. This is analogous to the learning process in the brain. Quantum neural networks (QNNs) combine artificial neural networks with quantum principles \cite{Abbas2021}. NNs are highly nonlinear models, whereas QNNs obey the laws of quantum mechanics and contain linear operators acting on quantum states. QNNs can offload nonlinearities to classical computers, or use quantum kernels. The latter option encodes classical data into Hilbert space using a nonlinear transformation, the quantum feature map \cite{Schuld2019}.

There are many different proposals for QNN architectures, such as the fully quantum, feed-forward neural networks \cite{Farhi2018}, and convolutional neural networks \cite{Cong2019}. Most QNNs follow similar steps \cite{Beer2020}: initialize a neural network architecture; specify a learning task; implement a training algorithm; and simulate the learning task. 

\begin{figure}[ht!]
\centering
\includegraphics[width=\linewidth]{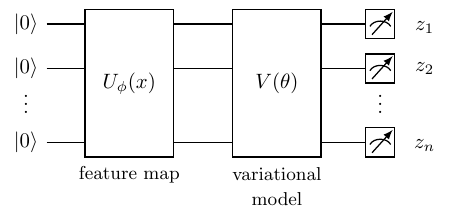}
\caption{Example of quantum neural network using variational quantum circuit.}
\label{fig:vqnn}
\end{figure}
There are similarities between QNNs and kernel methods \cite{Schuld2021,Schuld2019}, which are well-established machine learning techniques. Kernel methods map the features of a dataset into a high-dimensional space through function $\phi(\vec{x})$, then uses a kernel $K(x_1,x_2)$ to measure the distance between two data points in  high-dimensional space. We observe the same behavior in QNNs: when the features are mapped on to a high-dimensional Hilbert space, we find the level of similarity between two points in the space by calculating the overlap of two encoded states.

For QNNs constructed using quantum circuits, the parameters that perform an accurate mapping are defined by a set of gate operations \cite{Schuld2020}, corresponding to $V(\theta)$ (Figure \ref{fig:vqnn}). There are typically three steps. First, encode the input data, which is usually classical, into a quantum state using a quantum feature map \cite{Schuld2021v} $\phi$, applied to the initialized qubits through unitary operation $U_\phi(x)$ (Figure \ref{fig:vqnn}). The choice of feature map aims to enhance the QNN performance. 
\begin{equation}
\hat{U}_\phi(x)\ket{0}^{\otimes n} = \ket{\phi(x)}. 
\end{equation}
Second, we use a variational model parameterized by vector $\vec{\theta}$ to transform the quantum state
\begin{equation}
\ket{\psi(x,\vec{\theta})} = V(\vec{\theta})\ket{\phi(x)}.
\end{equation}
It contains parameterized gate operations that are optimized for a particular task, analogous to classical ML techniques \cite{Dunjko2018}. The aim is to choose parameters $\vec{\theta}$ to minimize a loss function. 
Third, we measure an observable $\hat{M}$. Classical post-processing procedures on the measurement outcomes $\vec{z}=[z_1, z_2, ..., z_n]$ extract the output of the model,
\begin{equation}
f(\vec{z}) = 
\bra{\psi(x,\vec{\theta})} \hat{M} \ket{\psi(x,\vec{\theta})}.
\end{equation}
The main advantage of variational circuits actually arises from the feature map: when $\hat{U}_\phi(x)$ cannot be simulated efficiently on classical computers, the overall quantum speedup can be up to exponential. 

Many types of QNNs have been proposed for different types of data analysis, such as quantum Boltzmann machines \cite{Amin2018,Kieferova2017} which are a natural fit for quantum annealing \cite{Benedetti2016,Xu2021} hardware, as are quantum support vector machines \cite{Delilbasic2024,Willsch2020}. Constructing QNN architectures is currently an active area of research.  For a concise review of quantum machine learning, see \cite{Cerezo2022} and references therein.

\subsubsection{VQA limitations and enhancements}\label{sec:vqa-limitations}\hfill

\begin{figure}[ht!]
\centering
\includegraphics[width=\linewidth]{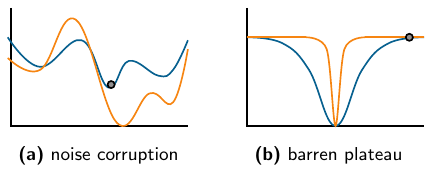}
\caption{Ideal solution landscape (blue) corrupted by (a) noise and (b) barren plateau, both orange.}
\label{fig:vqa-landscape}
\end{figure}

Variational quantum algorithms are a very active area of current research, and face a range of problems (see \cite{Cerezo2021,Cerezo2022} for an overview). In general, there is a trade-off between the expressivity and trainability of the circuit \cite{Holmes2022}. Performance decreases significantly when qubit number and circuit depth increase. Using more quantum resources to implement the ansatz results in greater expressivity of the ansatz, which increases the likelihood of finding the correct solution \cite{Benedetti2019,Du2020e,Du2022e}. However, large circuit depth means the classical optimizer would receive noisy and potentially unusable gradient information, which can lead to divergent optimization or barren plateaus \cite{McClean2018,Wang2021,Wang2021n}. 
We illustrate these two common problems in Figure \ref{fig:vqa-landscape}.

The barren plateau phenomenon occurs when the non-convex loss function landscape becomes exponentially flat (on average) as qubit number increases. At the same time, the valley containing the global minimum also shrinks exponentially with problem size, leading to a narrow gorge \cite{Arrasmith2022} (Figure \ref{fig:vqa-landscape}(b)). The cost function gradient is exponentially suppressed except in an exponentially small region. For cost functions in \eqref{eq:vqa-cost}, barren plateaus can be caused by circuit properties $\hat{U}(\vec{\theta})$ such as its structure and depth \cite{McClean2018,Larocca2022,Wang2021}, expressibility \cite{Holmes2021,Holmes2022} and entanglement power \cite{Marrero2021,Patti2021,Sharma2022}. The effects on parameter optimization can potentially destroy any quantum speedup. 

There have been considerable efforts to understand and mitigate the effects of barren plateaus, for a review, see \citeauthor{Larocca2024} \cite{Larocca2024}. 
These include building shallow circuits \cite{ZhangHK2024,Cerezo2021c,Wang2021n}, developing gradient free algorithms \cite{Koczor2022}, using error mitigation techniques \cite{Wang2024,Endo2021}, and developing more sophisticated state preparation strategies when initializing the variational parameters \cite{Grant2019,Liu2023b,Sack2022}.
Others have presented the mathematical foundations for a unified Lie algebraic theory to explain the sources of barren plateaus and quantify the ultimate expressiveness of parameterized circuits \cite{Ragone2024,Fontana2024}. There is also ongoing work to develop a more general barren plateau theory \cite{Diaz2023}. 

In addition, there have been considerable efforts in designing suitable ans{\"a}tze. Quantum architecture search (QAS) dynamically designs optimal ans{\"a}tze for VQAs \cite{ZhangSX2022,Du2022,Martyniuk2024}. Given a search space (possibly constrained by quantum hardware properties), performance criteria, and the underlying problem, QAS aims to automatically find an optimal quantum circuit with parameters $\vec{\theta}'$ that maximize the VQA performance. For VQEs, there are approaches which construct variational models constrained to a (smaller) symmetry-respecting solution space to preserve the problem Hamiltonian symmetries \cite{Sagastizabal2019,Ganzhorn2019,Cade2020}. This results in smaller circuits more suitable for NISQ hardware. This has been extended to quantum machine learning models \cite{Larocca2022g,Meyer2023,Nguyen2024}. 

Quantum optimal control (QOC) theory \cite{Koch2022,Ansel2024} has also inspired work into finding connections with VQAs \cite{Magann2021}. For example, the quantum-optimal-control-inspired ansatz (QOCA) \cite{Choquette2021} goes beyond symmetry-preserving methods by introducing a set of unitaries that break the symmetries of the problem Hamiltonian.

\subsection{Quantum software}\label{ssec:qsoftware}

As well as quantum hardware, making use of quantum computers also needs new software at various levels of the quantum computing stack, see Figure \ref{fig:stack2013}.  This includes programming languages, compilers, operating systems, and low level classical control software.
Quantum software start up companies are promising seamless integration into existing programming languages, such that you won't need to know you are using a quantum computer.  HPC users will be well aware that this is hype. Good performance is only achieved with some consideration of how best to map an application to the hardware.  For large scale HPC, this is increasingly becoming a professional team effort, requiring research software engineers specialized in tuning applications to the specific HPC hardware it is running on.  This is especially true for using GPUs, and for task scheduling to ensure all available compute nodes are kept busy.
Since adding QPUs to HPC increases the hardware complexity, we can expect to need the same level of professional team support to fully leverage future large scale quantum computers.  Nonetheless, current quantum computers that are available for use in the cloud have not yet reached this level of complexity, and writing and executing test bed applications is well within reach of the average computational scientist or engineer.  There are many tutorials available: each available platform provides its own, and there are guides and open source code for specific applications, see, for example, \cite{Camino2023} for optimizing crystal structures on D-Wave Systems.

Currently available commercial quantum computers tend to use Python modules for the user-level programming language, and provide simulators, compilers and optimizers specific to their platform.  In particular, Qiskit from IBM \cite{Qiskit} is open source and has been adapted for other platforms such as ion traps and neutral atoms.  As one of the first online services, IBM has the most comprehensive set of online resources, and also provide a visual drag and drop interface for creating quantum circuits, which is especially suitable for pedagogical uses.  Programming gate by gate is very low level, about equivalent to assembler code for classical computers, despite the high level interface.  This is indicative of the early stage of development of quantum hardware with only tens or hundreds of qubits, where circuit optimization by hand is still possible.  Indeed, hand optimization will tend to beat the automated compilers for specific applications (the automated compilers are generic), and is necessary to obtain optimal performance from the limited capabilities of NISQ systems.
On some platforms (including IBM), it is even possible to experiment at the hardware pulse control level to optimize quantum gate operations.  
Equivalently low level control is used for non-circuit model hardware, such as quantum annealers, where the programming consists of mapping the problem to the hardware settings (fields and couplers) and shaping the annealing schedule.  
Higher level tools to map the problem to the hardware graph across 5,000 qubits are available \cite{DWave_dimid,DWave_ocean}, but do not necessarily beat a hand-crafted assignment for specific problems.

There are are several software start ups specializing in operating systems and control software designed to operate equally well on multiple platforms and provide a similar interface regardless of the underlying hardware, even recommending the best choice if multiple platforms are available.  While we are still some way from this being viable for using quantum computers, the automated lower level control techniques are showing promise when integrated correctly
\cite{Mundada2023}.  Meanwhile, those who are ready to be quantum computing pioneers will find ample support and opportunities online, while those who prefer to wait for the next advances can be assured that software at an appropriate level will be available, either commercially or open source, when the hardware is ready.

\subsection{Performance benchmarks: quality, speed, scale}\label{ssec:benchmark}

Benchmarking for high performance computers is well established, and essential for managing HPC facilities and optimizing their throughput. For example, the High Performance Linpack (HPL) benchmark \cite{Petitet2018} solves a dense linear system to measure the throughput of a computing system. Running the same computational tasks on different HPC hardware can provide some level of side-by-side comparison of different facilities. However, the results from standard benchmarks cannot fully predict performance for a specific application, and bespoke tests are often run to tune performance on a per-application basis.

Benchmarks for quantum computers are under active development. Fair comparisons of quantum hardware are even more challenging than for classical hardware, given the diverse types of qubits under development.
Hardware based on superconducting qubits \cite{Krantz2019,Blais2021} is arguably the most popular choice \cite{Bravyi2022,Blais2020}, followed by ion-trap-based systems \cite{Bruzewicz2019,Malinowski2023,Moses2023}. Semiconductor-based spin qubits \cite{Chatterjee2021,Burkhard2023} are less advanced but potentially offer more scalability in future. Photonic platforms \cite{Pelucchi2022,Moody2022} require less stringent cooling regimes but other aspects of the engineering (e.g., single photon sources and photon loss) are challenging. Neutral atoms, especially using highly excited Rydberg states \cite{Adams2020}, are versatile and already showing impressive results for directly simulating quantum many-body systems \cite{Surace2020,Ebadi2021}.
Further types of hardware are under active development, and it is not clear yet which will prove the most useful and scalable in the long term.

The criteria for good quantum computer hardware are fairly straightforward \cite{DiVincenzo2000}. It must have as many high quality qubits as possible. Qubits must be individually controlled to generate complex (entangled) states. We must be able to apply a large number of sequential operations before the qubits lose coherence. The control and gate operations are characterized by the quality of single- and two-qubit gates. Scaling up to obtain good performance on large numbers of qubits requires error correction \cite{Terhal2015,Roffe2019} to be implemented in the hardware. This needs mid-circuit measurements to be possible, as well as measuring the qubit states at the end of the computation.

\subsubsection{Low- and high-level benchmarks}\label{sssec:lowhighbenchmarks}\hfill

Low-level benchmarks measure the performance of individual qubits, single- and two-qubit gate fidelities, state preparation and measurement operations. We can often directly measure the $T_{1,2}$ decoherence times, gate speed or gate operation reliability (gate fidelity). However these characteristics do not provide an accurate benchmark of quantum processor performance. Using low-level benchmarks alone can label a slow high-fidelity quantum computer as equivalent to a fast low-fidelity quantum computer \cite{Campbell2017a}. 

For more thorough testing of gate operations, there is randomized benchmarking \cite{Knill2008,Magesan2011}, cycle benchmarking \cite{Erhard2019a}, or gate set tomography \cite{Nielsen2021}. Efficient methods to verify the output state are now available \cite{Huang2024}, when the expected state is known. 
Randomized benchmarking protocols typically characterize the quality of Clifford operations, which are classically simulable (see section \ref{ssec:universal}). For non-Clifford operations, random circuit sampling and direct randomized benchmarking can be used \cite{Liu2022b,Proctor2019}. These methods may not be scalable to large numbers of qubits as it is difficult to classically simulate non-Clifford operations to check the test outcomes. A class of methods that aims to circumvent this issue is mirror benchmarking \cite{Proctor2021, Mayer2021}. The trivial ideal output of the circuits involved in these benchmarks makes the results easily verifiable and scalable to large numbers of qubits.

At a slightly higher level, specific quantum circuits can be used to test hardware performance for gate sequences. The circuits can be random, application-agnostic circuits \cite{Cross2019,Proctor2022}, or circuits run as part of certain algorithms \cite{Mills2021,Quetschlich2023,Li2022b}. Other types of gate sequences can be used including identity operations and entanglement generation circuits \cite{Michielsen2017}. For non-gate-based hardware, a generalized form of randomized benchmarking can be used, for example in programmable analogue quantum simulators \cite{Derbyshire2020}.

Qubit layout and connectivity constrain the implementation of gate sequences and hence affect algorithm efficiency. Some types of hardware (e.g., neutral atoms \cite{Adams2020}) can easily be reconfigured between runs. Other platforms, such as superconducting or semiconductor qubits have connectivity baked in at the manufacturing stage. 
Algorithm- \cite{Georgopoulos2021} or application-level \cite{Lubinski2023,Donkers2022} benchmarks are thus important for a full understanding of hardware performance.  Such benchmarks require knowing the ideal output in order to evaluate the solution quality. In some cases, analytical solutions are available for suitable problems. For example, the one-dimensional Fermi-Hubbard model is exactly solvable using the Bethe ansatz \cite{DallaireDemers2020}.  For a comprehensive assessment of performance, there are benchmarks comprising of a suite of many different applications (e.g., \cite{Cornelissen2021,Tomesh2022,Murali2019}). Most include well-known algorithms such as the quantum Fourier transform (section \ref{sssec:QFT}), quantum phase estimation (section \ref{sssec:QPE}), Grover's search algorithm (section \ref{sssec:search}), VQE (section \ref{sssec:VQE}), the Bernstein-Vazirani algorithm \cite{Bernstein1993}, and the hidden shift algorithm \cite{vanDam2006}. In general, device performance on simple problems or circuits may not scale \cite{Lubinski2023}. It is also necessary to define suitable high-level success criteria, e.g., for quality \cite{BlumeKohout2020,Mills2021}. 

It is instructive to test at both low and high levels because extrapolating from low level performance to higher level is unreliable. Cross-talk errors can dominate and these are not captured by lower level tests using only one- or two- qubit gates.  Even at a higher level, there may be large differences between performance on random and structured circuits \cite{Lubinski2023,Proctor2022}. 

Determining whether an actual computation has run correctly requires verification.  Classical verification methods are becoming increasingly sophisticated for larger, more complex problems (see, for example, \cite{Groen2022}).  Verifying quantum computations \cite{Carrasco2021}, especially for NISQ hardware, requires some extra tools, such the accreditation scheme of \citeauthor{Ferracin2019} \cite{Ferracin2019} for gate-based computations, which has been experimentally implemented \cite{Ferracin2021}.  Accreditation makes use of similar tools to low-level benchmarking, like mirror circuits, to quantify the errors and hence assign a confidence level to the correctness of the output of the computation, which has been run alongside the test circuits.  Accreditation has been extended beyond the circuit model to certain types of quantum simulation and quantum annealing \cite{Jackson2024}.

\subsubsection{Quantifying quantum device performance}\label{sssec:quantifyperformance}\hfill

Various combinations of benchmarks have been proposed as metrics to capture the performance of quantum computers in a single number or a few parameters.  Most of these use fairly low level benchmarks. For example, quantum volume is a benchmark developed by IBM \cite{Cross2019} to test hardware performance for gate sequences. Quantum volume is calculated by running randomized square circuits with different numbers of qubits. The quantum volume is given by the largest square circuit that can be run on the device which passes an acceptance criterion that quantifies the output quality. It has been generalized to more realistic non-square circuits \cite{BlumeKohout2020,Miller2022}.

Metrics such as \emph{circuit layer operations per second} (CLOPS) \cite{Wack2021}, layer fidelity \cite{McKay2023}, and qubit number have been proposed to collectively quantify the quality, speed, and scale of quantum computers. Single-number metrics are useful to make cross-device comparisons straightforward, but similar to classical metrics, they may not accurately describe quantum computer performance for all kinds of algorithms.  In a non-gate model setting, \citeauthor{McGeoch2019} \cite{McGeoch2019} has proposed guidelines for reporting and analysing quantum annealer performance. 

As well as accuracy, it is also important to measure how quickly the quantum processor can execute layers of a parameterized model circuit similar to those used to measure quantum volume. Increased quantum processor speed is critical to support near-term variational algorithms, which require thousands of iterations. Some benchmarks evaluate speed as a sub-score and combine it with other factors to determine a final metric \cite{Donkers2022}. There also exist standalone metrics for speed \cite{Wack2021}. \citeauthor{Lubinski2023} \cite{Lubinski2023} note that evaluating measures of speed for comparing different platforms is difficult. Timing information (when available at all) is reported differently by different hardware vendors, and gate speeds naturally vary by several orders of magnitude between hardware types (see section \ref{sec:hpc}). Another time-related factor is consistency, or how the performance varies over time. This determines how often hardware must be recalibrated. Consistency can be captured by repeating the same tests, and is an important consideration for current hardware which requires significant downtime for regular recalibration.

The above methods (section \ref{sssec:lowhighbenchmarks}) and metrics (section \ref{sssec:quantifyperformance}) are useful for benchmarking hardware dependent characteristics.  On top of this, the compilers and software stack efficiency can significantly affect performance, in much the same way as classical HPC requires highly tuned software to optimize performance.  Quantum software benchmarking is still at an early stage of development, but there are indications it will be critical for making quantum hardware useful, see for example, \cite{Mundada2023}.

\subsubsection{Quantum circuit simulation methods
}\label{sec:TNbenchmark}\hfill

In the current NISQ era, simulating quantum circuits is important for multiple purposes.  Simulations of ideal quantum circuits allow quantum algorithm development and testing in a more reliable environment than quantum hardware currently provides, albeit limited to about 45 qubits in full generality  due to the exponential size of Hilbert space in qubit number.
Several benchmarking tools (e.g., \cite{Liu2022,Proctor2019}) require knowledge of the ideal circuit outputs produced by such simulations, to verify the outputs of the quantum hardware.   Further examples of tensor network-based benchmarking methods include the holoQUADS algorithm \cite{FossFeig2021,FossFeig2022} and qFlex \cite{Villalonga2019}.

Classical simulation also establishes a classical computational bar that quantum computers must pass to demonstrate a quantum advantage.  It is important to carry out classical simulations as efficiently as possible, so that reports of quantum advantage are not exaggerated.  One of the main tests used to claim a quantum advantage from a NISQ device involves sampling the bit-strings from a random quantum circuit (RQC) \cite{Boixo2018}. The result must be within some variational distance of the output distribution defined by the circuit \cite{Villalonga2019,Bouland2019} to pass the test. Challenges to Google's supremacy claims \cite{Arute2019} used tensor network simulation methods \cite{Pan2022,Pan2022s,ZhangSX2023} to develop efficient classical simulations for imperfect hardware.  Decoherence and other errors in the hardware means it can only generate entanglement over a limited range \cite{Zhou2020x}.  This allows sufficiently accurate classical simulations of larger numbers of qubits to be performed.
The tensor network contraction method is one of the best classical methods for simulating RQCs for sizes close to the quantum advantage regime \cite{Wahl2023,Huang2021}; for recent reviews and tutorials, see, e.g., \cite{Orus2014,Orus2019t,Ran2021}.

\subsubsection{Measuring quantum advantage}\hfill

It is not straightforward to determine exactly when a quantum computer provides a practical advantage over a classical computer. The latter are benchmarked using FLOPS, which measure the number of floating-point operations per second. The equivalent speed benchmark for quantum processors is CLOPS (see section \ref{sssec:quantifyperformance}) which measures the number of circuit layer operations per second. It does not make sense to convert between the two metrics, since they do not correspond to the same unit of computation. 

Demonstrating a quantum advantage at the level of whole algorithms also highlights the difference between theoretical computational complexity and practical implementation.  Many quantum machine learning (QML) algorithms make different assumptions from the classical algorithms they are compared with \cite{Aaronson2015}. In some cases, it is possible to find ``dequantized'' versions of QML algorithms \cite{Tang2022}. These are fully classical algorithms that process classical data and perform only polynomially slower than their quantum counterparts. If a dequantized algorithm exists, then its quantum counterpart cannot give an exponential speedup on classical data.   
Many QML algorithms rely on pre-processed data via quantum encoding. Breakthroughs that claim a quantum speed-up often neglect the quantum encoding step in the algorithm runtime, while making comparisons to classical algorithms that operate on raw, unprocessed data. This means it is often possible to achieve equivalent speed-ups using only classical resources, if we perform the same pre-processing for classical algorithms. For example, the pre-processing in \cite{Tang2019} allows us to apply ``dequantized'' classical sampling techniques to the data, and achieve a significant speedup over previous classical methods.  We discuss the issues around encoding classical data into quantum systems in more depth in section \ref{ssec:encoding}.
On the other hand, dequantization shows how replacing quantum linear algebra algorithms with classical sampling techniques can potentially create a classical algorithm that runs exponentially faster than any other known classical algorithm \cite{Tang2019}. Quantum-inspired classical algorithms that provide a significant improvement on current best methods are a useful benefit from quantum algorithm research, if they turn out to be practical to implement on classical HPC. 

Comparing state-of-the-art quantum computers with the best classical algorithm for the same problem is not enough to establish a quantum advantage.  For the RQC used in Google's claim of quantum supremacy \cite{Arute2019}, classical algorithms had not been optimized for the problem: it is not a useful problem and there was little research on it.  Subsequent work brought down the classical runtime by some orders of magnitude \cite{Pan2022,Pan2022s,ZhangSX2023}.  Similar advances have occurred with boson sampling \cite{Zhong2020,Bulmer2022,Oh2024}.

Current research predicts a narrow window in terms of the number of qubits where circuit model NISQ hardware can potentially outperform classical hardware for certain problems \cite{Bouland2021}.  Larger numbers of qubits need longer gate sequences for the same algorithm. Errors inevitably build up and render the computation unreliable without active error correction. Recent proof-of-concept studies of quantum error correction \cite{RyanAnderson2021,Sivak2023} begin to show errors suppressed for extended gate sequences, heralding a new era of fault tolerant quantum computing.


\section{Quantum simulation for quantum systems}\label{sec:qsim}

Simulating quantum systems on classical computers has long been known for the exponential explosion problem, where the size of the quantum state space (Hilbert space) increases exponentially with the number of qubits.  This limits the size of full classical simulations to a state space equivalent to around 45 qubits ($2^{45}$ complex numbers). For quantum systems with a larger Hilbert space, this means the simulation is constrained to a subspace that may not be large enough to reveal the full behavior.
This can be overcome by using quantum simulation \cite{Feynman1982}: we use some controllable quantum system to study the model of another less controllable or accessible system. The method is especially relevant for studying quantum many-body problems \cite{Weimer2021} in condensed matter physics, atomic physics, and quantum chemistry. For authoritative reviews, see \cite{Georgescu2014,Altman2021,Daley2022}.
We first provide a brief overview of methods for quantum simulation, then, to provide a focus for a more in depth assessment, we provide a more detailed account of quantum computational chemistry and the prospects for quantum computers to advance this and related fields.

\subsection{Quantum simulation methods}\label{ssec:qsimtools}

\begin{table*}[ht!]
\centering
\caption{Compare digital vs analogue quantum simulation.}
\bgroup
\def\arraystretch{1.5}
\begin{tabular}{p{.8in}p{.1in}p{2.4in}p{.1in}p{2.4in}}
&&\textbf{Analogue} && \textbf{Digital} \\
\hline
Hamiltonian
 && 
Directly implement system Hamiltonian by experimentally building a controllable quantum system (simulator).
 &&  
Express Hamiltonian in second quantized form, then construct Hamiltonian out of quantum logic gates. 
 \\
System vs simulator
 && 
System may be completely different from simulator, e.g., using atoms in optical lattices to simulate electrons in a crystal.
 &&  
Implements circuits containing quantum gates, therefore has same level of technical difficulty as quantum computer.
 \\
Physical model
 && 
Needs to match the simulator Hamiltonian within the available experimental controls.
 && 
In principle, any quantum system can be simulated using digital (universal) quantum simulation.
 \\
Hardware
 && 
Easier to implement for certain problems well matched to the simulator hardware. Likely to become practicable before universal quantum computers become available.
 &&  
Requires high depth circuits (many quantum gates) therefore unlikely to be practical before fully error corrected quantum computers are available.
 \\
 \hline
\end{tabular}
\egroup
\label{tab:quantumsim}
\end{table*}

Quantum simulation can be implemented using quantum computers (digital quantum simulation), or with analogue devices (analogue quantum simulation) that are not necessarily as powerful as universal quantum computers and therefore are easier to build. An example of an analogue quantum simulator is the ultracold atomic systems often used to simulate condensed matter physics \cite{Trotzky2012,Choi2016} and quantum chemistry \cite{ArguelloLuengo2019}. A comparison of the relative merits of the different methods is summarized in Table \ref{tab:quantumsim}.  

The basic task underlying most quantum simulations involves time evolution under the system Hamiltonian. For Hamiltonian $\hat{H}$, the unitary operator $\hat{U}(t)$ evolves the dynamics for a time $t$,
\begin{equation}
\hat{U}(t) = \exp \left( -i \int_0^t \hat{H}(t) dt \right), 
\end{equation}
using units where $\hbar=1$.
For many physical systems, the system Hamiltonian $\hat{H}$ can be written as a sum of terms that describe local interactions,
\begin{equation}
\hat{H} = \sum_{j=1}^n \hat{H}_j.
\end{equation}
In analogue quantum simulation, the system Hamiltonian $\hat{H}_\text{sys}$ is directly mapped \cite{Somaroo1999} on to the simulator Hamiltonian $\hat{H}_\text{sys} \mapsto \hat{H}_\text{sim}$, and the time evolution proceeds under the natural Hamiltonian evolution of the simulator.

In contrast, implementing $\hat{U}(t)$ on a digital quantum computer usually relies on numerical integration, i.e., breaking up the evolution time into small steps of duration $\Delta t$, taking care to preserve the order of non-commuting operators. The first-order Trotter-Suzuki formula gives  
\begin{equation}\label{eq:STformula}
\hat{U}(\Delta t) 
= e^{-i \sum_j \hat{H}_j \Delta t}
= \prod_j e^{-i\hat{H}_j\Delta t} + O((\Delta t)^2).
\end{equation}
and higher order approximations can also be used to improve accuracy \cite{Hatano2005}. Digitizing Hamiltonian time evolution in this way is often referred to as ``Trotterization''.
Acceptable precision comes at the expense of small $\Delta t$, which requires many quantum gates and hence extremely deep circuits, well beyond NISQ capabilities for useful problem sizes.

\begin{figure}[ht!]
\centering
\includegraphics[width=\linewidth]{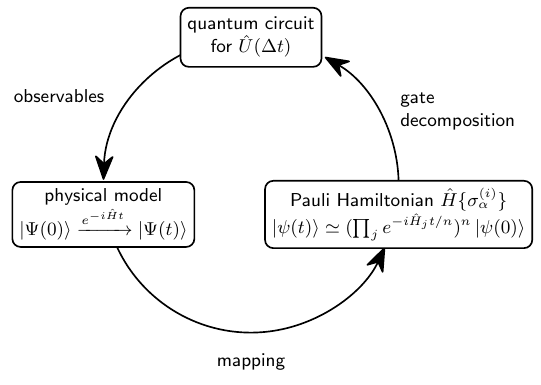}
\caption{Conceptual diagram of universal simulator on digital quantum computing device \cite{Tacchino2020}.}
\label{fig:universalsimulator}
\end{figure}

Figure \ref{fig:universalsimulator} gives a conceptual illustration of how to perform quantum simulations on a digital quantum computer. A physical model describes the quantum state evolution of interest $\Psi(t)$; this evolution can be approximated to arbitrary precision by mapping the given model to a spin-type Hamiltonian which is easily encoded onto a qubit register.  The sequence of unitary operations to implement the time evolution using \eqref{eq:STformula} or higher order variants can be programmed as a quantum circuit, giving the approximated evolved state as output $\psi(t)$ starting from a given input state $\ket{\psi(0)}$.  The mapping from the physical system to qubits can then, in principle, be reversed to obtain the desired physical output state.  However, since the output is a quantum state, not a classical description of it, in practice various types of post-processing are usually required to compute properties of interest (e.g., see section \ref{sssec:QPE}).  

Although time evolution can be readily realized using Trotter product formulas, accuracy requires many small time steps, leading to very deep circuits. Recent work has demonstrated that variational approaches can speed up the calculation of Trotter terms \cite{Benedetti2021}. Moreover, tensor network methods (see section \ref{sssec:quantifyperformance}) can be used to generate quantum circuits that can be more efficient than Trotterization \cite{McKeever2023}.

One of the main classical methods for simulating quantum systems beyond the constraints imposed by the exponential size of Hilbert space is quantum Monte Carlo (QMC) \cite{Foulkes2001,Becca2017,Mazzola2024}.  Instead of a full classical simulation, which would be beyond the reach of HPC, individual quantum trajectories are sampled to build up a statistical average of the properties of interest.

The method requires large numbers of samples to achieve reasonable accuracy, often on the order of millions of samples, which can take hours or even days on current HPC.
As well as computational chemistry \cite{Mazzola2024}, QMC methods are widely used in materials science \cite{Foulkes2001}, 
nuclear physics \cite{Carlson2015}, and condensed matter physics \cite{Cohen2015,Li2019}.  

In all fermionic quantum systems, QMC can suffer from the ``sign problem'' where enforcing fermion exchange statistics leads to a lack of convergence in the QMC sampling \cite{Alexandru2022}. 
This is where quantum simulators that can naturally represent fermionic properties are particularly promising \cite{Bakr2009,Cheuk2015}.
Quantum field theories are used to describe condensed matter many-body systems.  Work on quantum simulation in this setting, e.g.,
quantum algorithms \cite{Byrnes2006}, and analogue \cite{Zohar2011} and digital simulators \cite{Buchler2005}, has evolved into an experimental field that has attracted the attention of particle and nuclear physicists \cite{Cloet2019,Bauer2023}. 
In particular, after the first quantum degenerate Bose and Fermi gases were experimentally created \cite{Anderson1995,DeMarco1999}, these became a versatile analogue simulator to study the behavior of condensed matter systems \cite{Schafer2020,Daley2023}. The quantum gas microscope is another significant advance, allowing research into quantum-degenerate gas phenomena where particle indistinguishability plays a key role \cite{Bakr2009,Cheuk2015}.

Quantum field theory is also the main model used in high energy physics (HEP).
In 2016, the first full quantum simulation of a high-energy physics (HEP) experiment demonstrated the creation of electron-positron pairs from energy \cite{Martinez2016}. 
Gauge field theories are the underlying formalism describing interactions among elementary particles in the Standard Model, and can also be extended to physics beyond the Standard Model. Quantum simulators can be used to study gauge field theories and investigate non-perturbative dynamics at strong coupling. 
Rapid advances in atomic, optical, molecular, and solid-state platforms \cite{Altman2021} have given us analogue quantum simulators to explore HEP models. Better HEP simulations may eventually help us understand difficult problems like quark confinement and the properties of dense nuclear matter, e.g., neutron stars. 
\citeauthor{Bauer2023} \cite{Bauer2023} provide a roadmap of quantum simulations for HEP. 

In the NISQ era, while we wait for digital quantum hardware to develop, variational quantum algorithms (section \ref{ssec:VQAs}) have been shown to produce meaningful results with shallow circuits and without active error-correction procedures. There has been work on simulating field theories, including finding the lowest lying energy spectra of low-dimensional Abelian and non-Abelian lattice gauge theories \cite{Kokail2019,Atas2021,Irmejs2023}. It is important to note that many lattice gauge theory simulations on NISQ hardware require fully or partially removing redundant degrees of freedom or imposing symmetries. Simplifications like these are essential in the resource-limited NISQ era.

In the next sections, we provide a more detailed account of the potential of quantum computers to advance quantum simulation in quantum computational chemistry.  Many of the methods are more widely applicable, and can provide insight for adjacent and related fields.

\subsection{Quantum computational chemistry}

One of the original motivations for the development of quantum computers was to be able to accurately simulate and characterize systems of interacting fermions \cite{Abrams1997,ortiz2001quantum}. 
Obtaining accurate solutions of the electronic structure of many-body systems is a major challenge in computational chemistry. Research has been carried out to develop algorithms to solve the electronic structure problem using quantum computers \cite{Lin2022,Huggins2022,Somma2019}. Fault-tolerant quantum computers are predicted to solve the electronic structure problem for many-body systems \cite{Cao2019} in polynomial time, for example, using the quantum phase estimation algorithm \cite{AspuruGuzik2005,Abrams1997,Kitaev1996} (section \ref{sssec:QPE}). It is possible that a quantum advantage exists even in the absence of a clear polynomial scaling. A detailed overview of the state-of-the-art methods developed in this field can be found in \cite{Cao2019} and \cite{McArdle2020}. In practice, the possibility of achieving an exponential quantum advantage for quantum chemistry problems remains controversial \cite{Lee2023c}.  It is challenging for quantum simulations to achieve quantum advantage over the best classical computations, given the decades of work already done on classical methods for quantum systems. As classical algorithms continue to improve, quantum information processing has been useful in developing advanced ``quantum-inspired'' classical techniques \cite{Jia2021}. 
In this regard, we review new techniques for classical and quantum hardware from different perspectives. We cover a range of methods applicable to problems such as \textit{ab initio} simulations, Hubbard models, spin Hamiltonians, interacting Fermi liquids, solid-state systems, and molecular systems. 

Modern computational chemistry aims to simulate molecules and materials within 1 kcal/mol accuracy, which is required to predict chemical phenomena at room temperature. Despite the tremendous progress in quantum chemistry algorithms in the past century, such a level of precision can only be achieved for small systems due to the polynomial scaling (higher than O($x^{4}$) with system size $x$) of advanced computational chemistry methods \cite{Grimme2018}. Hence, current research has a strong emphasis on developing hybrid quantum-classical and quantum algorithms; estimating quantum resources; and designing quantum hardware architectures for simulations of practical interest. 

Another central quantity in statistical mechanics and materials science is the partition function. All thermodynamic quantities of interest can be derived from the partition function, which then help us understand the behavior of many-body systems. Calculating partition functions of physical systems is generally a \#P-hard complexity class problem \cite{Lidar2004,Jaeger1990} -- both classical and quantum methods are unlikely to provide exact solutions. Quantum algorithmic attempts to estimate partition functions involve quantum phase estimation \cite{Temme2011} combined with a quantum version of the Monte Carlo Metropolis algorithm, and algorithms based on quantum walks (section \ref{sssec:qw}) \cite{Wocjan2009,Arunachalam2022}. Hybrid quantum-classical algorithms \cite{Wu2021g} and quantum-inspired algorithms \cite{Jackson2023} have also been proposed, with quantum algorithms potentially capable of performing better than classical ones.

Even with noisy hardware, the long-term impact of using quantum simulation for quantum chemistry and materials is promising. 
For example, Google's Sycamore quantum processor used a hybrid quantum-classical Monte Carlo method \cite{Arute2019} to calculate the atomization energy of the strongly correlated square H$_{\text{4}}$ molecule. The results are competitive with those of the state-of-the-art classical methods. The algorithm relies on being able to efficiently prepare a good initial guess of the ground state on the quantum hardware, which is a non-trivial step.

Understanding the strengths and weaknesses of classical algorithms can indicate where we may see a quantum advantage. It also indicates where issues may arise in quantum simulation approaches in the near and long term. For example, \citeauthor{Tubman2018} \cite{Tubman2018} used classical simulations to provide resource estimates for crucially important but often neglected aspects of quantum chemistry simulations. These include efficient state preparation for quantum methods such as phase estimation. 
Sections \ref{sec:ManyElectronProblem} to \ref{sec:PostHF} present an overview of techniques in computational chemistry along with difficulties encountered in modelling chemical systems. Sections \ref{sssec:PostHFQC} and \ref{sec:quantumoptimizationchemistry} cover early applications of quantum computing within computational chemistry. Section \ref{sec:IndustryApplications} discusses two examples of chemical systems with industrial applications that are beyond the capabilities of classical computational chemistry methods.

\subsubsection{Quantum chemistry's many-electron problem}\hfill
\label{sec:ManyElectronProblem}

The core objective of quantum chemistry is to solve the stationary Schr{\"o}dinger equation for an interacting electronic system in the static external field of the nuclei, which describes isolated molecules in the Born-Oppenheimer approximation. The problem can be defined as follows. There are $N$ electrons in the field of stationary $M$ point charges (nuclei). Electrons interact only through electrostatic Coulomb terms and are collectively described by a wave function with $4N$ coordinates $\vec{x}_j = (\vec{r}_j, \vec{\omega}_j)$, i.e., $3N$ spatial coordinates $\vec{r}_j$ and $N$ spin coordinates $\omega_j$. The antisymmetric electronic wave functions are solutions of the time-independent Schr{\"o}dinger equation
\begin{gather}
\hat{H} \Psi(\vec{r}) = E \Psi(\vec{r})
\end{gather}
where we describe the motion of the $N$ electrons ($j$, $k$ indices) in the field of the $M$ nuclei ($A$ index) with Hamiltonian \cite{McArdle2020}
\begin{gather}
\hat{H} = 
\underbrace{- \sum_j \frac{\nabla_j^2}{2}}_{=\hat{H}_K}
\underbrace{-\sum_{jA}\frac{z_A}{\vert\vec{r}_j-\vec{r}_A\vert}}_{=\hat{H}_{e\text{-}n}} +
\underbrace{\frac{1}{2} \sum_{k \neq j} \frac{1}{\vert\vec{r}_j-\vec{r}_k\vert}}_{=\hat{H}_{e\text{-}e}}
\label{schroedinger_equation}
\end{gather}
in atomic units $z$. The Hamiltonian contains the kinetic term $\hat{H}_K$, the electron-electron interaction $\hat{H}_{e\text{-}e}$, and the electron-nuclei interaction $\hat{H}_{e\text{-}n}$. When there are two or more electrons, we cannot solve the Schr{\"o}dinger equation exactly. This is the fundamental challenge of computational chemistry.

There are two main families of approximations to solve this problem: \textit{Hartree-Fock} (HF) \cite{slater1951} methods with refinements and \textit{density functional theory} (DFT) \cite{Hohenberg1964,Kohn1965}.  Each has its own opportunities for quantum enhancements.  We summarize the main methods and their limitations before discussing potential quantum enhancements.

\subsubsection{HF and the electron correlation problem}\hfill

The Hartree-Fock (mean-field) method \cite{Echenique2007} describes the potential experienced by each electron resulting from the average field of all electrons. This means that the $\hat{H}_{e\text{-}e}$ term in (\ref{schroedinger_equation}) is replaced by a mean-field potential to make the equation solvable. However, this approximation does not accurately describe the electron-electron correlation.  Electron correlation accounts for the fact that electrons do not move independently in the average field of other electrons rather, their movement is influenced by the instantaneous positions and motions of other electrons. The particle exchange energy resulting from the anti-symmetric wave function (i.e., Pauli principle)
\begin{equation}
\Psi(\vec{x}_1, ..., \vec{x}_j, \vec{x}_k, ..., \vec{x}_N) = 
-\Psi(\vec{x}_1, ..., \vec{x}_k, \vec{x}_j, ..., \vec{x}_N),
\end{equation}
means we cannot express the wave function as a separable product of single-particle wave functions,
\begin{equation}
\Psi(\vec{x}_1, \vec{x}_2, ..., \vec{x}_N) \neq \psi_1(\vec{x}_1) \psi_2(\vec{x}_2) ... \psi_N(\vec{x}_N)
\end{equation}
where the spatial and spin parts are defined as $\psi_j(\vec{x}_j) = \phi_j(\vec{r}_j) \chi_j(\vec{\omega}_j)$. Instead, the HF method approximates the many-body wave function with a Slater determinant
\begin{multline}
\Psi(\vec{x}_1, \vec{x}_2, ..., \vec{x}_N) =\\ 
\frac{1}{\sqrt{N!}}
\begin{vmatrix}
\psi_1(\vec{x}_1) & \psi_2(\vec{x}_1) & ... & \psi_N(\vec{x}_1) \\
\psi_1(\vec{x}_2) & \psi_2(\vec{x}_2) & ... & \psi_N(\vec{x}_2) \\
\vdots & \vdots & \ddots & \vdots \\
\psi_1(\vec{x}_N) & \psi_2(\vec{x}_N) & ... & \psi_N(\vec{x}_N)
\end{vmatrix},
\label{slater_determinant}
\end{multline}
where the set of $\{\psi_j(\vec{x}_k)\}$ are chosen to minimize the ground state Hartree-Fock energy 
\begin{equation}
E_0 = \bra{\Psi_0} \hat{H} \ket{\Psi_0}.
\end{equation}
The wave function depends on the Hamiltonian, and, in turn, the mean-field part of the Hamiltonian depends on the wave function. This means that the HF method must be solved iteratively using the self-consistent field (SCF) method. We start from an initial guess for the wave function, build an initial Hamiltonian, and then calculate the initial energy. We change the wave function, update the Hamiltonian, and calculate the energy until the difference between cycles is lower than a defined threshold. The SCF method relies on the variational principle, which states that the energy of any approximate wave function is higher than or equal to the exact energy. 
The final energy is the exact HF energy $E_0$. This value is always higher than the true ground-state energy because the formalism cannot fully account for electron correlation effects. Recovering the correlation energy is the main goal of post-HF methods.

For small systems, such as atoms, the unknown $\psi$ functions in (\ref{slater_determinant}) can be solved numerically on a three-dimensional grid of points. Larger systems require grids that are too dense to reach convergence. Hence we express the $\psi$ functions as a linear combination of a set of known functions, the basis set:
\begin{equation}
f(x) = \sum_{j}c_{j}\chi_{j}(x).
\end{equation}
In practice, the basis set could be built from any type of function $\chi_{j}$ following two criteria: they should reproduce the behavior of the electron density (the highest density should be located at the atom positions, and they should monotonically decrease to zero at an infinite distance from the nuclei) and they should be easy to integrate. Among the most widely used basis sets, there are Gaussian and Slater-type basis sets for molecules \cite{basisset} and plane waves for periodic systems \cite{Singh2006}. 

Selecting the appropriate basis set is a critical consideration. It is a balance between achieving the desired accuracy and managing computational resources. Classical computer simulations often employ extensive basis sets, which can provide high levels of accuracy, including up to 30 functions per atom \cite{Peterson2007,Jensen2016}. The computational capacity of classical computers typically means that the basis set size is not the primary limitation on accuracy. In contrast, NISQ devices impose limitations on the basis set size due to hardware constraints \cite{Chien2022}, making the basis set size a more significant factor affecting accuracy as \citeauthor{McArdle2020} discuss in their comprehensive review \cite{McArdle2020}.

\subsubsection{Density functional theory (DFT)}\hfill

Density functional theory (DFT) \cite{Capelle2006} focuses on the electronic density 
\begin{gather}
n(\vec{r}) = N \int d^1 r_1 \int d^2 r_2 ... \int d^N r_N \vert\Psi(\vec{r}, \vec{r}_2, ..., \vec{r}_N)\vert^2
\end{gather}
rather than electron coordinates. DFT systematically maps the many-body problem to a single-body problem through the particle density $n(\vec{r})$ from which we calculate all other observables.

The Hohenberg-Kohn (HK) theorem is at the heart of DFT. It states that the system's ground-state energy is uniquely determined by the ground-state density $n_0(\vec{r})$. The ground-state wave function $\Psi_0(\vec{r}_1, \vec{r}_2, ..., \vec{r}_N)$ must both reproduce the ground-state density $n_0(\vec{r})$ and minimize the energy. Hence the aim is to minimize the sum of kinetic, interaction and potential energies
\begin{align}\label{HK}
&E_0[n(\vec{r})] \nonumber\\
&= \min_{\Psi \to n} \bra{\Psi} \hat{H}_K + \hat{H}_{e-e} + \hat{H}_{e-n} \ket{\Psi} \\
&= \underbrace{\min_{\Psi \to n} \bra{\Psi} \hat{H}_K + \hat{H}_{e-e} \ket{\Psi}}_{=: F[n]} + \int d^3 r \; n(\vec{r}) v(\vec{r})
\end{align}
with HK density functional $F[n]$. We then minimize $E_0$ with a variational method using Lagrange multipliers to obtain the Kohn-Sham equations \cite{Baseden2014},
\begin{equation}
-\frac{\nabla^2}{2} \psi_j(\vec{r}) + v_{\text{eff}}(\vec{r})\psi_j(\vec{r}) = \epsilon_j \psi_j(\vec{r}).
\end{equation}
The effective potential $v_{\text{eff}}(\vec{r})$ is defined as the sum of the external potential (the interaction with the nuclei) $v_{\text{ext}}(\vec{r})$, the Coulomb interaction among the electrons $v_{\text{H}}(\vec{r})$, and the exchange-correlation potential $v_{\text{xc}}(\vec{r})$:
\begin{equation}
v_{\text{eff}}(\vec{r}) = v_{\text{ext}}(\vec{r}) + v_{\text{H}}(\vec{r}) + v_{\text{xc}}(\vec{r}).
\end{equation}
The exchange-correlation potential takes into account non-classical effects such as the Pauli principle and the electron correlation. This functional, which must be parameterized, determines the quality of the result.
DFT is system-specific and does not reach the desired chemical accuracy. Similarly to the Hartree-Fock method, it does not take into account direct electron correlation, although this can be compensated for by the functional. Additionally, this method has the disadvantage of self-interaction, which occurs when each electron interacts with the electron density generated by all the electrons, including itself. 

Because DFT is based on the three-dimensional distribution of the electron density, it is possible to directly calculate it on a three-dimensional grid around the atoms and bonds that represent the electron density at different points in space. However, in practical implementations, DFT frequently uses auxiliary functions to discretize the electron density on the grid. This approach ensures precise numerical integration, computational efficiency, management of exchange-correlation functionals, and allows us to apply linear scaling methods for larger systems. Consequently, DFT uses basis set functions similar to those in HF theory.

\subsubsection{Post Hartree-Fock methods}\hfill
\label{sec:PostHF}

Since neither HF nor DFT fully take into account the electron-electron correlation, the repulsive force between the electrons must be added \textit{a posteriori}. Post-Hartree-Fock (post-HF) methods are more accurate than HF, but at greater computational cost. For example, the full configuration interaction (FCI) method considers all possible electron configurations (excitations) within the chosen basis set and gives the exact solution of the Schr{\"o}dinger equation within a basis set expansion \cite{Eriksen2021}. 
The number of all possible occupancy combinations scales as $O(N^M)$ with $N$ electrons and $M$ basis functions. It can only be performed on classical computers for small systems. Another example is the M{\o}ller-Plesset perturbation theory \cite{Moller1934} which takes the core terms $\hat{H}_K+\hat{H}_{e-n}$ as an unperturbed Hamiltonian and treats the electron repulsion terms $\hat{H}_{e-e}$ as a perturbation. Applying the Rayleigh-Schr{\"o}dinger perturbation theory gives an approximate correction of the correlation energy to the HF energy. In comparison, coupled cluster techniques \cite{Cizek1966} apply an exponential cluster operator $\hat{T} = \sum_k \hat{T}_k$ containing $k$-fold excitation operators $\hat{T}_k$ to the HF wave function $\ket{\Psi_0}$. We obtain the exact solution $e^{\hat{T}} \ket{\Psi_0}$ through perturbation theory. 
These methods can be used for larger systems in their truncated form. For example, instead of the FCI approach, we could consider a single excitation to reduce the computational cost. However, this will also decrease the accuracy of the final energy. Another approach is to reduce the number of electrons and orbitals considered in the CI expansion by selecting a complete active space (CAS) where all excitations are considered. This approach relies on the premise that chemical reactions primarily involve higher-energy electrons. Although we may not obtain the exact total energy, energy differences (such as those between reactants and products) should account for the electron correlation. 

The discussion above focuses on the simulation of molecules. The same concepts apply when simulating crystalline materials when the basis set consists of plane waves \cite{Singh2006} or Bloch functions built from local orbitals \cite{Hoffmann1987}. Crystalline orbitals, which are the periodic equivalent of molecular orbitals, are delocalized over the cell and, therefore, are difficult to use in a CAS approach. One method to solve this problem is to use the Wannier single-particle electron basis \cite{Wannier1937}. A detailed analysis of how these functions can be used in the context of quantum algorithms can be found in \cite{Clinton2024}.

\subsubsection{Quantum algorithms for computational chemistry}\hfill
\label{sssec:PostHFQC}

Quantum computing holds significant promise for enhancing two primary categories of chemistry challenges: post-Hartree-Fock (post-HF) methods and optimization/global minimum search tasks. Both categories deal with the challenges of combinatorial explosion. The former arises when encompassing all possible excitations, as seen in the FCI approach, while the latter emerges in scenarios where a multitude of configurations need to be evaluated, such as in determining the thermodynamic properties of complex materials or the optimum folded structure of particular proteins. Efforts have been made to map chemical problems to quantum computing-based simulations \cite{Cao2019,Kassal2011}; understand the possible benefits of using quantum hardware \cite{Bauer2020}; and explore the possibility of QML \cite{Bernal2022}. It is worth noting that the FCI method, while being a significant improvement over HF or DFT for a given basis set, is still constrained in its accuracy by the choice of basis set.

An additional challenge when simulating electrons arises from their fermionic nature (half-integer spin). The need for their wavefunction to be antisymmetric introduces non-locality, affecting the mapping to qubits. The mapping process starts from the second quantized form of the wavefunction, where the computational basis state is represented by a Slater determinant in Fock space. The most widely used methods to map it to a qubit system are the Jordan-Wigner \cite{Jordan1928} and Bravyi-Kitaev \cite{Bravyi2002} mappings \cite{Tranter2018}. Both methods ensure a one-to-one correspondence between electrons and qubits. Therefore, the non-locality of the wavefunction manifests itself in the operators, resulting in long strings of spin operators that increase gate depth and potentially degrade performance on NISQ devices. Although the Bravyi-Kitaev method aims to reduce the length of these operators, it is more challenging to implement and yields more complex quantum circuits, which may be harder to optimize. Recent and innovative mapping techniques, as discussed in \cite{OBrien2024} and \cite{Derby2021}, achieve circuit lengths that scale logarithmically with the number of fermions in the system.

One promising research direction is to improve existing quantum methods for estimating electronic ground-state energies, such as the quantum phase estimation (QPE) \cite{AspuruGuzik2005} (section \ref{sssec:QPE}) and VQEs \cite{Kandala2017} (section \ref{ssec:VQAs}) methods.  
The key difference between VQE and QPE-based approaches is that QPE requires encoding the full Hamiltonian into the unitary gates for the energy read-out method. Hence, QPE requires a large number of qubits and quantum gates \cite{Wecker2014,Babbush2018}. This complicated procedure is not needed in VQE. In addition, while QPE assumes the ansatz has a reasonable overlap with the exact ground state, VQE aims to approximate the ground state by variationally optimizing a set of parameters.

Ans{\"a}tze widely used in quantum chemistry problems include the unitary coupled clustered (UCC) ansatz \cite{Bartlett1989,Lee2019}, for calculating the ground-state energy of a fermionic molecular Hamiltonian. In the UCC method, we expand a reference wavefunction on the basis of molecular orbitals mapped to the qubit basis in a quantum circuit. Then, we apply a parameterized unitary operator to the circuit, which produces a trial wavefunction. This procedure encodes complicated electronic states in a small number of qubits. 

Variants of the UCC ansatz can reduce the circuit depth by compiling the fermionic operators more efficiently \cite{Motta2021,Matsuzawa2020,Setia2019}. The variational Hamiltonian ansatz prepares a trial ground state for a given Hamiltonian by Trotterizing an adiabatic state preparation process \cite{Wecker2015,Wiersema2020}. The variable structure ansatz optimizes the circuit structure itself by adding or removing unnecessary circuit elements \cite{Grimsley2019,Tang2021a,Yordanov2021}. Hybrid ans{\"a}tze offload some of the complexity from the quantum ansatz to a classical device \cite{Takeshita2020,Sokolov2020}. They are especially useful for exploiting the classical simulability of the free-fermion dynamics. 

The adaptive derivative-assembled pseudo-Trotter VQE (ADAPT-VQE) approach \cite{Grimsley2019,Tang2021a,VaqueroSabater2024} is another promising way to construct accurate wavefunctions and accurately predict the ground state properties of chemical systems \cite{Grimsley2023,Feniou2023}. ADAPT-VQE follows a dynamic strategy for selecting and refining the ansatz. It systematically expands the electronic wavefunction to include the operators that contribute significantly to the energy. This makes it well suited for strongly correlated electronic systems. ADAPT-VQE produces hardware-friendly circuits with few entangling gates, but the present technological limitations make it necessary to consider variants of ADAPT-VQE to enhance its performance.

Since quantum circuits can be simulated by tensor networks \cite{Markov2008}, it is also possible to combine tensor network techniques with a quantum ansatz \cite{Liu2019,Barratt2021}. Alternatively, the deep variational quantum eigensolver divides the system into smaller subsystems and sequentially solves each subsystem and the interaction between subsystems \cite{Fujii2022}.

So far, gate-based quantum hardware has been the technology of choice for performing post-HF simulations. On NISQ devices, FCI is suitable for small molecules such as H$_2$ and LiH \cite{McArdle2020}, while, for larger chemical systems, an embedding scheme is generally preferred. Embedding techniques are well-established in classical computing for quantum chemical systems, see for example ~\cite{catlow2017quantum} for an overview of common approaches. The model system is typically divided into two regions, an active region that is treated at a high level of theory, using DFT or methods beyond DFT, embedded within an environment or inactive (frozen) region. Different schemes are available for embedding via systems of point charges, molecular mechanical models \cite{catlow2017quantum}, electronic density \cite{berger2014embedded}, a density matrix \cite{knizia2012density} or Green's function-based method \cite{anisimov1997first}. This approach has high potential to be used with quantum computing systems, where quantum algorithms are applied to the active region in the embedding model. For example, \citeauthor{Gao2021} \cite{Gao2021} selected an orbital active space derived from an initial Hartree-Fock (HF) wavefunction to study the lithium superoxide dimer rearrangement. The method incorporates relevant orbitals present in reactants, products, and those involved in the transition state. In another instance, \citeauthor{Gujarati2022} \cite{Gujarati2022} chose active space orbitals from DFT orbitals to study the water splitting reaction on magnesium surfaces.

Performing calculations of the electronic excited-state properties of molecules is another challenging area of research \cite{Westermayr2021}. This is relevant when we want to simulate spectroscopic experiments, non-equilibrium dynamics, light-matter interactions and other excited-state based phenomena. Accurate simulations of these phenomena depend on precise calculations of the ground and excited states for the system, specifically the eigenvalues of the Hamiltonian operator. As discussed above, the VQE can achieve this with high precision \cite{Ollitrault2020}. Furthermore, the quantum subspace expansion (QSE) method \cite{Colless2018}, which is an extension of the VQE, allows for the definition of a subspace of low-energy excited states once the ground state of the system is determined.

Quantum computers hold significant potential for the simulation of dynamic processes \cite{Ollitrault2021} such as chemical reactions, the study of transition states, and molecular dynamics. The most widely used approach relies on the Suzuki-Trotter decomposition \cite{Hatano2005,Babbush2015} to break down the time evolution into small steps that can be implemented on quantum computers. This approach requires deep circuits and is therefore limited by the size of current NISQ devices. A different approach \cite{Miessen2021}, based on the VQE, results in a significant reduction in the gate count, but requires a large number of measurements, which are only possible for small systems.

\subsubsection{Quantum optimization for chemical systems}\label{sec:quantumoptimizationchemistry}\hfill

Quantum annealing (section \ref{ssec:qopt}) has been demonstrated to be a valuable tool for solving optimization problems in chemistry, biochemistry \cite{Fedorov2021} and materials science \cite{Camino2023}. Given a set of variables (for example, atom position in a solid solution lattice or bond angle in protein folding), the aim is to find the configurations of those variables that correspond to the lowest energy. The configuration space to explore is proportional to $\binom{n}{k}$ with $n$ variables and $k$ possible values that each variable can take. This configurational space rapidly becomes impossible to sample using classical methods for large values of $n$. Quantum annealing can offer an alternative means of finding the lowest energy configurations by exploiting quantum superposition and tunnelling (see \citeauthor{Camino2023}'s \cite{Camino2023} tutorial for more details).

Quantum annealers have also been used to solve the electronic Hamiltonian eigenvalue-eigenvector problem \cite{Teplukhin2020} and to calculate excited electronic states \cite{Teplukhin2021} for biatomic and triatomic molecules. In industry, researchers at Volkswagen used D-Wave to calculate the ground state energies of H$_2$ and LiH as a proof-of-principle \cite{Streif2018}, where their mixed discrete-continuous optimization algorithm found the lowest eigenstate of the qubit coupled cluster method and used a quantum annealer to solve the discrete part of the problem \cite{Genin2019}. Menten AI used D-Wave to address the rotamer optimization problem for protein design without major simplifications or a decrease in accuracy \cite{Mulligan2019}. GlaxoSmithKline studied the mRNA codon optimization problem to help drug discovery \cite{Fox2021} using quantum annealers. Fujitsu developed a quantum-inspired annealing machine to screen chemicals for materials discovery \cite{Hatakeyama-Sato}. 

Optimization can also be performed on gate-based hardware when the problem is formulated as a QAOA problem (section \ref{sssec:QAOA}). For example, the design of deuterated molecules for OLED applications has been studied using an approach that combines DFT-derived energies and VQE \cite{Gao2023}.

\subsection{Important applications in industry}
\label{sec:IndustryApplications}

Quantum simulation will potentially solve many industrial problems and bring about significant societal and environmental benefits. We give two examples in this section. Researchers are using the Fermi-Hubbard model as an analogue quantum simulator for high-temperature superconductors, specifically copper oxide compounds. On the other hand, digital quantum simulation may help us understand how to produce ammonia fertilizers using less energy intensive processes.

\subsubsection{High-temperature superconductivity}\hfill

Theoretical high-energy and condensed matter physics share common fundamental concepts such as symmetry breaking, renormalization group, and Feynman diagrams. Interesting phenomena occur in strongly correlated electronic systems where several physical interactions (spin, charge, lattice, and/or orbital) are simultaneously active. In particular, the discovery of high-temperature superconductors (HTS) in the 1980s \cite{Bednorz1986,Sheng1988} launched decades of diligent efforts to understand and use these compounds \cite{Zhou2021,Lilia2022}. 

\begin{figure}[t!]
\centering
\includegraphics[width=1\linewidth]{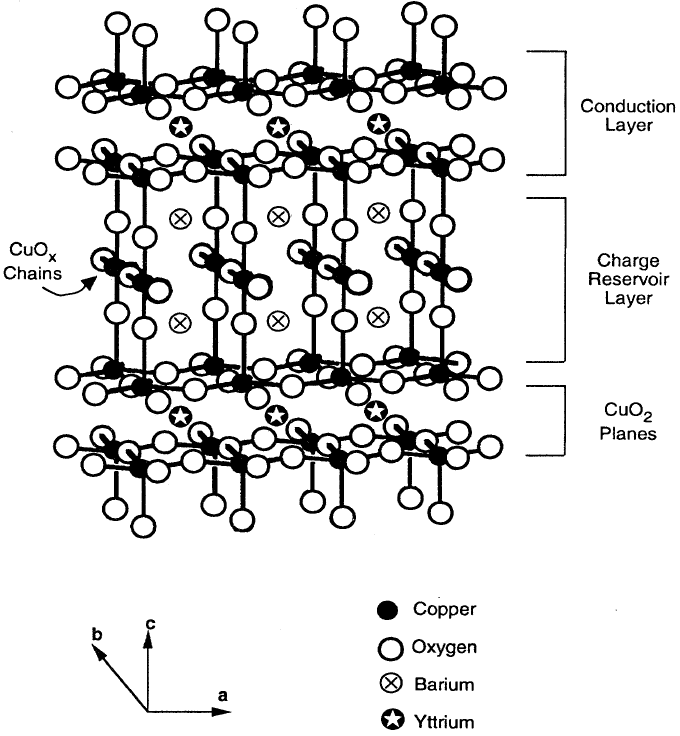}
\caption{Crystal structure of cuprate superconductor YBa$_2$Cu$_{3}$O$_{6+x}$ \cite{Dagotto1994}.}
\label{fig:YBCO}
\end{figure}

The Bardeen-Cooper-Schrieffer (BCS) theory \cite{Bardeen1957,Bardeen1957a} introduced the idea of electron-hole Cooper pairs to explain conventional low temperature superconductivity. However, the pairing mechanism in unconventional (high temperature) superconductors is still unknown. These materials do not conform to conventional BCS theory or its extensions. Conventional superconductors are used in many application areas such as MRI machines and high speed trains, but they can only operate at relatively low temperatures. This severely limits their wider use. One of the holy grails of materials science is to develop superconducting materials that operate at room temperature. This would revolutionize many technologies \cite{Lilia2022}. In particular, room-temperature superconductors would decrease the heat wasted from electronic devices and allow them to run more efficiently. On a larger scale, HTS can help achieve the International Energy Agency's roadmap to carbon-free economies \cite{IEA2021} via nuclear fusion-generated electricity \cite{Ball2021}.

Researchers have focused on the two-dimensional Hubbard model as it is believed to capture the important behavior of HTS \cite{Dagotto1994,Fradkin2015}, specifically the cuprate superconductors (a popular HTS) in the copper-oxygen planes (``CuO$_2$ planes'' in Figure \ref{fig:YBCO}). Despite the Hubbard model's apparent simplicity, its theoretical properties are far from fully understood. It is difficult to solve accurately as the model exhibits competing orders in its phase diagram where it is most relevant to cuprates. 
The Hubbard model is also widely used to benchmark numerical methods for strongly correlated systems \cite{SimonsCollaboration2015}. 
Quantum efforts include experimental work building analogue quantum simulators of the model using quantum dots \cite{Hensgens2017,Wang2022} and ultracold atoms in optical lattices \cite{Mazurenko2017,Hirthe2023}.
Digital quantum approaches include VQE variants \cite{Cade2020,Cai2020,Stanisic2022}, while Google simulated the Fermi-Hubbard model using digital superconducting quantum processors \cite{Google2020}. 

A state-of-the-art circuit-model algorithm for simulating the two-dimensional Hubbard time dynamics on an $8 \times 8$ lattice requires roughly $10^7$ Toffoli (CNOT) gates \cite{Kivlichan2020}. This includes the overhead for fault tolerance, given the gate fidelities of current and near-term hardware. A significant contribution to the gate count comes from the phase estimation procedure. These large gate counts are corroborated by \citeauthor{Clinton2021} \cite{Clinton2021}, who estimate the resources to simulate the time dynamics on a $5 \times 5$ square lattice to $\leq 10$\% accuracy require 50 qubits and 1,243,586 standard two-qubit gates. This is an optimistic estimate that assumes the effects of decoherence and errors in the circuit can be neglected; the source of inaccuracy is the approximations in the gate sequences.

There is much work to be done before we have a complete microscopic theory of HTS \cite{Lilia2022,Zhou2021}. The combination of theory, simulation, materials synthesis and experiment has been crucial to progress in the last two decades. For example, the breakthrough discovery of hydrogen-rich superconductors (super-hydrides) \cite{Drozdov2015} may not have happened for another century \cite{Castelvecchi2023s} had there not been significant advances in simulation and algorithms for chemical structure prediction \cite{Li2014}. Similarly, numerical work to understand the two-dimensional doped Hubbard model has relied on the lastest advances in Monte Carlo methods \cite{Xiao2023,Xu2024}. Experiments with ultracold atoms in optical lattices have also proved promising \cite{Mazurenko2017,Xu2023f,Homeier2024} due to advances in quantum control with quantum microscopes and other techniques \cite{Bohrdt2021}. Despite the lack of a clear timeline on when we can expect a fundamental theory for HTS, researchers remain optimistic \cite{Castelvecchi2023s}.

\subsubsection{Fertilizers}\label{sec:fertilizers}\hfill 

Another strong motivation for quantum simulation is to develop new materials and processes to make significant environmental contributions. For example, the Haber-Bosch process for producing ammonia (NH$_3$) fertilizers is one of the world's most CO$_2$-intensive chemical processes, consuming up to 2\% of global energy output \cite{Lv2021} and 3-5\% of all natural gas generated globally \cite{BCG2020}. 

The Haber-Bosch process turns nitrogen in the air into ammonia-based fertilizer for crops, which in turn provide food for about 40\% of the world's population. The natural biological process, nitrogen fixation, is a much more efficient process: microorganisms that contain the biological enzyme \textit{nitrogenase} convert atmospheric dinitrogen (N$_2$) to ammonia under ambient conditions. Despite almost a century of research, the reaction mechanism is still unexplained \cite{Burgess1996,Vogiatzis2019}. Understanding how this enzyme works would be an important step towards replacing the Haber-Bosch process and creating less energy-intensive synthetic fertilizers.

\begin{figure}[ht!]
\centering
\includegraphics[width=1\linewidth]{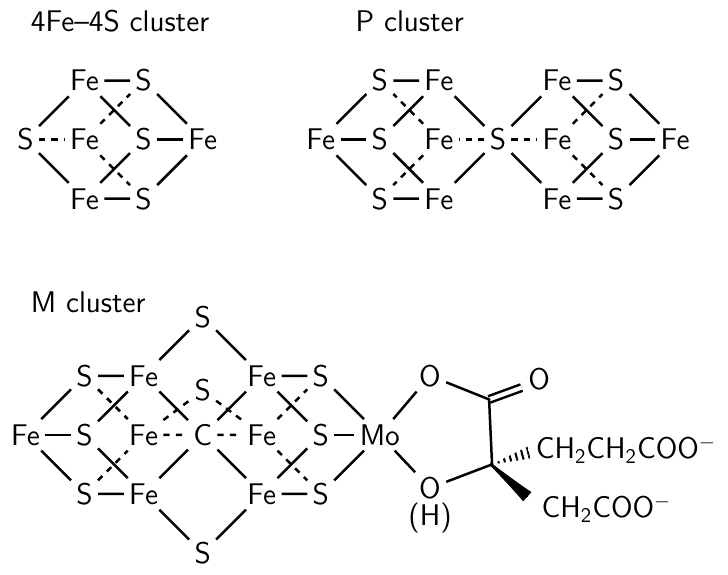}
\caption{Nitrogenase enzymes convert atmospheric dinitrogen into ammonia. The process involves transferring multiple electrons and protons to dinitrogen and uses multiple metalloclusters found in the nitrogenase enzyme, including the 4Fe--4S cluster (top left), P cluster (top right) and Fe-Mo cofactor M cluster (bottom).}
\label{fig:femoco}
\end{figure}

The transition metal compounds within the enzyme potentially hold the answers. These are the ``4Fe--4S'' cluster (containing iron and sulfur), ``P cluster'' and the iron molybdenum cofactor ``M cluster'' (FeMoco, containing iron, molybdenum, carbon, hydrogen, and oxygen) \cite{Hoffman2014} (Figure \ref{fig:femoco}). The FeMoco active space contains 54 electrons in 108 spin orbitals.  Proposed computational models of FeMoco are beyond the reach of current classical methods but are possible with small error-corrected quantum computers. 

As an initial study to develop simulation methods, \citeauthor{Reiher2017} \cite{Reiher2017} used a Trotterization approach to simulate the active-space model of FeMoCo. Their algorithm had T-gate complexity scaling as approximately $O(N^2S/\epsilon^{3/2})$, with Hamiltonian sparsity $S$ and error rate $\epsilon$, and required over $10^{14}$ T gates. This corresponds to roughly $10^8$ physical qubits if implemented in the surface code \cite{Fowler2012} with gates at $10^{-3}$ error rate. Their work focused on counting and reducing the required number of T gates. In practical error-correcting codes such as the surface code, these gates require significantly more time to implement than any other gate and require a large number of physical qubits for implementation.
This was followed by \citeauthor{low2019hamiltonian}'s \cite{low2019hamiltonian} a qubitization approach which reduced the gate cost estimates by several orders of magnitude, to roughly four days on a fault-tolerant quantum computer equipped with $4 \times 10^6$ physical qubits \cite{Lee2021}. 

Qubitization is particularly promising for reducing the computational costs of digital quantum simulation \cite{Babbush2018,Low2024}, which largely depends on how the input Hamiltonians are accessed by quantum computers. While simulations typically use sparse matrices \cite{Aharonov2003} or linear combination of unitaries (LCU) \cite{Childs2012,Berry2015}, \citeauthor{low2019hamiltonian}'s standard-form encoding   \cite{low2019hamiltonian} is much more general and includes sparse matrices and LCU as special cases. 
Qubitization uses the fact that whenever the encoded Hamiltonian $\hat{H}$ contains an eigenvalue $\lambda$, the standard-form encoding operation contains a $2 \times 2$ block 
\[
\begin{bmatrix} \lambda & -\sqrt{1-\lambda^2} \\ \pm \sqrt{1-\lambda^2} & \pm \lambda \end{bmatrix}
\]
on the diagonal, with respect to a basis determined jointly by $\hat{H}$ eigenstates and the encoding \cite{low2019hamiltonian}. Different $\lambda$ values produce different $2 \times 2$ blocks, which are analogous to single-qubit rotations or reflections, hence the name ``qubitization''. This spectral relation builds on earlier work such as Szegedy's quantum walk \cite{Szegedy2004,Childs2010} and quantum Merlin-Arthur amplification \cite{Marriott2005}, but the formulation by \citeauthor{low2019hamiltonian} is more suitable for quantum simulation.

Google and collaborators \cite{Tazhigulov2022} recently attempted to calculate the energy states of the nitrogenase 4Fe--4S cluster on their Sycamore quantum computer. 
However, simulations containing over 300 gates were overwhelmed by noise. 
The authors concluded that quantum circuit based quantum computers may not provide much advantage over classical computers until they incorporate noise reduction and/or full quantum error correction methods. These results are both exciting and daunting: quantum computational chemistry and materials science has made great progress over the last decades, but \cite{Tazhigulov2022} also shows how much work is still ahead.



\section{Quantum algorithms for classical simulation}\label{sec:qfluid}

High-performance computing is used for modelling large-scale systems in many socially and economically important areas, including weather forecasting, cosmology, plasmas, coastal engineering, real time traffic management, and many other applications. These models often have analytically intractable non-linear partial differential equations (NLPDEs) at their core, with a large number of variables in two or more dimensions.  
Classical numerical solvers for NLPDEs have been an active area of research for decades \cite{Debnath2012,Tadmore2012,Bartels2015}. Finite difference, volume, and element methods; spectral Galerkin methods; and neural networks are commonly used. The latter are particularly suitable for (approximately) solving high-dimensional PDEs \cite{Han2018,Muther2023}. 
Given the large amounts of HPC resources used to solve NLPDEs, improved algorithms are highly important and the subject of significant research investment. For example, NASA's Quantum Artificial Intelligence Laboratory (QuAIL) aims to show how quantum algorithms may dramatically improve the agency's computational problem-solving ability \cite{Biswas2017,Rieffel2019,Rieffel2024}. 
The HHL algorithm \cite{Harrow2009} (section \ref{sssec:HHL}) can be used in regimes where it is sufficient to take a linear approximation of the NLPDEs. For many applications, it is crucial to model the nonlinear effects more accurately. In this section, we focus on one of the most common NLPDEs, the Navier-Stokes equation for computational fluid dynamics (CFD), but the quantum algorithms can be readily adapted to other NLPDEs.

\subsection{Solving the Navier-Stokes equations}

In CFD, the aim is to solve the Navier-Stokes (NS) equations which describe a flow of pure gas or liquid (single- or multi-phase flows)
\begin{equation}\label{eq:navierstokes}
\frac{\partial\vec{u}}{\partial t} = \underbrace{-\vec{u}\cdot\nabla\vec{u}}_\text{advection} - \nabla p + \underbrace{\frac{1}{\mathcal{R}}\nabla^2\vec{u}}_\text{viscous diffusion}
\end{equation}
with continuity equation 
\begin{equation}
\nabla\cdot\vec{u}
\begin{cases}
= 0, & \text{incompressible flow,} \\
\neq 0, & \text{compressible flow.} 
\end{cases}
\end{equation}
Velocity $\vec{u}(\vec{r},t)$ and pressure $p(\vec{r},t)$ fields are functions of time $t$ and spatial coordinates in continuous space $\vec{r}$. The Reynolds number $\mathcal{R}$ describes the ratio of inertial force to viscous force (viscous diffusion). 

The Navier-Stokes equations consider the contribution of all forces acting on an infinitesimal element of volume and its surface. Given a certain mass of fluid in a region of space, two types of forces act on it: volume (forces outside the region) and surface (internal forces arising from fluid interactions via the boundary surfaces). The Navier-Stokes equations are a system of three balance equations (PDEs) of continuum mechanics which describe a linear viscous fluid. Under this umbrella, Stokes' law (kinematic balance) refers to the specific case of force on a moving sphere in fluid, and Fourier's law (energy balance) is a fundamental law of the material. 

\begin{figure*}[ht!]
\centering
\includegraphics[width=\linewidth]{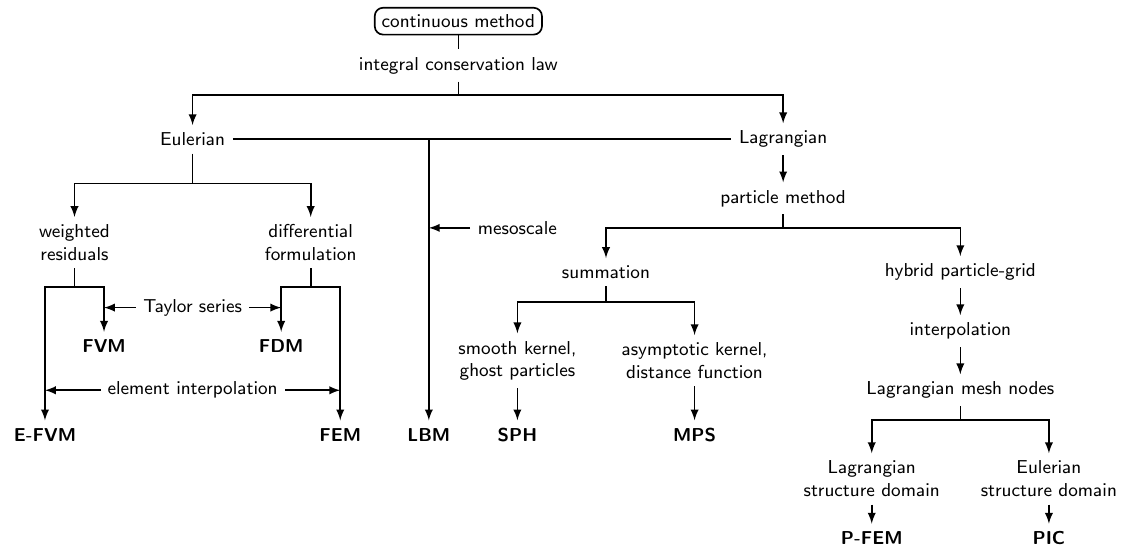}
\caption{Examples of numerical methods for simulating CFD problems: finite volume method (FVM), element-based finite volume method (E-FVM), finite element method (FEM), finite difference method (FDM), particle-in-cell (PIC), moving-particle semi-implicit method (MPS), particle-based FEM (P-FEM), smooth particle hydrodynamics (SPH), and lattice Boltzmann (LBM).}
\label{fig:cfd-methods}
\end{figure*}

A straightforward approach to solving the NS equations is direct numerical simulation (DNS) \cite{Moin1998}. DNS directly discretizes the NS equations and relies on using a mesh size that provides resolution at all scales of turbulent motion, including Kolmogorov length scales. Thus, there is no need for any subgrid modelling in order to capture turbulent flow dynamics, and an exhaustive description of the fluid is available throughout the domain. However, this understanding comes at high computational cost due to the requirement of a sufficiently fine mesh. DNS is too costly to use in most industrial problems and instead is generally relegated to more fundamental research. Hence, this fuels the need for alternatives, better suited to real world applications. 

Recent progress in machine learning techniques includes methods to replace some of the expensive time evolution in fluid simulation. Physics-informed neural networks (PINNs) \cite{Raissi2019,Karniadakis2021} are trained to solve supervised learning tasks constrained by the PDEs' physical laws, while respecting the conservation laws in fluid mechanics \cite{Brunton2020,Raissi2020}. This allows us to develop a more general predictive model for different problems. See \cite{Huang2022,Fallah2024,Hafiz2024} for a pedagogical introduction.

Figure \ref{fig:cfd-methods} summarizes some methods for solving the NS equations which form the basis of many commercial software packages \cite{Chung2002}. These include mesh-based methods such as the finite difference methods (FDM), finite volume methods (FVM) and finite element methods (FEM). Mesh-based models divide a continuum domain into smaller subdomains where the subdomain size is varied depending on the level of detail required around that region. 
Despite their success in science and engineering problems, mesh-based methods encounter difficulties at free surfaces, deformed boundaries, moving interfaces, and extremely large deformation and crack propagation. It is time-consuming and expensive to generate quality meshes for complicated geometries. For example, in FDM, irregular or complex geometries usually require additional transformations (e.g., mesh-rezoning \cite{Liu2003m}) that are more expensive than directly solving the problem itself, and may introduce numerical inaccuracies. 

In contrast to more established and mostly mesh-based codes, the smoothed particle hydrodynamics (SPH) method offers a viable alternative, especially for flows with a free surface \cite{Monaghan2005}. SPH is a purely Lagrangian approach that discretizes a continuum using a set of point particles. It has become popular in the last few decades as a way to simulate a wide range of applications, e.g., in engineering \cite{Monaghan2012} and astrophysics \cite{Lucy1977,Springel2010}. Another relatively new approach, the lattice-Boltzmann (LB) method \cite{Succi2001,Kruger2017}, combines both mesh- and particle-based methods.  LB is an advance on lattice gas automata that overcomes the convergence problems of directly using particles on grids.
Both SPH and LB have interesting potential for quantum algorithms or quantum-classical hybrid enhancements.  We discuss both in turn, first summarizing the classical method, then explaining the proposed quantum algorithms.

\subsection{Lattice Boltzmann method }\label{ssec:LB}

LB is a mesoscale approach based on the Boltzmann equation. It naturally bridges microscopic phenomena with the continuum macroscopic equations and accommodates a wide variety of boundary conditions \cite{Kruger2017,Succi2001}. LB has good scalability for solving PDEs due to its algorithmic locality, hence is inherently suited for large-scale parallelization on HPC systems, including use of GPUs \cite{Feichtinger2015,Liu2017l,Tran2015}.

\begin{figure*}[ht!]
\centering
\includegraphics[width=\linewidth]{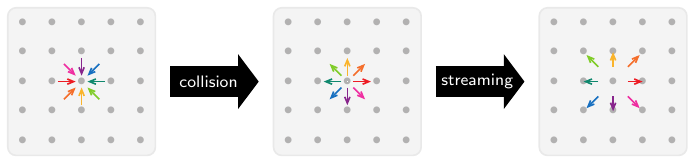}
\caption{Schematics of two-dimensional $D2Q9$ LB model showing nodes (grey), collision and streaming processes under temporal evolution.  Colors represent direction vectors $\vec{e}_k$ where $k$ corresponds to different directions in $D2Q9$.
}
\label{fig:lbm-grid-2}
\end{figure*}

The LB method considers a typical volume element of fluid that contains a collection of particles \cite{Chen1998}. These are characterized by a particle velocity distribution function for each fluid component at each grid point. At discrete time steps, fluid particles can collide with each other as they move, possibly under applied forces. The particle collision behavior is designed such that their time-average motion is consistent with the Navier-Stokes equation. The fluid is treated as a group of particles that have only mass and no volume. Particles flow in several directions of the lattice and collide with the particles around them. LB uses the collective motion of microscopic particles to describe the macroscopic parameters such as velocity, pressure, and temperature according to kinetic theory.

We use the Lattice Bhatnagar-Gross-Krook (LBGK) notation to describe lattice structures: the \textit{DnQb} classification indicates an $n$-dimensional space where each particle collides with $b$ surrounding particles (including itself). For example, the $D2Q9$ lattice in two-dimensional space contains a particle labelled $c_0$ which collides with the surrounding particles in eight directions (Figure \ref{fig:lbm-grid-2}). Particle movement is described in two steps: collision and streaming. In the collision step, particles collide with the opposite particles along each axis, changing the velocity. In the streaming step, particles move in the velocity direction to the neighboring lattice. The evolution equation is
\begin{equation}\label{eq:lbm}
f_k(\vec{x}+\vec{e}_k, t+\Delta t) - f_k(\vec{x}, t) = 
\Omega(f_k^\text{eq}(\vec{x}, t) - f_k(\vec{x}, t))
\end{equation}
with collision term $\Omega$, fluid distribution $f_k(\vec{x}, t)$ at point $\vec{x}$ at time $t$, and equilibrium distribution $f_k^\text{eq}$. At the next timestep $t+\Delta t$, the fluid would be at point $\vec{x}+\vec{e}_k$. The direction vector is $\vec{e}_k$ where $k$ represents the nine different directions in $D2Q9$. Each node is connected to its nearest neighbors with different colored vectors, denoted as $\vec{e}_k$. (Not shown in the figure is one $\vec{e}_k$ pointing to the node itself.) Every $f_k$ moves along its $\vec{e}_k$ vector to its neighbor and replaces the neighbor's distribution $f_{k'}$, except the one pointing to the node itself. In practice, the $D2Q9$ lattice requires up to several million nodes to generate an accurate flow field. 

\begin{figure*}
\centering
\begin{subfigure}[b]{\textwidth}
\centering
\includegraphics[width=\textwidth]{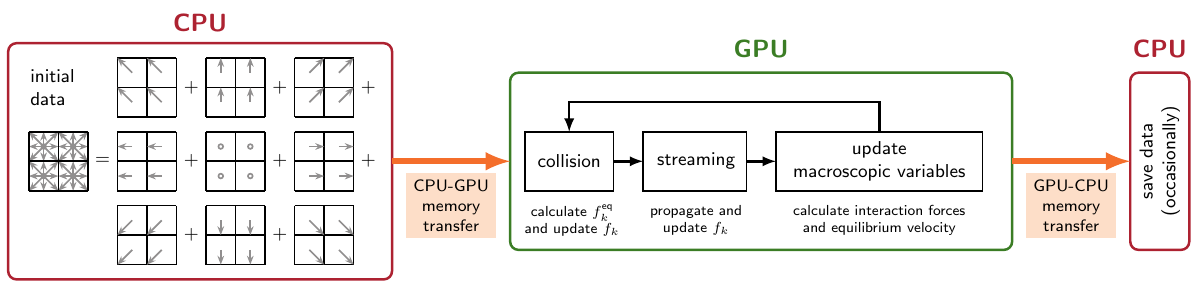}
\caption{Parallelized LB.}
\label{fig:lbm-gpu}
\end{subfigure}
\begin{subfigure}[b]{\textwidth}
\centering
\includegraphics[width=.8\textwidth]{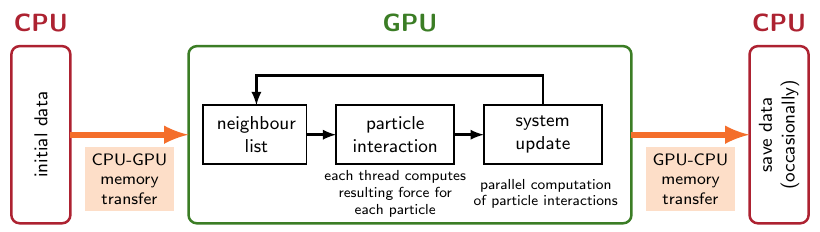}
\caption{Parallelized SPH.}
\label{fig:sph-gpu}
\end{subfigure}
\caption{Sketch of parallelized LB and SPH numerical processes with CPU and GPU.}
\label{fig:lbu-sph-gpu}
\end{figure*}

A common theme in LB applications is their suitability for parallel computing \cite{Alowayyed2017,Succi2019}. This is primarily due to the intrinsic locality of the method. For each time step, both the collision and streaming operators only require communication between neighboring cells at most. Recent research has focused on parallelizing and optimizing the computation using heterogeneous acceleration devices such as GPUs. The idea is to divide the LB lattice and group the distributions $f_k$ into arrays according to their velocity vectors $\mathbf{e}_k$ (Figure \ref{fig:lbm-gpu}). The first iteration starts with the distributions as inputs to calculate the density and velocity outputs. These quantities allow us to find the equilibrium distributions. New distributions are then computed from the input distributions and the equilibrium distributions according to the collision and the streaming operations. Finally, the boundary and outflow conditions are used to update the distributions. The updated distributions are then used as inputs for the next simulation step. For further details, see \cite{Wei2004}. Example open-source software platforms include OpenLB \cite{Krause2021} and Palabos \cite{Latt2021}.

\subsection{Quantum lattice Boltzmann}

Early exploration of quantum methods for CFD used quantum lattice gas models \cite{yepez2001quantum} and ``type II quantum computers'' \cite{Berman2002}. Both contain several quantum nodes connected by classical channels that carry bits instead of qubits \cite{yepez2001type}, while the latter are mathematically equivalent to a classical LB formulation \cite{love2006type}. These lattice models are susceptible to noise, non-isotropic advection and violation of Galilean invariance. Recently, a revised quantum algorithm for lattice gas automata \cite{Kocherla2024} managed to eliminate the need for repeated measurements at each time step. However, the size of the required quantum register scales linearly with the number of lattice sites. This makes it difficult to scale to realistic use cases.

There is a direct correspondence between the LB streaming step and quantum walks \cite{succi2015quantum}. \citeauthor{Todorova2020} \cite{Todorova2020} build on the latter and present the first fully quantum (as opposed to hybrid classical-quantum) LB method. The authors consider the simplified case of a collisionless Boltzmann system which reduces to
\begin{equation}\label{e:lbm}
\frac{\partial F (\vec{x},\vec{c};t)}{\partial t} + \vec{c} \frac{\partial F(\vec{x},\vec{c};t)}{\partial\vec{x}}=0
\end{equation}
with single-particle distribution function $F(\vec{x},\vec{c};t)$ defined in physical $\vec{x}$ and velocity $\vec{c}$ spaces. Classically, the discrete velocity method discretizes the solution state space into potentially three spatial and flow dimensions in \eqref{e:lbm}, leading to a six dimensional solution space which can be computationally expensive. Therefore, a key stepping stone for any appreciable quantum advantage is to represent this higher dimension space as a quantum state using a limited number of qubits. As shown by \citeauthor{Todorova2020}, this can provide an exponential reduction in memory, particularly for the extension to fluid mixtures. Classically, where doubling the number of fluid components would require doubling the memory, this was shown to be achievable with one extra qubit. Note also that the assumption of a collisionless system essentially decouples \eqref{e:lbm} for each fluid, a condition which is not true for real fluids. 

A primary obstacle to quantum LB algorithms is the need to take measurements during the simulation. This results in costly initialization routines when re-preparing the quantum state.
\citeauthor{Todorova2020} \cite{Todorova2020} address this issue by proposing that the purpose of their QLB is different to conventional CFD. The idea is to forego acquiring the complete flow field, i.e., the algorithm does not obtain a complete picture of the classical state. Instead, it efficiently obtains specific information such as particle number densities and concentrations for multi-fluid configurations. Thus, the algorithm allows an uninterrupted temporal iteration with measurement postponed until the end, at the cost of reducing the obtainable information. Similar approaches from quantum simulation and quantum chemistry are discussed in \cite{ortiz2001quantum,somma2002simulating} or techniques to obtain a single amplitude in \cite{Brassard2002}.

Including the collision operator in LB adds a nonlinear complexity, but is essential for using the method as a general PDE solver. Being able to choose a particular collision operator provides flexibility in which equation is actually being solved at the continuum level. One approach relies on Carleman linearization where an equivalent linear system in a higher dimensional space essentially replaces the nonlinear system. 
Recent work \cite{itani2022analysis} explored this in the context of nearly incompressible LB to find a quantum formulation for the collision term. Building on their findings, the authors propose a fully unitary streaming and collision operator \cite{Itani2024}. However, their collision step has a sub-optimal dependence on the number of time steps. This makes it unclear whether it is possible to achieve a practical quantum advantage, particularly at large Reynolds numbers. 

Alternatively, \citeauthor{Budinski2021} \cite{Budinski2021} formulates a fully quantum algorithm for the complete LB, i.e., including non-unitary collision operators using the standard form encoding approach \cite{low2019hamiltonian} and a quantum walk for the propagation step. Having initially applied this to an advection-diffusion equation, it was later extended to the streamfunction-vorticity formulation of the Navier-Stokes \cite{ljubomir2022quantum}. However, because the collision operator is retained in the equilibrium distribution function, the resulting nonlinearity restricts the algorithm to a single time-step before the state must be reinitialized. 
It has been shown \cite{Schalkers2024} that the conventional amplitude and basis encodings do not allow unitary operators for both streaming and collision.  This means future work needs to focus on alternative encodings, or, on ways to implement nonunitary operations as part of a quantum algorithm. 

\subsection{Smoothed particle hydrodynamics}\label{ssec:SPH}

SPH is a particle-based method with no mesh or grid, the particles move according to classical mechanics, and the fluid velocity is obtained as an average over the particle distributions.  Like LB, it is a more general method than just for solving Navier Stokes equations.
Mathematically, the SPH method expresses a function in terms of its values at locations of the virtual SPH particles. The integral interpolant of any function $A(\mathrm{r})$ is an integral $A(\mathbf{r}) = \int_\Gamma A(\mathbf{r'}) W(\mathbf{r}-\mathbf{r'},h) d\mathbf{r'}$ over the entire space $\Gamma$ for any point $\mathbf{r}$ in space and smoothing kernel $W$ with width $h$. This can be approximated by a summation interpolant,
\begin{equation}\label{eq:sph-interpolant}
A_S(\vec{r}) = \sum_i m_i \frac{A(\vec{r}_i)}{\rho_i} W(\vec{r}-\vec{r}_i,h)
\end{equation}
that sums over the set of all SPH particles $\{i\}$. Each particle $i$ has mass $m_i$, position $\vec{r}_i$, density $\rho_i$, and velocity $\vec{v}_i$. Hence, we can construct a differentiable interpolant of a function from its values at the particle level (interpolation points) by using a differentiable kernel \cite{Monaghan2005}.

\begin{figure}[ht!]
\centering
\includegraphics[width=\linewidth]{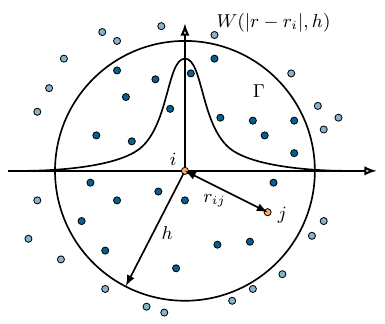}
\caption{Schematics of two-dimensional SPH model with SPH particle $i$, its neighbor $j$, kernel $W$, smoothing length $h$ and compact support domain $\Gamma$. Here, $h$ is equal to radius of support domain (support length). Full support length can typically be between $2h$ and $4h$.}
\label{fig:sph-kernel}
\end{figure}

The general SPH workflow contains three main stages that are repeated at each time step. First, the algorithm creates a neighbor list. Then it calculates the particle interactions for momentum and continuity conservation equations. This involves calculating the smoothing length and SPH kernels. The final step is time integration (or system update). The SPH particle distribution follows the mass density of the fluid while their evolution relies on a weighted interpolation over nearest neighboring particles. This has several implications. The SPH kernel smoothes the physical properties within the range of interpolation (Figure \ref{fig:sph-kernel}). This is characterized by the smoothing length $h$ which also determines the local spatial resolution. The identity of the neighboring particles change as the simulated system evolves. There is no computationally efficient method to predict which particles will be neighbors over time, hence we must identify the neighbors at each timestep.

Parallelization strategies for SPH are strikingly different from those of mesh-based methods. The computational domain is divided into a grid of cells where each SPH particle is assigned a cell. We build a list of its neighbors by searching for particles only in nearby cells. Dynamical neighbor lists require specialized methods for data packing and communicating. Particles migrating between adjacent domains can cause difficulties on memory management especially in distributed-memory architectures, as can large variations in particle density and domain size. The total number of particles can narrow the choice of hardware, as memory space is one of the main limitations of shared memory architectures. Using both shared and distributed memories, most SPH parallelization schemes are scalable \cite{Ferrari2009,Dominguez2013}. However, if communication latencies increase, scalability decreases rapidly \cite{Oger2016}.

Diverse SPH applications demonstrate the versatility of SPH methods in multiple fields. SPHysics is an SPH code tailored for simulating free surface flows, which are challenging for other methods but effectively handled by SPH. DualSPHysics \cite{Dominguez2022} runs on a hybrid architecture of CPUs and GPU accelerators for enhanced performance. The astrophysics community also extensively uses SPH codes, as evidenced by the popularity of SWIFT \cite{Schaller2024} and GADGET \cite{springel2005cosmological}. These tools have significantly contributed to advancements in cosmological and astrophysical research.

Figures \ref{fig:lbm-gpu} and \ref{fig:sph-gpu} compare the typical workflow for combined CPU-GPU implementation of LB and SPH, illustrating the commonalities in the processes. We break down the algorithm into parts that are suitable for different types of classical hardware. This illustrates how to incorporate accelerators into large scale simulations. Extending this to incorporate quantum computers used as accelerators can hopefully provide further efficiencies.

\subsection{Quantum SPH algorithm}

Recent reviews \cite{Lind2020,Vacondio2021} discuss the computational bottlenecks and grand challenges we face before SPH becomes more widely used for practical problems (e.g., in engineering). The SPH method has traditionally been considered computationally expensive \cite{Lind2020} due to two major factors: a large number of SPH particles are needed for good solution accuracy and time steps must be small enough to obey empirical stability criteria. One strategy to address these bottlenecks involves using quantum subroutines \cite{AuYeung2024}. Below we list examples of how to substitute classical SPH procedures with quantum algorithms.

Rewriting the SPH interpolant \eqref{eq:sph-interpolant} as an inner product of the form
\begin{equation}
\sum_i m_i \frac{A(\vec{r}_i)}{\rho_i} W(\vec{r}-\vec{r}_i,h) \to
\langle m A(\vec{r})/\rho
\vert W(\vec{r},h)
\rangle
\end{equation}
requires efficiently loading the classical data (floating point numbers) into quantum processors using quantum encoding techniques. There are several possible methods to calculate the inner product: the swap test \cite{Buhrman2001} or one of its variations \cite{Fanizza2020}; the Bell-basis algorithm \cite{Cincio2018} which is efficient on NISQ devices; quantum mean estimation and support vector machines \cite{Liu2022}; or quantum counting algorithm \cite{Brassard1998,Aaronson2019}.

The summation in \eqref{eq:sph-interpolant} means that the algorithm requires a search algorithm to find all the neighboring particles inside the compact support domain $\Gamma$ (Figure \ref{fig:sph-kernel}). This is another major bottleneck that we can, in principle, address using Grover's search algorithm \cite{Grover1996} implemented with a quantum walk (sections \ref{sssec:search}, \ref{sssec:qw}). This can be an effective search method when combined with existing SPH neighbor-list approaches, eg., cell-linked or Verlet lists \cite{Dominguez2011}.

The timestepping procedure is subject to empirical stability criteria like the Courant-Friedrichs-Lewy convergence condition \cite{Courant1967}. Physically, in the widely-used weakly-compressible \cite{Violeau2014} and incompressible \cite{Violeau2015} forms of SPH, the timestep is also limited by the speed of sound and maximum velocity respectively. This leads to timesteps of the order of $\leq 10^{-5}$ seconds \cite{Lind2020}, or typically one million time steps to simulate one second of physical time. Most three-dimensional applications require 10-100 million SPH particles. High computational costs have motivated research into timestepping procedures such as the Runge-Kutta Chebyshev scheme \cite{He2021}. This is where quantum algorithms may potentially provide an even greater speed up, since we do not need to read out the data every time step.  A quantum algorithm that can evolve for many time steps between measurements has the best chance of providing real advantages.

\subsection{Other quantum CFD algorithms}

There are quantum algorithms that directly perform numerical integration for linear differential equations and provide some quantum advantages \cite{Childs2021p,Linden2022}. Efforts to develop a quantum solver for the Poisson equation have applied variational quantum algorithms (section \ref{ssec:VQAs}) \cite{Liu2021v,Sato2021} and HHL framework (section \ref{sssec:HHL}) \cite{Robson2022,Saha2024}. However, specialized quantum algorithms are less well developed, and methods for NLPDEs remain open \cite{Balducci2022}. NLPDE problems are some of the most computationally demanding calculations in CFD. The nonlinearities may be due to the physics (one or more of the following: convection, diffusion, forcing, turbulence source terms, reacting flows) or the numerics. Compared to linear problems, there have been few advances in developing quantum algorithms for NLPDEs due to difficulties in expressing nonlinearities with unitary gates. Up until recently, the only notable work was an HHL-based method that uses post-selection to implement the nonlinearity \cite{Leyton2008}. Unfortunately, it is an exponential-time algorithm that also needs an exponential number of copies of the initial state.

This calls for another way to linearize the nonlinear equations (e.g., with Carleman linearization \cite{Liu2021p,itani2022analysis}). One approach is to discretize the NLPDEs then directly apply a quantum solver to the resulting nonlinear ordinary differential equation (ODE). This approach was adopted in \cite{Gaitan2020,Gaitan2021} using an ODE solver that had gone largely unnoticed until recently \cite{kacewicz1987optimal}. The method comes with stringent requirements on the underlying flow field and its rate of change. Despite this, a regime exists where there may be a quadratic speedup over classical random methods, and an exponential speedup over classical deterministic methods. Replacing the ODE solver with alternative formulations may also provide some benefits within the overall method.

While most discussions of quantum computing focus on when quantum computers will be able to consistently beat their classical counterparts, the reality is more likely to involve hybrid quantum-classical devices. For example, if it is possible, solve the non-linear parts of the calculation on conventional HPC, then send the linear part to a quantum processing unit which uses the HHL algorithm. However, if the classical part then requires consistent feedback from the quantum part, the cost of measurement and reloading input states may significantly reduce any quantum advantage. Some recent works include using a quantum algorithm as a predictor-corrector to accelerate the classical components of the algorithm without necessitating a complete solution readout \cite{rathore2024integrating}. Alternatively, in \cite{bharadwaj2023hybrid,ingelmann2024two} the authors discuss a framework for simulating simplified fluid dynamic configurations described by linear PDEs, using quantum linear equation solvers in conjunction with classical protocols. Instead of simply measuring out the complete velocity field, the authors propose a novel quantum postprocessing protocol for efficiently computing in situ nonlinear functions of the resulting flow field, namely the average viscous dissipation rate. 

An advantage of employing PINNs for complex flow simulations is their robustness to parameter changes. For instance, in classical CFD altering the geometrical boundaries typically necessitates a restart of the simulation in order to determine the new flow profile. However, by applying transfer learning methods \cite{weiss2016survey} a PINN model trained on a slightly different geometry can be adapted for the new configuration without requiring a complete restart. For PINNs to provide reliable outputs, it is imperative to have high expressibility and this is already known to be a key strength of quantum computers \cite{Schuld2021v,kordzanganeh2023exponentially}. Recent work \cite{sedykh2024hybrid} in the field includes a hybrid quantum PINN based on a quantum depth infused layer implemented using a VQA and coupled with transfer learning. The authors reported a $21\%$ reduction in loss compared with its purely classical counterpart.  

More general VQA methods (see section \ref{ssec:VQAs}) are promising for solving NLPDEs on NISQ devices \cite{Jaksch2023}.  \citeauthor{Lubasch2020} applied their method to the one-dimensional heat equation and Burger's equation. It is possible to solve the Schr{\"o}dinger-Poisson equation by combining this method with a subroutine to train a state to represent a nonlinear potential \cite{Mocz2021}. The reaction-diffusion equations can be solved similarly \cite{Leong2022}. Further work presented a general version of this quantum method to treat a large class of nonlinearities and inhomogeneous terms \cite{Sarma2024}, making the method applicable to most NLPDEs and multidimensional PDEs which are first-order in time, such as the Navier-Stokes equations. It can also be used to adapt many explicit and semi-implicit numerical schemes, such as the fourth-order Runge-Kutta algorithm. 

The vortex-in-cell method is an example of how the quantum Fourier transform can be used to build a Poisson solver \cite{Steijl2018}. However, the main drawback is the assumption that multiplication with the wavenumbers can be done efficiently on classical hardware. There is also the added complexity of accommodating non-periodic boundary conditions with any Fourier-based CFD method, meaning its translation into a quantum setting may not be straightforward. 

While there is an encouraging diversity of different quantum approaches, many existing demonstrations of quantum algorithms for CFD are limited to relatively simple flow cases. The effectiveness of these algorithms at scaling with the Reynolds number is either yet to be established or appears limited. While configurations involving low Reynolds numbers are valuable in areas like geophysics and biophysics, the study of turbulence at high Reynolds numbers is crucial for many disciplines. In this regime, a quantum-inspired algorithm using Schmidt decomposition combined with tensor network methods can successfully account for the interscale correlations in turbulence \cite{Gourianov2022}.  Even if useful quantum simulations of turbulence at scale are some way in the future, quantum-inspired algorithms like this can provide benefits to current classical simulations.



\section{Interfacing with high performance computing}\label{sec:hpc}

Compared to classical computerss, quantum hardware operates on different intrinsic timescales (clock speeds), and uses different methods for data encoding -- some of which are necessary to achieve a quantum advantage.  This makes interfacing between classical and quantum hardware highly nontrivial.  Despite the significant practical hurdles, using quantum computers as accelerators for HPC is the logical way to maximise overall computing power, and test beds offering coupled QPU and CPU configurations are being set up \cite{HPCQS2023,Innsbruck2024}.  In this section, we briefly review these interface issues, and the progress required to overcome them.

\subsection{Clock speed mismatch}\label{ssec:clock}

Every computation on a classical computer processor corresponds to a sequence of layers of logic gates. The time taken to execute one layer is the \textit{clock cycle}. Modern CPUs contain several billion transistors operating at around 3 to 4 GHz or 1/3 nanosecond per cycle. When quantifying the speed of quantum computers, we can break down quantum algorithms into a sequence of quantum gate layers. However we cannot define the \textit{quantum clock speed} as an inverse of the quantum gate layer time because of two major variables:
\begin{enumerate}
\item Gate choice. Quantum algorithms can be decomposed into elemental gates, but the choice of gates is flexible. Hence, the quantum gate duration can vary, even for different gate layers in the same hardware.
\item Quantum error correction. Quantum computers make many more errors than classical CPUs. Large-scale quantum computers will likely have a runtime dominated by quantum error correction (QEC) overheads. Each QEC method consists of different routines with different runtimes, that can be dominated by mid-circuit measurements, rather than gate times.
\end{enumerate}

\begin{table}[ht!]
\centering
\caption{Quantum hardware platforms ordered by decreasing gate speed (conservative estimates). 
}
\bgroup
\def\arraystretch{2}
\footnotesize
\begin{tabular}{p{1.2in}p{.8in}p{.9in}}
\textbf{Platform} & \textbf{two-qubit gate speed} & \textbf{measurement time} \\[4pt]
\hline
superconducting qubits$^\text{\cite{Google2023}}$ &
10 MHz & 660 ns
\\
\makecell{photons \\ (boson sampling)$^\text{\cite{Madsen2022}}$} &
6 MHz & 36 $\mu$s
\\
silicon spin qubits$^\text{\cite{Blumoff2022}}$ &
750 kHz & 1.3 $\mu$s
\\
trapped ions$^\text{\cite{Harty2014,Todaro2021}}$ &
6 kHz & 100 $\mu$s
\\
Rydberg arrays$^\text{\cite{Kwon2017}}$ &
170 Hz & 6 ms
\\
\makecell{photons \\ (fusion-based)$^\text{\cite{Istrati2020}}$} &
10 Hz & 100 ms
\\
\hline
\end{tabular}
\egroup
\label{tab:hardware_clock_speed}
\end{table}
Table \ref{tab:hardware_clock_speed} shows typical gate operation speeds and measurement times for different hardware platforms.  Note that these are physical gate speeds that error correction overheads will significantly modify.  Slower, but higher quality, trapped ions or Rydberg atoms will need less error correction than more noisy, but faster, superconducting qubits. 
Fusion-based photonic platforms are slow because they have high overheads to generate the heralded resource states required for fully universal quantum computing \cite{Pelucchi2022,Cogan2023}.  
In general, state preparation can be done ``offline'' in parallel with other operations, to avoid this step slowing down the overall compute time.  This works well for trapped ions or atoms \cite{Nickerson2014}, but for photonic fusion gates the overheads are extreme and dominate the scaling.
Photons can also be used for boson sampling \cite{Zhong2020,Zhong2021}, which does not require post-selection and can therefore operate much faster. \citeauthor{Madsen2022} \cite{Madsen2022} performed boson sampling in just 36 $\mu$s via a 216-mode squeezed photon state. Raw photon generation and detection runs at 6 MHz, putting this platform up near the top of the table for processor speed, although it cannot perform universal quantum computing.

Classical CPU clock speeds have remained around 3 GHz over the last decade, which is significantly faster than any current quantum hardware. The steady increase in processor speeds that occurred during several decades of Moore's law scaling \cite{Moore1965} (Figure \ref{fig:microprocessor_trends}) is not possible for quantum hardware platforms. Individual atoms, ions or electrons have their own intrinsic physical frequencies which cannot be made smaller.  We therefore need to develop programming models that can leverage processors with vastly different effective clock speeds.  A hybrid algorithm may need to queue 100 or $10^4$ operations on the classical processor for every operation on the quantum processor.  This should be viable for some of the large, multi-scale simulations that dominate today's HPC use, where many parts of the computation can be done classically while the quantum processor solves key components that are expensive for classical computation.

\subsection{Quantum data encoding}\label{ssec:encoding}

Theoretical frameworks that address the interface between classical memory and quantum devices are critical for implementing quantum algorithms. In classical computing, floating point numbers are ubiquitous for numerical data in scientific applications.  However, they are not the natural way to encode data into qubits, and the choice of data encoding can have a big impact on quantum algorithm efficiency.  In this section, we summarise the available data encoding approaches for quantum computing, and the challenges they present for interfacing with classical HPC.

\paragraph{Basis (digital) encoding.} Basis encoding is the simplest encoding. It corresponds to binary encoding in classical bits.  Multi-qubit basis states are often written in a single ``ket'' as
$\{ \ket{0...00}, \ket{0...01}, ... \ket{1...11} \}$ or $\ket{j}$ where it is understood that the integer $j$ corresponds in binary to the qubit values. The bitstring $b_{n-1} ... b_0$ is thus encoded by the $\ket{ b_{n-1} ... b_0}$ state.  Single computational basis states are easy to prepare from the all zero state, by applying bit flips to the appropriate qubits.

\paragraph{Angle encoding.} Angle encoding extends basis state encoding to use the ability of qubits to be in any superposition of zero and one.  A convenient parameterization of this using angles $\vartheta$ is
\[
\cos\vartheta\ket{0} + \sin{\vartheta}\ket{1}.
\]
An $n$-qubit register can thus encode $n$ angle variables $\{\vartheta_k\}$.  This is efficient to prepare if arbitrary single qubit rotation gates are available. We can extend this by encoding a second angle $\varphi$ in the phase,
\[
\cos\vartheta\ket{0} + e^{i\varphi}\sin{\vartheta}\ket{1},
\]
and noting that qubits correspond to unit vectors on a Bloch sphere (Figure \ref{fig:bloch-sphere}).  However, it is not possible to read out the value of the angles using a single measurement on a single copy of the qubit.  While the state preparation is easy, readout is only straightforward for results encoded as basis states.

\paragraph{Amplitude (analogue) encoding.} Amplitude encoding loads a vector of real numbers $\vec{X}=(x_0,x_1\dots,x_{N-1})$, with $x_k\in[0,1]$, into the amplitudes of the quantum state, so they are stored in superposition 
\[
\vec{X} \to \sum_{k=0}^{N-1} x_k \ket{k}.
\]
The overall normalization factor is stored separately if it is not equal to one.  An $n$-qubit register stores up to $N=2^n$ real values, so this is an efficient encoding. However, to manipulate individual amplitudes, the number of gates grows exponentially with qubit number. Preparing an arbitrary state is thus an expensive operation \cite{Long2001,Plesch2011,Sanders2019}. Refinements include ``approximate amplitude encoding'' \cite{Nakaji2022,Mitsuda2024} which reduces the preparation cost to $O(\text{poly}(n))$ gates for $n$ qubits at a cost of reduced accuracy. It is also possible to encode data more efficiently for wavepacket dynamical simulations in quantum chemistry \cite{Ollitrault2020e,Chan2023}.

\paragraph{QRAM encoding.} 
Quantum random access memory (QRAM) is a general-purpose architecture for quantum oracles (see \cite{Jaques2023,Hann2021} for excellent introduction). It is a generalization of classical RAM \cite{Giovannetti2008,Giovannetti2008a}: given an address $k$ as input, the RAM returns memory contents $j_k$. Analogously, QRAM takes a superposition of addresses $\ket{\psi_\text{in}}$ and returns an entangled state $\ket{\psi_\text{out}}$ where each address is correlated with the corresponding memory element:
\begin{gather}
\ket{\psi_\text{in}} =
\sum_{k=0}^{N-1} \ket{k}^A \ket{0}^B \nonumber\\
\xrightarrow{\text{QRAM}}
\sum_{k=0}^{N-1} \ket{k}^A \ket{j_k}^B
= \ket{\psi_\text{out}} \label{eq:qram}
\end{gather}
for memory size $N$. Input and output registers are denoted by $A$ and $B$ respectively. Note we may alternatively denote the memory size by $2^n$, where register $A$ contains $n \equiv \log N$ qubits. The principle behind QRAM is that, if a query state is a quantum superposition over all addresses, then the circuit responds by performing a memory access operation over all addresses simultaneously. If we imagine the memory laid out in space, a QRAM access must transfer some information to each location in memory for it to correctly perform a superposition of memory accesses. 
QRAM performs the operation in \eqref{eq:qram} in $O(\log N)$ time at the cost of $O(N)$ ancillary qubits. This makes QRAM especially appealing for quantum algorithms that require $O(\log N)$ query times to claim exponential speedup. However, it is still uncertain whether QRAM can be used to achieve quantum speedups, in principle or in practice \cite{Aaronson2015,Ciliberto2018}. 
QRAM is sensitive to decoherence \cite{Arunachalam2015} and is subject to an overhead associated with error correction of ancillary qubits \cite{DiMatteo2020}. Recent efforts have led to developments in highly noise-resilient QRAM using bucket-brigade QRAM architecture \cite{Hann2021}, and this continues to be an active open area of research.

\paragraph{Floating point encodings.}  Classical digital computers encode floating-point numbers using the IEEE 754 international standard, established in 1985 by the Institute of Electrical and Electronics Engineers (IEEE) \cite{IEEE754}.  
Work is currently underway to develop the IEEE P7130 Standard for Quantum Computing Definitions \cite{IEEE7130}.
Basis state encoding is one option for floating point numbers. It uses a mantissa and an exponent in direct analogy to the binary representation of classical floating point numbers. This is not the only (or the best) method, depending on the application. 
For example, quantum algorithms that rely on amplitude encoding for their efficiency gains cannot use basis state encoding.
\citeauthor{Seidel2022} \cite{Seidel2022} explore an alternative way to design and handle various types of arithmetic operations. They formulate arithmetic ring operations as semiboolean polynomials, and extend their encoding from intergers to floating point numbers, with significant efficiency gains over basis state encodings.

\paragraph{Block encodings.}  Block encodings of matrices play an important role in quantum algorithms derived from the quantum singular value transformation \cite{Gilyen2019,Camps2020b,Sunderhauf2024}, such as Hamiltonian simulation (section \ref{sec:qsim}) and quantum machine learning \cite{Chakraborty2019}. Block encoding involves embedding a (scaled) non-unitary operator $\hat{A}=\sum_j\alpha_j\ket{w_k}\bra{v_k}$ as the leading principal block in a larger unitary matrix,
\begin{gather}
\hat{U} = \begin{bmatrix} \hat{A}/\alpha & * \\ * & * \end{bmatrix}
\Leftrightarrow
\hat{A} = \hat{\Pi}' \hat{U} \hat{\Pi},
\end{gather}
with arbitrary matrix elements $*$. The projectors $\hat{\Pi}'=\sum_j\ket{w_k}\bra{w_k}$ and $\hat{\Pi}=\sum_j\ket{v_k}\bra{v_k}$ locate $\hat{A}$ within $\hat{U}$: respectively, $\hat{\Pi}$ and $\hat{\Pi}'$ selects the columns and rows in which $\hat{A}$ is encoded. See \cite{Martyn2021,Sunderhauf2024} for a pedagogical introduction.

\paragraph{Hamiltonian encoding.}  Hamiltonian encoding is suitable for quantum optimization.  The problem constraints are encoded in a QUBO or the fields and couplings of an Ising Hamiltonian (section \ref{ssec:qopt}). Loading the classical data on to a quantum device involves setting the values of the couplers and fields through classical control lines. Although straightforward to implement, the required precision increases with the problem size. Additionally, not every problem can be easily mapped on to the Ising Hamiltonian parameters. 

\paragraph{One-hot and domain-wall encoding. } One-hot and domain-wall encodings are unary encodings originally designed for higher dimensional variables in quantum optimization \cite{Chancellor2019,Berwald2023}. 
One-hot encoding sets a single bit in a position that determines the value of the variable.
Domain-wall encoding is based on a topological defect in an Ising spin chain. One side of the chain is in a spin-down state, whereas the other is spin-up. The domain-wall (topological defect) is where the system jumps from spin-up to -down.  The location of the domain wall represents the value of the variable.  The number of qubits in the spin chain determines the number of available values.  Domain wall encoding is efficient in NISQ devices for variables with a small fixed range of values \cite{Plewa2021}.

\subsection{Data interconnections}

As HPC users know well, exchanging data between different processing units is where many bottlenecks occur in large scale computations. The physical interconnections between quantum and classical hardware are already the focus of significant engineering development, since they are necessary to control the quantum hardware.  A significant issue is transferring data and instructions from room temperature to the cryogenic conditions required for operating many types of quantum hardware.  Electronic and electromagnetic signals carry unwanted heat into the cryogenic environment that must be minimised and removed to maintain ideal operating conditions.  This introduces another constraint for optimizing code across both platforms: minimizing data exchange to reduce cooling costs.  It takes time to run the procedures to re-cool cryogenic systems, and like calibration, this is time when the quantum processors are not available for computation.

Slower quantum ``clock speeds'' (section \ref{ssec:clock}) are accompanied by even slower timescales for measurements to read out the qubit state and convert into classical data signals.  Table \ref{tab:hardware_clock_speed} lists typical measurement times for different hardware.
These factors all need to be taken into account when designing algorithms, and significantly affect the wall-clock time required \cite{Barnes2023}.

The most important next step for integrating quantum computers with HPC is making test bed hardware widely available \cite{HPCQS2023,Innsbruck2024}. Exploration of their combined capabilities can guide the engineering required to integrate them more efficiently at both the hardware and software levels.



\section{Outlook}\label{sec:outlook}

\begin{figure*}[ht!]
\centering
\includegraphics[width=\linewidth]{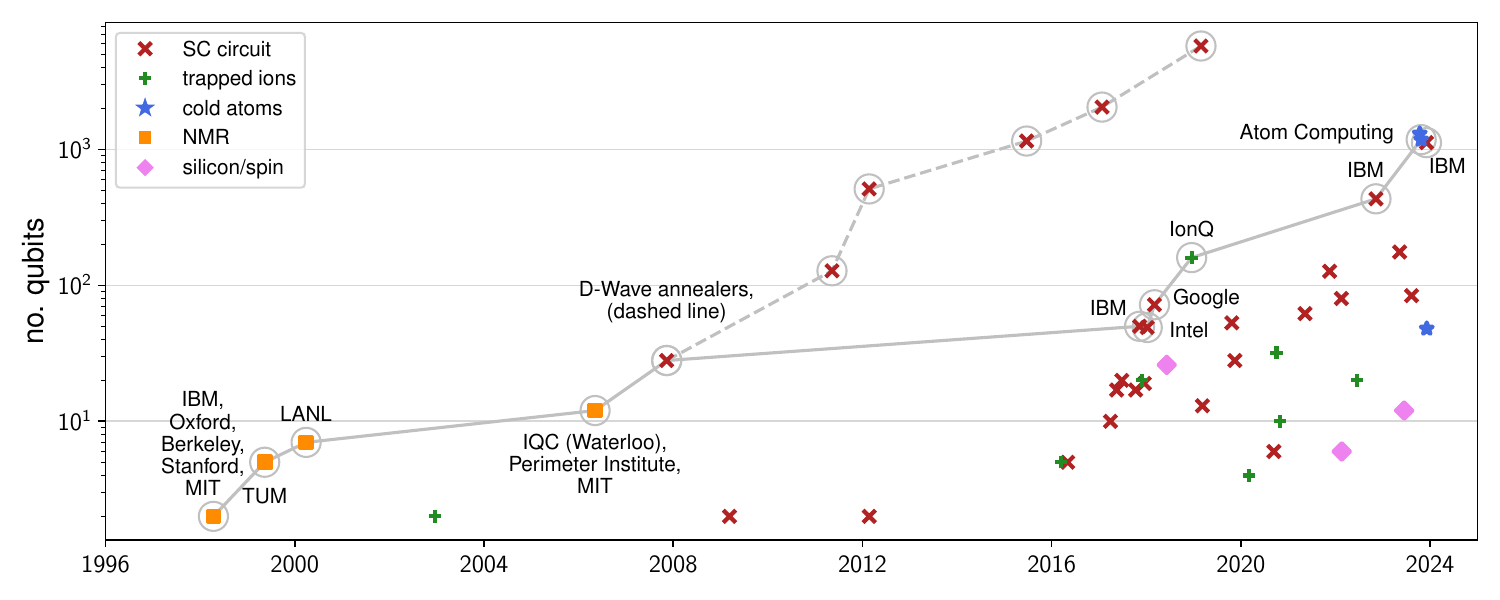}
\caption{Growth in number of qubits per device from 1998 to 2024. Showing SC circuits (red $\times$), trapped ions (green $+$), cold atoms (blue stars), NMR (orange squares) and silicon/spin (pink diamonds) platforms. Results from selected teams annotated with team name and circled in grey. Compiled by R. Au-Yeung from Statista \cite{Statista} and press releases.}
\label{fig:qubits-progress}
\end{figure*}

Looking back over the past two decades, it is already possible to identify a quantum equivalent of Moore's Law through the steady improvements in the qubit number and quality (Figure \ref{fig:qubits-progress}). This is most evident in platforms that have been under development for the longest, such as superconducting qubits and ion traps. Early quantum computers are available now through cloud computing services, such as Amazon Braket, enabling researchers to develop and test of proof-of-concept algorithms. Cloud accessible quantum computing was pioneered by IBM, who released their first cloud accessible five-qubit superconducting quantum computer in 2016 \cite{Mandelbaum2021}. IBM recently announced their newest quantum chip, which provides over a thousand physical qubits \cite{Castelvecchi2023}, an impressive development trajectory.  However, further progress requires error correction \cite{Terhal2015,Roffe2019} to reduce the accumulated errors that occur during computation, and hence enable long enough computations for useful applications. This will produce high quality qubits, but each of IBM's thousand-qubit chips will contain only a few of these logical qubits, which will then need to be connected through sophisticated quantum network interfaces \cite{Nickerson2014}.

Ion trap based systems have made similar progress, prioritizing qubit quality to achieve the highest fidelity quantum gates of any current platform \cite{Leu2023}, in an academic research lab setting.  IonQ's latest systems have up to 32 qubits, with 64 qubit systems due in 2025.  This is enough qubits for simulating quantum systems beyond the reach of classical HPC. Again, significantly larger systems will require error correction. Newer entrants, such as Rydberg \cite{Cong2022,Wintersperger2023} (neutral atoms), are catching up fast in performance and flexibility. Despite the current leading position of superconducting qubits and ion traps, it is by no means clear they will turn out to be the main type of quantum hardware in the longer term. Also firmly in the running are photonic and silicon or other semiconductor-based systems. An ecosystem that supports several types of hardware for different applications is both plausible and promising. Despite the significant engineering challenges, there is growing confidence that the hardware will deliver on the anticipated timescales. Companies are forecasting fully error corrected quantum processors on the scale of hundreds or thousands of qubits by 2030, and are investing the resources necessary to achieve this.

However, standalone quantum hardware will support only niche applications.  Some of these are significant: simulating many-body quantum systems (see section \ref{sec:qsim}) is a natural problem for quantum hardware that can potentially deliver many new scientific results with valuable impact in a wide range of commercial sectors.  Nonetheless, to leverage the full potential of quantum computing, interfacing with classical HPC is essential, and already underway \cite{HPCQS2023,Innsbruck2024}. As discussed in section \ref{sec:hpc}, there are major challenges here that require new science and engineering to be developed.  The most important are related to the different data encodings that are natural for quantum computers, making it highly non-trivial to transfer data between quantum and classical processors.  This is compounded by clock speed mismatches that, depending on the type of quantum hardware, can be up to six orders of magnitude.  While data conversion and transfer are largely an engineering challenge, we do not yet have good programming models for how to handle asynchronous computing on these timescales. Commercial promises of seamless software integration are plausible for test bed applications, but do not yet have an assured path for scaling up.

In addition, quantum computing is being developed in a context of significant changes in classical computing.  Physical limits have been reached for scaling up conventional CMOS CPUs: the energy used by a typical large data centre or HPC facility is equivalent to a small town. More compute power requires new hardware that is less power-hungry for the same amount of computation.  GPUs have provided this for the past decade, but they have also shown us that redesigning codes and algorithms to use different types of hardware is extremely challenging and time-consuming for large scale applications \cite{Betcke2022}.  Quantum processors will be significantly more challenging still to deploy alongside classical HPC at scale.  The investment of time and resources necessary to leverage quantum computing is a significant barrier to widespread adoption as quantum computers become more available.  Applications developed and tested in an academic setting are important for opening up areas with commercial potential.

Artificial intelligence (AI) is also accelerating the pace of change in classical computing.  In recent years, AI has come under scrutiny over its high energy consumption \cite{Dhar2020,Debus2023} and carbon emissions. For example, the carbon footprint of training a large natural language processing (NLP) model equals roughly 300 tonnes of carbon dioxide emissions \cite{Strubell2019}, or 125 round-trip flights between New York City and Beijing \cite{Dhar2020}. This increases the pressure to develop energy-efficient hardware optimized for the massive data processing required for AI applications.  This will significantly reshape the application areas where quantum computing can deliver real benefits.  While AI may overtake quantum in predicting solutions to scientific computing applications, quantum may in turn provide advantages \cite{Dunjko2018, Cerezo2022} for other types of AI applications, as quantum hardware becomes more capable.

Despite the remaining hurdles on the path to useful quantum computing, this is an exciting time to be doing computational science and engineering. There are multiple potential routes to significantly larger simulations and the scientific and technical breakthroughs they will produce.  Achieving these goals requires interdisciplinary collaborations to identify where and how quantum computing can provide the most benefits. 
There are multiple projects aiming to accelerate this process.  Using the UK for examples, we have the
\begin{itemize}
\item National Quantum Computing Centre's SPARQ program and hackathons \url{https://www.nqcc.ac.uk/engage/sparq-programme/},
\item Quantum Software Lab in Edinburgh \url{https://www.quantumsoftwarelab.com/},
\item Collaborative Computational Project on Quantum Computing \url{https://ccp-qc.ac.uk/}, 
\item ExCALIBUR cross-cutting project QEVEC \url{https://excalibur.ac.uk/projects/qevec/}, 
\item multiple projects funded by the Software for Quantum Computing call \url{https://gow.epsrc.ukri.org/NGBOViewPanel.aspx?PanelId=1-FIN46S}, 
\item and the Quantum Technology Hub for Quantum Computing and Simulation \url{https://www.qcshub.org/}, and its recently announced successor, the Quantum Technology Hub for Quantum Computing via Integrated and Interconnected Implementations, due to start in December 2024.
\end{itemize}

The Royal Academy of Engineering (RAEng) 2024 Quantum Infrastructure Review outlines a range of recommendations for developing the UK's quantum tech sector over the next decade \cite{RAEng2024}.  This is motivated by the need to transition from a successful program of fundamental research to commercialization and deployment in real-world applications. As a result, industry will require a workforce that is knowledgeable about quantum mechanics \cite{Chen2023}. However, there are fears that unless we enhance accessibility to quantum education \cite{Goorney2024,Heidt2024} and persuade students that it is a realistic career path \cite{Meyer2024,Rosenberg2024}, current inequities will persist into the future quantum workforce. Initiatives such as QWorld \url{https://qworld.net/} allows researchers from any country with internet access to participate in quantum application development. Since many countries do not have their own quantum computing hardware development programs, this enlarges the overall effort and ensures that a wide range of relevant problems are tackled.  

In the global race to develop quantum computers, there has been growing debate on its ethical use \cite{TenHolter2023,WEF2024}, especially given the discussion on potential military applications \cite{Neumann2021}. 
Even as governments move to restrict technology access for national security reasons, it is essential for researchers to have cloud access to compute resources, to ensure the benefits are available to all.


\vspace{1em}

\ack
RA, BC, OR, and VK are supported by UK Research and Innovation (UKRI) Grants EP/W00772X/2 (QEVEC). RA, OR, and VK are supported by EP/Y004566/1 (QuANDiE). RA, and VK are supported by EP/T001062/1 (QCS Hub). VK is supported by EP/T026715/2 (CCP-QC). The authors thank John Buckeridge, Nick Chancellor, Animesh Datta, Steph Foulds and Steve Lind for useful discussions and proofreading the manuscript. 

\section*{ORCID}
R. Au-Yeung 
\orcidlink{0000-0002-0082-5382} 
\href{https://orcid.org/0000-0002-0082-5382}{0000-0002-0082-5382} 
\\ 
B. Camino 
\orcidlink{0000-0002-9569-2219} 
\href{https://orcid.org/0000-0002-9569-2219}{0000-0002-9569-2219} 
\\ 
O. Rathore 
\orcidlink{0009-0002-2087-6715}
\href{https://orcid.org/0009-0002-2087-6715}{0009-0002-2087-6715}
\\
V. Kendon 
\orcidlink{0000-0002-6551-3056} 
\href{https://orcid.org/0000-0002-6551-3056}{0000-0002-6551-3056}


\addcontentsline{toc}{section}{References}
\providecommand{\noopsort}[1]{}\providecommand{\singleletter}[1]{#1}%


\begin{thebibliography}{601}
\providecommand{\natexlab}[1]{#1}
\providecommand{\url}[1]{\texttt{#1}}
\expandafter\ifx\csname urlstyle\endcsname\relax
  \providecommand{\doi}[1]{doi: #1}\else
  \providecommand{\doi}{doi: \begingroup \urlstyle{rm}\Url}\fi

\bibitem[Arute et~al.(2019)Arute, Arya, Babbush, Bacon, Bardin, Barends,
  Biswas, Boixo, Brandao, Buell, Burkett, Chen, Chen, Chiaro, Collins,
  Courtney, Dunsworth, Farhi, Foxen, Fowler, Gidney, Giustina, Graff, Guerin,
  Habegger, Harrigan, Hartmann, Ho, Hoffmann, Huang, Humble, Isakov, Jeffrey,
  Jiang, Kafri, Kechedzhi, Kelly, Klimov, Knysh, Korotkov, Kostritsa, Landhuis,
  Lindmark, Lucero, Lyakh, Mandr{\`a}, McClean, McEwen, Megrant, Mi,
  Michielsen, Mohseni, Mutus, Naaman, Neeley, Neill, Niu, Ostby, Petukhov,
  Platt, Quintana, Rieffel, Roushan, Rubin, Sank, Satzinger, Smelyanskiy, Sung,
  Trevithick, Vainsencher, Villalonga, White, Yao, Yeh, Zalcman, Neven, and
  Martinis]{Arute2019}
F.~Arute, K.~Arya, R.~Babbush, D.~Bacon, J.~C. Bardin, R.~Barends, R.~Biswas,
  S.~Boixo, F.~G. S.~L. Brandao, D.~A. Buell, B.~Burkett, Y.~Chen, Z.~Chen,
  B.~Chiaro, R.~Collins, W.~Courtney, A.~Dunsworth, E.~Farhi, B.~Foxen,
  A.~Fowler, C.~Gidney, M.~Giustina, R.~Graff, K.~Guerin, S.~Habegger, M.~P.
  Harrigan, M.~J. Hartmann, A.~Ho, M.~Hoffmann, T.~Huang, T.~S. Humble, S.~V.
  Isakov, E.~Jeffrey, Z.~Jiang, D.~Kafri, K.~Kechedzhi, J.~Kelly, P.~V. Klimov,
  S.~Knysh, A.~Korotkov, F.~Kostritsa, D.~Landhuis, M.~Lindmark, E.~Lucero,
  D.~Lyakh, S.~Mandr{\`a}, J.~R. McClean, M.~McEwen, A.~Megrant, X.~Mi,
  K.~Michielsen, M.~Mohseni, J.~Mutus, O.~Naaman, M.~Neeley, C.~Neill, M.~Y.
  Niu, E.~Ostby, A.~Petukhov, J.~C. Platt, C.~Quintana, E.~G. Rieffel,
  P.~Roushan, N.~C. Rubin, D.~Sank, K.~J. Satzinger, V.~Smelyanskiy, K.~J.
  Sung, M.~D. Trevithick, A.~Vainsencher, B.~Villalonga, T.~White, Z.~J. Yao,
  P.~Yeh, A.~Zalcman, H.~Neven, and J.~M. Martinis.
\newblock Quantum supremacy using a programmable superconducting processor.
\newblock \emph{Nature}, 574:\penalty0 505--510, 2019.
\newblock \doi{10.1038/s41586-019-1666-5}.

\bibitem[Li et~al.(2022{\natexlab{a}})Li, Gan, Chen, Chen, Lu, Lu, Pan, Fu, and
  Yang]{Li2022g}
Y.~Li, L.~Gan, M.~Chen, Y.~Chen, H.~Lu, C.~Lu, J.~Pan, H.~Fu, and G.~Yang.
\newblock Benchmarking 50-photon {Gaussian} boson sampling on the {Sunway
  TaihuLight}.
\newblock \emph{IEEE Trans. Parallel Distrib. Syst.}, 33:\penalty0 1357--1372,
  2022{\natexlab{a}}.
\newblock \doi{10.1109/TPDS.2021.3111185}.

\bibitem[Rupp(2022)]{Rupp2022}
K.~Rupp.
\newblock Microprocessor trend data.
\newblock \url{https://github.com/karlrupp/microprocessor-trend-data}, 2022.

\bibitem[Mann(2020)]{Mann2020}
A.~Mann.
\newblock Nascent exascale supercomputers offer promise, present challenges.
\newblock \emph{Proc. Natl. Acad. Sci.}, 117:\penalty0 22623--22625, 2020.
\newblock \doi{10.1073/pnas.2015968117}.

\bibitem[Betcke et~al.(2022)Betcke, Bungartz, Wells, and Woodley]{Betcke2022}
T.~Betcke, H.~J. Bungartz, G.~Wells, and S.~Woodley.
\newblock {ExCALIBUR}--{U.K.}'s preparation for the arrival of the next
  generation of {HPC}.
\newblock \emph{Comput. Sci. Eng.}, 24:\penalty0 5--7, 2022.
\newblock \doi{10.1109/MCSE.2022.3144927}.

\bibitem[Moore(1965)]{Moore1965}
G.~E. Moore.
\newblock Cramming more components onto integrated circuits.
\newblock \emph{Electron.}, 38:\penalty0 114, 1965.

\bibitem[Khan et~al.(2018)Khan, Hounshell, and Fuchs]{Khan2018}
H.~N. Khan, D.~A. Hounshell, and E.~R.~H. Fuchs.
\newblock Science and research policy at the end of {Moore's} law.
\newblock \emph{Nat. Electron.}, 1:\penalty0 14--21, 2018.
\newblock \doi{10.1038/s41928-017-0005-9}.

\bibitem[Leiserson et~al.(2020)Leiserson, Thompson, Emer, Kuszmaul, Lampson,
  Sanchez, and Schardl]{Leiserson2020}
C.~E. Leiserson, N.~C. Thompson, J.~S. Emer, B.~C. Kuszmaul, B.~W. Lampson,
  D.~Sanchez, and T.~B. Schardl.
\newblock There's plenty of room at the top: What will drive computer
  performance after {Moore's} law?
\newblock \emph{Science}, 368:\penalty0 eaam9744, 2020.
\newblock \doi{10.1126/science.aam9744}.

\bibitem[Strohmaier et~al.(2023{\natexlab{a}})Strohmaier, Dongarra, Simon, and
  Meuer]{top500}
E.~Strohmaier, J.~Dongarra, H.~Simon, and M.~Meuer.
\newblock Top500.
\newblock \url{https://top500.org/}, 2023{\natexlab{a}}.
\newblock Accessed: 2023-08-01, updated every six months.

\bibitem[Strohmaier et~al.(2023{\natexlab{b}})Strohmaier, Dongarra, Simon, and
  Meuer]{green500}
E.~Strohmaier, J.~Dongarra, H.~Simon, and M.~Meuer.
\newblock Green500.
\newblock \url{https://www.top500.org/lists/green500/}, 2023{\natexlab{b}}.
\newblock Accessed: 2023-12-03, updated every six months.

\bibitem[Feynman(1982)]{Feynman1982}
R.~P. Feynman.
\newblock Simulating physics with computers.
\newblock \emph{Int. J. Theor. Phys.}, 21:\penalty0 467--488, 1982.
\newblock \doi{10.1007/BF02650179}.

\bibitem[Deutsch(1985)]{Deutsch1985}
D.~Deutsch.
\newblock Quantum theory, the {Church-Turing} principle and the universal
  quantum computer.
\newblock \emph{Proc. R. Soc. A}, 400:\penalty0 97--117, 1985.
\newblock \doi{10.1098/rspa.1985.0070}.

\bibitem[Shor(1994)]{Shor1994}
P.~W. Shor.
\newblock Polynomial time algorithms for discrete logarithms and factoring on a
  quantum computer.
\newblock In L.~M. Adleman and M.-D. Huang, editors, \emph{Algorithmic Number
  Theory}, volume 877 of \emph{Lecture Notes in Computer Science}, page 289.
  Springer, Berlin, Heidelberg, 1994.
\newblock \doi{10.1007/3-540-58691-1_68}.

\bibitem[Grover(1996)]{Grover1996}
L.~K. Grover.
\newblock A fast quantum mechanical algorithm for database search.
\newblock In \emph{Proc. 28th STOC}, pages 212--219. ACM, New York, NY, 1996.
\newblock \doi{10.1145/237814.237866}.

\bibitem[Grover(1997)]{Grover1997}
L.~K. Grover.
\newblock Quantum mechanics helps in searching for a needle in a haystack.
\newblock \emph{Phys. Rev. Lett.}, 79:\penalty0 325--328, 1997.
\newblock \doi{10.1103/PhysRevLett.79.325}.

\bibitem[Georgescu et~al.(2014)Georgescu, Ashhab, and Nori]{Georgescu2014}
I.~M. Georgescu, S.~Ashhab, and F.~Nori.
\newblock Quantum simulation.
\newblock \emph{Rev. Mod. Phys.}, 86:\penalty0 153--185, 2014.
\newblock \doi{10.1103/RevModPhys.86.153}.

\bibitem[Altman et~al.(2021)Altman, Brown, Carleo, Carr, Demler, Chin, DeMarco,
  Economou, Eriksson, Fu, Greiner, Hazzard, Hulet, Koll{\'a}r, Lev, Lukin, Ma,
  Mi, Misra, Monroe, Murch, Nazario, Ni, Potter, Roushan, Saffman,
  Schleier-Smith, Siddiqi, Simmonds, Singh, Spielman, Temme, Weiss, Vu{\v
  c}kovi{\'c}, Vuleti{\'c}, Ye, and Zwierlein]{Altman2021}
E.~Altman, K.~R. Brown, G.~Carleo, L.~D. Carr, E.~Demler, C.~Chin, B.~DeMarco,
  S.~E. Economou, M.~A. Eriksson, K.-M.~C. Fu, M.~Greiner, K.~R.~A. Hazzard,
  R.~G. Hulet, A.~J. Koll{\'a}r, B.~L. Lev, M.~D. Lukin, R.~Ma, X.~Mi,
  S.~Misra, C.~Monroe, K.~Murch, Z.~Nazario, K.-K. Ni, A.~C. Potter,
  P.~Roushan, M.~Saffman, M.~Schleier-Smith, I.~Siddiqi, R.~Simmonds, M.~Singh,
  I.~B. Spielman, K.~Temme, D.~S. Weiss, J.~Vu{\v c}kovi{\'c}, V.~Vuleti{\'c},
  J.~Ye, and M.~Zwierlein.
\newblock Quantum simulators: Architectures and opportunities.
\newblock \emph{PRX Quantum}, 2:\penalty0 017003, 2021.
\newblock \doi{10.1103/PRXQuantum.2.017003}.

\bibitem[{Department for Science, Innovation and Technology}(2023)]{DSIT2023}
{Department for Science, Innovation and Technology}.
\newblock National quantum strategy.
\newblock Technical report, UK Government, 2023.
\newblock
  \url{https://www.gov.uk/government/publications/national-quantum-strategy}.

\bibitem[Ac{\'i}n et~al.(2018)Ac{\'i}n, Bloch, Buhrman, Calarco, Eichler,
  Eisert, Esteve, Gisin, Glaser, Jelezko, Kuhr, Lewenstein, Riedel, Schmidt,
  Thew, Wallraff, Walmsley, and Wilhelm]{Acin2018}
A.~Ac{\'i}n, I.~Bloch, H.~Buhrman, T.~Calarco, C.~Eichler, J.~Eisert,
  D.~Esteve, N.~Gisin, S.~J. Glaser, F.~Jelezko, S.~Kuhr, M.~Lewenstein, M.~F.
  Riedel, P.~O. Schmidt, R.~Thew, A.~Wallraff, I.~Walmsley, and F.~K. Wilhelm.
\newblock The quantum technologies roadmap: a {European} community view.
\newblock \emph{New J. Phys.}, 20:\penalty0 080201, 2018.
\newblock \doi{10.1088/1367-2630/aad1ea}.

\bibitem[Monroe et~al.(2019)Monroe, Raymer, and Taylor]{Monroe2019}
C.~Monroe, M.~G. Raymer, and J.~Taylor.
\newblock The {U.S. National Quantum Initiative}: From act to action.
\newblock \emph{Science}, 364:\penalty0 440--442, 2019.
\newblock \doi{10.1126/science.aax0578}.

\bibitem[{Department of Industry, Science and Resources}(2023)]{Australia2023}
{Department of Industry, Science and Resources}.
\newblock National quantum strategy.
\newblock Technical report, {Australian Government}, 2023.
\newblock
  \url{https://www.industry.gov.au/publications/national-quantum-strategy}.

\bibitem[{UNESCO}(2024)]{UN_IYQ}
{UNESCO}.
\newblock International year of quantum science and technology.
\newblock \url{https://quantum2025.org}, 2024.
\newblock Accessed: 2024-06-15.

\bibitem[MacQuarrie et~al.(2020)MacQuarrie, Simon, Simmons, and
  Maine]{MacQuarrie2020}
E.~R. MacQuarrie, C.~Simon, S.~Simmons, and E.~Maine.
\newblock The emerging commercial landscape of quantum computing.
\newblock \emph{Nat. Rev. Phys.}, 2:\penalty0 596--598, 2020.
\newblock \doi{10.1038/s42254-020-00247-5}.

\bibitem[Gibney(2019)]{Gibney2019}
E.~Gibney.
\newblock Quantum gold rush: the private funding pouring into quantum
  start-ups.
\newblock \emph{Nature}, 574:\penalty0 22--24, 2019.
\newblock \doi{10.1038/d41586-019-02935-4}.

\bibitem[Bova et~al.(2021)Bova, Goldfarb, and Melko]{Bova2021}
F.~Bova, A.~Goldfarb, and R.~G. Melko.
\newblock Commercial applications of quantum computing.
\newblock \emph{EPJ Quantum Technol.}, 8:\penalty0 2, 2021.
\newblock \doi{10.1140/epjqt/s40507-021-00091-1}.

\bibitem[Preskill(2018)]{Preskill2018}
J.~Preskill.
\newblock Quantum computing in the {NISQ} era and beyond.
\newblock \emph{Quantum}, 2:\penalty0 79, 2018.
\newblock \doi{10.22331/q-2018-08-06-79}.

\bibitem[Bharti et~al.(2022)Bharti, Cervera-Lierta, Kyaw, Haug, Alperin-Lea,
  Anand, Degroote, Heimonen, Kottmann, Menke, Mok, Sim, Kwek, and
  Aspuru-Guzik]{Bharti2022}
K.~Bharti, A.~Cervera-Lierta, T.~H. Kyaw, T.~Haug, S.~Alperin-Lea, A.~Anand,
  M.~Degroote, H.~Heimonen, J.~S. Kottmann, T.~Menke, W.-K. Mok, S.~Sim, L.-C.
  Kwek, and A.~Aspuru-Guzik.
\newblock Noisy intermediate-scale quantum algorithms.
\newblock \emph{Rev. Mod. Phys.}, 94:\penalty0 015004, 2022.
\newblock \doi{10.1103/RevModPhys.94.015004}.

\bibitem[{IQM}(2023)]{IQMtech}
{IQM}.
\newblock Technology.
\newblock \url{https://meetiqm.com/technology/}, 2023.
\newblock Accessed: 2023-12-01.

\bibitem[Atos(2020)]{Atos2020}
Atos.
\newblock Atos joins forces with start-up {Pasqal} to accelerate high
  performance computing using quantum neutral atom technology {[Press release,
  4 Nov 2020]}.
\newblock
  \url{https://atos.net/en/2020/press-release_2020_11_04/atos-joins-forces-with-start-up-pasqal-to-accelerate-high-performance-computing-using-quantum-neutral-atom-technology},
  2020.
\newblock Accessed: 2023-12-01.

\bibitem[Atos(2021)]{Atos2021}
Atos.
\newblock Atos confirms role as global leader in quantum hybridization
  technologies at its {8th Quantum Advisory Board} {[Press release, 3 Dec
  2021]}.
\newblock
  \url{https://atos.net/en/2021/press-release_2021_12_03/atos-confirms-role-as-global-leader-in-quantum-hybridization-technologies-at-its-8th-quantum-advisory-board},
  2021.
\newblock Accessed: 2024-08-01.

\bibitem[{HPCQS}(2021)]{HPCQS2021}
{HPCQS}.
\newblock Towards a world-class supercomputing ecosystem - {HPCQS} pioneers
  federated quantum-super-computing in {Europe} {[Press release, 1 Dec 2021].}
\newblock Archived
  \url{https://web.archive.org/web/20211201181208/https://www.hpcqs.eu/}.,
  2021.
\newblock Accessed: 2024-08-01.

\bibitem[{IQM}(2024)]{IQM2024}
{IQM}.
\newblock {Germany} launches its first hybrid quantum computer at {Leibniz
  Supercomputing Centre} {[Press release, 18 Jun 2024]}.
\newblock
  \url{https://www.meetiqm.com/newsroom/press-releases/germany-launches-its-first-hybrid-quantum-computer},
  2024.
\newblock Accessed: 2024-08-01.

\bibitem[McCaskey(2022)]{NVIDIAcuQuantum2022}
A.~McCaskey.
\newblock The road to the hybrid quantum-{HPC} data center starts here {[Press
  release, 30 May 2022]}.
\newblock \url{https://blogs.nvidia.com/blog/quantum-computing-hpc-isc2022},
  2022.
\newblock Accessed: 2023-12-01.

\bibitem[{Zapata AI}(2023)]{ZapataOrquestra}
{Zapata AI}.
\newblock Build and deploy industrial generative {AI} applications on
  {Orquestra}.
\newblock \url{https://zapata.ai/platform-orquestra/}, 2023.
\newblock Accessed: 2023-12-01.

\bibitem[Herr et~al.(2019)Herr, Duke, Godel, Natraj, Makowska, Williams,
  Perkins, and Visser]{LondonEconomics2019}
D.~Herr, C.~Duke, M.~Godel, A.~Natraj, A.~Makowska, R.~Williams, R.~Perkins,
  and C.~Visser.
\newblock The impact of {EPSRC}'s investments in high performance computing
  infrastructure: Final report.
\newblock Technical report, London Economics, 2019.
\newblock
  \url{https://www.ukri.org/publications/impact-of-epsrcs-investments-in-hpc-infrastructure/}.

\bibitem[{Department for Science, Innovation and Technology}(2022)]{DSIT2022}
{Department for Science, Innovation and Technology}.
\newblock Independent review of the future of compute: Final report and
  recommendations.
\newblock Technical report, UK Government, 2022.
\newblock
  \url{https://www.gov.uk/government/publications/future-of-compute-review/the-future-of-compute-report-of-the-review-of-independent-panel-of-experts}.

\bibitem[{European Commission}(2023)]{EuroHPC}
{European Commission}.
\newblock The {European} high performance computing joint undertaking.
\newblock
  \url{https://digital-strategy.ec.europa.eu/en/policies/high-performance-computing-joint-undertaking},
  2023.
\newblock Accessed: 2023-12-01.

\bibitem[{UK Research and Innovation}(2024)]{UKRI2024}
{UK Research and Innovation}.
\newblock Digital research infrastructure.
\newblock
  \url{https://www.ukri.org/what-we-do/creating-world-class-research-and-innovation-infrastructure/digital-research-infrastructure/},
  2024.
\newblock Accessed: 2024-08-01.

\bibitem[Dunjko and Briegel(2018)]{Dunjko2018}
V.~Dunjko and H.~J. Briegel.
\newblock Machine learning \& artificial intelligence in the quantum domain: a
  review of recent progress.
\newblock \emph{Rep. Prog. Phys.}, 81:\penalty0 074001, 2018.
\newblock \doi{10.1088/1361-6633/aab406}.

\bibitem[Nielsen and Chuang(2010)]{Nielsen2010}
M.~A. Nielsen and I.~L. Chuang.
\newblock \emph{Quantum Computation and Quantum Information: 10th Anniversary
  Edition}.
\newblock Cambridge University Press, 2010.
\newblock \doi{10.1017/CBO9780511976667}.

\bibitem[Abhijith et~al.(2022)Abhijith, Adedoyin, Ambrosiano, Anisimov, Casper,
  Chennupati, Coffrin, Djidjev, Gunter, Karra, Lemons, Lin, Malyzhenkov,
  Mascarenas, Mniszewski, Nadiga, O’malley, Oyen, Pakin, Prasad, Roberts,
  Romero, Santhi, Sinitsyn, Swart, Wendelberger, Yoon, Zamora, Zhu, Eidenbenz,
  B{\"a}rtschi, Coles, Vuffray, and Lokhov]{Abhijith2022}
J.~Abhijith, A.~Adedoyin, J.~Ambrosiano, P.~Anisimov, W.~Casper, G.~Chennupati,
  C.~Coffrin, H.~Djidjev, D.~Gunter, S.~Karra, N.~Lemons, S.~Lin,
  A.~Malyzhenkov, D.~Mascarenas, S.~Mniszewski, B.~Nadiga, D.~O’malley,
  D.~Oyen, S.~Pakin, L.~Prasad, R.~Roberts, P.~Romero, N.~Santhi, N.~Sinitsyn,
  P.~J. Swart, J.~G. Wendelberger, B.~Yoon, R.~Zamora, W.~Zhu, S.~Eidenbenz,
  A.~B{\"a}rtschi, P.~J. Coles, M.~Vuffray, and A.~Y. Lokhov.
\newblock Quantum algorithm implementations for beginners.
\newblock \emph{ACM Trans. Quantum Comput.}, 3:\penalty0 18, 2022.
\newblock \doi{10.1145/3517340}.

\bibitem[Montanaro(2016)]{Montanaro2016}
A.~Montanaro.
\newblock Quantum algorithms: an overview.
\newblock \emph{npj Quantum Inf.}, 2:\penalty0 15023, 2016.
\newblock \doi{10.1038/npjqi.2015.23}.

\bibitem[Albash and Lidar(2018)]{Albash2018}
T.~Albash and D.~A. Lidar.
\newblock Adiabatic quantum computation.
\newblock \emph{Rev. Mod. Phys.}, 90:\penalty0 015002, 2018.
\newblock \doi{10.1103/RevModPhys.90.015002}.

\bibitem[Childs and van Dam(2010)]{Childs2010r}
A.~M. Childs and W.~van Dam.
\newblock Quantum algorithms for algebraic problems.
\newblock \emph{Rev. Mod. Phys.}, 82:\penalty0 1--52, 2010.
\newblock \doi{10.1103/RevModPhys.82.1}.

\bibitem[Jordan(2021)]{QuantumAlgorithmZoo}
S.~Jordan.
\newblock {Quantum Algorithm Zoo}.
\newblock \url{https://quantumalgorithmzoo.org}, 2021.
\newblock Accessed: 2022-08-01.

\bibitem[Bernstein and Vazirani(1993)]{Bernstein1993}
E.~Bernstein and U.~Vazirani.
\newblock Quantum complexity theory.
\newblock In \emph{Proc. 25th STOC}, page 11–20. ACM, New York, NY, 1993.
\newblock \doi{10.1145/167088.167097}.

\bibitem[Vazirani(2002)]{Vazirani2002}
U.~Vazirani.
\newblock A survey of quantum complexity theory.
\newblock In \emph{Proc. Sympos. Appl. Math.}, volume~58, pages 193--220. Amer.
  Math. Soc., Providence, RI, 2002.
\newblock \doi{10.1090/psapm/058/1922899}.

\bibitem[Terhal(2015)]{Terhal2015}
B.~M. Terhal.
\newblock Quantum error correction for quantum memories.
\newblock \emph{Rev. Mod. Phys.}, 87:\penalty0 307--346, 2015.
\newblock \doi{10.1103/RevModPhys.87.307}.

\bibitem[Roffe(2019)]{Roffe2019}
J.~Roffe.
\newblock Quantum error correction: an introductory guide.
\newblock \emph{Contemp. Phys.}, 60:\penalty0 226--245, 2019.
\newblock \doi{10.1080/00107514.2019.1667078}.

\bibitem[Eisert et~al.(2020)Eisert, Hangleiter, Walk, Roth, Markham, Parekh,
  Chabaud, and Kashefi]{Eisert2020}
J.~Eisert, D.~Hangleiter, N.~Walk, I.~Roth, D.~Markham, R.~Parekh, U.~Chabaud,
  and E.~Kashefi.
\newblock Quantum certification and benchmarking.
\newblock \emph{Nat. Rev. Phys.}, 2:\penalty0 382--390, 2020.
\newblock \doi{10.1038/s42254-020-0186-4}.

\bibitem[Carrasco et~al.(2021)Carrasco, Elben, Kokail, Kraus, and
  Zoller]{Carrasco2021}
J.~Carrasco, A.~Elben, C.~Kokail, B.~Kraus, and P.~Zoller.
\newblock Theoretical and experimental perspectives of quantum verification.
\newblock \emph{PRX Quantum}, 2:\penalty0 010102, 2021.
\newblock \doi{10.1103/PRXQuantum.2.010102}.

\bibitem[Hauke et~al.(2020)Hauke, Katzgraber, Lechner, Nishimori, and
  Oliver]{Hauke2020}
P.~Hauke, H.~G. Katzgraber, W.~Lechner, H.~Nishimori, and W.~D. Oliver.
\newblock Perspectives of quantum annealing: methods and implementations.
\newblock \emph{Rep. Prog. Phys.}, 83:\penalty0 054401, 2020.
\newblock \doi{10.1088/1361-6633/ab85b8}.

\bibitem[Yarkoni et~al.(2022)Yarkoni, Raponi, B{\"a}ck, and
  Schmitt]{Yarkoni2022}
S.~Yarkoni, E.~Raponi, T.~B{\"a}ck, and S.~Schmitt.
\newblock Quantum annealing for industry applications: introduction and review.
\newblock \emph{Rep. Prog. Phys.}, 85:\penalty0 104001, 2022.
\newblock \doi{10.1088/1361-6633/ac8c54}.

\bibitem[Wu et~al.(2021)Wu, Bao, Cao, Chen, Chen, Chen, Chung, Deng, Du, Fan,
  Gong, Guo, Guo, Guo, Han, Hong, Huang, Huo, Li, Li, Li, Li, Liang, Lin, Lin,
  Qian, Qiao, Rong, Su, Sun, Wang, Wang, Wu, Xu, Yan, Yang, Yang, Ye, Yin,
  Ying, Yu, Zha, Zhang, Zhang, Zhang, Zhang, Zhao, Zhao, Zhou, Zhu, Lu, Peng,
  Zhu, and Pan]{Wu2021}
Y.~Wu, W.-S. Bao, S.~Cao, F.~Chen, M.~C. Chen, X.~Chen, T.-H. Chung, H.~Deng,
  Y.~Du, D.~Fan, M.~Gong, C.~Guo, C.~Guo, S.~Guo, L.~Han, L.~Hong, H.-L. Huang,
  Y.-H. Huo, L.~Li, N.~Li, S.~Li, Y.~Li, F.~Liang, C.~Lin, J.~Lin, H.~Qian,
  D.~Qiao, H.~Rong, H.~Su, L.~Sun, L.~Wang, S.~Wang, D.~Wu, Y.~Xu, K.~Yan,
  W.~Yang, Y.~Yang, Y.~Ye, J.~Yin, C.~Ying, J.~Yu, C.~Zha, C.~Zhang, H.~Zhang,
  K.~Zhang, Y.~Zhang, H.~Zhao, Y.~Zhao, L.~Zhou, Q.~Zhu, C.-Y. Lu, C.~Z. Peng,
  X.~Zhu, and J.~W. Pan.
\newblock Strong quantum computational advantage using a superconducting
  quantum processor.
\newblock \emph{Phys. Rev. Lett.}, 127:\penalty0 180501, 2021.
\newblock \doi{10.1103/PhysRevLett.127.180501}.

\bibitem[Kim et~al.(2023)Kim, Eddins, Anand, Wei, van~den Berg, Rosenblatt,
  Nayfeh, Wu, Zaletel, Temme, and Kandala]{Kim2023a}
Y.~Kim, A.~Eddins, S.~Anand, K.~X. Wei, E.~van~den Berg, S.~Rosenblatt,
  H.~Nayfeh, Y.~Wu, M.~Zaletel, K.~Temme, and A.~Kandala.
\newblock Evidence for the utility of quantum computing before fault tolerance.
\newblock \emph{Nature}, 618:\penalty0 500--505, 2023.
\newblock \doi{10.1038/s41586-023-06096-3}.

\bibitem[Leymann and Barzen(2020)]{Leymann2020}
F.~Leymann and J.~Barzen.
\newblock The bitter truth about gate-based quantum algorithms in the {NISQ}
  era.
\newblock \emph{Quantum Sci. Technol.}, 5:\penalty0 044007, 2020.
\newblock \doi{10.1088/2058-9565/abae7d}.

\bibitem[Coveney and Highfield(2020)]{Coveney2020}
P.~V. Coveney and R.~R. Highfield.
\newblock From digital hype to analogue reality: Universal simulation beyond
  the quantum and exascale eras.
\newblock \emph{J. Comput. Sci.}, 46:\penalty0 101093, 2020.
\newblock \doi{10.1016/j.jocs.2020.101093}.

\bibitem[Aaronson(2015)]{Aaronson2015}
S.~Aaronson.
\newblock Read the fine print.
\newblock \emph{Nat. Phys.}, 11:\penalty0 291--293, 2015.
\newblock \doi{10.1038/nphys3272}.

\bibitem[Chancellor et~al.(2020)Chancellor, Cumming, and
  Thomas]{Chancellor2020}
N.~Chancellor, R.~Cumming, and T.~Thomas.
\newblock Toward a standardized methodology for constructing quantum computing
  use cases, 2020.
\newblock \href{https://doi.org/10.48550/arxiv.2006.05846}{arXiv:2006.05846}.

\bibitem[Perini and Ciarletta(2021)]{IQM2021}
S.~Perini and F.~Ciarletta.
\newblock Untangling the {HPC} innovation dilemma through quantum computing.
\newblock Technical report, IQM and Atos, 2021.
\newblock
  \url{https://www.meetiqm.com/technology/iqm-atos-state-of-quantum-research-2021/}.

\bibitem[Wootters and Zurek(1982)]{Wootters1982}
W.~Wootters and W.~Zurek.
\newblock A single quantum cannot be cloned.
\newblock \emph{Nature}, 299:\penalty0 802--803, 1982.
\newblock \doi{10.1038/299802a0}.

\bibitem[van Meter and Horsman(2013)]{vanMeter2013}
R.~van Meter and D.~Horsman.
\newblock A blueprint for building a quantum computer.
\newblock \emph{Commun. ACM}, 56:\penalty0 84–93, 2013.
\newblock \doi{10.1145/2494568}.

\bibitem[DiVincenzo(2000)]{DiVincenzo2000}
D.~P. DiVincenzo.
\newblock The physical implementation of quantum computation.
\newblock \emph{Fortschr. Phys.}, 48:\penalty0 771--783, 2000.
\newblock \doi{10.1002/1521-3978(200009)48:9/11<771::AID-PROP771>3.0.CO;2-E}.

\bibitem[Reiher et~al.(2017)Reiher, Wiebe, Svore, Wecker, and
  Troyer]{Reiher2017}
M.~Reiher, N.~Wiebe, K.~M. Svore, D.~Wecker, and M.~Troyer.
\newblock Elucidating reaction mechanisms on quantum computers.
\newblock \emph{Proc. Natl. Acad. Sci.}, 114:\penalty0 7555--7560, 2017.
\newblock \doi{10.1073/pnas.1619152114}.

\bibitem[Portugal(2022)]{Portugal2022}
R.~Portugal.
\newblock Basic quantum algorithms, 2022.
\newblock \href{https://doi.org/10.48550/arxiv.2201.10574}{arXiv:2201.10574}.

\bibitem[{Qiskit}(2023)]{Qiskit}
{Qiskit}.
\newblock Qiskit is the open-source toolkit for useful quantum.
\newblock \url{https://www.ibm.com/quantum/qiskit}, 2023.
\newblock Accessed: 2023-10-01.

\bibitem[Camps et~al.(2020)Camps, Beeumen, and Yang]{Camps2020}
D.~Camps, R.~Van Beeumen, and C.~Yang.
\newblock Quantum {Fourier} transform revisited.
\newblock \emph{Numer. Lin. Algebra Appl.}, 28:\penalty0 e2331, 2020.
\newblock \doi{10.1002/nla.2331}.

\bibitem[Griffiths and Niu(1996)]{Griffiths1996}
R.~B. Griffiths and C.-S. Niu.
\newblock Semiclassical {Fourier} transform for quantum computation.
\newblock \emph{Phys. Rev. Lett.}, 76:\penalty0 3228--3231, 1996.
\newblock \doi{10.1103/PhysRevLett.76.3228}.

\bibitem[Browne(2007)]{Browne2007}
D.~Browne.
\newblock Efficient classical simulation of the semi-classical quantum
  {Fourier} transform.
\newblock \emph{New J. Phys.}, 9:\penalty0 146, 2007.
\newblock \doi{10.1088/1367-2630/9/5/146}.

\bibitem[Kitaev(1996)]{Kitaev1996}
A.~Y. Kitaev.
\newblock Quantum measurements and the {Abelian} stabilizer problem.
\newblock \emph{Electron. Colloquium Comput. Complex.}, 3:\penalty0 22, 1996.

\bibitem[Ahmadi and Chiang(2012)]{Ahmadi2012}
H.~Ahmadi and C.-F. Chiang.
\newblock Quantum phase estimation with arbitrary constant-precision phase
  shift operators.
\newblock \emph{Quantum Inf. Comput.}, 12:\penalty0 864–875, 2012.

\bibitem[Ni et~al.(2023)Ni, Li, and Ying]{Ni2023}
H.~Ni, H.~Li, and L.~Ying.
\newblock On low-depth algorithms for quantum phase estimation.
\newblock \emph{Quantum}, 7:\penalty0 2265, 2023.
\newblock \doi{10.22331/q-2023-11-06-1165}.

\bibitem[Aspuru-Guzik et~al.(2005)Aspuru-Guzik, Dutoi, Love, and
  Head-Gordon]{AspuruGuzik2005}
A.~Aspuru-Guzik, A.~D. Dutoi, P.~J. Love, and M.~Head-Gordon.
\newblock Simulated quantum computation of molecular energies.
\newblock \emph{Science}, 309:\penalty0 1704--1707, 2005.
\newblock \doi{10.1126/science.1113479}.

\bibitem[Ekert and Jozsa(1996)]{Ekert1996}
A.~Ekert and R.~Jozsa.
\newblock Quantum computation and {Shor's} factoring algorithm.
\newblock \emph{Rev. Mod. Phys.}, 68:\penalty0 733--753, 1996.
\newblock \doi{10.1103/RevModPhys.68.733}.

\bibitem[Rivest et~al.(1978)Rivest, Shamir, and Adleman]{Rivest1978}
R.~L. Rivest, A.~Shamir, and L.~Adleman.
\newblock A method for obtaining digital signatures and public-key
  cryptosystems.
\newblock \emph{Commun. ACM}, 21:\penalty0 120--126, 1978.
\newblock \doi{10.1145/359340.359342}.

\bibitem[Zimmermann(2019)]{classicalfactoring}
P.~Zimmermann.
\newblock Factorization of {RSA-250}, 2019.
\newblock URL
  \url{https://sympa.inria.fr/sympa/arc/cado-nfs/2020-02/msg00001.html}.
\newblock {Previous records: Kleinjung, Thomas (18 Feb 2010). "Factorization of
  a 768-bit RSA modulus". Thom{\'e}, Emmanuel (December 2, 2019). "795-bit
  factoring and discrete logarithms". cado-nfs-discuss (Mailing list).}

\bibitem[van Meter and Devitt(2016)]{vanMeter2016}
R.~van Meter and S.~J. Devitt.
\newblock The path to scalable distributed quantum computing.
\newblock \emph{IEEE Comput.}, 49:\penalty0 31--42, 2016.
\newblock \doi{10.1109/MC.2016.291}.

\bibitem[Ambainis and {\v S}palek(2006)]{Ambainis2006}
A.~Ambainis and R.~{\v S}palek.
\newblock Quantum algorithms for matching and network flows.
\newblock In B.~Durand and W.~Thomas, editors, \emph{STACS 2006}, pages
  172--183. Springer, Berlin, Heidelberg, 2006.
\newblock \doi{10.1007/11672142_13}.

\bibitem[Shenvi et~al.(2003)Shenvi, Kempe, and {Birgitta Whaley}]{Shenvi2002}
N.~Shenvi, J.~Kempe, and K.~{Birgitta Whaley}.
\newblock Quantum random-walk search algorithm.
\newblock \emph{Phys. Rev. A}, 67:\penalty0 052307, 2003.
\newblock \doi{10.1103/PhysRevA.67.052307}.

\bibitem[Childs and Eisenberg(2005)]{Childs2003}
A.~Childs and J.~M. Eisenberg.
\newblock Quantum algorithms for subset finding.
\newblock \emph{Quantum Inf. Comput.}, 5:\penalty0 593--604, 2005.
\newblock \doi{10.26421/QIC5.7-7}.

\bibitem[Brassard et~al.(2002)Brassard, H{\o}yer, Mosca, and
  Tapp]{Brassard2002}
G.~Brassard, P.~H{\o}yer, M.~Mosca, and A.~Tapp.
\newblock Quantum amplitude amplification and estimation.
\newblock \emph{Contemp. Math.}, 305:\penalty0 53--74, 2002.
\newblock \doi{10.1090/conm/305/05215}.

\bibitem[Ambainis(2012)]{Ambainis2012}
A.~Ambainis.
\newblock Variable time amplitude amplification and quantum algorithms for
  linear algebra problems.
\newblock In C.~D{\"u}rr and T.~Wilke, editors, \emph{29th Int. STACS 2012},
  volume~14 of \emph{LIPIcs}, pages 636--647. Schloss Dagstuhl--Leibniz-Zentrum
  fuer Informatik, Dagstuhl, Germany, 2012.
\newblock \doi{10.4230/LIPIcs.STACS.2012.636}.

\bibitem[Hogg and Portnov(2000)]{Hogg2000}
T.~Hogg and D.~Portnov.
\newblock Quantum optimization.
\newblock \emph{Inf. Sci.}, 128:\penalty0 181--197, 2000.
\newblock \doi{10.1016/S0020-0255(00)00052-9}.

\bibitem[D{\"u}rr and H{\o}yer(1996)]{Durr1996}
C.~D{\"u}rr and P.~H{\o}yer.
\newblock A quantum algorithm for finding the minimum, 1996.
\newblock
  \href{https://doi.org/10.48550/arxiv.quant-ph/9607014}{arXiv.quant-ph/9607014}.

\bibitem[D{\"u}rr et~al.(2004)D{\"u}rr, Heiligman, H{\o}yer, and
  Mhalla]{Durr2004}
C.~D{\"u}rr, M.~Heiligman, P.~H{\o}yer, and M.~Mhalla.
\newblock Quantum query complexity of some graph problems.
\newblock In \emph{ICALP 2004}, pages 481--493, Berlin, Heidelberg, 2004.
  Springer.
\newblock \doi{10.1007/978-3-540-27836-8_42}.

\bibitem[Ramesh and Vinay(2003)]{Ramesh2003}
H.~Ramesh and V.~Vinay.
\newblock String matching in {$\tilde{O}(\sqrt{n}+\sqrt{m})$} quantum time.
\newblock \emph{J. Discrete Algorithms}, 1:\penalty0 103--110, 2003.
\newblock \doi{10.1016/s1570-8667(03)00010-8}.

\bibitem[Brassard et~al.(1998)Brassard, H{\o}yer, and Tapp]{Brassard1998}
G.~Brassard, P.~H{\o}yer, and A.~Tapp.
\newblock Quantum counting.
\newblock In K.~G. Larsen, S.~Skyum, and G.~Winskel, editors, \emph{Automata,
  Languages and Programming}, volume 1443 of \emph{LNCS}, pages 820--831.
  Springer, Berlin, Heidelberg, 1998.
\newblock \doi{10.1007/BFb0055105}.

\bibitem[Davenport and Pring(2020)]{Davenport2020}
J.~H. Davenport and B.~Pring.
\newblock Improvements to quantum search techniques for block-ciphers, with
  applications to {AES}.
\newblock In \emph{Selected Areas in Cryptography: 27th International
  Conference}, pages 360--384, Berlin, Heidelberg, 2020. Springer-Verlag.
\newblock \doi{10.1007/978-3-030-81652-0_14}.

\bibitem[Motwani and Raghavan(1995)]{Motwani1995}
R.~Motwani and P.~Raghavan.
\newblock \emph{Randomized Algorithms}.
\newblock Cambridge University Press, Cambridge, 1995.
\newblock \doi{10.1017/CBO9780511814075}.

\bibitem[Aharonov et~al.(2001)Aharonov, Ambainis, Kempe, and
  Vazirani]{Aharonov2000}
D.~Aharonov, A.~Ambainis, J.~Kempe, and U.~Vazirani.
\newblock Quantum walks on graphs.
\newblock In \emph{Proc. 33rd ACM STOC}, pages 50--59, New York, NY, 2001. ACM.
\newblock \doi{10.1145/380752.380758}.

\bibitem[Ambainis et~al.(2001)Ambainis, Bach, Nayak, Vishwanath, and
  Watrous]{Ambainis2001}
A.~Ambainis, E.~Bach, A.~Nayak, A.~Vishwanath, and J.~Watrous.
\newblock One-dimensional quantum walks.
\newblock In \emph{Proc. 33rd ACM STOC}, pages 60--69, New York, NY, 2001. ACM.

\bibitem[Farhi and Gutmann(1998)]{Farhi1998}
E.~Farhi and S.~Gutmann.
\newblock Quantum computation and decision trees.
\newblock \emph{Phys. Rev. A}, 58:\penalty0 915--928, 1998.
\newblock \doi{10.1103/PhysRevA.58.915}.

\bibitem[Childs et~al.(2003)Childs, Cleve, Deotto, Farhi, Gutmann, and
  Spielman]{Childs2003b}
A.~M. Childs, R.~Cleve, E.~Deotto, E.~Farhi, S.~Gutmann, and D.~A. Spielman.
\newblock Exponential algorithmic speedup by a quantum walk.
\newblock In \emph{Proc. 35th STOC}, pages 59--68, New York, NY, 2003. ACM.
\newblock \doi{10.1145/780542.780552}.

\bibitem[Venegas-Andraca(2012)]{VenegasAndraca2012}
S.~E. Venegas-Andraca.
\newblock Quantum walks: a comprehensive review.
\newblock \emph{Quantum Inf. Process.}, 11:\penalty0 1015--1106, 2012.
\newblock \doi{10.1007/s11128-012-0432-5}.

\bibitem[Kadian et~al.(2021)Kadian, Garhwal, and Kumar]{Kadian2021}
K.~Kadian, S.~Garhwal, and A.~Kumar.
\newblock Quantum walk and its application domains: A systematic review.
\newblock \emph{Comput. Sci. Rev.}, 41:\penalty0 100419, 2021.
\newblock \doi{10.1016/j.cosrev.2021.100419}.

\bibitem[Callison et~al.(2019)Callison, Chancellor, Mintert, and
  Kendon]{Callison2019}
A.~Callison, N.~Chancellor, F.~Mintert, and V.~Kendon.
\newblock Finding spin glass ground states using quantum walks.
\newblock \emph{New J. Phys.}, 21:\penalty0 123022, 2019.
\newblock \doi{10.1088/1367-2630/ab5ca2}.

\bibitem[Childs(2009)]{Childs2009}
A.~M. Childs.
\newblock Universal computation by quantum walk.
\newblock \emph{Phys. Rev. Lett.}, 102:\penalty0 180501, 2009.
\newblock \doi{10.1103/PhysRevLett.102.180501}.

\bibitem[Lovett et~al.(2010)Lovett, Cooper, Everitt, Trevers, and
  Kendon]{Lovett2010}
N.~B. Lovett, S.~Cooper, M.~Everitt, M.~Trevers, and V.~Kendon.
\newblock Universal quantum computation using the discrete-time quantum walk.
\newblock \emph{Phys. Rev. A}, 81:\penalty0 042330, 2010.
\newblock \doi{10.1103/PhysRevA.81.042330}.

\bibitem[Aaronson and Arkhipov(2011)]{Aaronson2011}
S.~Aaronson and A.~Arkhipov.
\newblock The computational complexity of linear optics.
\newblock In \emph{Proc. 43rd STOC}, page 333–342, New York, NY, 2011. ACM.
\newblock \doi{10.1145/1993636.1993682}.

\bibitem[Childs et~al.(2013)Childs, Gosset, and Webb]{Childs2013}
A.~M. Childs, D.~Gosset, and Z.~Webb.
\newblock Universal computation by multiparticle quantum walk.
\newblock \emph{Science}, 339:\penalty0 791--794, 2013.
\newblock \doi{10.1126/science.1229957}.

\bibitem[Schumacher and Werner(2004)]{Schumacher2004}
B.~Schumacher and R.~F. Werner.
\newblock Reversible quantum cellular automata, 2004.
\newblock
  \href{https://doi.org/10.48550/arXiv.quant-ph/0405174}{arXiv:quant-ph/0405174}.

\bibitem[Karski et~al.(2009)Karski, Forster, Choi, Steffen, Alt, Meschede, and
  Widera]{Karski2009}
M.~Karski, L.~Forster, J.-M. Choi, A.~Steffen, W.~Alt, D.~Meschede, and
  A.~Widera.
\newblock Quantum walk in position space with single optically trapped atoms.
\newblock \emph{Science}, 325:\penalty0 174--177, 2009.
\newblock \doi{10.1126/science.1174436}.

\bibitem[Harrow et~al.(2009)Harrow, Hassidim, and Lloyd]{Harrow2009}
A.~W. Harrow, A.~Hassidim, and S.~Lloyd.
\newblock Quantum algorithm for linear systems of equations.
\newblock \emph{Phys. Rev. Lett.}, 103:\penalty0 150502, 2009.
\newblock \doi{10.1103/PhysRevLett.103.150502}.

\bibitem[Dervovic et~al.(2018)Dervovic, Herbster, Mountney, Severini, Usher,
  and Wossnig]{Dervovic2018}
D.~Dervovic, M.~Herbster, P.~Mountney, S.~Severini, N.~Usher, and L.~Wossnig.
\newblock Quantum linear systems algorithms: a primer, 2018.
\newblock \href{https://arxiv.org/abs/1802.08227}{arXiv:1802.08227}.

\bibitem[Childs et~al.(2017)Childs, Kothari, and Somma]{Childs2017}
A.~M. Childs, R.~Kothari, and R.~D. Somma.
\newblock Quantum algorithm for systems of linear equations with exponentially
  improved dependence on precision.
\newblock \emph{SIAM J. Comput.}, 46:\penalty0 1920--1950, 2017.
\newblock \doi{10.1137/16M1087072}.

\bibitem[Montanaro and Pallister(2016)]{Montanaro2016a}
A.~Montanaro and S.~Pallister.
\newblock Quantum algorithms and the finite element method.
\newblock \emph{Phys. Rev. A}, 93:\penalty0 032324, 2016.
\newblock \doi{10.1103/PhysRevA.93.032324}.

\bibitem[Chowdhury and Somma(2017)]{Chowdhury2017}
A.~N. Chowdhury and R.~D. Somma.
\newblock Quantum algorithms for {Gibbs} sampling and hitting-time estimation.
\newblock \emph{Quantum Inf. Comput.}, 17:\penalty0 41--64, 2017.
\newblock \doi{10.26421/QIC17.1-2-3}.

\bibitem[Saito et~al.(2021)Saito, Lee, Cai, and Asai]{Saito2021}
Y.~Saito, X.~Lee, D.~Cai, and N.~Asai.
\newblock An iterative improvement method for {HHL} algorithm for solving
  linear system of equations, 2021.
\newblock \href{https://doi.org/10.48550/arxiv.2108.07744}{arXiv:2108.07744}.

\bibitem[Lee et~al.(2019{\natexlab{a}})Lee, Joo, and Lee]{lee2019hybrid}
Y.~Lee, J.~Joo, and S.~Lee.
\newblock Hybrid quantum linear equation algorithm and its experimental test on
  {IBM Quantum Experience}.
\newblock \emph{Sci. Rep.}, 9:\penalty0 4778, 2019{\natexlab{a}}.
\newblock \doi{10.1038/s41598-019-41324-9}.

\bibitem[Gao et~al.(2023{\natexlab{a}})Gao, Wu, Guo, Dai, and Shuang]{Gao2023h}
F.~Gao, G.~Wu, S.~Guo, W.~Dai, and F.~Shuang.
\newblock Solving {DC} power flow problems using quantum and hybrid algorithms.
\newblock \emph{Appl. Soft. Comput.}, 137:\penalty0 110147, 2023{\natexlab{a}}.
\newblock \doi{10.1016/j.asoc.2023.110147}.

\bibitem[Angara et~al.(2020)Angara, Stege, M{\"u}ller, and
  Bozzo-Rey]{angara2020hybrid}
P.~P. Angara, U.~Stege, H.~A. M{\"u}ller, and M.~Bozzo-Rey.
\newblock Hybrid quantum-classical problem solving in the {NISQ} era.
\newblock In \emph{CASCON '20}, pages 247--252, USA, 2020. IBM.

\bibitem[Wilson et~al.(2015)Wilson, Dann, and Nickisch]{Wilson2015}
A.~G. Wilson, C.~Dann, and H.~Nickisch.
\newblock Thoughts on massively scalable {Gaussian} processes, 2015.
\newblock \href{https://doi.org/10.48550/arxiv.1511.01870}{arXiv:1511.01870}.

\bibitem[Wossnig et~al.(2018)Wossnig, Zhao, and Prakash]{Wossnig2018}
L.~Wossnig, Z.~Zhao, and A.~Prakash.
\newblock Quantum linear system algorithm for dense matrices.
\newblock \emph{Phys. Rev. Lett.}, 120:\penalty0 050502, 2018.
\newblock \doi{10.1103/PhysRevLett.120.050502}.

\bibitem[Bravo-Prieto et~al.(2020)Bravo-Prieto, Garc{\'i}a-Mart{\'i}n, and
  Latorre]{BravoPrieto2020}
C.~Bravo-Prieto, D.~Garc{\'i}a-Mart{\'i}n, and J.~I. Latorre.
\newblock Quantum singular value decomposer.
\newblock \emph{Phys. Rev. A}, 101:\penalty0 062310, 2020.
\newblock \doi{10.1103/PhysRevA.101.062310}.

\bibitem[Rebentrost et~al.(2014)Rebentrost, Mohseni, and Lloyd]{Rebentrost2014}
P.~Rebentrost, M.~Mohseni, and S.~Lloyd.
\newblock Quantum support vector machine for big data classification.
\newblock \emph{Phys. Rev. Lett.}, 113:\penalty0 130503, 2014.
\newblock \doi{10.1103/PhysRevLett.113.130503}.

\bibitem[Lloyd et~al.(2014)Lloyd, Mohseni, and Rebentrost]{Lloyd2014}
S.~Lloyd, M.~Mohseni, and P.~Rebentrost.
\newblock Quantum principal component analysis.
\newblock \emph{Nat. Phys.}, 10:\penalty0 631--633, 2014.
\newblock \doi{10.1038/nphys3029}.

\bibitem[Cong and Duan(2016)]{Cong2016}
I.~Cong and L.~Duan.
\newblock Quantum discriminant analysis for dimensionality reduction and
  classification.
\newblock \emph{New J. Phys.}, 18:\penalty0 073011, 2016.
\newblock \doi{10.1088/1367-2630/18/7/073011}.

\bibitem[Wiebe et~al.(2012)Wiebe, Braun, and Lloyd]{Wiebe2012}
N.~Wiebe, D.~Braun, and S.~Lloyd.
\newblock Quantum algorithm for data fitting.
\newblock \emph{Phys. Rev. Lett.}, 109:\penalty0 050505, 2012.
\newblock \doi{10.1103/PhysRevLett.109.050505}.

\bibitem[Wang(2017)]{Wang2017a}
G.~Wang.
\newblock Quantum algorithm for linear regression.
\newblock \emph{Phys. Rev. A}, 96:\penalty0 012335, 2017.
\newblock \doi{10.1103/PhysRevA.96.012335}.

\bibitem[Schuld et~al.(2016)Schuld, Sinayskiy, and Petruccione]{Schuld2016}
M.~Schuld, I.~Sinayskiy, and F.~Petruccione.
\newblock Prediction by linear regression on a quantum computer.
\newblock \emph{Phys. Rev. A}, 94:\penalty0 022342, 2016.
\newblock \doi{10.1103/PhysRevA.94.022342}.

\bibitem[Yu et~al.(2021)Yu, Gao, and Wen]{Yu2021a}
C.-H. Yu, F.~Gao, and Q.-Y. Wen.
\newblock An improved quantum algorithm for ridge regression.
\newblock \emph{IEEE Trans. Knowl. Data Eng.}, 33:\penalty0 858--866, 2021.
\newblock \doi{10.1109/TKDE.2019.2937491}.

\bibitem[Kerenidis and Prakash(2017)]{Kerenidis2017}
I.~Kerenidis and A.~Prakash.
\newblock Quantum recommendation systems.
\newblock In C.~H. Papadimitriou, editor, \emph{8th ITCS 2017}, volume~67 of
  \emph{LIPIcs}, pages 49:1--49:21. Schloss Dagstuhl--Leibniz-Zentrum fuer
  Informatik, Dagstuhl, Germany, 2017.
\newblock \doi{10.4230/LIPIcs.ITCS.2017.49}.

\bibitem[Duan et~al.(2018)Duan, Yuan, Liu, and Li]{Duan2018}
B.~Duan, J.~Yuan, Y.~Liu, and D.~Li.
\newblock Efficient quantum circuit for singular-value thresholding.
\newblock \emph{Phys. Rev. A}, 98:\penalty0 012308, 2018.
\newblock \doi{10.1103/PhysRevA.98.012308}.

\bibitem[Rebentrost et~al.(2018)Rebentrost, Bromley, Weedbrook, and
  Lloyd]{Rebentrost2018}
P.~Rebentrost, T.~R. Bromley, C.~Weedbrook, and S.~Lloyd.
\newblock Quantum {Hopfield} neural network.
\newblock \emph{Phys. Rev. A}, 98:\penalty0 042308, 2018.
\newblock \doi{10.1103/PhysRevA.98.042308}.

\bibitem[Rathore et~al.(2024)Rathore, Basden, Chancellor, and
  Kusumaatmaja]{rathore2024integrating}
O.~Rathore, A.~Basden, N.~Chancellor, and H.~Kusumaatmaja.
\newblock Integrating quantum algorithms into classical frameworks: A
  predictor-corrector approach using {HHL}, 2024.
\newblock \href{https://arxiv.org/abs/2406.19996}{arXiv:2406.19996}.

\bibitem[Bharadwaj and Sreenivasan(2023)]{bharadwaj2023hybrid}
S.~S. Bharadwaj and K.~R. Sreenivasan.
\newblock Hybrid quantum algorithms for flow problems.
\newblock \emph{Proc. Natl. Acad. Sci.}, 120:\penalty0 e2311014120, 2023.
\newblock \doi{10.1073/pnas.2311014120}.

\bibitem[Au-Yeung et~al.(2023)Au-Yeung, Chancellor, and Halffmann]{AuYeung2023}
R.~Au-Yeung, N.~Chancellor, and P.~Halffmann.
\newblock {NP}-hard but no longer hard to solve? {Using} quantum computing to
  tackle optimization problems.
\newblock \emph{Front. Quantum Sci. Technol.}, 2:\penalty0 1128576, 2023.
\newblock \doi{10.3389/frqst.2023.1128576}.

\bibitem[Emani et~al.(2021)Emani, Warrell, Anticevic, Bekiranov, Gandal,
  McConnell, Sapiro, Aspuru-Guzik, Baker, Bastiani, Murray, Sotiropoulos,
  Taylor, Senthil, Lehner, Gerstein, and Harrow]{Emani2021}
P.~S. Emani, J.~Warrell, A.~Anticevic, S.~Bekiranov, M.~Gandal, M.~J.
  McConnell, G.~Sapiro, A.~Aspuru-Guzik, J.~T. Baker, M.~Bastiani, J.~D.
  Murray, S.~N. Sotiropoulos, J.~Taylor, G.~Senthil, T.~Lehner, M.~B. Gerstein,
  and A.~W. Harrow.
\newblock Quantum computing at the frontiers of biological sciences.
\newblock \emph{Nat. Methods}, 18:\penalty0 701--709, 2021.
\newblock \doi{10.1038/s41592-020-01004-3}.

\bibitem[Kirkpatrick et~al.(1983)Kirkpatrick, Gelatt, and
  Vecchi]{Kirkpatrick1983}
S.~Kirkpatrick, C.~D. Gelatt, and M.~P. Vecchi.
\newblock Optimization by simulated annealing.
\newblock \emph{Science}, 220:\penalty0 671--680, 1983.
\newblock \doi{10.1126/science.220.4598.671}.

\bibitem[Finnila et~al.(1994)Finnila, Gomez, Sebenik, and Doll]{Finnila1994}
A.~B. Finnila, M.~A. Gomez, C.~Sebenik, and J.~D. Doll.
\newblock Quantum annealing: A new method for minimizing multidimensional
  functions.
\newblock \emph{Chem. Phys. Lett.}, 219:\penalty0 343, 1994.
\newblock \doi{10.1016/0009-2614(94)00117-0}.

\bibitem[Kadowaki and Nishimori(1998)]{Kadowaki1998}
T.~Kadowaki and H.~Nishimori.
\newblock Quantum annealing in the transverse {Ising} model.
\newblock \emph{Phys. Rev. E}, 58:\penalty0 5355--5363, 1998.
\newblock \doi{10.1103/PhysRevE.58.5355}.

\bibitem[Lucas(2014)]{Lucas2014}
A.~Lucas.
\newblock Ising formulations of many {NP} problems.
\newblock \emph{Front. Phys.}, 2, 2014.
\newblock \doi{10.3389/fphy.2014.00005}.

\bibitem[Choi(2011)]{Choi2011}
V.~Choi.
\newblock Different adiabatic quantum optimization algorithms for the
  {NP}-complete exact cover and {3SAT} problems.
\newblock \emph{Quantum Inf. Comput.}, 11:\penalty0 638--648, 2011.
\newblock \doi{10.26421/QIC11.7-8-7}.

\bibitem[Farhi et~al.(2001)Farhi, Goldstone, Gutmann, Lapan, Lundgren, and
  Preda]{Farhi2001}
E.~Farhi, J.~Goldstone, S.~Gutmann, J.~Lapan, A.~Lundgren, and D.~Preda.
\newblock A quantum adiabatic evolution algorithm applied to random instances
  of an {NP}-complete problem.
\newblock \emph{Science}, 292:\penalty0 472--475, 2001.
\newblock \doi{10.1126/science.1057726}.

\bibitem[Childs et~al.(2001)Childs, Farhi, and Preskill]{Childs2001}
A.~M. Childs, E.~Farhi, and J.~Preskill.
\newblock Robustness of adiabatic quantum computation.
\newblock \emph{Phys. Rev. A}, 65:\penalty0 012322, 2001.
\newblock \doi{10.1103/PhysRevA.65.012322}.

\bibitem[Aharonov et~al.(2004)Aharonov, van Dam, Kempe, Landau, Lloyd, and
  Regev]{Aharonov2004}
D.~Aharonov, W.~van Dam, J.~Kempe, Z.~Landau, S.~Lloyd, and O.~Regev.
\newblock Adiabatic quantum computation is equivalent to standard quantum
  computation.
\newblock In \emph{45th IEEE SFCS}, pages 42--51. IEEE, USA, 2004.
\newblock \doi{10.1109/FOCS.2004.8}.

\bibitem[Aharonov et~al.(2008)Aharonov, van Dam, Kempe, Landau, Lloyd, and
  Regev]{Aharonov2008}
D.~Aharonov, W.~van Dam, J.~Kempe, Z.~Landau, S.~Lloyd, and O.~Regev.
\newblock Adiabatic quantum computation is equivalent to standard quantum
  computation.
\newblock \emph{SIAM Rev.}, 50:\penalty0 755--787, 2008.
\newblock \doi{10.1137/080734479}.

\bibitem[Crosson and Lidar(2021)]{Crosson2020}
E.~J. Crosson and D.~A. Lidar.
\newblock Prospects for quantum enhancement with diabatic quantum annealing.
\newblock \emph{Nat. Rev. Phys.}, 3:\penalty0 466--489, 2021.
\newblock \doi{10.1038/s42254-021-00313-6}.

\bibitem[Farhi et~al.(2014)Farhi, Goldstone, and Gutmann]{Farhi2014}
E.~Farhi, J.~Goldstone, and S.~Gutmann.
\newblock A quantum approximate optimization algorithm, 2014.
\newblock \href{https://doi.org/10.48550/arxiv.1411.4028}{arXiv:1411.4028}.

\bibitem[Hadfield et~al.(2019)Hadfield, Wang, O'Gorman, Rieffel, Venturelli,
  and Biswas]{Hadfield2019}
S.~Hadfield, Z.~Wang, B.~O'Gorman, E.~G. Rieffel, D.~Venturelli, and R.~Biswas.
\newblock From the quantum approximate optimization algorithm to a quantum
  alternating operator ansatz.
\newblock \emph{Algorithms}, 12:\penalty0 34, 2019.
\newblock \doi{10.3390/a12020034}.

\bibitem[Banks et~al.(2024)Banks, Haque, Nazef, Fethallah, Ruqaya, Ahsan, Vora,
  Tahir, Ahmad, Hewins, Shah, Baranwal, Arora, Asad, Khan, Hasan, Azad,
  Fedaiee, Majeed, Bhuyan, Tarannum, Ali, Browne, and Warburton]{Banks2024}
R.~J. Banks, E.~Haque, F.~Nazef, F.~Fethallah, F.~Ruqaya, H.~Ahsan, H.~Vora,
  H.~Tahir, I.~Ahmad, I.~Hewins, I.~Shah, K.~Baranwal, M.~Arora, M.~Asad,
  M.~Khan, N.~Hasan, N.~Azad, S.~Fedaiee, S.~Majeed, S.~Bhuyan, T.~Tarannum,
  Y.~Ali, D.~E. Browne, and P.~A. Warburton.
\newblock Continuous-time quantum walks for {MAX-CUT} are hot.
\newblock \emph{Quantum}, 8:\penalty0 1254, 2024.
\newblock \doi{10.22331/q-2024-02-13-1254}.

\bibitem[Schulz et~al.(2024)Schulz, Willsch, and Michielsen]{Schulz2024}
S.~Schulz, D.~Willsch, and K.~Michielsen.
\newblock Guided quantum walk.
\newblock \emph{Phys. Rev. Res.}, 6:\penalty0 013312, 2024.
\newblock \doi{10.1103/PhysRevResearch.6.013312}.

\bibitem[Cerezo et~al.(2021{\natexlab{a}})Cerezo, Arrasmith, Babbush, Benjamin,
  Endo, Fujii, McClean, Mitarai, Yuan, Cincio, and Coles]{Cerezo2021}
M.~Cerezo, A.~Arrasmith, R.~Babbush, S.~C. Benjamin, S.~Endo, K.~Fujii, J.~R.
  McClean, K.~Mitarai, X.~Yuan, L.~Cincio, and P.~J. Coles.
\newblock Variational quantum algorithms.
\newblock \emph{Nat. Rev. Phys.}, 3:\penalty0 625--644, 2021{\natexlab{a}}.
\newblock \doi{10.1038/s42254-021-00348-9}.

\bibitem[Lubasch et~al.(2020)Lubasch, Joo, Moinier, Kiffner, and
  Jaksch]{Lubasch2020}
M.~Lubasch, J.~Joo, P.~Moinier, M.~Kiffner, and D.~Jaksch.
\newblock Variational quantum algorithms for nonlinear problems.
\newblock \emph{Phys. Rev. A}, 101:\penalty0 010301, 2020.
\newblock \doi{10.1103/PhysRevA.101.010301}.

\bibitem[Kyriienko et~al.(2021)Kyriienko, Paine, and Elfving]{Kyriienko2021}
O.~Kyriienko, A.~E. Paine, and V.~E. Elfving.
\newblock Solving nonlinear differential equations with differentiable quantum
  circuits.
\newblock \emph{Phys. Rev. A}, 103:\penalty0 052416, 2021.
\newblock \doi{10.1103/PhysRevA.103.052416}.

\bibitem[Kokail et~al.(2019)Kokail, Maier, van Bijnen, Brydges, Joshi,
  Jurcevic, Muschik, Silvi, Blatt, Roos, and Zoller]{Kokail2019}
C.~Kokail, C.~Maier, R.~van Bijnen, T.~Brydges, M.~K. Joshi, P.~Jurcevic, C.~A.
  Muschik, P.~Silvi, R.~Blatt, C.~F. Roos, and P.~Zoller.
\newblock Self-verifying variational quantum simulation of lattice models.
\newblock \emph{Nature}, 569:\penalty0 355--360, 2019.
\newblock \doi{10.1038/s41586-019-1177-4}.

\bibitem[Kandala et~al.(2017)Kandala, Mezzacapo, Temme, Takita, Brink, Chow,
  and Gambetta]{Kandala2017}
A.~Kandala, A.~Mezzacapo, K.~Temme, M.~Takita, M.~Brink, J.~M. Chow, and J.~M.
  Gambetta.
\newblock Hardware-efficient variational quantum eigensolver for small
  molecules and quantum magnets.
\newblock \emph{Nature}, 549:\penalty0 242--246, 2017.
\newblock \doi{10.1038/nature23879}.

\bibitem[Botelho et~al.(2022)Botelho, Glos, Kundu, Miszczak, Salehi, and
  Zimbor{\'a}s]{Botelho2022}
L.~Botelho, A.~Glos, A.~Kundu, J.~A. Miszczak, {\"O}.~Salehi, and
  Z.~Zimbor{\'a}s.
\newblock Error mitigation for variational quantum algorithms through
  mid-circuit measurements.
\newblock \emph{Phys. Rev. A}, 105:\penalty0 022441, 2022.
\newblock \doi{10.1103/PhysRevA.105.022441}.

\bibitem[Cai et~al.(2023)Cai, Babbush, Benjamin, Endo, Huggins, Li, McClean,
  and O'Brien]{Cai2023}
Z.~Cai, R.~Babbush, S.~C. Benjamin, S.~Endo, W.~J. Huggins, Y.~Li, J.~R.
  McClean, and T.~E. O'Brien.
\newblock Quantum error mitigation.
\newblock \emph{Rev. Mod. Phys.}, 95:\penalty0 045005, 2023.
\newblock \doi{10.1103/RevModPhys.95.045005}.

\bibitem[Zhou et~al.(2020{\natexlab{a}})Zhou, Wang, Choi, Pichler, and
  Lukin]{Zhou2020}
L.~Zhou, S.-T. Wang, S.~Choi, H.~Pichler, and M.~D. Lukin.
\newblock Quantum approximate optimization algorithm: Performance, mechanism,
  and implementation on near-term devices.
\newblock \emph{Phys. Rev. X}, 10:\penalty0 021067, 2020{\natexlab{a}}.
\newblock \doi{10.1103/PhysRevX.10.021067}.

\bibitem[Wang et~al.(2020)Wang, Rubin, Dominy, and Rieffel]{Wang2020}
Z.~Wang, N.~C. Rubin, J.~M. Dominy, and E.~G. Rieffel.
\newblock {$XY$} mixers: Analytical and numerical results for the quantum
  alternating operator ansatz.
\newblock \emph{Phys. Rev. A}, 101:\penalty0 012320, 2020.
\newblock \doi{10.1103/PhysRevA.101.012320}.

\bibitem[Blekos et~al.(2024)Blekos, Brand, Ceschini, Chou, Li, Pandya, and
  Summer]{Blekos2024}
K.~Blekos, D.~Brand, A.~Ceschini, C.-H. Chou, R.-H. Li, K.~Pandya, and
  A.~Summer.
\newblock A review on quantum approximate optimization algorithm and its
  variants.
\newblock \emph{Phys. Rep.}, 1068:\penalty0 1--66, 2024.
\newblock \doi{10.1016/j.physrep.2024.03.002}.

\bibitem[Jain et~al.(2022)Jain, Coyle, Kashefi, and Kumar]{Jain2022}
N.~Jain, B.~Coyle, E.~Kashefi, and N.~Kumar.
\newblock Graph neural network initialisation of quantum approximate
  optimisation.
\newblock \emph{Quantum}, 6:\penalty0 861, 2022.
\newblock \doi{10.22331/q-2022-11-17-861}.

\bibitem[Galda et~al.(2021)Galda, Liu, Lykov, Alexeev, and Safro]{Galda2021}
A.~Galda, X.~Liu, D.~Lykov, Y.~Alexeev, and I.~Safro.
\newblock Transferability of optimal {QAOA} parameters between random graphs.
\newblock In \emph{2021 QCE}, pages 171--180. IEEE, 2021.
\newblock \doi{10.1109/QCE52317.2021.00034}.

\bibitem[Moussa et~al.(2022)Moussa, Wang, B{\"a}ck, and Dunjko]{Moussa2022}
C.~Moussa, H.~Wang, T.~B{\"a}ck, and V.~Dunjko.
\newblock Unsupervised strategies for identifying optimal parameters in quantum
  approximate optimization algorithm.
\newblock \emph{EPJ Quantum Technol.}, 9:\penalty0 11, 2022.
\newblock \doi{10.1140/epjqt/s40507-022-00131-4}.

\bibitem[Cheng et~al.(2024)Cheng, Chen, Zhang, and Zhang]{Cheng2024}
L.~Cheng, Y.-Q. Chen, S.-X. Zhang, and S.~Zhang.
\newblock Quantum approximate optimization via learning-based adaptive
  optimization.
\newblock \emph{Commun. Phys.}, 7:\penalty0 83, 2024.
\newblock \doi{10.1038/s42005-024-01577-x}.

\bibitem[Alam et~al.(2020)Alam, Ash-Saki, and Ghosh]{Alam2020}
M.~Alam, A.~Ash-Saki, and S.~Ghosh.
\newblock Accelerating quantum approximate optimization algorithm using machine
  learning.
\newblock In \emph{2020 DATE}, pages 686--689. IEEE, 2020.
\newblock \doi{10.23919/DATE48585.2020.9116348}.

\bibitem[Dong et~al.(2020)Dong, Meng, Lin, Kosut, and Whaley]{Dong2020}
T.~Dong, X.~Meng, L.~Lin, R.~Kosut, and K.~B. Whaley.
\newblock Robust control optimization for quantum approximate optimization
  algorithms.
\newblock \emph{IFAC-PapersOnLine}, 53:\penalty0 242--249, 2020.
\newblock \doi{10.1016/j.ifacol.2020.12.130}.

\bibitem[Shaydulin et~al.(2021)Shaydulin, Hadfield, Hogg, and
  Safro]{Shaydulin2021}
R.~Shaydulin, S.~Hadfield, T.~Hogg, and I.~Safro.
\newblock Classical symmetries and the quantum approximate optimization
  algorithm.
\newblock \emph{Quantum Inf. Process.}, 20:\penalty0 359, 2021.
\newblock \doi{10.1007/s11128-021-03298-4}.

\bibitem[Shaydulin and Wild(2021)]{Shaydulin2021s}
R.~Shaydulin and S.~M. Wild.
\newblock Exploiting symmetry reduces the cost of training {QAOA}.
\newblock \emph{IEEE Trans. Quantum Eng.}, 2:\penalty0 1--0, 2021.
\newblock \doi{10.1109/TQE.2021.3066275}.

\bibitem[Brady et~al.(2021)Brady, Baldwin, Bapat, Kharkov, and
  Gorshkov]{Brady2021}
L.~T. Brady, C.~L. Baldwin, A.~Bapat, Y.~Kharkov, and A.~V. Gorshkov.
\newblock Optimal protocols in quantum annealing and quantum approximate
  optimization algorithm problems.
\newblock \emph{Phys. Rev. Lett.}, 126:\penalty0 070505, 2021.
\newblock \doi{10.1103/PhysRevLett.126.070505}.

\bibitem[Yang et~al.(2017)Yang, Rahmani, Shabani, Neven, and Chamon]{Yang2017}
Z.-C. Yang, A.~Rahmani, A.~Shabani, H.~Neven, and C.~Chamon.
\newblock Optimizing variational quantum algorithms using {Pontryagin's}
  minimum principle.
\newblock \emph{Phys. Rev. X}, 7:\penalty0 021027, 2017.
\newblock \doi{10.1103/PhysRevX.7.021027}.

\bibitem[Gerblich et~al.(2024)Gerblich, Dasanjh, Wong, Ross, Novo, Chancellor,
  and Kendon]{Gerblich2024}
L.~Gerblich, T.~Dasanjh, H.~Q.~X. Wong, D.~Ross, L.~Novo, N.~Chancellor, and
  V.~Kendon.
\newblock Advantages of multistage quantum walks over {QAOA}, 2024.
\newblock \href{https://arxiv.org/abs/2407.06663}{arXiv:2407.06663}.

\bibitem[McClean et~al.(2016)McClean, Romero, Babbush, and
  Aspuru-Guzik]{McClean2016}
J.~R. McClean, J.~Romero, R.~Babbush, and A.~Aspuru-Guzik.
\newblock The theory of variational hybrid quantum-classical algorithms.
\newblock \emph{New J. Phys.}, 18:\penalty0 023023, 2016.
\newblock \doi{10.1088/1367-2630/18/2/023023}.

\bibitem[Tilly et~al.(2022)Tilly, Chen, Cao, Picozzi, Setia, Li, Grant,
  Wossnig, Rungger, Booth, and Tennyson]{Tilly2022}
J.~Tilly, H.~Chen, S.~Cao, D.~Picozzi, K.~Setia, Y.~Li, E.~Grant, L.~Wossnig,
  I.~Rungger, G.~H. Booth, and J.~Tennyson.
\newblock The variational quantum eigensolver: A review of methods and best
  practices.
\newblock \emph{Phys. Rep.}, 986:\penalty0 1--128, 2022.
\newblock \doi{10.1016/j.physrep.2022.08.003}.

\bibitem[Amaro et~al.(2022)Amaro, Modica, Rosenkranz, Fiorentini, Benedetti,
  and Lubasch]{Amaro2022}
D.~Amaro, C.~Modica, M.~Rosenkranz, M.~Fiorentini, M.~Benedetti, and
  M.~Lubasch.
\newblock Filtering variational quantum algorithms for combinatorial
  optimization.
\newblock \emph{Quantum Sci. Technol.}, 7:\penalty0 015021, 2022.
\newblock \doi{10.1088/2058-9565/ac3e54}.

\bibitem[Lee et~al.(2019{\natexlab{b}})Lee, Huggins, Head-Gordon, and
  Whaley]{Lee2019}
J.~Lee, W.~J. Huggins, M.~Head-Gordon, and K.~B. Whaley.
\newblock Generalized unitary coupled cluster wave functions for quantum
  computation.
\newblock \emph{J. Chem. Theory Comput.}, 13:\penalty0 311--324,
  2019{\natexlab{b}}.
\newblock \doi{10.1021/acs.jctc.8b01004}.

\bibitem[Low and Chuang(2019)]{low2019hamiltonian}
G.~H. Low and I.~L. Chuang.
\newblock Hamiltonian simulation by qubitization.
\newblock \emph{Quantum}, 3:\penalty0 163, 2019.
\newblock \doi{10.22331/q-2019-07-12-163}.

\bibitem[Yuan et~al.(2019)Yuan, Endo, Zhao, Li, and Benjamin]{Yuan2019}
X.~Yuan, S.~Endo, Q.~Zhao, Y.~Li, and S.~C. Benjamin.
\newblock Theory of variational quantum simulation.
\newblock \emph{Quantum}, 3:\penalty0 191, 2019.
\newblock \doi{10.22331/q-2019-10-07-191}.

\bibitem[Endo et~al.(2020)Endo, Sun, Li, Benjamin, and Yuan]{Endo2020}
S.~Endo, J.~Sun, Y.~Li, S.~C. Benjamin, and X.~Yuan.
\newblock Variational quantum simulation of general processes.
\newblock \emph{Phys. Rev. Lett.}, 125:\penalty0 010501, 2020.
\newblock \doi{10.1103/PhysRevLett.125.010501}.

\bibitem[Xu et~al.(2021)Xu, Sun, Endo, Li, Benjamin, and Yuan]{Xu2021v}
X.~Xu, J.~Sun, S.~Endo, Y.~Li, S.~C. Benjamin, and X.~Yuan.
\newblock Variational algorithms for linear algebra.
\newblock \emph{Sci. Bull.}, 66:\penalty0 2181--2188, 2021.
\newblock \doi{10.1016/j.scib.2021.06.023}.

\bibitem[LeCun et~al.(2015)LeCun, Bengio, and Hinton]{LeCun2015}
Y.~LeCun, Y.~Bengio, and G.~Hinton.
\newblock Deep learning.
\newblock \emph{Nature}, 521:\penalty0 436--444, 2015.
\newblock \doi{10.1038/nature14539}.

\bibitem[Abbas et~al.(2021)Abbas, Sutter, Zoufal, Lucchi, Figalli, and
  Woerner]{Abbas2021}
A.~Abbas, D.~Sutter, C.~Zoufal, A.~Lucchi, A.~Figalli, and S.~Woerner.
\newblock The power of quantum neural networks.
\newblock \emph{Nat. Comput. Sci.}, 1:\penalty0 403--409, 2021.
\newblock \doi{10.1038/s43588-021-00084-1}.

\bibitem[Schuld and Killoran(2019)]{Schuld2019}
M.~Schuld and N.~Killoran.
\newblock Quantum machine learning in feature {Hilbert} spaces.
\newblock \emph{Phys. Rev. Lett.}, 122:\penalty0 040504, 2019.
\newblock \doi{10.1103/PhysRevLett.122.040504}.

\bibitem[Farhi and Neven(2018)]{Farhi2018}
E.~Farhi and H.~Neven.
\newblock Classification with quantum neural networks on near term processors,
  2018.
\newblock \href{https://doi.org/10.48550/arxiv.1802.06002}{arXiv:1802.06002}.

\bibitem[Cong et~al.(2019)Cong, Choi, and Lukin]{Cong2019}
I.~Cong, S.~Choi, and M.~D. Lukin.
\newblock Quantum convolutional neural networks.
\newblock \emph{Nat. Phys.}, 15:\penalty0 1273--1278, 2019.
\newblock \doi{10.1038/s41567-019-0648-8}.

\bibitem[Beer et~al.(2020)Beer, Bondarenk, Farrelly, Osborne, Salzmann,
  Scheiermann, and Wolf]{Beer2020}
K.~Beer, D.~Bondarenk, T.~Farrelly, T.~J. Osborne, R.~Salzmann, D.~Scheiermann,
  and R.~Wolf.
\newblock Training deep quantum neural networks.
\newblock \emph{Nat. Commun.}, 11:\penalty0 808, 2020.
\newblock \doi{10.1038/s41467-020-14454-2}.

\bibitem[Schuld and Petruccione(2021)]{Schuld2021}
M.~Schuld and F.~Petruccione.
\newblock \emph{Quantum Models as Kernel Methods}, chapter~6, pages 217--245.
\newblock Springer, Cham, 2021.
\newblock \doi{10.1007/978-3-030-83098-4_6}.

\bibitem[Schuld et~al.(2020)Schuld, Bocharov, Svore, and Wiebe]{Schuld2020}
M.~Schuld, A.~Bocharov, K.~M. Svore, and N.~Wiebe.
\newblock Circuit-centric quantum classifiers.
\newblock \emph{Phys. Rev. A}, 101:\penalty0 032308, 2020.
\newblock \doi{10.1103/PhysRevA.101.032308}.

\bibitem[Schuld et~al.(2021)Schuld, Sweke, and Meyer]{Schuld2021v}
M.~Schuld, R.~Sweke, and J.~J. Meyer.
\newblock Effect of data encoding on the expressive power of variational
  quantum-machine-learning models.
\newblock \emph{Phys. Rev. A}, 103:\penalty0 032430, 2021.
\newblock \doi{10.1103/PhysRevA.103.032430}.

\bibitem[Amin et~al.(2018)Amin, Andriyash, Rolfe, Kulchytskyy, and
  Melko]{Amin2018}
M.~H. Amin, E.~Andriyash, J.~Rolfe, B.~Kulchytskyy, and R.~Melko.
\newblock Quantum {Boltzmann} machine.
\newblock \emph{Phys. Rev. X}, 8:\penalty0 021050, 2018.
\newblock \doi{10.1103/PhysRevX.8.021050}.

\bibitem[Kieferov{\'a} and Wiebe(2017)]{Kieferova2017}
M.~Kieferov{\'a} and N.~Wiebe.
\newblock Tomography and generative training with quantum {Boltzmann} machines.
\newblock \emph{Phys. Rev. A}, 96:\penalty0 062327, 2017.
\newblock \doi{10.1103/PhysRevA.96.062327}.

\bibitem[Benedetti et~al.(2016)Benedetti, Realpe-G{\'o}mez, Biswas, and
  Perdomo-Ortiz]{Benedetti2016}
M.~Benedetti, J.~Realpe-G{\'o}mez, R.~Biswas, and A.~Perdomo-Ortiz.
\newblock Estimation of effective temperatures in quantum annealers for
  sampling applications: A case study with possible applications in deep
  learning.
\newblock \emph{Phys. Rev. A}, 94:\penalty0 022308, 2016.
\newblock \doi{10.1103/PhysRevA.94.022308}.

\bibitem[Xu and Oates(2021)]{Xu2021}
G.~Xu and W.~S. Oates.
\newblock Adaptive hyperparameter updating for training restricted {Boltzmann}
  machines on quantum annealers.
\newblock \emph{Sci. Rep.}, 11:\penalty0 2727, 2021.
\newblock \doi{10.1038/s41598-021-82197-1}.

\bibitem[Delilbasic et~al.(2024)Delilbasic, Saux, Riedel, Michielsen, and
  Cavallaro]{Delilbasic2024}
A.~Delilbasic, B.~Le Saux, M.~Riedel, K.~Michielsen, and G.~Cavallaro.
\newblock A single-step multiclass {SVM} based on quantum annealing for remote
  sensing data classification.
\newblock \emph{JSTARS}, 17:\penalty0 1434--1445, 2024.
\newblock \doi{10.1109/JSTARS.2023.3336926}.

\bibitem[Willsch et~al.(2020)Willsch, Willsch, Raedt, and
  Michielsen]{Willsch2020}
D.~Willsch, M.~Willsch, H.~De Raedt, and K.~Michielsen.
\newblock Support vector machines on the {D-Wave} quantum annealer.
\newblock \emph{Comput. Phys. Commun.}, 248:\penalty0 107006, 2020.
\newblock \doi{10.1016/j.cpc.2019.107006}.

\bibitem[Cerezo et~al.(2022)Cerezo, Verdon, Huang, Cincio, and
  Coles]{Cerezo2022}
M.~Cerezo, G.~Verdon, H.-Y. Huang, L.~Cincio, and P.~J. Coles.
\newblock Challenges and opportunities in quantum machine learning.
\newblock \emph{Nat. Comput. Sci.}, 2:\penalty0 567--576, 2022.
\newblock \doi{10.1038/s43588-022-00311-3}.

\bibitem[Holmes et~al.(2022)Holmes, Sharma, Cerezo, and Coles]{Holmes2022}
Z.~Holmes, K.~Sharma, M.~Cerezo, and P.~J. Coles.
\newblock Connecting ansatz expressibility to gradient magnitudes and barren
  plateaus.
\newblock \emph{PRX Quantum}, 3:\penalty0 010313, 2022.
\newblock \doi{10.1103/PRXQuantum.3.010313}.

\bibitem[Benedetti et~al.(2019)Benedetti, Lloyd, Sack, and
  Fiorentini]{Benedetti2019}
M.~Benedetti, E.~Lloyd, S.~Sack, and M.~Fiorentini.
\newblock Parameterized quantum circuits as machine learning models.
\newblock \emph{Quantum Sci. Technol.}, 4:\penalty0 043001, 2019.
\newblock \doi{10.1088/2058-9565/ab4eb5}.

\bibitem[Du et~al.(2020)Du, Hsieh, Liu, and Tao]{Du2020e}
Y.~Du, M.-H. Hsieh, T.~Liu, and D.~Tao.
\newblock Expressive power of parametrized quantum circuits.
\newblock \emph{Phys. Rev. Res.}, 2:\penalty0 033125, 2020.
\newblock \doi{10.1103/PhysRevResearch.2.033125}.

\bibitem[Du et~al.(2022{\natexlab{a}})Du, Tu, Yuan, and Tao]{Du2022e}
Y.~Du, Z.~Tu, X.~Yuan, and D.~Tao.
\newblock Efficient measure for the expressivity of variational quantum
  algorithms.
\newblock \emph{Phys. Rev. Lett.}, 128:\penalty0 080506, 2022{\natexlab{a}}.
\newblock \doi{10.1103/PhysRevLett.128.080506}.

\bibitem[McClean et~al.(2018)McClean, Boixo, Smelyanskiy, Babbush, and
  Neven]{McClean2018}
J.~R. McClean, S.~Boixo, V.~N. Smelyanskiy, R.~Babbush, and H.~Neven.
\newblock Barren plateaus in quantum neural network training landscapes.
\newblock \emph{Nat. Commun.}, 9:\penalty0 4812, 2018.
\newblock \doi{10.1038/s41467-018-07090-4}.

\bibitem[Wang et~al.(2021{\natexlab{a}})Wang, Fontana, Cerezo, Sharma, Sone,
  Cincio, and Coles]{Wang2021}
S.~Wang, E.~Fontana, M.~Cerezo, K.~Sharma, A.~Sone, L.~Cincio, and P.~J. Coles.
\newblock Noise-induced barren plateaus in variational quantum algorithms.
\newblock \emph{Nat. Commun.}, 12:\penalty0 6961, 2021{\natexlab{a}}.
\newblock \doi{10.1038/s41467-021-27045-6}.

\bibitem[Wang et~al.(2021{\natexlab{b}})Wang, Fontana, Cerezo, Sharma, Sone,
  Cincio, and Coles]{Wang2021n}
S.~Wang, E.~Fontana, M.~Cerezo, K.~Sharma, A.~Sone, L.~Cincio, and P.~J. Coles.
\newblock Noise-induced barren plateaus in variational quantum algorithms.
\newblock \emph{Nat. Commun.}, 12:\penalty0 6961, 2021{\natexlab{b}}.
\newblock \doi{10.1038/s41467-021-27045-6}.

\bibitem[Arrasmith et~al.(2022)Arrasmith, Holmes, Cerezo, and
  Coles]{Arrasmith2022}
A.~Arrasmith, Z.~Holmes, M.~Cerezo, and P.~J. Coles.
\newblock Equivalence of quantum barren plateaus to cost concentration and
  narrow gorges.
\newblock \emph{Quantum Sci. Technol.}, 7:\penalty0 045015, 2022.
\newblock \doi{10.1088/2058-9565/ac7d06}.

\bibitem[Larocca et~al.(2022{\natexlab{a}})Larocca, Czarnik, Sharma,
  Muraleedharan, Coles, and Cerezo]{Larocca2022}
M.~Larocca, P.~Czarnik, K.~Sharma, G.~Muraleedharan, P.~J. Coles, and
  M.~Cerezo.
\newblock Diagnosing barren plateaus with tools from quantum optimal control.
\newblock \emph{Quantum}, 6:\penalty0 824, 2022{\natexlab{a}}.
\newblock \doi{10.22331/q-2022-09-29-824}.

\bibitem[Holmes et~al.(2021)Holmes, Arrasmith, Yan, Coles, Albrecht, and
  Sornborger]{Holmes2021}
Z.~Holmes, A.~Arrasmith, B.~Yan, P.~J. Coles, A.~Albrecht, and A.~T.
  Sornborger.
\newblock Barren plateaus preclude learning scramblers.
\newblock \emph{Phys. Rev. Lett.}, 126:\penalty0 190501, 2021.
\newblock \doi{10.1103/PhysRevLett.126.190501}.

\bibitem[Marrero et~al.(2021)Marrero, Kieferov{\'a}, and Wiebe]{Marrero2021}
C.~O. Marrero, M.~Kieferov{\'a}, and N.~Wiebe.
\newblock Entanglement-induced barren plateaus.
\newblock \emph{PRX Quantum}, 2:\penalty0 040316, 2021.
\newblock \doi{10.1103/PRXQuantum.2.040316}.

\bibitem[Patti et~al.(2021)Patti, Najafi, Gao, and Yelin]{Patti2021}
T.~L. Patti, K.~Najafi, X.~Gao, and S.~F. Yelin.
\newblock Entanglement devised barren plateau mitigation.
\newblock \emph{Phys. Rev. Res.}, 3:\penalty0 033090, 2021.
\newblock \doi{10.1103/PhysRevResearch.3.033090}.

\bibitem[Sharma et~al.(2022)Sharma, Cerezo, Cincio, and Coles]{Sharma2022}
K.~Sharma, M.~Cerezo, L.~Cincio, and P.~J. Coles.
\newblock Trainability of dissipative perceptron-based quantum neural networks.
\newblock \emph{Phys. Rev. Lett.}, 128:\penalty0 180505, 2022.
\newblock \doi{10.1103/PhysRevLett.128.180505}.

\bibitem[Larocca et~al.(2024)Larocca, Thanasilp, Wang, Sharma, Biamonte, Coles,
  Cincio, McClean, Holmes, and Cerezo]{Larocca2024}
M.~Larocca, S.~Thanasilp, S.~Wang, K.~Sharma, J.~Biamonte, P.~J. Coles,
  L.~Cincio, J.~R. McClean, Z.~Holmes, and M.~Cerezo.
\newblock A review of barren plateaus in variational quantum computing, 2024.
\newblock \href{https://arxiv.org/abs/2405.00781}{arXiv:2405.00781}.

\bibitem[Zhang et~al.(2024)Zhang, Liu, and Zhang]{ZhangHK2024}
H.-K. Zhang, S.~Liu, and S.-X. Zhang.
\newblock Absence of barren plateaus in finite local-depth circuits with
  long-range entanglement.
\newblock \emph{Phys. Rev. Lett.}, 132:\penalty0 150603, 2024.
\newblock \doi{10.1103/PhysRevLett.132.150603}.

\bibitem[Cerezo et~al.(2021{\natexlab{b}})Cerezo, Sone, Volkoff, Cincio, and
  Coles]{Cerezo2021c}
M.~Cerezo, A.~Sone, T.~Volkoff, L.~Cincio, and P.~J. Coles.
\newblock Cost function dependent barren plateaus in shallow parametrized
  quantum circuits.
\newblock \emph{Nat. Commun.}, 12:\penalty0 1791, 2021{\natexlab{b}}.
\newblock \doi{10.1038/s41467-021-21728-w}.

\bibitem[Koczor and Benjamin(2022)]{Koczor2022}
B.~Koczor and S.~C. Benjamin.
\newblock Quantum analytic descent.
\newblock \emph{Phys. Rev. Res.}, 4:\penalty0 023017, 2022.
\newblock \doi{10.1103/PhysRevResearch.4.023017}.

\bibitem[Wang et~al.(2024)Wang, Czarnik, Arrasmith, Cerezo, Cincio, and
  Coles]{Wang2024}
S.~Wang, P.~Czarnik, A.~Arrasmith, M.~Cerezo, L.~Cincio, and P.~J. Coles.
\newblock Can error mitigation improve trainability of noisy variational
  quantum algorithms?
\newblock \emph{Quantum}, 8:\penalty0 1287, 2024.
\newblock \doi{10.22331/q-2024-03-14-1287}.

\bibitem[Endo et~al.(2021)Endo, Cai, Benjamin, and Yuan]{Endo2021}
S.~Endo, Z.~Cai, S.~C. Benjamin, and X.~Yuan.
\newblock Hybrid quantum-classical algorithms and quantum error mitigation.
\newblock \emph{J. Phys. Soc. Japan}, 90:\penalty0 032001, 2021.
\newblock \doi{10.7566/JPSJ.90.032001}.

\bibitem[Grant et~al.(2019)Grant, Wossnig, Ostaszewski, and
  Benedetti]{Grant2019}
E.~Grant, L.~Wossnig, M.~Ostaszewski, and M.~Benedetti.
\newblock An initialization strategy for addressing barren plateaus in
  parametrized quantum circuits.
\newblock \emph{Quantum}, 3:\penalty0 214, 2019.
\newblock \doi{10.22331/q-2019-12-09-214}.

\bibitem[Liu et~al.(2023)Liu, Sun, Wu, Han, and Guo]{Liu2023b}
H.-Y. Liu, T.-P. Sun, Y.-C. Wu, Y.-J. Han, and G.-P. Guo.
\newblock Mitigating barren plateaus with transfer-learning-inspired parameter
  initializations.
\newblock \emph{New J. Phys.}, 25:\penalty0 013039, 2023.
\newblock \doi{10.1088/1367-2630/acb58e}.

\bibitem[Sack et~al.(2022)Sack, Medina, Michailidis, Kueng, and
  Serbyn]{Sack2022}
S.~H. Sack, R.~A. Medina, A.~A. Michailidis, R.~Kueng, and M.~Serbyn.
\newblock Avoiding barren plateaus using classical shadows.
\newblock \emph{PRX Quantum}, 3:\penalty0 020365, 2022.
\newblock \doi{10.1103/PRXQuantum.3.020365}.

\bibitem[Ragone et~al.(2024)Ragone, Bakalov, Sauvage, Kemper, Marrero, Larocca,
  and Cerezo]{Ragone2024}
M.~Ragone, B.~N. Bakalov, F.~Sauvage, A.~F. Kemper, C.~O. Marrero, M.~Larocca,
  and M.~Cerezo.
\newblock A {Lie} algebraic theory of barren plateaus for deep parameterized
  quantum circuits.
\newblock \emph{Nat. Commun.}, 15:\penalty0 7172, 2024.
\newblock \doi{10.1038/s41467-024-49909-3}.

\bibitem[Fontana et~al.(2024)Fontana, Herman, Chakrabarti, Kumar, Yalovetzky,
  Heredge, Sureshbabu, and Pistoia]{Fontana2024}
E.~Fontana, D.~Herman, S.~Chakrabarti, N.~Kumar, R.~Yalovetzky, J.~Heredge,
  S.~H. Sureshbabu, and M.~Pistoia.
\newblock Characterizing barren plateaus in quantum ans{\"a}tze with the
  adjoint representation.
\newblock \emph{Nat. Commun.}, 15:\penalty0 7171, 2024.
\newblock \doi{10.1038/s41467-024-49910-w}.

\bibitem[Diaz et~al.(2023)Diaz, Garc{\'i}a-Mart{\'i}n, Kazi, Larocca, and
  Cerezo]{Diaz2023}
N.~L. Diaz, D.~Garc{\'i}a-Mart{\'i}n, S.~Kazi, M.~Larocca, and M.~Cerezo.
\newblock Showcasing a barren plateau theory beyond the dynamical {Lie}
  algebra, 2023.
\newblock \href{https://arxiv.org/abs/2310.11505}{arXiv:2310.11505}.

\bibitem[Zhang et~al.(2022)Zhang, Hsieh, Zhang, and Yao]{ZhangSX2022}
S.-X. Zhang, C.-Y. Hsieh, S.~Zhang, and H.~Yao.
\newblock Differentiable quantum architecture search.
\newblock \emph{Quantum Sci. Technol.}, 7:\penalty0 045023, 2022.
\newblock \doi{10.1088/2058-9565/ac87cd}.

\bibitem[Du et~al.(2022{\natexlab{b}})Du, Huang, You, Hsieh, and Tao]{Du2022}
Y.~Du, T.~Huang, S.~You, M.-H. Hsieh, and D.~Tao.
\newblock Quantum circuit architecture search for variational quantum
  algorithms.
\newblock \emph{npj Quantum Inf.}, 8:\penalty0 62, 2022{\natexlab{b}}.
\newblock \doi{10.1038/s41534-022-00570-y}.

\bibitem[Martyniuk et~al.(2024)Martyniuk, Jung, and Paschke]{Martyniuk2024}
D.~Martyniuk, J.~Jung, and A.~Paschke.
\newblock Quantum architecture search: A survey, 2024.
\newblock \href{https://arxiv.org/abs/2406.06210}{arXiv:2406.06210}.

\bibitem[Sagastizabal et~al.(2019)Sagastizabal, Bonet-Monroig, Singh, Rol,
  Bultink, Fu, Price, Ostroukh, Muthusubramanian, Bruno, Beekman, Haider,
  O'Brien, and DiCarlo]{Sagastizabal2019}
R.~Sagastizabal, X.~Bonet-Monroig, M.~Singh, M.~A. Rol, C.~C. Bultink, X.~Fu,
  C.~H. Price, V.~P. Ostroukh, N.~Muthusubramanian, A.~Bruno, M.~Beekman,
  N.~Haider, T.~E. O'Brien, and L.~DiCarlo.
\newblock Experimental error mitigation via symmetry verification in a
  variational quantum eigensolver.
\newblock \emph{Phys. Rev. A}, 100:\penalty0 010302, 2019.
\newblock \doi{10.1103/PhysRevA.100.010302}.

\bibitem[Ganzhorn et~al.(2019)Ganzhorn, Egger, Barkoutsos, Ollitrault, Salis,
  Moll, Roth, Fuhrer, Mueller, Woerner, Tavernelli, and Filipp]{Ganzhorn2019}
M.~Ganzhorn, D.~J. Egger, P.~Barkoutsos, P.~Ollitrault, G.~Salis, N.~Moll,
  M.~Roth, A.~Fuhrer, P.~Mueller, S.~Woerner, I.~Tavernelli, and S.~Filipp.
\newblock Gate-efficient simulation of molecular eigenstates on a quantum
  computer.
\newblock \emph{Phys. Rev. Appl.}, 11:\penalty0 044092, 2019.
\newblock \doi{10.1103/PhysRevApplied.11.044092}.

\bibitem[Cade et~al.(2020)Cade, Mineh, Montanaro, and Stanisic]{Cade2020}
C.~Cade, L.~Mineh, A.~Montanaro, and S.~Stanisic.
\newblock Strategies for solving the {Fermi-Hubbard} model on near-term quantum
  computers.
\newblock \emph{Phys. Rev. B}, 102:\penalty0 235122, 2020.
\newblock \doi{10.1103/PhysRevB.102.235122}.

\bibitem[Larocca et~al.(2022{\natexlab{b}})Larocca, Sauvage, Sbahi, Verdon,
  Coles, and Cerezo]{Larocca2022g}
M.~Larocca, F.~Sauvage, F.~M. Sbahi, G.~Verdon, P.~J. Coles, and M.~Cerezo.
\newblock Group-invariant quantum machine learning.
\newblock \emph{PRX Quantum}, 3:\penalty0 030341, 2022{\natexlab{b}}.
\newblock \doi{10.1103/PRXQuantum.3.030341}.

\bibitem[Meyer et~al.(2023)Meyer, Mularski, Gil-Fuster, Mele, Arzani, Wilms,
  and Eisert]{Meyer2023}
J.~J. Meyer, M.~Mularski, E.~Gil-Fuster, A.~A. Mele, F.~Arzani, A.~Wilms, and
  J.~Eisert.
\newblock Exploiting symmetry in variational quantum machine learning.
\newblock \emph{PRX Quantum}, 4:\penalty0 010328, 2023.
\newblock \doi{10.1103/PRXQuantum.4.010328}.

\bibitem[Nguyen et~al.(2024)Nguyen, Schatzki, Braccia, Ragone, Coles, Sauvage,
  Larocca, and Cerezo]{Nguyen2024}
Q.~T. Nguyen, L.~Schatzki, P.~Braccia, M.~Ragone, P.~J. Coles, F.~Sauvage,
  M.~Larocca, and M.~Cerezo.
\newblock Theory for equivariant quantum neural networks.
\newblock \emph{PRX Quantum}, 5:\penalty0 020328, 2024.
\newblock \doi{10.1103/PRXQuantum.5.020328}.

\bibitem[Koch et~al.(2022)Koch, Boscain, Calarco, Dirr, Filipp, Glaser,
  Kosloff, Montangero, Schulte-Herbr{\"u}ggen, Sugny, and Wilhelm]{Koch2022}
C.~P. Koch, U.~Boscain, T.~Calarco, G.~Dirr, S.~Filipp, S.~J. Glaser,
  R.~Kosloff, S.~Montangero, T.~Schulte-Herbr{\"u}ggen, D.~Sugny, and F.~K.
  Wilhelm.
\newblock Quantum optimal control in quantum technologies. strategic report on
  current status, visions and goals for research in {Europe}.
\newblock \emph{EPJ Quantum Technol.}, 9:\penalty0 19, 2022.
\newblock \doi{10.1140/epjqt/s40507-022-00138-x}.

\bibitem[Ansel et~al.(2024)Ansel, Dionis, Arrouas, Peaudecerf, Gu{\'e}rin,
  Gu{\'e}ry-Odelin, and Sugny]{Ansel2024}
Q.~Ansel, E.~Dionis, F.~Arrouas, B.~Peaudecerf, S.~Gu{\'e}rin,
  D.~Gu{\'e}ry-Odelin, and D.~Sugny.
\newblock Introduction to theoretical and experimental aspects of quantum
  optimal control.
\newblock \emph{J. Phys. B: At. Mol. Opt. Phys.}, 57:\penalty0 133001, 2024.
\newblock \doi{10.1088/1361-6455/ad46a5}.

\bibitem[Magann et~al.(2021)Magann, Arenz, Grace, Ho, Kosut, McClean, Rabitz,
  and Sarovar]{Magann2021}
A.~B. Magann, C.~Arenz, M.~D. Grace, T.-S. Ho, R.~L. Kosut, J.~R. McClean,
  H.~A. Rabitz, and M.~Sarovar.
\newblock From pulses to circuits and back again: A quantum optimal control
  perspective on variational quantum algorithms.
\newblock \emph{PRX Quantum}, 2:\penalty0 010101, 2021.
\newblock \doi{10.1103/PRXQuantum.2.010101}.

\bibitem[Choquette et~al.(2021)Choquette, Paolo, Barkoutsos, S{\'e}n{\'e}chal,
  Tavernelli, and Blais]{Choquette2021}
A.~Choquette, A.~Di Paolo, P.~K. Barkoutsos, D.~S{\'e}n{\'e}chal,
  I.~Tavernelli, and A.~Blais.
\newblock Quantum-optimal-control-inspired ansatz for variational quantum
  algorithms.
\newblock \emph{Phys. Rev. Res.}, 3:\penalty0 023092, 2021.
\newblock \doi{10.1103/PhysRevResearch.3.023092}.

\bibitem[Camino et~al.(2023)Camino, Buckeridge, Warburton, Kendon, and
  Woodley]{Camino2023}
B.~Camino, J.~Buckeridge, P.~A. Warburton, V.~Kendon, and S.~M. Woodley.
\newblock Quantum computing and materials science: A practical guide to
  applying quantum annealing to the configurational analysis of materials.
\newblock \emph{J. Appl. Phys.}, 133:\penalty0 221102, 2023.
\newblock \doi{10.1063/5.0151346}.

\bibitem[{D-Wave}(2024{\natexlab{a}})]{DWave_dimid}
{D-Wave}.
\newblock {D-Wave systems: Dimod}.
\newblock \url{https://docs.ocean.dwavesys.com/en/stable/docs_dimod/},
  2024{\natexlab{a}}.
\newblock Accessed: 2024-08-01.

\bibitem[{D-Wave}(2024{\natexlab{b}})]{DWave_ocean}
{D-Wave}.
\newblock {D-Wave systems: Ocean}.
\newblock \url{https://docs.ocean.dwavesys.com/}, 2024{\natexlab{b}}.
\newblock Accessed: 2024-08-01.

\bibitem[Mundada et~al.(2023)Mundada, Barbosa, Maity, Wang, Merkh, Stace,
  Nielson, Carvalho, Hush, Biercuk, and Baum]{Mundada2023}
P.~S. Mundada, A.~Barbosa, S.~Maity, Y.~Wang, T.~Merkh, T.~M. Stace,
  F.~Nielson, A.~R.~R. Carvalho, M.~Hush, M.~J. Biercuk, and Y.~Baum.
\newblock Experimental benchmarking of an automated deterministic
  error-suppression workflow for quantum algorithms.
\newblock \emph{Phys. Rev. Appl.}, 20:\penalty0 024034, 2023.
\newblock \doi{10.1103/PhysRevApplied.20.024034}.

\bibitem[Petitet et~al.(2018)Petitet, Whaley, Dongarra, and
  Cleary]{Petitet2018}
A.~Petitet, R.~C. Whaley, J.~Dongarra, and A.~Cleary.
\newblock {HPL} - a portable implementation of the high-performance {Linpack}
  benchmark for distributed-memory computers, 2018.
\newblock \url{https://netlib.org/benchmark/hpl/}.

\bibitem[Krantz et~al.(2019)Krantz, Kjaergaard, Yan, Orlando, Gustavsson, and
  Oliver]{Krantz2019}
P.~Krantz, M.~Kjaergaard, F.~Yan, T.~P. Orlando, S.~Gustavsson, and W.~D.
  Oliver.
\newblock A quantum engineer's guide to superconducting qubits.
\newblock \emph{Appl. Phys. Rev.}, 6:\penalty0 021318, 2019.
\newblock \doi{10.1063/1.5089550}.

\bibitem[Blais et~al.(2021)Blais, Grimsmo, Girvin, and Wallraff]{Blais2021}
A.~Blais, A.~L. Grimsmo, S.~M. Girvin, and A.~Wallraff.
\newblock Circuit quantum electrodynamics.
\newblock \emph{Rev. Mod. Phys.}, 93:\penalty0 025005, 2021.
\newblock \doi{10.1103/RevModPhys.93.025005}.

\bibitem[Bravyi et~al.(2022)Bravyi, Dial, Gambetta, Gil, and
  Nazarioa]{Bravyi2022}
S.~Bravyi, O.~Dial, J.~M. Gambetta, D.~Gil, and Z.~Nazarioa.
\newblock The future of quantum computing with superconducting qubits.
\newblock \emph{J. Appl. Phys.}, 132:\penalty0 160902, 2022.
\newblock \doi{10.1063/5.0082975}.

\bibitem[Blais et~al.(2020)Blais, Girvin, and Oliver]{Blais2020}
A.~Blais, S.~M. Girvin, and W.~D. Oliver.
\newblock Quantum information processing and quantum optics with circuit
  quantum electrodynamics.
\newblock \emph{Nat. Phys.}, 16:\penalty0 247--256, 2020.
\newblock \doi{10.1038/s41567-020-0806-z}.

\bibitem[Bruzewicz et~al.(2019)Bruzewicz, Chiaverini, McConnell, and
  Sage]{Bruzewicz2019}
C.~D. Bruzewicz, J.~Chiaverini, R.~McConnell, and J.~M. Sage.
\newblock Trapped-ion quantum computing: Progress and challenges.
\newblock \emph{Appl. Phys. Rev.}, 6:\penalty0 021314, 2019.
\newblock \doi{10.1063/1.5088164}.

\bibitem[Malinowski et~al.(2023)Malinowski, Allcock, and
  Ballance]{Malinowski2023}
M.~Malinowski, D.~T.~C. Allcock, and C.~J. Ballance.
\newblock How to wire a 1000-qubit trapped-ion quantum computer.
\newblock \emph{PRX Quantum}, 4:\penalty0 040313, 2023.
\newblock \doi{10.1103/PRXQuantum.4.040313}.

\bibitem[Moses et~al.(2023)Moses, Baldwin, Allman, Ancona, Ascarrunz, Barnes,
  Bartolotta, Bjork, Blanchard, Bohn, Bohnet, Brown, Burdick, Burton, Campbell,
  Campora, Carron, Chambers, Chan, Chen, Chernoguzov, Chertkov, Colina, Curtis,
  Daniel, DeCross, Deen, Delaney, Dreiling, Ertsgaard, Esposito, Estey,
  Fabrikant, Figgatt, Foltz, Foss-Feig, Francois, Gaebler, Gatterman, Gilbreth,
  Giles, Glynn, Hall, Hankin, Hansen, Hayes, Higashi, Hoffman, Horning, Hout,
  Jacobs, Johansen, Jones, Karcz, Klein, Lauria, Lee, Liefer, Lu, Lucchetti,
  Lytle, Malm, Matheny, Mathewson, Mayer, Miller, Mills, Neyenhuis, Nugent,
  Olson, Parks, Price, Price, Pugh, Ransford, Reed, Roman, Rowe, Ryan-Anderson,
  Sanders, Sedlacek, Shevchuk, Siegfried, Skripka, Spaun, Sprenkle, Stutz,
  Swallows, Tobey, Tran, Tran, Vogt, Volin, Walker, Zolot, and Pino]{Moses2023}
S.~A. Moses, C.~H. Baldwin, M.~S. Allman, R.~Ancona, L.~Ascarrunz, C.~Barnes,
  J.~Bartolotta, B.~Bjork, P.~Blanchard, M.~Bohn, J.~G. Bohnet, N.~C. Brown,
  N.~Q. Burdick, W.~C. Burton, S.~L. Campbell, J.~P. Campora, C.~Carron,
  J.~Chambers, J.~W. Chan, Y.~H. Chen, A.~Chernoguzov, E.~Chertkov, J.~Colina,
  J.~P. Curtis, R.~Daniel, M.~DeCross, D.~Deen, C.~Delaney, J.~M. Dreiling,
  C.~T. Ertsgaard, J.~Esposito, B.~Estey, M.~Fabrikant, C.~Figgatt, C.~Foltz,
  M.~Foss-Feig, D.~Francois, J.~P. Gaebler, T.~M. Gatterman, C.~N. Gilbreth,
  J.~Giles, E.~Glynn, A.~Hall, A.~M. Hankin, A.~Hansen, D.~Hayes, B.~Higashi,
  I.~M. Hoffman, B.~Horning, J.~J. Hout, R.~Jacobs, J.~Johansen, L.~Jones,
  J.~Karcz, T.~Klein, P.~Lauria, P.~Lee, D.~Liefer, S.~T. Lu, D.~Lucchetti,
  C.~Lytle, A.~Malm, M.~Matheny, B.~Mathewson, K.~Mayer, D.~B. Miller,
  M.~Mills, B.~Neyenhuis, L.~Nugent, S.~Olson, J.~Parks, G.~N. Price, Z.~Price,
  M.~Pugh, A.~Ransford, A.~P. Reed, C.~Roman, M.~Rowe, C.~Ryan-Anderson,
  S.~Sanders, J.~Sedlacek, P.~Shevchuk, P.~Siegfried, T.~Skripka, B.~Spaun,
  R.~T. Sprenkle, R.~P. Stutz, M.~Swallows, R.~I. Tobey, A.~Tran, T.~Tran,
  E.~Vogt, C.~Volin, J.~Walker, A.~M. Zolot, and J.~M. Pino.
\newblock A race-track trapped-ion quantum processor.
\newblock \emph{Phys. Rev. X}, 13:\penalty0 041052, 2023.
\newblock \doi{10.1103/PhysRevX.13.041052}.

\bibitem[Chatterjee et~al.(2021)Chatterjee, Stevenson, Franceschi, Morello,
  de~Leon, and Kuemmeth]{Chatterjee2021}
A.~Chatterjee, P.~Stevenson, S.~De Franceschi, A.~Morello, N.~P. de~Leon, and
  F.~Kuemmeth.
\newblock Semiconductor qubits in practice.
\newblock \emph{Nat. Rev. Phys.}, 3:\penalty0 157--177, 2021.
\newblock \doi{10.1038/s42254-021-00283-9}.

\bibitem[Burkard et~al.(2023)Burkard, Ladd, Pan, Nichol, and
  Petta]{Burkhard2023}
G.~Burkard, T.~D. Ladd, A.~Pan, J.~M. Nichol, and J.~R. Petta.
\newblock Semiconductor spin qubits.
\newblock \emph{Rev. Mod. Phys.}, 95:\penalty0 025003, 2023.
\newblock \doi{10.1103/RevModPhys.95.025003}.

\bibitem[Pelucchi et~al.(2022)Pelucchi, Fagas, Aharonovich, Englund, Figueroa,
  Gong, Hannes, Liu, Lu, Matsuda, Pan, Schreck, Sciarrino, Silberhorn, Wang,
  and J{\"o}ns]{Pelucchi2022}
E.~Pelucchi, G.~Fagas, I.~Aharonovich, D.~Englund, E.~Figueroa, Q.~Gong,
  H.~Hannes, J.~Liu, C.-Y. Lu, N.~Matsuda, J.-W. Pan, F.~Schreck, F.~Sciarrino,
  C.~Silberhorn, J.~Wang, and K.~D. J{\"o}ns.
\newblock The potential and global outlook of integrated photonics for quantum
  technologies.
\newblock \emph{Nat. Rev. Phys.}, 4:\penalty0 194--208, 2022.
\newblock \doi{10.1038/s42254-021-00398-z}.

\bibitem[Moody et~al.(2022)Moody, Sorger, Blumenthal, Juodawlkis, Loh,
  Sorace-Agaskar, Jones, Balram, Matthews, Laing, Davanco, Chang, Bowers,
  Quack, Galland, Aharonovich, Wolff, Schuck, Sinclair, Lon{\v c}ar,
  Komljenovic, Weld, Mookherjea, Buckley, Radulaski, Reitzenstein, Pingault,
  Machielse, Mukhopadhyay, Akimov, Zheltikov, Agarwal, Srinivasan, Lu, Tang,
  Jiang, McKenna, Safavi-Naeini, Steinhauer, Elshaari, Zwiller, Davids,
  Martinez, Gehl, Chiaverini, Mehta, Romero, Lingaraju, Weiner, Peace,
  Cernansky, Lobino, Diamanti, Vidarte, and Camacho]{Moody2022}
G.~Moody, V.~J. Sorger, D.~J. Blumenthal, P.~W. Juodawlkis, W.~Loh,
  C.~Sorace-Agaskar, A.~E. Jones, K.~C. Balram, J.~C.~F. Matthews, A.~Laing,
  M.~Davanco, L.~Chang, J.~E. Bowers, N.~Quack, C.~Galland, I.~Aharonovich,
  M.~A. Wolff, C.~Schuck, N.~Sinclair, M.~Lon{\v c}ar, T.~Komljenovic, D.~Weld,
  S.~Mookherjea, S.~Buckley, M.~Radulaski, S.~Reitzenstein, B.~Pingault,
  B.~Machielse, D.~Mukhopadhyay, A.~Akimov, A.~Zheltikov, G.~S. Agarwal,
  K.~Srinivasan, J.~Lu, H.~X. Tang, W.~Jiang, T.~P. McKenna, A.~H.
  Safavi-Naeini, S.~Steinhauer, A.~W. Elshaari, V.~Zwiller, P.~S. Davids,
  N.~Martinez, M.~Gehl, J.~Chiaverini, K.~K. Mehta, J.~Romero, N.~B. Lingaraju,
  A.~M. Weiner, D.~Peace, R.~Cernansky, M.~Lobino, E.~Diamanti, L.~T. Vidarte,
  and R.~M. Camacho.
\newblock 2022 roadmap on integrated quantum photonics.
\newblock \emph{J. Phys. Photon.}, 4:\penalty0 012501, 2022.
\newblock \doi{10.1088/2515-7647/ac1ef4}.

\bibitem[Adams et~al.(2020)Adams, Pritchard, and Shaffer]{Adams2020}
C.~S. Adams, J.~D. Pritchard, and J.~P. Shaffer.
\newblock Rydberg atom quantum technologies.
\newblock \emph{J. Phys. B: At. Mol. Opt. Phys.}, 53:\penalty0 012002, 2020.
\newblock \doi{10.1088/1361-6455/ab52ef}.

\bibitem[Surace et~al.(2020)Surace, Mazza, Giudici, Lerose, Gambassi, and
  Dalmonte]{Surace2020}
F.~M. Surace, P.~P. Mazza, G.~Giudici, A.~Lerose, A.~Gambassi, and M.~Dalmonte.
\newblock Lattice gauge theories and string dynamics in {Rydberg} atom quantum
  simulators.
\newblock \emph{Phys. Rev. X}, 10:\penalty0 021041, 2020.
\newblock \doi{10.1103/PhysRevX.10.021041}.

\bibitem[Ebadi et~al.(2021)Ebadi, Wang, Levine, Keesling, Semeghini, Omran,
  Bluvstein, Samajdar, Pichler, Ho, Choi, Sachdev, Greiner, Vuleti{\'c}, and
  Lukin]{Ebadi2021}
S.~Ebadi, T.~T. Wang, H.~Levine, A.~Keesling, G.~Semeghini, A.~Omran,
  D.~Bluvstein, R.~Samajdar, H.~Pichler, W.~W. Ho, S.~Choi, S.~Sachdev,
  M.~Greiner, V.~Vuleti{\'c}, and M.~D. Lukin.
\newblock Quantum phases of matter on a 256-atom programmable quantum
  simulator.
\newblock \emph{Nature}, 595:\penalty0 227--232, 2021.
\newblock \doi{10.1038/s41586-021-03582-4}.

\bibitem[Campbell and Deffner(2017)]{Campbell2017a}
S.~Campbell and S.~Deffner.
\newblock Trade-off between speed and cost in shortcuts to adiabaticity.
\newblock \emph{Phys. Rev. Lett.}, 118:\penalty0 100601, 2017.
\newblock \doi{10.1103/PhysRevLett.118.100601}.

\bibitem[Knill et~al.(2008)Knill, Leibfried, Reichle, Britton, Blakestad, Jost,
  Langer, Ozeri, Seidelin, and Wineland]{Knill2008}
E.~Knill, D.~Leibfried, R.~Reichle, J.~Britton, R.~B. Blakestad, J.~D. Jost,
  C.~Langer, R.~Ozeri, S.~Seidelin, and D.~J. Wineland.
\newblock Randomized benchmarking of quantum gates.
\newblock \emph{Phys. Rev. A}, 77:\penalty0 012307, 2008.
\newblock \doi{10.1103/PhysRevA.77.012307}.

\bibitem[Magesan et~al.(2011)Magesan, Gambetta, and Emerson]{Magesan2011}
E.~Magesan, J.~M. Gambetta, and J.~Emerson.
\newblock Scalable and robust randomized benchmarking of quantum processes.
\newblock \emph{Phys. Rev. Lett.}, 106:\penalty0 180504, 2011.
\newblock \doi{10.1103/PhysRevLett.106.180504}.

\bibitem[Erhard et~al.(2019)Erhard, Wallman, Postler, Meth, Stricker, Martinez,
  Schindler, Monz, Emerson, and Blatt]{Erhard2019a}
A.~Erhard, J.~J. Wallman, L.~Postler, M.~Meth, R.~Stricker, E.~A. Martinez,
  P.~Schindler, T.~Monz, J.~Emerson, and R.~Blatt.
\newblock Characterizing large-scale quantum computers via cycle benchmarking.
\newblock \emph{Nat. Commun.}, 10:\penalty0 5347, 2019.
\newblock \doi{10.1038/s41467-019-13068-7}.

\bibitem[Nielsen et~al.(2021)Nielsen, Gamble, Rudinger, Scholten, Young, and
  Blume-Kohout]{Nielsen2021}
E.~Nielsen, J.~K. Gamble, K.~Rudinger, T.~Scholten, K.~Young, and
  R.~Blume-Kohout.
\newblock Gate set tomography.
\newblock \emph{Quantum}, 5:\penalty0 557, 2021.
\newblock \doi{10.22331/q-2021-10-05-557}.

\bibitem[Huang et~al.(2024)Huang, Preskill, and Soleimanifar]{Huang2024}
H.-Y. Huang, J.~Preskill, and M.~Soleimanifar.
\newblock Certifying almost all quantum states with few single-qubit
  measurements, 2024.
\newblock \href{https://arxiv.org/abs/2404.07281}{arXiv:2404.07281}.

\bibitem[Liu et~al.(2022)Liu, Otten, Bassirianjahromi, Jiang, and
  Fefferman]{Liu2022b}
Y.~Liu, M.~Otten, R.~Bassirianjahromi, L.~Jiang, and B.~Fefferman.
\newblock Benchmarking near-term quantum computers via random circuit sampling,
  2022.
\newblock \href{https://arxiv.org/abs/2105.05232}{arXiv:2105.05232}.

\bibitem[Proctor et~al.(2019)Proctor, Carignan-Dugas, Rudinger, Nielsen,
  Blume-Kohout, and Young]{Proctor2019}
T.~J. Proctor, A.~Carignan-Dugas, K.~Rudinger, E.~Nielsen, R.~Blume-Kohout, and
  K.~Young.
\newblock Direct randomized benchmarking for multiqubit devices.
\newblock \emph{Phys. Rev. Lett.}, 123:\penalty0 030503, 2019.
\newblock \doi{10.1103/PhysRevLett.123.030503}.

\bibitem[Proctor et~al.(2022{\natexlab{a}})Proctor, Seritan, Rudinger, Nielsen,
  Blume-Kohout, and Young]{Proctor2021}
T.~Proctor, S.~Seritan, K.~Rudinger, E.~Nielsen, R.~Blume-Kohout, and K.~Young.
\newblock Scalable randomized benchmarking of quantum computers using mirror
  circuits.
\newblock \emph{Phys. Rev. Lett.}, 129:\penalty0 150502, 2022{\natexlab{a}}.
\newblock \doi{10.1103/PhysRevLett.129.150502}.

\bibitem[Mayer et~al.(2021)Mayer, Hall, Gatterman, Halit, Lee, Bohnet, Gresh,
  Hankin, Gilmore, Gerber, and Gaebler]{Mayer2021}
K.~Mayer, A.~Hall, T.~Gatterman, S.~K. Halit, K.~Lee, J.~Bohnet, D.~Gresh,
  A.~Hankin, K.~Gilmore, J.~Gerber, and J.~Gaebler.
\newblock Theory of mirror benchmarking and demonstration on a quantum
  computer, 2021.
\newblock \href{https://arxiv.org/abs/2108.10431}{arXiv:2108.10431}.

\bibitem[Cross et~al.(2019)Cross, Bishop, Sheldon, Nation, and
  Gambetta]{Cross2019}
A.~W. Cross, L.~S. Bishop, S.~Sheldon, P.~D. Nation, and J.~M. Gambetta.
\newblock Validating quantum computers using randomized model circuits.
\newblock \emph{Phys. Rev. A}, 100:\penalty0 032328, 2019.
\newblock \doi{10.1103/PhysRevA.100.032328}.

\bibitem[Proctor et~al.(2022{\natexlab{b}})Proctor, Rudinger, Young, Nielsen,
  and Blume-Kohout]{Proctor2022}
T.~Proctor, K.~Rudinger, K.~Young, E.~Nielsen, and R.~Blume-Kohout.
\newblock Measuring the capabilities of quantum computers.
\newblock \emph{Nat. Phys.}, 18:\penalty0 75--79, 2022{\natexlab{b}}.
\newblock \doi{10.1038/s41567-021-01409-7}.

\bibitem[Mills et~al.(2021)Mills, Sivarajah, Scholten, and Duncan]{Mills2021}
D.~Mills, S.~Sivarajah, T.~L. Scholten, and R.~Duncan.
\newblock Application-motivated, holistic benchmarking of a full quantum
  computing stack.
\newblock \emph{Quantum}, 5:\penalty0 415, 2021.
\newblock \doi{10.22331/q-2021-03-22-415}.

\bibitem[Quetschlich et~al.(2023)Quetschlich, Burgholzer, and
  Wille]{Quetschlich2023}
N.~Quetschlich, L.~Burgholzer, and R.~Wille.
\newblock {MQT Bench}: Benchmarking software and design automation tools for
  quantum computing.
\newblock \emph{Quantum}, 7:\penalty0 1062, 2023.
\newblock \doi{10.22331/q-2023-07-20-1062}.

\bibitem[Li et~al.(2022{\natexlab{b}})Li, Stein, Krishnamoorthy, and
  Ang]{Li2022b}
A.~Li, S.~Stein, S.~Krishnamoorthy, and J.~Ang.
\newblock {QASMBench}: A low-level {QASM} benchmark suite for {NISQ} evaluation
  and simulation.
\newblock \emph{ACM Trans. Quantum Comput.}, 4:\penalty0 1--26,
  2022{\natexlab{b}}.
\newblock \doi{10.1145/3550488}.

\bibitem[Michielsen et~al.(2017)Michielsen, Nocon, Willsch, Jin, Lippert, and
  Raedt]{Michielsen2017}
K.~Michielsen, M.~Nocon, D.~Willsch, F.~Jin, T.~Lippert, and H.~De Raedt.
\newblock Benchmarking gate-based quantum computers.
\newblock \emph{Comput. Phys. Commun.}, 220:\penalty0 44--55, 2017.
\newblock \doi{10.1016/j.cpc.2017.06.011}.

\bibitem[Derbyshire et~al.(2020)Derbyshire, Malo, Daley, Kashefi, and
  Wallden]{Derbyshire2020}
E~Derbyshire, J.~Yago Malo, A.~J. Daley, E.~Kashefi, and P.~Wallden.
\newblock Randomized benchmarking in the analogue setting.
\newblock \emph{Quantum Sci. Technol.}, 5:\penalty0 034001, 2020.
\newblock \doi{10.1088/2058-9565/ab7eec}.

\bibitem[Georgopoulos et~al.(2021)Georgopoulos, Emary, and
  Zuliani]{Georgopoulos2021}
K.~Georgopoulos, C.~Emary, and P.~Zuliani.
\newblock Quantum computer benchmarking via quantum algorithms, 2021.
\newblock \href{https://arxiv.org/abs/2112.09457}{arXiv:2112.09457}.

\bibitem[Lubinski et~al.(2023)Lubinski, Johri, Varosy, Coleman, Zhao, Necaise,
  Baldwin, Mayer, and Proctor]{Lubinski2023}
T.~Lubinski, S.~Johri, P.~Varosy, J.~Coleman, L.~Zhao, J.~Necaise, C.~H.
  Baldwin, K.~Mayer, and T.~Proctor.
\newblock Application-oriented performance benchmarks for quantum computing.
\newblock \emph{IEEE Trans. Quantum Eng.}, 4:\penalty0 1--32, 2023.
\newblock \doi{10.1109/TQE.2023.3253761}.

\bibitem[Donkers et~al.(2022)Donkers, Mesman, Al-Ars, and
  M{\"o}ller]{Donkers2022}
H.~Donkers, K.~Mesman, Z.~Al-Ars, and M.~M{\"o}ller.
\newblock {QPack Scores}: Quantitative performance metrics for
  application-oriented quantum computer benchmarking, 2022.
\newblock \href{https://arxiv.org/abs/2205.12142}{arXiv:2205.12142}.

\bibitem[Dallaire-Demers et~al.(2020)Dallaire-Demers, St{\k{e}}ch{\l}y,
  Gonthier, Bashige, Romero, and Cao]{DallaireDemers2020}
P.~L. Dallaire-Demers, M.~St{\k{e}}ch{\l}y, J.~F. Gonthier, N.~T. Bashige,
  J.~Romero, and Y.~Cao.
\newblock An application benchmark for fermionic quantum simulations, 2020.
\newblock \href{https://arxiv.org/abs/2003.01862}{arXiv:2003.01862}.

\bibitem[amd J.~Bausch and Gily{\'e}n(2021)]{Cornelissen2021}
A.~Cornelissen amd J.~Bausch and A.~Gily{\'e}n.
\newblock Scalable benchmarks for gate-based quantum computers, 2021.
\newblock \href{https://arxiv.org/abs/2104.10698}{arXiv:2104.10698}.

\bibitem[Tomesh et~al.(2022)Tomesh, Gokhale, Omole, Ravi, Smith, Viszlai, Wu,
  Hardavellas, Martonosi, and Chong]{Tomesh2022}
T.~Tomesh, P.~Gokhale, V.~Omole, G.~S. Ravi, K.~N. Smith, J.~Viszlai, X.~C. Wu,
  N.~Hardavellas, M.~R. Martonosi, and F.~T. Chong.
\newblock {SupermarQ}: A scalable quantum benchmark suite.
\newblock In \emph{2022 IEEE HPCA}, pages 587--603, 2022.
\newblock \doi{10.1109/HPCA53966.2022.00050}.

\bibitem[Murali et~al.(2019)Murali, Linke, Martonosi, Abhari, Nguyen, and
  Alderete]{Murali2019}
P.~Murali, N.~M. Linke, M.~Martonosi, A.~J. Abhari, N.~H. Nguyen, and C.~H.
  Alderete.
\newblock Full-stack, real-system quantum computer studies: architectural
  comparisons and design insights.
\newblock In \emph{ISCA 2019}, pages 527--540, New York, NY, 2019. ACM.
\newblock \doi{10.1145/3307650.3322273}.

\bibitem[van Dam et~al.(2006)van Dam, Hallgren, and Ip]{vanDam2006}
W.~van Dam, S.~Hallgren, and L.~Ip.
\newblock Quantum algorithms for some hidden shift problems.
\newblock \emph{SIAM J. Comput.}, 36:\penalty0 763--778, 2006.
\newblock \doi{10.1137/S009753970343141X}.

\bibitem[Blume-Kohout and Young(2020)]{BlumeKohout2020}
R.~Blume-Kohout and K.~C. Young.
\newblock A volumetric framework for quantum computer benchmarks.
\newblock \emph{Quantum}, 4:\penalty0 362, 2020.
\newblock \doi{10.22331/q-2020-11-15-362}.

\bibitem[Groen et~al.(2022)Groen, Arabnejad, Suleimenova, Edeling, Raffin, Xue,
  Bronik, Monnier, and Coveney]{Groen2022}
D.~Groen, H.~Arabnejad, D.~Suleimenova, W.~Edeling, E.~Raffin, Y.~Xue,
  K.~Bronik, N.~Monnier, and P.~V. Coveney.
\newblock {FabSim3}: An automation toolkit for verified simulations using high
  performance computing.
\newblock \emph{Comput. Phys. Commun.}, 283:\penalty0 108596, 2022.
\newblock \doi{10.1016/j.cpc.2022.108596}.

\bibitem[Ferracin et~al.(2019)Ferracin, Kapourniotis, and Datta]{Ferracin2019}
S.~Ferracin, T.~Kapourniotis, and A.~Datta.
\newblock Accrediting outputs of noisy intermediate-scale quantum computing
  devices.
\newblock \emph{New J. Phys.}, 21:\penalty0 113038, 2019.
\newblock \doi{10.1088/1367-2630/ab4fd6}.

\bibitem[Ferracin et~al.(2021)Ferracin, Merkel, McKay, and Datta]{Ferracin2021}
S.~Ferracin, S.~T. Merkel, D.~McKay, and A.~Datta.
\newblock Experimental accreditation of outputs of noisy quantum computers.
\newblock \emph{Phys. Rev. A}, 104:\penalty0 042603, 2021.
\newblock \doi{10.1103/PhysRevA.104.042603}.

\bibitem[Jackson et~al.(2024)Jackson, Kapourniotis, and Datta]{Jackson2024}
A.~Jackson, T.~Kapourniotis, and A.~Datta.
\newblock Accreditation of analogue quantum simulators.
\newblock \emph{Proc. Natl. Acad. Sci.}, 121:\penalty0 e2309627121, 2024.
\newblock \doi{10.1073/pnas.2309627121}.

\bibitem[Miller et~al.(2022)Miller, Broomfield, Cox, Kinast, and
  Rodenburg]{Miller2022}
K.~Miller, C.~Broomfield, A.~Cox, J.~Kinast, and B.~Rodenburg.
\newblock An improved volumetric metric for quantum computers via more
  representative quantum circuit shapes, 2022.
\newblock \href{https://arxiv.org/abs/2207.02315}{arXiv:2207.02315}.

\bibitem[Wack et~al.(2021)Wack, Paik, Javadi-Abhari, Jurcevic, Faro, Gambetta,
  and Johnson]{Wack2021}
A.~Wack, H.~Paik, A.~Javadi-Abhari, P.~Jurcevic, I.~Faro, J.~M. Gambetta, and
  B.~R. Johnson.
\newblock Quality, speed, and scale: three key attributes to measure the
  performance of near-term quantum computers, 2021.
\newblock \href{https://doi.org/10.48550/arxiv.2110.14108}{arXiv:2110.14108}.

\bibitem[McKay et~al.(2023)McKay, Hincks, Pritchett, Carroll, Govia, and
  Merkel]{McKay2023}
D.~C. McKay, I.~Hincks, E.~J. Pritchett, M.~Carroll, L.~C.~G. Govia, and S.~T.
  Merkel.
\newblock Benchmarking quantum processor performance at scale, 2023.
\newblock \href{https://arxiv.org/abs/2311.05933}{arXiv.2311.05933}.

\bibitem[McGeoch(2019)]{McGeoch2019}
C.~C. McGeoch.
\newblock Principles and guidelines for quantum performance analysis.
\newblock In S.~Feld and C.~Linnhoff-Popien, editors, \emph{QTOP 2019}, pages
  36--48, Cham, 2019. Springer.
\newblock \doi{10.1007/978-3-030-14082-3_4}.

\bibitem[Liu and Yin(2022)]{Liu2022}
W.~Liu and H.-W. Yin.
\newblock A quantum scheme of state overlap based on quantum mean estimation
  and support vector machine.
\newblock \emph{Phys. A: Stat. Mech. Appl.}, 606:\penalty0 128117, 2022.
\newblock \doi{10.1016/j.physa.2022.128117}.

\bibitem[Foss-Feig et~al.(2021)Foss-Feig, Hayes, Dreiling, Figgatt, Gaebler,
  Moses, Pino, and Potter]{FossFeig2021}
M.~Foss-Feig, D.~Hayes, J.~M. Dreiling, C.~Figgatt, J.~P. Gaebler, S.~A. Moses,
  J.~M. Pino, and A.~C. Potter.
\newblock Holographic quantum algorithms for simulating correlated spin
  systems.
\newblock \emph{Phys. Rev. Research}, 3:\penalty0 033002, 2021.
\newblock \doi{10.1103/PhysRevResearch.3.033002}.

\bibitem[Foss-Feig et~al.(2022)Foss-Feig, Ragole, Potter, Dreiling, Figgatt,
  Gaebler, Hall, Moses, Pino, Spaun, Neyenhuis, and Hayes]{FossFeig2022}
M.~Foss-Feig, S.~Ragole, A.~Potter, J.~Dreiling, C.~Figgatt, J.~Gaebler,
  A.~Hall, S.~Moses, J.~Pino, B.~Spaun, B.~Neyenhuis, and D.~Hayes.
\newblock Entanglement from tensor networks on a trapped-ion quantum computer.
\newblock \emph{Phys. Rev. Lett.}, 128:\penalty0 150504, 2022.
\newblock \doi{10.1103/PhysRevLett.128.150504}.

\bibitem[Villalonga et~al.(2019)Villalonga, Boixo, Nelson, Henze, Rieffel,
  Biswas, and Mandr{\`a}]{Villalonga2019}
B.~Villalonga, S.~Boixo, B.~Nelson, C.~Henze, E.~Rieffel, R.~Biswas, and
  S.~Mandr{\`a}.
\newblock A flexible high-performance simulator for verifying and benchmarking
  quantum circuits implemented on real hardware.
\newblock \emph{npj Quantum Inf.}, 5:\penalty0 86, 2019.
\newblock \doi{10.1038/s41534-019-0196-1}.

\bibitem[Boixo et~al.(2018)Boixo, Isakov, Smelyanskiy, Babbush, Ding, Jiang,
  Bremner, Martinis, and Neven]{Boixo2018}
S.~Boixo, S.~V. Isakov, V.~N. Smelyanskiy, R.~Babbush, N.~Ding, Z.~Jiang, M.~J.
  Bremner, J.~M. Martinis, and H.~Neven.
\newblock Characterizing quantum supremacy in near-term devices.
\newblock \emph{Nat. Phys.}, 14:\penalty0 595--600, 2018.
\newblock \doi{10.1038/s41567-018-0124-x}.

\bibitem[Bouland et~al.(2019)Bouland, Fefferman, Nirkhe, and
  Vazirani]{Bouland2019}
A.~Bouland, B.~Fefferman, C.~Nirkhe, and U.~Vazirani.
\newblock On the complexity and verification of quantum random circuit
  sampling.
\newblock \emph{Nat. Phys.}, 15:\penalty0 159--163, 2019.
\newblock \doi{10.1038/s41567-018-0318-2}.

\bibitem[Pan and Zhang(2022)]{Pan2022}
F.~Pan and P.~Zhang.
\newblock Simulation of quantum circuits using the big-batch tensor network
  method.
\newblock \emph{Phys. Rev. Lett.}, 128:\penalty0 030501, 2022.
\newblock \doi{10.1103/PhysRevLett.128.030501}.

\bibitem[Pan et~al.(2022)Pan, Chen, and Zhang]{Pan2022s}
F.~Pan, K.~Chen, and P.~Zhang.
\newblock Solving the sampling problem of the {Sycamore} quantum circuits.
\newblock \emph{Phys. Rev. Lett.}, 129:\penalty0 090502, 2022.
\newblock \doi{10.1103/PhysRevLett.129.090502}.

\bibitem[Zhang et~al.(2023)Zhang, Allcock, Wan, Liu, Sun, Yu, Yang, Qiu, Ye,
  Chen, Lee, Zheng, Jian, Yao, Hsieh, and Zhang]{ZhangSX2023}
S.-X. Zhang, J.~Allcock, Z.-Q. Wan, S.~Liu, J.~Sun, H.~Yu, X.-H. Yang, J.~Qiu,
  Z.~Ye, Y.-Q. Chen, C.-K. Lee, Y.-C. Zheng, S.-K. Jian, H.~Yao, C.-Y. Hsieh,
  and S.~Zhang.
\newblock {TensorCircuit}: a quantum software framework for the {NISQ} era.
\newblock \emph{Quantum}, 7:\penalty0 912, 2023.
\newblock \doi{10.22331/q-2023-02-02-912}.

\bibitem[Zhou et~al.(2020{\natexlab{b}})Zhou, Stoudenmire, and
  Waintal]{Zhou2020x}
Y.~Zhou, E.~M. Stoudenmire, and X.~Waintal.
\newblock What limits the simulation of quantum computers?
\newblock \emph{Phys. Rev. X}, 10:\penalty0 041038, 2020{\natexlab{b}}.
\newblock \doi{10.1103/PhysRevX.10.041038}.

\bibitem[Wahl and Strelchuk(2023)]{Wahl2023}
T.~B. Wahl and S.~Strelchuk.
\newblock Simulating quantum circuits using efficient tensor network
  contraction algorithms with subexponential upper bound.
\newblock \emph{Phys. Rev. Lett.}, 131:\penalty0 180601, 2023.
\newblock \doi{10.1103/PhysRevLett.131.180601}.

\bibitem[Huang et~al.(2021)Huang, Zhang, Newman, Ni, Ding, Cai, Gao, Wang, Wu,
  Zhang, Ku, Tian, Wu, Xu, Yu, Yuan, Szegedy, Shi, Zhao, Deng, and
  Chen]{Huang2021}
C.~Huang, F.~Zhang, M.~Newman, X.~Ni, D.~Ding, J.~Cai, X.~Gao, T.~Wang, F.~Wu,
  G.~Zhang, H.-S. Ku, Z.~Tian, J.~Wu, H.~Xu, H.~Yu, B.~Yuan, M.~Szegedy,
  Y.~Shi, H.-H. Zhao, C.~Deng, and J.~Chen.
\newblock Efficient parallelization of tensor network contraction for
  simulating quantum computation.
\newblock \emph{Nat. Comput. Sci.}, 1:\penalty0 578--587, 2021.
\newblock \doi{10.1038/s43588-021-00119-7}.

\bibitem[Or{\'u}s(2014)]{Orus2014}
R.~Or{\'u}s.
\newblock A practical introduction to tensor networks: Matrix product states
  and projected entangled pair states.
\newblock \emph{Ann. Phys.}, 349:\penalty0 117--158, 2014.
\newblock \doi{10.1016/j.aop.2014.06.013}.

\bibitem[Or{\'u}s(2019)]{Orus2019t}
R.~Or{\'u}s.
\newblock Tensor networks for complex quantum systems.
\newblock \emph{Nat. Rev. Phys.}, 1:\penalty0 538--550, 2019.
\newblock \doi{10.1038/s42254-019-0086-7}.

\bibitem[Ran et~al.(2021)Ran, Tirrito, Peng, Chen, Tagliacozzo, Su, and
  Lewenstein]{Ran2021}
S.-J. Ran, E.~Tirrito, C.~Peng, X.~Chen, L.~Tagliacozzo, G.~Su, and
  M.~Lewenstein.
\newblock \emph{Tensor Network Contractions: Methods and Applications to
  Quantum Many-Body Systems}.
\newblock Springer, Cham, 2021.
\newblock \doi{10.1007/978-3-030-34489-4}.

\bibitem[Tang(2022)]{Tang2022}
E.~Tang.
\newblock Dequantizing algorithms to understand quantum advantage in machine
  learning.
\newblock \emph{Nat. Rev. Phys.}, 4:\penalty0 692--693, 2022.
\newblock \doi{10.1038/s42254-022-00511-w}.

\bibitem[Tang(2019)]{Tang2019}
E.~Tang.
\newblock A quantum-inspired classical algorithm for recommendation systems.
\newblock In \emph{Proc. 51st ACM SIGACT STOC}, pages 217--228. ACM, New York,
  NY, 2019.
\newblock \doi{10.1145/3313276.3316310}.

\bibitem[Zhong et~al.(2020)Zhong, Wang, Deng, Chen, Peng, Luo, Qin, Wu, Ding,
  Hu, Hu, Yang, Zhang, Li, Li, Jiang, Gan, Yang, You, Wang, Li, Liu, Lu, and
  Pan]{Zhong2020}
H.-S. Zhong, H.~Wang, Y.-H. Deng, M.-C. Chen, L.-C. Peng, Y.-H. Luo, J.~Qin,
  D.~Wu, X.~Ding, Y.~Hu, P.~Hu, X.-Y. Yang, W.-J. Zhang, H.~Li, Y.~Li,
  X.~Jiang, L.~Gan, G.~Yang, L.~You, Z.~Wang, L.~Li, N.-L. Liu, C.-Y. Lu, and
  J.-W. Pan.
\newblock Quantum computational advantage using photons.
\newblock \emph{Science}, 370:\penalty0 1460--1463, 2020.
\newblock \doi{10.1126/science.abe8770}.

\bibitem[Bulmer et~al.(2022)Bulmer, Bell, Chadwick, Jones, Moise, Rigazzi,
  Thorbecke, Haus, Vaerenbergh, Patel, Walmsley, and Laing]{Bulmer2022}
J.~F.~F. Bulmer, B.~A. Bell, R.~S. Chadwick, A.~E. Jones, D.~Moise, A.~Rigazzi,
  J.~Thorbecke, U.-U. Haus, T.~Van Vaerenbergh, R.~B. Patel, I.~A. Walmsley,
  and A.~Laing.
\newblock The boundary for quantum advantage in {Gaussian} boson sampling.
\newblock \emph{Sci. Adv.}, 8:\penalty0 eabl9236, 2022.
\newblock \doi{10.1126/sciadv.abl9236}.

\bibitem[Oh et~al.(2024)Oh, Liu, Alexeev, Fefferman, and Jiang]{Oh2024}
C.~Oh, M.~Liu, Y.~Alexeev, B.~Fefferman, and L.~Jiang.
\newblock Classical algorithm for simulating experimental {Gaussian} boson
  sampling.
\newblock \emph{Nat. Phys.}, 20:\penalty0 1461--1468, 2024.
\newblock \doi{10.1038/s41567-024-02535-8}.

\bibitem[Bouland et~al.(2022)Bouland, Fefferman, Landau, and Liu]{Bouland2021}
A.~Bouland, B.~Fefferman, Z.~Landau, and Y.~Liu.
\newblock Noise and the frontier of quantum supremacy.
\newblock In \emph{IEEE 62nd FOCS}, pages 1308--1317, 2022.
\newblock \doi{10.1109/FOCS52979.2021.00127}.

\bibitem[Ryan-Anderson et~al.(2021)Ryan-Anderson, Bohnet, Lee, Gresh, Hankin,
  Gaebler, Francois, Chernoguzov, Lucchetti, Brown, Gatterman, Halit, Gilmore,
  Gerber, Neyenhuis, Hayes, and Stutz]{RyanAnderson2021}
C.~Ryan-Anderson, J.~G. Bohnet, K.~Lee, D.~Gresh, A.~Hankin, J.~P. Gaebler,
  D.~Francois, A.~Chernoguzov, D.~Lucchetti, N.~C. Brown, T.~M. Gatterman,
  S.~K. Halit, K.~Gilmore, J.~A. Gerber, B.~Neyenhuis, D.~Hayes, and R.~P.
  Stutz.
\newblock Realization of real-time fault-tolerant quantum error correction.
\newblock \emph{Phys. Rev. X}, 11:\penalty0 041058, 2021.
\newblock \doi{10.1103/PhysRevX.11.041058}.

\bibitem[Sivak et~al.(2023)Sivak, Eickbusch, Royer, Singh, Tsioutsios, Ganjam,
  Miano, Brock, Ding, Frunzio, Girvin, Schoelkopf, and Devoret]{Sivak2023}
V.~V. Sivak, A.~Eickbusch, B.~Royer, S.~Singh, I.~Tsioutsios, S.~Ganjam,
  A.~Miano, B.~L. Brock, A.~Z. Ding, L.~Frunzio, S.~M. Girvin, R.~J.
  Schoelkopf, and M.~H. Devoret.
\newblock Real-time quantum error correction beyond break-even.
\newblock \emph{Nature}, 616:\penalty0 50--55, 2023.
\newblock \doi{10.1038/s41586-023-05782-6}.

\bibitem[Weimer et~al.(2021)Weimer, Kshetrimayum, and Or{\'u}s]{Weimer2021}
H.~Weimer, A.~Kshetrimayum, and R.~Or{\'u}s.
\newblock Simulation methods for open quantum many-body systems.
\newblock \emph{Rev. Mod. Phys.}, 93:\penalty0 015008, 2021.
\newblock \doi{10.1103/RevModPhys.93.015008}.

\bibitem[Daley et~al.(2022)Daley, Bloch, Kokail, Flannigan, Pearson, Troyer,
  and Zoller]{Daley2022}
A.~J. Daley, I.~Bloch, C.~Kokail, S.~Flannigan, N.~Pearson, M.~Troyer, and
  P.~Zoller.
\newblock Practical quantum advantage in quantum simulation.
\newblock \emph{Nature}, 607:\penalty0 667--676, 2022.
\newblock \doi{10.1038/s41586-022-04940-6}.

\bibitem[Trotzky et~al.(2012)Trotzky, Chen, Flesch, McCulloch, Schollw{\"o}ck,
  Eisert, and Bloch]{Trotzky2012}
S.~Trotzky, Y.~A. Chen, A.~Flesch, I.~P. McCulloch, U.~Schollw{\"o}ck,
  J.~Eisert, and I.~Bloch.
\newblock Probing the relaxation towards equilibrium in an isolated strongly
  correlated one-dimensional {Bose} gas.
\newblock \emph{Nat. Phys.}, 8:\penalty0 325--330, 2012.
\newblock \doi{10.1038/nphys2232}.

\bibitem[Choi et~al.(2016)Choi, Hild, Zeiher, Schau\ss, Rubio-Abadal, Yefsah,
  Khemani, Huse, Bloch, and Gross]{Choi2016}
J.-Y. Choi, S.~Hild, J.~Zeiher, P.~Schau\ss, A.~Rubio-Abadal, T.~Yefsah,
  V.~Khemani, D.~A. Huse, I.~Bloch, and C.~Gross.
\newblock Exploring the many-body localization transition in two dimensions.
\newblock \emph{Science}, 352:\penalty0 1547--1552, 2016.
\newblock \doi{10.1126/science.aaf8834}.

\bibitem[Arg{\"u}ello-Luengo et~al.(2019)Arg{\"u}ello-Luengo,
  Gonz{\'a}lez-Tudela, Shi, Zoller, and Cirac]{ArguelloLuengo2019}
J.~Arg{\"u}ello-Luengo, A.~Gonz{\'a}lez-Tudela, T.~Shi, P.~Zoller, and J.~I.
  Cirac.
\newblock Analogue quantum chemistry simulation.
\newblock \emph{Nature}, 574:\penalty0 215--218, 2019.
\newblock \doi{10.1038/s41586-019-1614-4}.

\bibitem[Somaroo et~al.(1999)Somaroo, Tseng, Havel, Laflamme, and
  Cory]{Somaroo1999}
S.~Somaroo, C.~H. Tseng, T.~F. Havel, R.~Laflamme, and D.~G. Cory.
\newblock Quantum simulations on a quantum computer.
\newblock \emph{Phys. Rev. Lett.}, 82:\penalty0 5381--5384, 1999.
\newblock \doi{10.1103/PhysRevLett.82.5381}.

\bibitem[Hatano and Suzuki(2005)]{Hatano2005}
N.~Hatano and M.~Suzuki.
\newblock Finding exponential product formulas of higher orders.
\newblock In A.~Das and B.~K. Chakrabarti, editors, \emph{Quantum Annealing and
  Other Optimization Methods}, chapter~2, pages 37--68. Springer, Berlin,
  Heidelberg, 2005.
\newblock \doi{10.1007/11526216_2}.

\bibitem[Tacchino et~al.(2020)Tacchino, Chiesa, Carretta, and
  Gerace]{Tacchino2020}
F.~Tacchino, A.~Chiesa, S.~Carretta, and D.~Gerace.
\newblock Quantum computers as universal quantum simulators: State-of-the-art
  and perspectives.
\newblock \emph{Adv. Quantum Technol.}, 3:\penalty0 1900052, 2020.
\newblock \doi{10.1002/qute.201900052}.

\bibitem[Benedetti et~al.(2021)Benedetti, Fiorentini, and
  Lubasch]{Benedetti2021}
M.~Benedetti, M.~Fiorentini, and M.~Lubasch.
\newblock Hardware-efficient variational quantum algorithms for time evolution.
\newblock \emph{Phys. Rev. Res.}, 3:\penalty0 033083, 2021.
\newblock \doi{10.1103/PhysRevResearch.3.033083}.

\bibitem[McKeever and Lubasch(2023)]{McKeever2023}
C.~McKeever and M.~Lubasch.
\newblock Classically optimized {Hamiltonian} simulation.
\newblock \emph{Phys. Rev. Res.}, 5:\penalty0 023146, 2023.
\newblock \doi{10.1103/PhysRevResearch.5.023146}.

\bibitem[Foulkes et~al.(2001)Foulkes, Mitas, Needs, and Rajagopal]{Foulkes2001}
W.~M.~C. Foulkes, L.~Mitas, R.~J. Needs, and G.~Rajagopal.
\newblock Quantum {Monte Carlo} simulations of solids.
\newblock \emph{Rev. Mod. Phys.}, 73:\penalty0 33--83, 2001.
\newblock \doi{10.1103/RevModPhys.73.33}.

\bibitem[Becca and Sorella(2017)]{Becca2017}
F.~Becca and S.~Sorella.
\newblock \emph{{Quantum Monte Carlo} Approaches for Correlated Systems}.
\newblock Cambridge University Press, Cambridge, 2017.
\newblock \doi{10.1017/9781316417041}.

\bibitem[Mazzola(2024)]{Mazzola2024}
G.~Mazzola.
\newblock Quantum computing for chemistry and physics applications from a
  {Monte Carlo} perspective.
\newblock \emph{J. Chem. Phys.}, 160:\penalty0 010901, 2024.
\newblock \doi{10.1063/5.0173591}.

\bibitem[Carlson et~al.(2015)Carlson, Gandolfi, Pederiva, Pieper, Schiavilla,
  Schmidt, and Wiringa]{Carlson2015}
J.~Carlson, S.~Gandolfi, F.~Pederiva, S.~C. Pieper, R.~Schiavilla, K.~E.
  Schmidt, and R.~B. Wiringa.
\newblock Quantum {Monte Carlo} methods for nuclear physics.
\newblock \emph{Rev. Mod. Phys.}, 87:\penalty0 1067--1118, 2015.
\newblock \doi{10.1103/RevModPhys.87.1067}.

\bibitem[Cohen et~al.(2015)Cohen, Gull, Reichman, and Millis]{Cohen2015}
G.~Cohen, E.~Gull, D.~R. Reichman, and A.~J. Millis.
\newblock Taming the dynamical sign problem in real-time evolution of quantum
  many-body problems.
\newblock \emph{Phys. Rev. Lett.}, 115:\penalty0 266802, 2015.
\newblock \doi{10.1103/PhysRevLett.115.266802}.

\bibitem[Li and Yao(2019)]{Li2019}
Z.-X. Li and H.~Yao.
\newblock Sign-problem-free fermionic quantum {Monte Carlo}: Developments and
  applications.
\newblock \emph{Annu. Rev. Condens. Matter Phys.}, 10:\penalty0 337--356, 2019.
\newblock \doi{10.1146/annurev-conmatphys-033117-054307}.

\bibitem[Alexandru et~al.(2022)Alexandru, Başar, Bedaque, and
  Warrington]{Alexandru2022}
A.~Alexandru, G.~Başar, P.~F. Bedaque, and N.~C. Warrington.
\newblock Complex paths around the sign problem.
\newblock \emph{Rev. Mod. Phys.}, 94:\penalty0 015006, 2022.
\newblock \doi{10.1103/RevModPhys.94.015006}.

\bibitem[Bakr et~al.(2009)Bakr, Gillen, Peng, F{\"o}lling, and
  Greiner]{Bakr2009}
W.~S. Bakr, J.~I. Gillen, A.~Peng, S.~F{\"o}lling, and M.~Greiner.
\newblock A quantum gas microscope for detecting single atoms in a
  {Hubbard}-regime optical lattice.
\newblock \emph{Nature}, 462:\penalty0 74--77, 2009.
\newblock \doi{10.1038/nature08482}.

\bibitem[Cheuk et~al.(2015)Cheuk, Nichols, Okan, Gersdorf, Ramasesh, Bakr,
  Lompe, and Zwierlein]{Cheuk2015}
L.~W. Cheuk, M.~A. Nichols, M.~Okan, T.~Gersdorf, V.~V. Ramasesh, W.~S. Bakr,
  T.~Lompe, and M.~W. Zwierlein.
\newblock Quantum-gas microscope for fermionic atoms.
\newblock \emph{Phys. Rev. Lett.}, 114:\penalty0 193001, 2015.
\newblock \doi{10.1103/PhysRevLett.114.193001}.

\bibitem[Byrnes and Yamamoto(2006)]{Byrnes2006}
T.~Byrnes and Y.~Yamamoto.
\newblock Simulating lattice gauge theories on a quantum computer.
\newblock \emph{Phys. Rev. A}, 73:\penalty0 022328, 2006.
\newblock \doi{10.1103/PhysRevA.73.022328}.

\bibitem[Zohar and Reznik(2011)]{Zohar2011}
E.~Zohar and B.~Reznik.
\newblock High-temperature superconductivity.
\newblock \emph{Phys. Rev. Lett.}, 107:\penalty0 275301, 2011.
\newblock \doi{10.1103/PhysRevLett.107.275301}.

\bibitem[B{\"u}chler et~al.(2005)B{\"u}chler, Hermele, Huber, Fisher, and
  Zoller]{Buchler2005}
H.~P. B{\"u}chler, M.~Hermele, S.~D. Huber, M.~P.~A. Fisher, and P.~Zoller.
\newblock Atomic quantum simulator for lattice gauge theories and ring exchange
  models.
\newblock \emph{Phys. Rev. Lett.}, 95:\penalty0 040402, 2005.
\newblock \doi{10.1103/PhysRevLett.95.040402}.

\bibitem[Clo{\"e}t et~al.(2019)Clo{\"e}t, Dietrich, Arrington, Bazavov, Bishof,
  Freese, Gorshkov, Grassellino, Hafidi, Jacob, McGuigan, Meurice, Meziani,
  Mueller, Muschik, Osborn, Otten, Petreczky, Polakovic, Poon, Pooser, Roggero,
  Saffman, VanDevender, Zhang, and Zohar]{Cloet2019}
I.~C. Clo{\"e}t, M.~R. Dietrich, J.~Arrington, A.~Bazavov, M.~Bishof,
  A.~Freese, A.~V. Gorshkov, A.~Grassellino, K.~Hafidi, Z.~Jacob, M.~McGuigan,
  Y.~Meurice, Z.-E. Meziani, P.~Mueller, C.~Muschik, J.~Osborn, M.~Otten,
  P.~Petreczky, T.~Polakovic, A.~Poon, R.~Pooser, A.~Roggero, M.~Saffman,
  B.~VanDevender, J.~Zhang, and E.~Zohar.
\newblock Opportunities for nuclear physics \& quantum information science,
  2019.
\newblock \href{https://arxiv.org/abs/1903.05453}{arxiv:1903.05453}.

\bibitem[Bauer et~al.(2023)Bauer, Davoudi, Balantekin, Bhattacharya, Carena,
  de~Jong, Draper, El-Khadra, Gemelke, Hanada, Kharzeev, Lamm, Li, Liu, Lukin,
  Meurice, Monroe, Nachman, Pagano, Preskill, Rinaldi, Roggero, Santiago,
  Savage, Siddiqi, Siopsis, Zanten, Wiebe, Yamauchi, Yeter-Aydeniz, and
  Zorzetti]{Bauer2023}
C.~W. Bauer, Z.~Davoudi, A.~B. Balantekin, T.~Bhattacharya, M.~Carena, W.~A.
  de~Jong, P.~Draper, A.~El-Khadra, N.~Gemelke, M.~Hanada, D.~Kharzeev,
  H.~Lamm, Y.-Y. Li, J.~Liu, M.~Lukin, Y.~Meurice, C.~Monroe, B.~Nachman,
  G.~Pagano, J.~Preskill, E.~Rinaldi, A.~Roggero, D.~I. Santiago, M.~J. Savage,
  I.~Siddiqi, G.~Siopsis, D.~Van Zanten, N.~Wiebe, Y.~Yamauchi,
  K.~Yeter-Aydeniz, and S.~Zorzetti.
\newblock Quantum simulation for high-energy physics.
\newblock \emph{PRX Quantum}, 4:\penalty0 027001, 2023.
\newblock \doi{10.1103/PRXQuantum.4.027001}.

\bibitem[Anderson et~al.(1995)Anderson, Ensher, Matthews, Wieman, and
  Cornell]{Anderson1995}
M.~H. Anderson, J.~R. Ensher, M.~R. Matthews, C.~E. Wieman, and E.~A. Cornell.
\newblock Observation of {Bose-Einstein} condensation in a dilute atomic vapor.
\newblock \emph{Science}, 269:\penalty0 198--201, 1995.
\newblock \doi{10.1126/science.269.5221.198}.

\bibitem[DeMarco and Jin(1999)]{DeMarco1999}
B.~DeMarco and D.~S. Jin.
\newblock Onset of {Fermi} degeneracy in a trapped atomic gas.
\newblock \emph{Science}, 285:\penalty0 1703--1706, 1999.
\newblock \doi{10.1126/science.285.5434.17}.

\bibitem[Sch{\"a}fer et~al.(2020)Sch{\"a}fer, Fukuhara, Sugawa, Takasu, and
  Takahashi]{Schafer2020}
F.~Sch{\"a}fer, T.~Fukuhara, S.~Sugawa, Y.~Takasu, and Y.~Takahashi.
\newblock Tools for quantum simulation with ultracold atoms in optical
  lattices.
\newblock \emph{Nat. Rev. Phys.}, 2:\penalty0 411--425, 2020.
\newblock \doi{10.1038/s42254-020-0195-3}.

\bibitem[Daley(2023)]{Daley2023}
A.~J. Daley.
\newblock Twenty-five years of analogue quantum simulation.
\newblock \emph{Nat. Rev. Phys.}, 5:\penalty0 702--703, 2023.
\newblock \doi{10.1038/s42254-023-00666-0}.

\bibitem[Martinez et~al.(2016)Martinez, Muschik, Schindler, Nigg, Erhard, Heyl,
  Hauke, Dalmonte, Monz, Zoller, and Blatt]{Martinez2016}
E.~A. Martinez, C.~A. Muschik, P.~Schindler, D.~Nigg, A.~Erhard, M.~Heyl,
  P.~Hauke, M.~Dalmonte, T.~Monz, P.~Zoller, and R.~Blatt.
\newblock Real-time dynamics of lattice gauge theories with a few-qubit quantum
  computer.
\newblock \emph{Nature}, 534:\penalty0 516--519, 2016.
\newblock \doi{10.1038/nature18318}.

\bibitem[Atas et~al.(2021)Atas, Zhang, Lewis, Jahanpour, Haase, and
  Muschik]{Atas2021}
Y.~Y. Atas, J.~Zhang, R.~Lewis, A.~Jahanpour, J.~F. Haase, and C.~A. Muschik.
\newblock {SU(2)} hadrons on a quantum computer via a variational approach.
\newblock \emph{Nat. Commun.}, 12:\penalty0 6499, 2021.
\newblock \doi{10.1038/s41467-021-26825-4}.

\bibitem[Irmejs et~al.(2023)Irmejs, Ba{\~n}uls, and Cirac]{Irmejs2023}
R.~Irmejs, M.~C. Ba{\~n}uls, and J.~I. Cirac.
\newblock Quantum simulation of {$\mathbb{Z}_2$} lattice gauge theory with
  minimal resources.
\newblock \emph{Phys. Rev. D}, 108:\penalty0 074503, 2023.
\newblock \doi{10.1103/PhysRevD.108.074503}.

\bibitem[Abrams and Lloyd(1997)]{Abrams1997}
D.~S. Abrams and S.~Lloyd.
\newblock Simulation of many-body {Fermi} systems on a universal quantum
  computer.
\newblock \emph{Phys. Rev. Lett.}, 79:\penalty0 2586, 1997.
\newblock \doi{10.1103/PhysRevLett.79.2586}.

\bibitem[Ortiz et~al.(2001)Ortiz, Gubernatis, Knill, and
  Laflamme]{ortiz2001quantum}
G.~Ortiz, J.~E. Gubernatis, E.~Knill, and R.~Laflamme.
\newblock Quantum algorithms for fermionic simulations.
\newblock \emph{Phys. Rev. A}, 64:\penalty0 022319, 2001.
\newblock \doi{10.1103/PhysRevA.64.022319}.

\bibitem[Lin and Tong(2022)]{Lin2022}
L.~Lin and Y.~Tong.
\newblock Heisenberg-limited ground-state energy estimation for early
  fault-tolerant quantum computers.
\newblock \emph{PRX Quantum}, 3:\penalty0 010318, 2022.
\newblock \doi{10.1103/PRXQuantum.3.010318}.

\bibitem[Huggins et~al.(2022)Huggins, O'Gorman, Rubin, Reichman, Babbush, and
  Lee]{Huggins2022}
W.~J. Huggins, B.~A. O'Gorman, N.~C. Rubin, D.~R. Reichman, R.~Babbush, and
  J.~Lee.
\newblock Unbiasing fermionic quantum {Monte Carlo} with a quantum computer.
\newblock \emph{Nature}, 603:\penalty0 416--420, 2022.
\newblock \doi{10.1038/s41586-021-04351-z}.

\bibitem[Somma(2019)]{Somma2019}
R.~D. Somma.
\newblock Quantum eigenvalue estimation via time series analysis.
\newblock \emph{New J. Phys.}, 21:\penalty0 123025, 2019.
\newblock \doi{10.1088/1367-2630/ab5c60}.

\bibitem[Cao et~al.(2019)Cao, Romero, Olson, Degroote, Johnson, Kieferov\'{a},
  Kivlichan, Menke, Peropadre, Sawaya, Sim, Veis, and Aspuru-Guzik]{Cao2019}
Y.~Cao, J.~Romero, J.~P. Olson, M.~Degroote, P.~D. Johnson, M.~Kieferov\'{a},
  I.~D. Kivlichan, T.~Menke, B.~Peropadre, N.~P.~D. Sawaya, S.~Sim, L.~Veis,
  and A.~Aspuru-Guzik.
\newblock Quantum chemistry in the age of quantum computing.
\newblock \emph{Chem. Rev.}, 119:\penalty0 10856, 2019.
\newblock \doi{10.1021/acs.chemrev.8b00803}.

\bibitem[McArdle et~al.(2020)McArdle, Endo, Aspuru-Guzik, Benjamin, and
  Yuan]{McArdle2020}
S.~McArdle, S.~Endo, A.~Aspuru-Guzik, S.~C. Benjamin, and X.~Yuan.
\newblock Quantum computational chemistry.
\newblock \emph{Rev. Mod. Phys.}, 92:\penalty0 015003, 2020.
\newblock \doi{10.1103/RevModPhys.92.015003}.

\bibitem[Lee et~al.(2023)Lee, Lee, Zhai, Tong, Dalzell, Kumar, Helms, Gray,
  Cui, Liu, Kastoryano, Babbush, Preskill, Reichman, Campbell, Valeev, Lin, and
  Chan]{Lee2023c}
S.~Lee, J.~Lee, H.~Zhai, Y.~Tong, A.~M. Dalzell, A.~Kumar, P.~Helms, J.~Gray,
  Z.-H. Cui, W.~Liu, M.~Kastoryano, R.~Babbush, J.~Preskill, D.~R. Reichman,
  E.~T. Campbell, E.~F. Valeev, L.~Lin, and G.~K.-L. Chan.
\newblock Evaluating the evidence for exponential quantum advantage in
  ground-state quantum chemistry.
\newblock \emph{Nat. Commun.}, 14:\penalty0 1952, 2023.
\newblock \doi{10.1038/s41467-023-37587-6}.

\bibitem[Chen et~al.(2021)Chen, Cheng, and Freericks]{Jia2021}
J.~Chen, H.-P. Cheng, and J.~K. Freericks.
\newblock Quantum-inspired algorithm for the factorized form of unitary coupled
  cluster theory.
\newblock \emph{J. Chem. Theory Comput.}, 17:\penalty0 841--847, 2021.
\newblock \doi{10.1021/acs.jctc.0c01052}.

\bibitem[Grimme and Schreiner(2018)]{Grimme2018}
S.~Grimme and P.~R. Schreiner.
\newblock Computational chemistry: The fate of current methods and future
  challenges.
\newblock \emph{Angew. Chem. Int. Ed.}, 57:\penalty0 4170--4176, 2018.
\newblock \doi{10.1002/anie.201709943}.

\bibitem[Lidar(2004)]{Lidar2004}
D.~Lidar.
\newblock On the quantum computational complexity of the {Ising} spin glass
  partition function and of knot invariants.
\newblock \emph{New J. Phys.}, 6:\penalty0 167, 2004.
\newblock \doi{10.1088/1367-2630/6/1/167}.

\bibitem[Jaeger et~al.(1990)Jaeger, Vertigan, and Welsh]{Jaeger1990}
F.~Jaeger, D.~L. Vertigan, and D.~J.~A. Welsh.
\newblock On the computational complexity of the {Jones} and {Tutte}
  polynomials.
\newblock \emph{Math. Proc. Cambridge Philos. Soc.}, 108:\penalty0 35--53,
  1990.
\newblock \doi{10.1017/S0305004100068936}.

\bibitem[Temme et~al.(2011)Temme, Osborne, Vollbrecht, Poulin, and
  Verstraete]{Temme2011}
K.~Temme, T.~J. Osborne, K.~G. Vollbrecht, D.~Poulin, and F.~Verstraete.
\newblock Quantum {Metropolis} sampling.
\newblock \emph{Nature}, 471:\penalty0 87--90, 2011.
\newblock \doi{10.1038/nature09770}.

\bibitem[Wocjan et~al.(2009)Wocjan, Chiang, Nagaj, and Abeyesinghe]{Wocjan2009}
P.~Wocjan, C.-F. Chiang, D.~Nagaj, and A.~Abeyesinghe.
\newblock Quantum algorithm for approximating partition functions.
\newblock \emph{Phys. Rev. A}, 80:\penalty0 022340, 2009.
\newblock \doi{10.1103/PhysRevA.80.022340}.

\bibitem[Arunachalam et~al.(2022)Arunachalam, Havlicek, Nannicini, Temme, and
  Wocjan]{Arunachalam2022}
S.~Arunachalam, V.~Havlicek, G.~Nannicini, K.~Temme, and P.~Wocjan.
\newblock Simpler (classical) and faster (quantum) algorithms for {Gibbs}
  partition functions.
\newblock \emph{Quantum}, 6:\penalty0 789, 2022.
\newblock \doi{10.22331/q-2022-09-01-789}.

\bibitem[Wu and Wang(2021)]{Wu2021g}
Y.~Wu and J.~B. Wang.
\newblock Estimating {Gibbs} partition function with quantum {Clifford}
  sampling.
\newblock \emph{Quantum Sci. Technol.}, 7:\penalty0 025006, 2021.
\newblock \doi{10.1088/2058-9565/ac47f0}.

\bibitem[Jackson et~al.(2023)Jackson, Kapourniotis, and Datta]{Jackson2023}
A.~Jackson, T.~Kapourniotis, and A.~Datta.
\newblock Partition-function estimation: Quantum and quantum-inspired
  algorithms.
\newblock \emph{Phys. Rev. A}, 107:\penalty0 012421, 2023.
\newblock \doi{10.1103/PhysRevA.107.012421}.

\bibitem[Tubman et~al.(2018)Tubman, Mejuto-Zaera, Epstein, Hait, Levine,
  Huggins, Jiang, McClean, Babbush, Head-Gordon, and Whaley]{Tubman2018}
N.~M. Tubman, C.~Mejuto-Zaera, J.~M. Epstein, D.~Hait, D.~S. Levine,
  W.~Huggins, Z.~Jiang, J.~R. McClean, R.~Babbush, M.~Head-Gordon, and
  K.~Birgitta Whaley.
\newblock Postponing the orthogonality catastrophe: efficient state preparation
  for electronic structure simulations on quantum devices, 2018.
\newblock \href{https://doi.org/10.48550/arxiv.1809.05523}{arXiv:1809.05523}.

\bibitem[Slater(1951)]{slater1951}
J.~C. Slater.
\newblock A simplification of the {Hartree-Fock} method.
\newblock \emph{Phys. Rev.}, 81:\penalty0 385, 1951.
\newblock \doi{10.1103/PhysRev.81.385}.

\bibitem[Hohenberg and Kohn(1964)]{Hohenberg1964}
P.~Hohenberg and W.~Kohn.
\newblock Inhomogeneous electron gas.
\newblock \emph{Phys. Rev.}, 136:\penalty0 B864, 1964.
\newblock \doi{10.1103/PhysRev.136.B864}.

\bibitem[Kohn and Sham(1965)]{Kohn1965}
W.~Kohn and L.~J. Sham.
\newblock Self-consistent equations including exchange and correlation effects.
\newblock \emph{Phys. Rev.}, 140:\penalty0 A1133, 1965.
\newblock \doi{10.1103/PhysRev.140.A1133}.

\bibitem[Echenique and Alonso(2007)]{Echenique2007}
P.~Echenique and J.~L. Alonso.
\newblock A mathematical and computational review of {Hartree-Fock SCF} methods
  in quantum chemistry.
\newblock \emph{Mol. Phys.}, 105:\penalty0 3057--3098, 2007.
\newblock \doi{10.1080/00268970701757875}.

\bibitem[{Molecular Sciences Software Institute (MolSSI)} and {Pacific
  Northwest National Lab/Environmental Molecular Sciences Laboratory
  (PNNL/EMSL)}(2023)]{basisset}
{Molecular Sciences Software Institute (MolSSI)} and {Pacific Northwest
  National Lab/Environmental Molecular Sciences Laboratory (PNNL/EMSL)}.
\newblock Basis set exchange.
\newblock \url{https://www.basissetexchange.org/}, 2023.
\newblock Accessed: 2023-09-01.

\bibitem[Singh and Nordstr{\"o}m(2006)]{Singh2006}
D.~J. Singh and L.~Nordstr{\"o}m.
\newblock \emph{Planewaves, Pseudopotentials, and the {LAPW} method}.
\newblock Springer, New York, NY, 2006.
\newblock \doi{10.1007/978-0-387-29684-5}.

\bibitem[Peterson(2007)]{Peterson2007}
K.~A. Peterson.
\newblock Gaussian basis sets exhibiting systematic convergence to the complete
  basis set limit.
\newblock \emph{Annu. Rep. Comput. Chem.}, 3:\penalty0 195--206, 2007.
\newblock \doi{10.1016/S1574-1400(07)03011-3}.

\bibitem[Jensen et~al.(2016)Jensen, Fl{\o a}, Jonsson, Monstad, Ruud, and
  Frediani]{Jensen2016}
S.~R. Jensen, T.~Fl{\o a}, D.~Jonsson, R.~S. Monstad, K.~Ruud, and L.~Frediani.
\newblock Magnetic properties with multiwavelets and {DFT}: the complete basis
  set limit achieved.
\newblock \emph{Phys. Chem. Chem. Phys.}, 18:\penalty0 21145--21161, 2016.
\newblock \doi{10.1039/C6CP01294A}.

\bibitem[Chien and Klassen(2022)]{Chien2022}
R.~W. Chien and J.~Klassen.
\newblock Optimizing fermionic encodings for both {Hamiltonian} and hardware,
  2022.
\newblock \href{https://doi.org/10.48550/arxiv.2210.05652}{arxiv:2210.05652}.

\bibitem[Capelle(2006)]{Capelle2006}
K.~Capelle.
\newblock A bird's-eye view of density-functional theory.
\newblock \emph{Brazilian J. Phys.}, 36:\penalty0 1318--1343, 2006.
\newblock \doi{10.1590/S0103-97332006000700035}.

\bibitem[Baseden and Tye(2014)]{Baseden2014}
K.~A. Baseden and J.~W. Tye.
\newblock Introduction to density functional theory: Calculations by hand on
  the helium atom.
\newblock \emph{J. Chem. Educ.}, 91:\penalty0 2116--2123, 2014.
\newblock \doi{10.1021/ed5004788}.

\bibitem[Eriksen and Gauss(2021)]{Eriksen2021}
J.~J. Eriksen and J.~Gauss.
\newblock Incremental treatments of the full configuration interaction problem.
\newblock \emph{WIREs Comput. Mol. Sci.}, 11:\penalty0 e1525, 2021.
\newblock \doi{10.1002/wcms.1525}.

\bibitem[M{\o}ller and Plesset(1934)]{Moller1934}
C.~M{\o}ller and M.~S. Plesset.
\newblock Note on an approximation treatment for many-electron systems.
\newblock \emph{Phys. Rev.}, 46:\penalty0 618--622, 1934.
\newblock \doi{10.1103/PhysRev.46.618}.

\bibitem[{\v C}{\'i}{\v z}ek(1966)]{Cizek1966}
J.~{\v C}{\'i}{\v z}ek.
\newblock On the correlation problem in atomic and molecular systems.
  calculation of wavefunction components in {Ursell}-type expansions using
  quantum-field theoretical methods.
\newblock \emph{J. Chem. Phys.}, 45:\penalty0 4256--4266, 1966.

\bibitem[Hoffmann(1987)]{Hoffmann1987}
R.~Hoffmann.
\newblock How chemistry and physics meet in the solid state.
\newblock \emph{Angew. Chem. Int. Ed. Engl.}, 26:\penalty0 846--878, 1987.
\newblock \doi{10.1002/anie.198708461}.

\bibitem[Wannier(1937)]{Wannier1937}
G.~H. Wannier.
\newblock The structure of electronic excitation levels in insulating crystals.
\newblock \emph{Phys. Rev.}, 52:\penalty0 191--197, 1937.
\newblock \doi{10.1103/PhysRev.52.191}.

\bibitem[Clinton et~al.(2024)Clinton, Cubitt, Flynn, Gambetta, Klassen,
  Montanaro, Piddock, Santos, and Sheridan]{Clinton2024}
L.~Clinton, T.~Cubitt, B.~Flynn, F.~M. Gambetta, J.~Klassen, A.~Montanaro,
  S.~Piddock, R.~A. Santos, and E.~Sheridan.
\newblock Towards near-term quantum simulation of materials.
\newblock \emph{Nat. Commun.}, 15:\penalty0 211, 2024.
\newblock \doi{10.1038/s41467-023-43479-6}.

\bibitem[Kassal et~al.(2011)Kassal, Whitfield, Perdomo-Ortiz, Yung, and
  Aspuru-Guzik]{Kassal2011}
I.~Kassal, J.~D. Whitfield, A.~Perdomo-Ortiz, M.-H. Yung, and A.~Aspuru-Guzik.
\newblock Simulating chemistry using quantum computers.
\newblock \emph{Annu. Rev. Phys. Chem.}, 62:\penalty0 185--207, 2011.
\newblock \doi{10.1146/annurev-physchem-032210-103512}.

\bibitem[Bauer et~al.(2020)Bauer, Bravyi, Motta, and Chan]{Bauer2020}
B.~Bauer, S.~Bravyi, M.~Motta, and G.~K.-L. Chan.
\newblock Quantum algorithms for quantum chemistry and quantum materials
  science.
\newblock \emph{Chem. Rev.}, 120:\penalty0 12685--12717, 2020.
\newblock \doi{10.1021/acs.chemrev.9b00829}.

\bibitem[Bernal et~al.(2022)Bernal, Ajagekar, Harwood, Stober, Trenev, and
  You]{Bernal2022}
D.~E. Bernal, A.~Ajagekar, S.~M. Harwood, S.~T. Stober, D.~Trenev, and F.~You.
\newblock Perspectives of quantum computing for chemical engineering.
\newblock \emph{AIChE J.}, 68:\penalty0 e17651, 2022.
\newblock \doi{10.1002/aic.17651}.

\bibitem[Jordan and Wigner(1928)]{Jordan1928}
P.~Jordan and E.~Wigner.
\newblock {\"U}ber das {Paulische} {\"a}quivalenzverbot.
\newblock \emph{Z. Phys.}, 47:\penalty0 631--651, 1928.
\newblock \doi{10.1007/BF01331938}.

\bibitem[Bravyi and Kitaev(2002)]{Bravyi2002}
S.~B. Bravyi and A.~Y. Kitaev.
\newblock Fermionic quantum computation.
\newblock \emph{Ann. Phys.}, 298:\penalty0 210--226, 2002.
\newblock \doi{10.1006/aphy.2002.6254}.

\bibitem[Tranter et~al.(2018)Tranter, Love, Mintert, and Coveney]{Tranter2018}
A.~Tranter, P.~J. Love, F.~Mintert, and P.~V. Coveney.
\newblock A comparison of the {Bravyi–Kitaev} and {Jordan–Wigner}
  transformations for the quantum simulation of quantum chemistry.
\newblock \emph{J. Chem. Theory Comput.}, 14:\penalty0 5617--5630, 2018.
\newblock \doi{10.1021/acs.jctc.8b00450}.

\bibitem[O'Brien and Strelchuk(2024)]{OBrien2024}
O.~O'Brien and S.~Strelchuk.
\newblock Ultrafast hybrid fermion-to-qubit mapping.
\newblock \emph{Phys. Rev. B}, 109:\penalty0 115149, 2024.
\newblock \doi{10.1103/PhysRevB.109.115149}.

\bibitem[Derby et~al.(2021)Derby, Klassen, Bausch, and Cubitt]{Derby2021}
C.~Derby, J.~Klassen, J.~Bausch, and T.~Cubitt.
\newblock Compact fermion to qubit mappings.
\newblock \emph{Phys. Rev. B}, 104:\penalty0 035118, 2021.
\newblock \doi{10.1103/PhysRevB.104.035118}.

\bibitem[Wecker et~al.(2014)Wecker, Bauer, Clark, Hastings, and
  Troyer]{Wecker2014}
D.~Wecker, B.~Bauer, B.~K. Clark, M.~B. Hastings, and M.~Troyer.
\newblock Gate-count estimates for performing quantum chemistry on small
  quantum computers.
\newblock \emph{Phys. Rev. A}, 90:\penalty0 022305, 2014.
\newblock \doi{10.1103/PhysRevA.90.022305}.

\bibitem[Babbush et~al.(2018)Babbush, Gidney, Berry, Wiebe, McClean, Paler,
  Fowler, and Neven]{Babbush2018}
R.~Babbush, C.~Gidney, D.~W. Berry, N.~Wiebe, J.~McClean, A.~Paler, A.~Fowler,
  and H.~Neven.
\newblock Encoding electronic spectra in quantum circuits with linear {T}
  complexity.
\newblock \emph{Phys. Rev. X}, 8:\penalty0 041015, 2018.
\newblock \doi{10.1103/PhysRevX.8.041015}.

\bibitem[Bartlett et~al.(1989)Bartlett, Kucharski, and Noga]{Bartlett1989}
R.~J. Bartlett, S.~A. Kucharski, and J.~Noga.
\newblock Alternative coupled-cluster ans{\"a}tze {II}. the unitary
  coupled-cluster method.
\newblock \emph{Chem. Phys. Lett.}, 155:\penalty0 133--140, 1989.
\newblock \doi{10.1016/S0009-2614(89)87372-5}.

\bibitem[Motta et~al.(2021)Motta, Ye, McClean, Li, Minnich, Babbush, and
  Chan]{Motta2021}
M.~Motta, E.~Ye, J.~R. McClean, Z.~Li, A.~J. Minnich, R.~Babbush, and G.~K.-L.
  Chan.
\newblock Low rank representations for quantum simulation of electronic
  structure.
\newblock \emph{npj Quantum Inf.}, 7:\penalty0 83, 2021.
\newblock \doi{10.1038/s41534-021-00416-z}.

\bibitem[Matsuzawa and Kurashige(2020)]{Matsuzawa2020}
Y.~Matsuzawa and Y.~Kurashige.
\newblock Jastrow-type decomposition in quantum chemistry for low-depth quantum
  circuits.
\newblock \emph{J. Chem. Theory Comput.}, 16:\penalty0 944--952, 2020.
\newblock \doi{10.1021/acs.jctc.9b00963}.

\bibitem[Setia et~al.(2019)Setia, Bravyi, Mezzacapo, and Whitfield]{Setia2019}
K.~Setia, S.~Bravyi, A.~Mezzacapo, and J.~D. Whitfield.
\newblock Superfast encodings for fermionic quantum simulation.
\newblock \emph{Phys. Rev. Res.}, 1:\penalty0 033033, 2019.
\newblock \doi{10.1103/PhysRevResearch.1.033033}.

\bibitem[Wecker et~al.(2015)Wecker, Hastings, and Troyer]{Wecker2015}
D.~Wecker, M.~B. Hastings, and M.~Troyer.
\newblock Progress towards practical quantum variational algorithms.
\newblock \emph{Phys. Rev. A}, 92:\penalty0 042303, 2015.
\newblock \doi{10.1103/PhysRevA.92.042303}.

\bibitem[Wiersema et~al.(2020)Wiersema, Zhou, de~Sereville, Carrasquilla, Kim,
  and Yuen]{Wiersema2020}
R.~Wiersema, C.~Zhou, Y.~de~Sereville, J.~F. Carrasquilla, Y.~B. Kim, and
  H.~Yuen.
\newblock Exploring entanglement and optimization within the {Hamiltonian}
  variational ansatz.
\newblock \emph{PRX Quantum}, 1:\penalty0 020319, 2020.
\newblock \doi{10.1103/PRXQuantum.1.020319}.

\bibitem[Grimsley et~al.(2019)Grimsley, Economou, Barnes, and
  Mayhall]{Grimsley2019}
H.~R. Grimsley, S.~E. Economou, E.~Barnes, and N.~J. Mayhall.
\newblock An adaptive variational algorithm for exact molecular simulations on
  a quantum computer.
\newblock \emph{Nat. Commun.}, 10:\penalty0 3007, 2019.
\newblock \doi{10.1038/s41467-019-10988-2}.

\bibitem[Tang et~al.(2021)Tang, Shkolnikov, Barron, Grimsley, Mayhall, Barnes,
  and Economou]{Tang2021a}
H.~L. Tang, V.~O. Shkolnikov, G.~S. Barron, H.~R. Grimsley, N.~J. Mayhall,
  E.~Barnes, and S.~E. Economou.
\newblock {Qubit-ADAPT-VQE}: An adaptive algorithm for constructing
  hardware-efficient ans{\"a}tze on a quantum processor.
\newblock \emph{PRX Quantum}, 2:\penalty0 020310, 2021.
\newblock \doi{10.1103/PRXQuantum.2.020310}.

\bibitem[Yordanov et~al.(2021)Yordanov, Armaos, Barnes, and
  Arvidsson-Shukur]{Yordanov2021}
Y.~S. Yordanov, V.~Armaos, C.~H.~W. Barnes, and D.~R.~M. Arvidsson-Shukur.
\newblock Qubit-excitation-based adaptive variational quantum eigensolver.
\newblock \emph{Commun. Phys.}, 4:\penalty0 228, 2021.
\newblock \doi{10.1038/s42005-021-00730-0}.

\bibitem[Takeshita et~al.(2020)Takeshita, Rubin, Jiang, Lee, Babbush, and
  McClean]{Takeshita2020}
T.~Takeshita, N.~C. Rubin, Z.~Jiang, E.~Lee, R.~Babbush, and J.~R. McClean.
\newblock Increasing the representation accuracy of quantum simulations of
  chemistry without extra quantum resources.
\newblock \emph{Phys. Rev. X}, 10:\penalty0 011004, 2020.
\newblock \doi{10.1103/PhysRevX.10.011004}.

\bibitem[Sokolov et~al.(2020)Sokolov, Barkoutsos, Ollitrault, Greenberg, Rice,
  Pistoia, and Tavernelli]{Sokolov2020}
I.~O. Sokolov, P.~K.~I. Barkoutsos, P.~J. Ollitrault, D.~Greenberg, J.~Rice,
  M.~Pistoia, and I.~Tavernelli.
\newblock Quantum orbital-optimized unitary coupled cluster methods in the
  strongly correlated regime: Can quantum algorithms outperform their classical
  equivalents?
\newblock \emph{J. Chem. Phys.}, 152:\penalty0 124107, 2020.
\newblock \doi{10.1063/1.5141835}.

\bibitem[Vaquero-Sabater et~al.(2024)Vaquero-Sabater, Carreras, Or{\'u}s,
  Mayhall, and Casanova]{VaqueroSabater2024}
N.~Vaquero-Sabater, A.~Carreras, R.~Or{\'u}s, N.~J. Mayhall, and D.~Casanova.
\newblock Physically motivated improvements of variational quantum
  eigensolvers.
\newblock \emph{J. Chem. Theory Comput.}, 20:\penalty0 5133--5144, 2024.
\newblock \doi{10.1021/acs.jctc.4c00329}.

\bibitem[Grimsley et~al.(2023)Grimsley, Barron, Barnes, Economou, and
  Mayhall]{Grimsley2023}
H.~R. Grimsley, G.~S. Barron, E.~Barnes, S.~E. Economou, and N.~J. Mayhall.
\newblock Adaptive, problem-tailored variational quantum eigensolver mitigates
  rough parameter landscapes and barren plateaus.
\newblock \emph{npj Quantum Inf.}, 9:\penalty0 19, 2023.
\newblock \doi{10.1038/s41534-023-00681-0}.

\bibitem[Feniou et~al.(2023)Feniou, Hassan, Traor{\'e}, Giner, Maday, and
  Piquemal]{Feniou2023}
C.~Feniou, M.~Hassan, D.~Traor{\'e}, E.~Giner, Y.~Maday, and J.-P. Piquemal.
\newblock {Overlap-ADAPT-VQE}: practical quantum chemistry on quantum computers
  via overlap-guided compact {Ans{\"a}tze}.
\newblock \emph{Commun. Phys.}, 6:\penalty0 192, 2023.
\newblock \doi{10.1038/s42005-023-01312-y}.

\bibitem[Markov and Shi(2008)]{Markov2008}
I.~L. Markov and Y.~Shi.
\newblock Simulating quantum computation by contracting tensor networks.
\newblock \emph{SIAM J. Comput.}, 38:\penalty0 963--981, 2008.
\newblock \doi{10.1137/050644756}.

\bibitem[Liu et~al.(2019)Liu, Zhang, Wan, and Wang]{Liu2019}
J.-G. Liu, Y.-H. Zhang, Y.~Wan, and L.~Wang.
\newblock Variational quantum eigensolver with fewer qubits.
\newblock \emph{Phys. Rev. Res.}, 1:\penalty0 023025, 2019.
\newblock \doi{10.1103/PhysRevResearch.1.023025}.

\bibitem[Barratt et~al.(2021)Barratt, Dborin, Bal, Stojevic, Pollmann, and
  Green]{Barratt2021}
F.~Barratt, J.~Dborin, M.~Bal, V.~Stojevic, F.~Pollmann, and A.~G. Green.
\newblock Parallel quantum simulation of large systems on small {NISQ}
  computers.
\newblock \emph{npj Quantum Inf.}, 7:\penalty0 79, 2021.
\newblock \doi{10.1038/s41534-021-00420-3}.

\bibitem[Fujii et~al.(2022)Fujii, Mizuta, Ueda, Mitarai, Mizukami, and
  Nakagawa]{Fujii2022}
K.~Fujii, K.~Mizuta, H.~Ueda, K.~Mitarai, W.~Mizukami, and Y.~O. Nakagawa.
\newblock Deep variational quantum eigensolver: A divide-and-conquer method for
  solving a larger problem with smaller size quantum computers.
\newblock \emph{PRX Quantum}, 3:\penalty0 010346, 2022.
\newblock \doi{10.1103/PRXQuantum.3.010346}.

\bibitem[Catlow et~al.(2017)Catlow, Buckeridge, Farrow, Logsdail, and
  Sokol]{catlow2017quantum}
C.~R.~A. Catlow, J.~Buckeridge, M.~R. Farrow, A.~J. Logsdail, and A.~A. Sokol.
\newblock Quantum mechanical/molecular mechanical ({QM/MM}) approaches.
\newblock In R.~Dronskowski, S.~Kikkawa, and A.~Stein, editors, \emph{Handbook
  of Solid State Chemistry}. Wiley, 2017.
\newblock \doi{10.1002/9783527691036.hsscvol5012}.

\bibitem[Berger et~al.(2014)Berger, Logsdail, Oberhofer, Farrow, Catlow,
  Sherwood, Sokol, Blum, and Reuter]{berger2014embedded}
D.~Berger, A.~J. Logsdail, H.~Oberhofer, M.~R. Farrow, C.~R.~A. Catlow,
  P.~Sherwood, A.~A. Sokol, V.~Blum, and K.~Reuter.
\newblock Embedded-cluster calculations in a numeric atomic orbital
  density-functional theory framework.
\newblock \emph{J. Chem. Phys.}, 141:\penalty0 024105, 2014.
\newblock \doi{10.1063/1.4885816}.

\bibitem[Knizia and Chan(2012)]{knizia2012density}
G.~Knizia and G.~K.-L. Chan.
\newblock Density matrix embedding: A simple alternative to dynamical
  mean-field theory.
\newblock \emph{Phys. Rev. Lett.}, 109:\penalty0 186404, 2012.
\newblock \doi{10.1103/PhysRevLett.109.186404}.

\bibitem[Anisimov et~al.(1997)Anisimov, Poteryaev, Korotin, Anokhin, and
  Kotliar]{anisimov1997first}
V.~I. Anisimov, A.~I. Poteryaev, M.~A. Korotin, A.~O. Anokhin, and G.~Kotliar.
\newblock First-principles calculations of the electronic structure and spectra
  of strongly correlated systems: dynamical mean-field theory.
\newblock \emph{J. Phys.: Condens. Matter}, 9:\penalty0 7359, 1997.
\newblock \doi{10.1088/0953-8984/9/4/002}.

\bibitem[Gao et~al.(2021)Gao, Nakamura, Gujarati, Jones, Rice, Wood, Pistoia,
  Garcia, and Yamamoto]{Gao2021}
Q.~Gao, H.~Nakamura, T.~P. Gujarati, G.~O. Jones, J.~E. Rice, S.~P. Wood,
  M.~Pistoia, J.~M. Garcia, and N.~Yamamoto.
\newblock Computational investigations of the lithium superoxide dimer
  rearrangement on noisy quantum devices.
\newblock \emph{J. Phys. Chem. A}, 125:\penalty0 1827--1836, 2021.
\newblock \doi{10.1021/acs.jpca.0c09530}.

\bibitem[Gujarati et~al.(2023)Gujarati, Motta, Friedhoff, Rice, Nguyen,
  Barkoutsos, Thompson, Smith, Kagele, Brei, Jones, and Williams]{Gujarati2022}
T.~P. Gujarati, M.~Motta, T.~N. Friedhoff, J.~E. Rice, N.~Nguyen, P.~K.
  Barkoutsos, R.~J. Thompson, T.~Smith, M.~Kagele, M.~Brei, B.~A. Jones, and
  K.~Williams.
\newblock Quantum computation of reactions on surfaces using local embedding.
\newblock \emph{npj Quantum Inf.}, 9:\penalty0 88, 2023.
\newblock \doi{10.1038/s41534-023-00753-1}.

\bibitem[Westermayr and Marquetand(2021)]{Westermayr2021}
J.~Westermayr and P.~Marquetand.
\newblock Machine learning for electronically excited states of molecules.
\newblock \emph{Chem. Rev.}, 121:\penalty0 9873--9926, 2021.
\newblock \doi{10.1021/acs.chemrev.0c00749}.

\bibitem[Ollitrault et~al.(2020{\natexlab{a}})Ollitrault, Mazzola, and
  Tavernelli]{Ollitrault2020}
P.~J. Ollitrault, G.~Mazzola, and I.~Tavernelli.
\newblock Nonadiabatic molecular quantum dynamics with quantum computers.
\newblock \emph{Phys. Rev. Lett.}, 125:\penalty0 260511, 2020{\natexlab{a}}.
\newblock \doi{10.1103/PhysRevLett.125.260511}.

\bibitem[Colless et~al.(2018)Colless, Ramasesh, Dahlen, Blok, Kimchi-Schwartz,
  McClean, Carter, de~Jong, and Siddiqi]{Colless2018}
J.~I. Colless, V.~V. Ramasesh, D.~Dahlen, M.~S. Blok, M.~E. Kimchi-Schwartz,
  J.~R. McClean, J.~Carter, W.~A. de~Jong, and I.~Siddiqi.
\newblock Computation of molecular spectra on a quantum processor with an
  error-resilient algorithm.
\newblock \emph{Phys. Rev. X}, 8:\penalty0 011021, 2018.
\newblock \doi{10.1103/PhysRevX.8.011021}.

\bibitem[Ollitrault et~al.(2021)Ollitrault, Miessen, and
  Tavernelli]{Ollitrault2021}
P.~J. Ollitrault, A.~Miessen, and I.~Tavernelli.
\newblock Molecular quantum dynamics: A quantum computing perspective.
\newblock \emph{Acc. Chem. Res.}, 54:\penalty0 4229--4238, 2021.
\newblock \doi{10.1021/acs.accounts.1c00514}.

\bibitem[Babbush et~al.(2015)Babbush, McClean, Wecker, Aspuru-Guzik, and
  Wiebe]{Babbush2015}
R.~Babbush, J.~McClean, D.~Wecker, A.~Aspuru-Guzik, and N.~Wiebe.
\newblock Chemical basis of {Trotter-Suzuki} errors in quantum chemistry
  simulation.
\newblock \emph{Phys. Rev. A}, 91:\penalty0 022311, 2015.
\newblock \doi{10.1103/PhysRevA.91.022311}.

\bibitem[Miessen et~al.(2021)Miessen, Ollitrault, and Tavernelli]{Miessen2021}
A.~Miessen, P.~J. Ollitrault, and I.~Tavernelli.
\newblock Quantum algorithms for quantum dynamics: A performance study on the
  spin-boson model.
\newblock \emph{Phys. Rev. Res.}, 3:\penalty0 043212, 2021.
\newblock \doi{10.1103/PhysRevResearch.3.043212}.

\bibitem[Fedorov and Gelfand(2021)]{Fedorov2021}
A.~K. Fedorov and M.~S. Gelfand.
\newblock Towards practical applications in quantum computational biology.
\newblock \emph{Nat. Comput. Sci.}, 1:\penalty0 114--119, 2021.
\newblock \doi{10.1038/s43588-021-00024-z}.

\bibitem[Teplukhin et~al.(2020)Teplukhin, Kendrick, Tretiak, and
  Dub]{Teplukhin2020}
A.~Teplukhin, B.~K. Kendrick, S.~Tretiak, and P.~A. Dub.
\newblock Electronic structure with direct diagonalization on a {D-Wave}
  quantum annealer.
\newblock \emph{Sci. Rep.}, 10:\penalty0 20753, 2020.
\newblock \doi{10.1038/s41598-020-77315-4}.

\bibitem[Teplukhin et~al.(2021)Teplukhin, Kendrick, Mniszewski, Zhang, Kumar,
  Negre, Anisimov, Tretiak, and Dub]{Teplukhin2021}
A.~Teplukhin, B.~K. Kendrick, S.~M. Mniszewski, Y.~Zhang, A.~Kumar, C.~F.~A.
  Negre, P.~M. Anisimov, S.~Tretiak, and P.~A. Dub.
\newblock Computing molecular excited states on a {D-Wave} quantum annealer.
\newblock \emph{Sci. Rep.}, 11:\penalty0 18796, 2021.
\newblock \doi{10.1038/s41598-021-98331-y}.

\bibitem[Streif et~al.(2019)Streif, Neukart, and Leib]{Streif2018}
M.~Streif, F.~Neukart, and M.~Leib.
\newblock Solving quantum chemistry problems with a {D-Wave} quantum annealer.
\newblock In \emph{QTOP 2019}, pages 111--122. Springer, Cham, 2019.
\newblock \doi{10.1007/978-3-030-14082-3_10}.

\bibitem[Genin et~al.(2019)Genin, Ryabinkin, and Izmaylov]{Genin2019}
S.~N. Genin, I.~G. Ryabinkin, and A.~F. Izmaylov.
\newblock Quantum chemistry on quantum annealers, 2019.
\newblock \href{https://doi.org/10.48550/arxiv.1901.04715}{arXiv:1901.04715}.

\bibitem[Mulligan et~al.(2019)Mulligan, Melo, Merritt, Slocum, Weitzner,
  Watkins, Renfrew, Pelissier, Arora, and Bonneau]{Mulligan2019}
V.~K. Mulligan, H.~Melo, H.~I. Merritt, S.~Slocum, B.~D. Weitzner, A.~M.
  Watkins, P.~D. Renfrew, C.~Pelissier, P.~S. Arora, and R.~Bonneau.
\newblock Designing peptides on a quantum computer, 2019.
\newblock \href{https://doi.org/10.1101/752485}{bioRxiv:752485,
  doi:10.1101/752485}.

\bibitem[Fox et~al.(2021)Fox, Branson, and Walker]{Fox2021}
D.~M. Fox, K.~M. Branson, and R.~C. Walker.
\newblock {mRNA} codon optimization with quantum computers.
\newblock \emph{PLoS One}, 16:\penalty0 e0259101, 2021.
\newblock \doi{10.1371/journal.pone.0259101}.

\bibitem[Hatakeyama-Sato et~al.(2021)Hatakeyama-Sato, Kashikawa, Kimura, and
  Oyaizu]{Hatakeyama-Sato}
K.~Hatakeyama-Sato, T.~Kashikawa, K.~Kimura, and K.~Oyaizu.
\newblock Tackling the challenge of a huge materials science search space with
  quantum-inspired annealing.
\newblock \emph{Adv. Intell. Syst.}, 3:\penalty0 2000209, 2021.
\newblock \doi{10.1002/aisy.202000209}.

\bibitem[Gao et~al.(2023{\natexlab{b}})Gao, Jones, Kobayashi, Sugawara,
  Yamashita, Kawaguchi, Tanaka, and Yamamoto]{Gao2023}
Q.~Gao, G.~O. Jones, T.~Kobayashi, M.~Sugawara, H.~Yamashita, H.~Kawaguchi,
  S.~Tanaka, and N.~Yamamoto.
\newblock Quantum-classical computational molecular design of deuterated
  high-efficiency {OLED} emitters.
\newblock \emph{Intell. Comput.}, 2:\penalty0 0037, 2023{\natexlab{b}}.
\newblock \doi{10.34133/icomputing.0037}.

\bibitem[Bednorz and M{\"u}ller(1986)]{Bednorz1986}
J.~G. Bednorz and K.~A. M{\"u}ller.
\newblock Possible high {$T_c$} superconductivity in the {Ba-La-Cu-O} system.
\newblock \emph{Z. Phys. B Condens. Matter}, 64:\penalty0 189--193, 1986.
\newblock \doi{10.1007/BF01303701}.

\bibitem[Sheng and Hermann(1988)]{Sheng1988}
Z.~Z. Sheng and A.~M. Hermann.
\newblock Bulk superconductivity at {120 K} in the {Tl-Ca/Ba-Cu-O} system.
\newblock \emph{Nature}, 332:\penalty0 138--139, 1988.
\newblock \doi{10.1038/332138a0}.

\bibitem[Zhou et~al.(2021)Zhou, Lee, Imada, Trivedi, Phillips, Kee,
  T{\"o}rm{\"a}, and Eremets]{Zhou2021}
X.~Zhou, W.-S. Lee, M.~Imada, N.~Trivedi, P.~Phillips, H.-Y. Kee,
  P.~T{\"o}rm{\"a}, and M.~Eremets.
\newblock High-temperature superconductivity.
\newblock \emph{Nat. Rev. Phys.}, 3:\penalty0 462--465, 2021.
\newblock \doi{10.1038/s42254-021-00324-3}.

\bibitem[Lilia et~al.(2022)Lilia, Hennig, Hirschfeld, Profeta, Sanna, Zurek,
  Pickett, Amsler, Dias, Eremets, Heil, Hemley, Liu, Ma, Pierleoni, Kolmogorov,
  Rybin, Novoselov, Anisimov, Oganov, Pickard, Bi, Arita, Errea, Pellegrini,
  Requist, Gross, Margine, Xie, Quan, Hire, Fanfarillo, Stewart, Hamlin,
  Stanev, Gonnelli, Piatti, Romanin, Daghero, and Valenti]{Lilia2022}
B.~Lilia, R.~Hennig, P.~Hirschfeld, G.~Profeta, A.~Sanna, E.~Zurek, W.~E.
  Pickett, M.~Amsler, R.~Dias, M.~I. Eremets, C.~Heil, R.~J. Hemley, H.~Liu,
  Y.~Ma, C.~Pierleoni, A.~N. Kolmogorov, N.~Rybin, D.~Novoselov, V.~Anisimov,
  A.~R. Oganov, C.~J. Pickard, T.~Bi, R.~Arita, I.~Errea, C.~Pellegrini,
  R.~Requist, E.~K.~U. Gross, E.~R. Margine, S.~R. Xie, Y.~Quan, A.~Hire,
  L.~Fanfarillo, G.~R. Stewart, J.~J. Hamlin, V.~Stanev, R.~S. Gonnelli,
  E.~Piatti, D.~Romanin, D.~Daghero, and R.~Valenti.
\newblock The 2021 room-temperature superconductivity roadmap.
\newblock \emph{J. Phys.: Condens. Matter}, 34:\penalty0 183002, 2022.
\newblock \doi{10.1088/1361-648x/ac2864}.

\bibitem[Dagotto(1994)]{Dagotto1994}
E.~Dagotto.
\newblock Correlated electrons in high-temperature superconductors.
\newblock \emph{Rev. Mod. Phys.}, 66:\penalty0 763--840, 1994.
\newblock \doi{10.1103/RevModPhys.66.763}.

\bibitem[Bardeen et~al.(1957{\natexlab{a}})Bardeen, Cooper, and
  Schrieffer]{Bardeen1957}
J.~Bardeen, L.~N. Cooper, and J.~R. Schrieffer.
\newblock Microscopic theory of superconductivity.
\newblock \emph{Phys. Rev.}, 106:\penalty0 162--164, 1957{\natexlab{a}}.
\newblock \doi{10.1103/PhysRev.106.162}.

\bibitem[Bardeen et~al.(1957{\natexlab{b}})Bardeen, Cooper, and
  Schrieffer]{Bardeen1957a}
J.~Bardeen, L.~N. Cooper, and J.~R. Schrieffer.
\newblock Theory of superconductivity.
\newblock \emph{Phys. Rev.}, 108:\penalty0 1175--1204, 1957{\natexlab{b}}.
\newblock \doi{10.1103/PhysRev.108.1175}.

\bibitem[{International Energy Agency}(2021)]{IEA2021}
{International Energy Agency}.
\newblock Net zero by 2050.
\newblock Flagship report, IEA, 2021.
\newblock Accessed: 2023-06-01.

\bibitem[Ball(2021)]{Ball2021}
P.~Ball.
\newblock The chase for fusion energy.
\newblock \emph{Nature}, 599:\penalty0 362--366, 2021.
\newblock \doi{10.1038/d41586-021-03401-w}.

\bibitem[Fradkin et~al.(2015)Fradkin, Kivelson, and Tranquada]{Fradkin2015}
E.~Fradkin, S.~A. Kivelson, and J.~M. Tranquada.
\newblock Colloquium: Theory of intertwined orders in high temperature
  superconductors.
\newblock \emph{Rev. Mod. Phys.}, 87:\penalty0 457--482, 2015.
\newblock \doi{10.1103/RevModPhys.87.457}.

\bibitem[LeBlanc et~al.(2015)LeBlanc, Antipov, Becca, Bulik, Chan, Chung, Deng,
  Ferrero, Henderson, Jim{\'e}nez-Hoyos, Kozik, Liu, Millis, Prokof'ev, Qin,
  Scuseria, Shi, Svistunov, Tocchio, Tupitsyn, White, Zhang, Zheng, Zhu, and
  Gull]{SimonsCollaboration2015}
J.~P.~F. LeBlanc, A.~E. Antipov, F.~Becca, I.~W. Bulik, G.~K.-L. Chan, C.-M.
  Chung, Y.~Deng, M.~Ferrero, T.~M. Henderson, C.~A. Jim{\'e}nez-Hoyos,
  E.~Kozik, X.-W. Liu, A.~J. Millis, N.~V. Prokof'ev, M.~Qin, G.~E. Scuseria,
  H.~Shi, B.~V. Svistunov, L.~F. Tocchio, I.~S. Tupitsyn, S.~R. White,
  S.~Zhang, B.-X. Zheng, Z.~Zhu, and E.~Gull.
\newblock Solutions of the two-dimensional {Hubbard} model: Benchmarks and
  results from a wide range of numerical algorithms.
\newblock \emph{Phys. Rev. X}, 5:\penalty0 041041, 2015.
\newblock \doi{10.1103/PhysRevX.5.041041}.

\bibitem[Hensgens et~al.(2017)Hensgens, Fujita, Janssen, Li, Diepen, Reichl,
  Wegscheider, Sarma, and Vandersypen]{Hensgens2017}
T.~Hensgens, T.~Fujita, L.~Janssen, X.~Li, C.~J.~Van Diepen, C.~Reichl,
  W.~Wegscheider, S.~Das Sarma, and L.~M.~K. Vandersypen.
\newblock Quantum simulation of a {Fermi-Hubbard} model using a semiconductor
  quantum dot array.
\newblock \emph{Nature}, 548:\penalty0 70--73, 2017.
\newblock \doi{10.1038/nature23022}.

\bibitem[Wang et~al.(2022)Wang, Khatami, Fei, Wyrick, Namboodiri, Kashid,
  Rigosi, Bryant, and Silver]{Wang2022}
X.~Wang, E.~Khatami, F.~Fei, J.~Wyrick, P.~Namboodiri, R.~Kashid, A.~F. Rigosi,
  G.~Bryant, and R.~Silver.
\newblock Experimental realization of an extended {Fermi-Hubbard} model using a
  {2D} lattice of dopant-based quantum dots.
\newblock \emph{Nat. Commun.}, 13:\penalty0 6824, 2022.
\newblock \doi{10.1038/s41467-022-34220-w}.

\bibitem[Mazurenko et~al.(2017)Mazurenko, Chiu, Ji, Parsons, Kan{\'a}sz-Nagy,
  Schmidt, Grusdt, Demler, Greif, and Greiner]{Mazurenko2017}
A.~Mazurenko, C.~S. Chiu, G.~Ji, M.~F. Parsons, M.~Kan{\'a}sz-Nagy, R.~Schmidt,
  F.~Grusdt, E.~Demler, D.~Greif, and M.~Greiner.
\newblock A cold-atom {Fermi-Hubbard} antiferromagnet.
\newblock \emph{Nature}, 545:\penalty0 462--466, 2017.
\newblock \doi{10.1038/nature22362}.

\bibitem[Hirthe et~al.(2023)Hirthe, Chalopin, Bourgund, Bojovi{\'c}, Bohrdt,
  Demler, Grusdt, Bloch, and Hilker]{Hirthe2023}
S.~Hirthe, T.~Chalopin, D.~Bourgund, P.~Bojovi{\'c}, A.~Bohrdt, E.~Demler,
  F.~Grusdt, I.~Bloch, and T.~A. Hilker.
\newblock Magnetically mediated hole pairing in fermionic ladders of ultracold
  atoms.
\newblock \emph{Nature}, 413:\penalty0 463--467, 2023.
\newblock \doi{10.1038/s41586-022-05437-y}.

\bibitem[Cai(2020)]{Cai2020}
Z.~Cai.
\newblock Resource estimation for quantum variational simulations of the
  {Hubbard} model.
\newblock \emph{Phys. Rev. Appl.}, 14:\penalty0 014059, 2020.
\newblock \doi{10.1103/PhysRevApplied.14.014059}.

\bibitem[Stanisic et~al.(2022)Stanisic, Bosse, Gambetta, Santos, Mruczkiewicz,
  O’Brien, Ostby, and Montanaro]{Stanisic2022}
S.~Stanisic, J.~L. Bosse, F.~M. Gambetta, R.~A. Santos, W.~Mruczkiewicz, T.~E.
  O’Brien, E.~Ostby, and A.~Montanaro.
\newblock Observing ground-state properties of the {Fermi-Hubbard} model using
  a scalable algorithm on a quantum computer.
\newblock \emph{Nat. Commun.}, 13:\penalty0 5743, 2022.
\newblock \doi{10.1038/s41467-022-33335-4}.

\bibitem[{Google Quantum AI and collaborators}(2020)]{Google2020}
{Google Quantum AI and collaborators}.
\newblock Observation of separated dynamics of charge and spin in the
  {Fermi-Hubbard} model, 2020.
\newblock \href{https://arxiv.org/abs/2010.07965}{arXiv:2010.07965}.

\bibitem[Kivlichan et~al.(2020)Kivlichan, Gidney, Berry, Wiebe, McClean, Sun,
  Jiang, Rubin, Fowler, Aspuru-Guzik, Neven, and Babbush]{Kivlichan2020}
I.~D. Kivlichan, C.~Gidney, D.~W. Berry, N.~Wiebe, J.~McClean, W.~Sun,
  Z.~Jiang, N.~Rubin, A.~Fowler, A.~Aspuru-Guzik, H.~Neven, and R.~Babbush.
\newblock Improved fault-tolerant quantum simulation of condensed-phase
  correlated electrons via {Trotterization}.
\newblock \emph{Quantum}, 4:\penalty0 296, 2020.
\newblock \doi{10.22331/q-2020-07-16-296}.

\bibitem[Clinton et~al.(2021)Clinton, Bausch, and Cubitt]{Clinton2021}
L.~Clinton, J.~Bausch, and T.~Cubitt.
\newblock Hamiltonian simulation algorithms for near-term quantum hardware.
\newblock \emph{Nat. Commun.}, 12:\penalty0 4989, 2021.
\newblock \doi{10.1038/s41467-021-25196-0}.

\bibitem[Drozdov et~al.(2015)Drozdov, Eremets, Troyan, Ksenofontov, and
  Shylin]{Drozdov2015}
A.~P. Drozdov, M.~I. Eremets, I.~A. Troyan, V.~Ksenofontov, and S.~I. Shylin.
\newblock Conventional superconductivity at 203 kelvin at high pressures in the
  sulfur hydride system.
\newblock \emph{Nature}, 525:\penalty0 73--76, 2015.
\newblock \doi{10.1038/nature14964}.

\bibitem[Castelvecchi(2023{\natexlab{a}})]{Castelvecchi2023s}
D.~Castelvecchi.
\newblock Why superconductor research is in a `golden age' - despite
  controversy, 2023{\natexlab{a}}.
\newblock
  \href{https://doi.org/10.1038/d41586-023-03551-z}{doi:10.1038/d41586-023-03551-z}.

\bibitem[Li et~al.(2014)Li, Hao, Liu, Li, and Ma]{Li2014}
Y.~Li, J.~Hao, H.~Liu, Y.~Li, and Y.~Ma.
\newblock The metallization and superconductivity of dense hydrogen sulfide.
\newblock \emph{J. Chem. Phys.}, 140:\penalty0 174712, 2014.
\newblock \doi{10.1063/1.4874158}.

\bibitem[Xiao et~al.(2023)Xiao, He, Georges, and Zhang]{Xiao2023}
B.~Xiao, Y.-Y. He, A.~Georges, and S.~Zhang.
\newblock Temperature dependence of spin and charge orders in the doped
  two-dimensional {Hubbard} model.
\newblock \emph{Phys. Rev. X}, 13:\penalty0 011007, 2023.
\newblock \doi{10.1103/PhysRevX.13.011007}.

\bibitem[Xu et~al.(2024)Xu, Chung, Qin, Schollw{\"o}ck, White, and
  Zhang]{Xu2024}
H.~Xu, C.-M. Chung, M.~Qin, U.~Schollw{\"o}ck, S.~R. White, and S.~Zhang.
\newblock Coexistence of superconductivity with partially filled stripes in the
  {Hubbard} model.
\newblock \emph{Science}, 384:\penalty0 eadh7691, 2024.
\newblock \doi{10.1126/science.adh7691}.

\bibitem[Xu et~al.(2023)Xu, Kendrick, Kale, Gang, Ji, Scalettar, Lebrat, and
  Greiner]{Xu2023f}
M.~Xu, L.~H. Kendrick, A.~Kale, Y.~Gang, G.~Ji, R.~T. Scalettar, M.~Lebrat, and
  M.~Greiner.
\newblock Frustration- and doping-induced magnetism in a {Fermi–Hubbard}
  simulator.
\newblock \emph{Nature}, 620:\penalty0 971--976, 2023.
\newblock \doi{10.1038/s41586-023-06280-5}.

\bibitem[Homeier et~al.(2024)Homeier, Harris, Blatz, Geier, Hollerith,
  Schollw{\"o}ck, Grusdt, and Bohrdt]{Homeier2024}
L.~Homeier, T.~J. Harris, T.~Blatz, S.~Geier, S.~Hollerith, U.~Schollw{\"o}ck,
  F.~Grusdt, and A.~Bohrdt.
\newblock Antiferromagnetic bosonic {$t-J$} models and their quantum simulation
  in tweezer arrays.
\newblock \emph{Phys. Rev. Lett.}, 132:\penalty0 230401, 2024.
\newblock \doi{10.1103/PhysRevLett.132.230401}.

\bibitem[Bohrdt et~al.(2021)Bohrdt, Homeier, Reinmoser, Demler, and
  Grusdt]{Bohrdt2021}
A.~Bohrdt, L.~Homeier, C.~Reinmoser, E.~Demler, and F.~Grusdt.
\newblock Exploration of doped quantum magnets with ultracold atoms.
\newblock \emph{Ann. Phys.}, 435:\penalty0 168651, 2021.
\newblock \doi{10.1016/j.aop.2021.168651}.

\bibitem[Lv et~al.(2021)Lv, Zhong, Liu, Fang, Yan, Chen, Kong, Lee, Liu, Li,
  Liu, Song, Chen, Yan, and Yu]{Lv2021}
C.~Lv, L.~Zhong, H.~Liu, Z.~Fang, C.~Yan, M.~Chen, Y.~Kong, C.~Lee, D.~Liu,
  S.~Li, J.~Liu, L.~Song, G.~Chen, Q.~Yan, and G.~Yu.
\newblock Selective electrocatalytic synthesis of urea with nitrate and carbon
  dioxide.
\newblock \emph{Nat. Sustain.}, 4:\penalty0 868--876, 2021.
\newblock \doi{10.1038/s41893-021-00741-3}.

\bibitem[{Boston Consulting Group}(2020)]{BCG2020}
{Boston Consulting Group}.
\newblock A quantum advantage in fighting climate change.
\newblock
  \url{https://www.bcg.com/publications/2020/quantum-advantage-fighting-climate-change},
  2020.
\newblock Accessed: 2022-12-01.

\bibitem[Burgess and Lowe(1996)]{Burgess1996}
B.~K. Burgess and D.~J. Lowe.
\newblock Mechanism of molybdenum nitrogenase.
\newblock \emph{Chem. Rev.}, 96:\penalty0 2983--3012, 1996.
\newblock \doi{10.1021/cr950055x}.

\bibitem[Vogiatzis et~al.(2019)Vogiatzis, Polynski, Kirkland, Townsend,
  Hashemi, Liu, and Pidko]{Vogiatzis2019}
K.~D. Vogiatzis, M.~V. Polynski, J.~K. Kirkland, J.~Townsend, A.~Hashemi,
  C.~Liu, and E.~A. Pidko.
\newblock Computational approach to molecular catalysis by 3d transition
  metals: Challenges and opportunities.
\newblock \emph{Chem. Rev.}, 119:\penalty0 2453--2523, 2019.
\newblock \doi{10.1021/acs.chemrev.8b00361}.

\bibitem[Hoffman et~al.(2014)Hoffman, Lukoyanov, Yang, Dean, and
  Seefeldt]{Hoffman2014}
B.~M. Hoffman, D.~Lukoyanov, Z.-Y. Yang, D.~R. Dean, and L.~C. Seefeldt.
\newblock Mechanism of nitrogen fixation by nitrogenase: The next stage.
\newblock \emph{Chem. Rev.}, 114:\penalty0 4041--4062, 2014.
\newblock \doi{10.1021/cr400641x}.

\bibitem[Fowler et~al.(2012)Fowler, Mariantoni, Martinis, and
  Cleland]{Fowler2012}
A.~G. Fowler, M.~Mariantoni, J.~M. Martinis, and A.~N. Cleland.
\newblock Surface codes: Towards practical large-scale quantum computation.
\newblock \emph{Phys. Rev. A}, 86:\penalty0 032324, 2012.
\newblock \doi{10.1103/PhysRevA.86.032324}.

\bibitem[Lee et~al.(2021)Lee, Berry, Gidney, Huggins, McClean, Wiebe, and
  Babbush]{Lee2021}
J.~Lee, D.~W. Berry, C.~Gidney, W.~J. Huggins, J.~R. McClean, N.~Wiebe, and
  R.~Babbush.
\newblock Even more efficient quantum computations of chemistry through tensor
  hypercontraction.
\newblock \emph{PRX Quantum}, 2:\penalty0 030305, 2021.
\newblock \doi{10.1103/PRXQuantum.2.030305}.

\bibitem[Low et~al.(2024)Low, Kliuchnikov, and Schaeffer]{Low2024}
G.~H. Low, V.~Kliuchnikov, and L.~Schaeffer.
\newblock Trading {T-gates} for dirty qubits in state preparation and unitary
  synthesis.
\newblock \emph{Quantum}, 8:\penalty0 1375, 2024.
\newblock \doi{10.22331/q-2024-06-17-1375}.

\bibitem[Aharonov and Ta-Shma(2003)]{Aharonov2003}
D.~Aharonov and A.~Ta-Shma.
\newblock Adiabatic quantum state generation and statistical zero knowledge.
\newblock In \emph{Proc. 35rd ACM STOC}, pages 20--29, New York, NY, 2003. ACM.
\newblock \doi{10.1145/780542.780546}.

\bibitem[Childs and Wiebe(2012)]{Childs2012}
A.~M. Childs and N.~Wiebe.
\newblock Hamiltonian simulation using linear combinations of unitary
  operations.
\newblock \emph{Quantum Inf. Comput.}, 12:\penalty0 901--924, 2012.
\newblock \doi{10.26421/QIC12.11-12}.

\bibitem[Berry et~al.(2015)Berry, Childs, Cleve, Kothari, and Somma]{Berry2015}
D.~M. Berry, A.~M. Childs, R.~Cleve, R.~Kothari, and R.~D. Somma.
\newblock Simulating {Hamiltonian} dynamics with a truncated {Taylor} series.
\newblock \emph{Phys. Rev. Lett.}, 114:\penalty0 090502, 2015.
\newblock \doi{10.1103/PhysRevLett.114.090502}.

\bibitem[Szegedy(2004)]{Szegedy2004}
M.~Szegedy.
\newblock Quantum speed-up of {Markov} chain based algorithms.
\newblock In \emph{45th IEEE SFCS}, pages 32--41, 2004.
\newblock \doi{10.1109/FOCS.2004.53}.

\bibitem[Childs(2010)]{Childs2010}
A.~M. Childs.
\newblock On the relationship between continuous-and discrete-time quantum
  walk.
\newblock \emph{Commun. Math. Phys.}, 294:\penalty0 581--603, 2010.
\newblock \doi{10.1007/s00220-009-0930-1}.

\bibitem[Marriott and Watrous(2005)]{Marriott2005}
C.~Marriott and J.~Watrous.
\newblock Quantum {Arthur-Merlin} games.
\newblock \emph{Comput. Complex.}, 14:\penalty0 122--152, 2005.
\newblock \doi{10.1007/s00037-005-0194-x}.

\bibitem[Tazhigulov et~al.(2022)Tazhigulov, Sun, Haghshenas, Zhai, Tan, Rubin,
  Babbush, Minnich, and Chan]{Tazhigulov2022}
R.~N. Tazhigulov, S.-N. Sun, R.~Haghshenas, H.~Zhai, A.~T.~K. Tan, N.~C. Rubin,
  R.~Babbush, A.~J. Minnich, and G.~K.-L. Chan.
\newblock Simulating models of challenging correlated molecules and materials
  on the {Sycamore} quantum processor.
\newblock \emph{PRX Quantum}, 3:\penalty0 040318, 2022.
\newblock \doi{10.1103/PRXQuantum.3.040318}.

\bibitem[Debnath(2012)]{Debnath2012}
L.~Debnath.
\newblock \emph{Nonlinear Partial Differential Equations for Scientists and
  Engineers}.
\newblock Springer, New York, NY, 3rd edition, 2012.
\newblock \doi{10.1007/978-0-8176-8265-1}.

\bibitem[Tadmore(2012)]{Tadmore2012}
E.~Tadmore.
\newblock A review of numerical methods for nonlinear partial differential
  equations.
\newblock \emph{Bull. Amer. Math. Soc.}, 49:\penalty0 507--554, 2012.
\newblock \doi{10.1090/S0273-0979-2012-01379-4}.

\bibitem[Bartels(2015)]{Bartels2015}
S.~Bartels.
\newblock \emph{Numerical Methods for Nonlinear Partial Differential
  Equations}.
\newblock Springer, Cham, 2015.
\newblock \doi{10.1007/978-3-319-13797-1}.

\bibitem[Han et~al.(2018)Han, Jentzen, and E]{Han2018}
J.~Han, A.~Jentzen, and W.~E.
\newblock Solving high-dimensional partial differential equations using deep
  learning.
\newblock \emph{Proc. Natl. Acad. Sci.}, 115:\penalty0 8505--8510, 2018.
\newblock \doi{10.1073/pnas.1718942115}.

\bibitem[Muther et~al.(2023)Muther, Dahaghi, Syed, and Pham]{Muther2023}
T.~Muther, A.~K. Dahaghi, F.~I. Syed, and V.~V. Pham.
\newblock Physical laws meet machine intelligence: current developments and
  future directions.
\newblock \emph{Artif. Intell. Rev.}, 56:\penalty0 6947--7013, 2023.
\newblock \doi{10.1007/s10462-022-10329-8}.

\bibitem[Biswas et~al.(2017)Biswas, Jiang, Kechezhi, Knysh, Mandr{\`a},
  O'Gorman, Perdomo-Ortiz, Petukhov, Realpe-G{\'o}mez, Rieffel, Venturelli,
  Vasko, and Wang]{Biswas2017}
R.~Biswas, Z.~Jiang, K.~Kechezhi, S.~Knysh, S.~Mandr{\`a}, B.~O'Gorman,
  A.~Perdomo-Ortiz, A.~Petukhov, J.~Realpe-G{\'o}mez, E.~Rieffel,
  D.~Venturelli, F.~Vasko, and Z.~Wang.
\newblock A {NASA} perspective on quantum computing: Opportunities and
  challenges.
\newblock \emph{Parallel Comput.}, 64:\penalty0 81--98, 2017.
\newblock \doi{10.1016/j.parco.2016.11.002}.

\bibitem[Rieffel et~al.(2019)Rieffel, Hadfield, Hogg, Mandr{\`a}, Marshall,
  Mossi, O'Gorman, Plamadeala, Tubman, Venturelli, Vinci, Wang, Wilson,
  Wudarski, and Biswas]{Rieffel2019}
E.~G. Rieffel, S.~Hadfield, T.~Hogg, S.~Mandr{\`a}, J.~Marshall, G.~Mossi,
  B.~O'Gorman, E.~Plamadeala, N.~M. Tubman, D.~Venturelli, W.~Vinci, Z.~Wang,
  M.~Wilson, F.~Wudarski, and R.~Biswas.
\newblock From {Ans{\"a}tze} to {Z}-gates: A {NASA} view of quantum computing.
\newblock In L.~Grandinetti, G.~R. Joubert, K.~Michielsen, S.~L. Mirtaheri,
  M.~Taufer, and R.~Yokota, editors, \emph{Future Trends of HPC in a Disruptive
  Scenario}, volume~34 of \emph{Advances in Parallel Computing}, pages
  133--160. IOS Press, Amsterdam, Netherlands, 2019.
\newblock \doi{10.3233/APC190010}.

\bibitem[Rieffel et~al.(2024)Rieffel, Asanjan, Alam, Anand, Neira, Block,
  Brady, Cotton, Izquierdo, Grabbe, Gustafson, Hadfield, Lott, Maciejewski,
  Mandr{\`a}, Marshall, Mossi, Bauza, Saied, Suri, Venturelli, Wang, and
  Biswas]{Rieffel2024}
E.~G. Rieffel, A.~A. Asanjan, M.~S. Alam, N.~Anand, D.~E.~Bernal Neira,
  S.~Block, L.~T. Brady, S.~Cotton, Z.~Gonzalez Izquierdo, S.~Grabbe,
  E.~Gustafson, S.~Hadfield, P.~A. Lott, F.~B. Maciejewski, S.~Mandr{\`a},
  J.~Marshall, G.~Mossi, H.~Munoz Bauza, J.~Saied, N.~Suri, D.~Venturelli,
  Z.~Wang, and R.~Biswas.
\newblock Assessing and advancing the potential of quantum computing: A {NASA}
  case study.
\newblock \emph{Future Gener. Comput. Syst.}, 160:\penalty0 598--618, 2024.
\newblock \doi{10.1016/j.future.2024.06.012}.

\bibitem[Moin and Mahesh(1998)]{Moin1998}
P.~Moin and K.~Mahesh.
\newblock Direct numerical simulation: A tool in turbulence research.
\newblock \emph{Annu. Rev. Fluid Mech.}, 30:\penalty0 539--578, 1998.
\newblock \doi{10.1146/annurev.fluid.30.1.539}.

\bibitem[Raissi et~al.(2019)Raissi, Perdikaris, and Karniadakis]{Raissi2019}
M.~Raissi, P.~Perdikaris, and G.~E. Karniadakis.
\newblock Physics-informed neural networks: A deep learning framework for
  solving forward and inverse problems involving nonlinear partial differential
  equations.
\newblock \emph{J. Comput. Phys.}, 379:\penalty0 686--707, 2019.
\newblock \doi{10.1016/j.jcp.2018.10.045}.

\bibitem[Karniadakis et~al.(2021)Karniadakis, Kevrekidis, Lu, Perdikaris, Wang,
  and Yang]{Karniadakis2021}
G.~E. Karniadakis, I.~G. Kevrekidis, L.~Lu, P.~Perdikaris, S.~Wang, and
  L.~Yang.
\newblock Physics-informed machine learning.
\newblock \emph{Nat. Rev. Phys.}, 3:\penalty0 422--440, 2021.
\newblock \doi{10.1038/s42254-021-00314-5}.

\bibitem[Brunton et~al.(2020)Brunton, Noack, and Koumoutsakos]{Brunton2020}
S.~L. Brunton, B.~R. Noack, and P.~Koumoutsakos.
\newblock Machine learning for fluid mechanics.
\newblock \emph{Annu. Rev. Fluid Mech.}, 52:\penalty0 477--508, 2020.
\newblock \doi{10.1146/annurev-fluid-010719-060214}.

\bibitem[Raissi et~al.(2020)Raissi, Yazdani, and Karniadakis]{Raissi2020}
M.~Raissi, A.~Yazdani, and G.~E. Karniadakis.
\newblock Hidden fluid mechanics: Learning velocity and pressure fields from
  flow visualizations.
\newblock \emph{Science}, 367:\penalty0 1026--1030, 2020.
\newblock \doi{10.1126/science.aaw4741}.

\bibitem[Huang et~al.(2022)Huang, Feng, Tang, and Lv]{Huang2022}
S.~Huang, W.~Feng, C.~Tang, and J.~Lv.
\newblock Partial differential equations meet deep neural networks: A survey,
  2022.
\newblock \href{https://arxiv.org/abs/2211.05567}{arXiv:2211.05567}.

\bibitem[Fallah and Aghdam(2024)]{Fallah2024}
A.~Fallah and M.~M. Aghdam.
\newblock Physics-informed neural network for solution of nonlinear
  differential equations.
\newblock In R.~N. Jazar and L.~Dai, editors, \emph{Nonlinear Approaches in
  Engineering Application: Automotive Engineering Problems}, chapter~5, pages
  163--178. Springer, Cham, 2024.
\newblock \doi{10.1007/978-3-031-53582-6_5}.

\bibitem[Hafiz et~al.(2024)Hafiz, Faiq, and Hassaballah]{Hafiz2024}
A.~M. Hafiz, I.~Faiq, and M.~Hassaballah.
\newblock Solving partial differential equations using large-data models: a
  literature review.
\newblock \emph{Artif. Intell. Rev.}, 57:\penalty0 152, 2024.
\newblock \doi{10.1007/s10462-024-10784-5}.

\bibitem[Chung(2002)]{Chung2002}
T.~J. Chung.
\newblock \emph{Computational Fluid Dynamics}.
\newblock Cambridge University Press, Cambridge, 2002.
\newblock \doi{10.1017/CBO9780511606205}.

\bibitem[Liu(2003)]{Liu2003m}
G.~R. Liu.
\newblock \emph{Mesh Free Methods: Moving Beyond the Finite Element Method}.
\newblock CRC Press, Boca Raton, FL, 2003.

\bibitem[Monaghan(2005)]{Monaghan2005}
J.~J. Monaghan.
\newblock Smoothed particle hydrodynamics.
\newblock \emph{Rep. Prog. Phys.}, 68:\penalty0 1703--1759, 2005.
\newblock \doi{10.1088/0034-4885/68/8/r01}.

\bibitem[Monaghan(2012)]{Monaghan2012}
J.~J. Monaghan.
\newblock Smoothed particle hydrodynamics and its diverse applications.
\newblock \emph{Annu. Rev. Fluid Mech.}, 44:\penalty0 323–346, 2012.
\newblock \doi{10.1146/annurev-fluid-120710-101220}.

\bibitem[Lucy(1977)]{Lucy1977}
L.~B. Lucy.
\newblock A numerical approach to the testing of the fission hypothesis.
\newblock \emph{Astron. J.}, 82:\penalty0 1013--1024, 1977.

\bibitem[Springel(2010)]{Springel2010}
V.~Springel.
\newblock Smoothed particle hydrodynamics in astrophysics.
\newblock \emph{Annu. Rev. Astron. Astrophys.}, 48:\penalty0 391–430, 2010.
\newblock \doi{10.1146/annurev-astro-081309-130914}.

\bibitem[Succi(2001)]{Succi2001}
S.~Succi.
\newblock \emph{The Lattice Boltzmann Equation for Fluid Dynamics and Beyond}.
\newblock Oxford University Press, New York, NY, 2001.

\bibitem[Kr{\"u}ger et~al.(2017)Kr{\"u}ger, Kusumaatmaja, Kuzmin, Shardt,
  Silva, and Viggen]{Kruger2017}
T.~Kr{\"u}ger, H.~Kusumaatmaja, A.~Kuzmin, O.~Shardt, G.~Silva, and E.~M.
  Viggen.
\newblock \emph{The Lattice Boltzmann Method}.
\newblock Springer, Cham, 2017.
\newblock \doi{10.1007/978-3-319-44649-3}.

\bibitem[Feichtinger et~al.(2015)Feichtinger, Habich, K{\"o}stler, R{\"u}de,
  and Aoki]{Feichtinger2015}
C.~Feichtinger, J.~Habich, H.~K{\"o}stler, U.~R{\"u}de, and T.~Aoki.
\newblock Performance modeling and analysis of heterogeneous lattice
  {Boltzmann} simulations on {CPU-GPU} clusters.
\newblock \emph{Parallel Comput.}, 46:\penalty0 1--13, 2015.
\newblock \doi{10.1016/j.parco.2014.12.003}.

\bibitem[Liu et~al.(2017)Liu, Zou, Cui, and Wu]{Liu2017l}
S.~Liu, N.~Zou, Y.~Cui, and W.~Wu.
\newblock Accelerating the parallelization of lattice {Boltzmann} method by
  exploiting the temporal locality.
\newblock In \emph{2017 IEEE ISPA/IUCC}, pages 1186--1193. IEEE, 2017.
\newblock \doi{10.1109/ISPA/IUCC.2017.00178}.

\bibitem[Tran et~al.(2015)Tran, Lee, and Choi]{Tran2015}
N.~P. Tran, M.~Lee, and D.~H. Choi.
\newblock Memory-efficient parallelization of {3D} lattice {Boltzmann} flow
  solver on a {GPU}.
\newblock In \emph{2015 IEEE 22nd Int. Conf. HiPC}, pages 315--324. IEEE, 2015.
\newblock \doi{10.1109/HiPC.2015.49}.

\bibitem[Chen and Doolen(1998)]{Chen1998}
S.~Chen and G.~D. Doolen.
\newblock Lattice {Boltzmann} method for fluid flows.
\newblock \emph{Annu. Rev. Fluid Mech.}, 30:\penalty0 329--364, 1998.
\newblock \doi{10.1146/annurev.fluid.30.1.329}.

\bibitem[Alowayyed et~al.(2017)Alowayyed, Groen, Coveney, and
  Hoekstra]{Alowayyed2017}
S.~Alowayyed, D.~Groen, P.~V. Coveney, and A.~G. Hoekstra.
\newblock Multiscale computing in the exascale era.
\newblock \emph{J. Comput. Sci.}, 22:\penalty0 15--25, 2017.
\newblock \doi{10.1016/j.jocs.2017.07.004}.

\bibitem[Succi et~al.(2019)Succi, Amati, Bernaschi, Falcucci, Lauricella, and
  Montessori]{Succi2019}
S.~Succi, G.~Amati, M.~Bernaschi, G.~Falcucci, M.~Lauricella, and
  A.~Montessori.
\newblock Towards exascale lattice {Boltzmann} computing.
\newblock \emph{Comput. Fluids}, 181:\penalty0 107--115, 2019.
\newblock \doi{10.1016/j.compfluid.2019.01.005}.

\bibitem[Wei et~al.(2004)Wei, Li, Mueller, and Kaufman]{Wei2004}
X.~Wei, W.~Li, K.~Mueller, and A.~E. Kaufman.
\newblock The lattice-{Boltzmann} method for simulating gaseous phenomena.
\newblock \emph{IEEE Trans Viz. Comput. Graph.}, 10:\penalty0 164--176, 2004.
\newblock \doi{10.1109/TVCG.2004.1260768}.

\bibitem[Krause et~al.(2021)Krause, Kummerl{\"a}nder, Avis, Kusumaatmaja,
  Dapelo, Klemens, Gaedtke, Hafen, Mink, Trunk, Marquardt, Maier, Haussmann,
  and Simonis]{Krause2021}
M.~J. Krause, A.~Kummerl{\"a}nder, S.~J. Avis, H.~Kusumaatmaja, D.~Dapelo,
  F.~Klemens, M.~Gaedtke, N.~Hafen, A.~Mink, R.~Trunk, J.~E. Marquardt, M.-L.
  Maier, M.~Haussmann, and S.~Simonis.
\newblock {OpenLB}--open source lattice {Boltzmann} code.
\newblock \emph{Comput. Math. Appl.}, 81:\penalty0 258--288, 2021.
\newblock \doi{10.1016/j.camwa.2020.04.033}.

\bibitem[Latt et~al.(2021)Latt, Malaspinas, Kontaxakis, Parmigiani, Lagrava,
  Brogi, Belgacem, Thorimbert, Leclaire, Li, Marson, Lemus, Kotsalos, Conradin,
  Coreixas, Petkantchin, Raynaud, Beny, and Chopard]{Latt2021}
J.~Latt, O.~Malaspinas, D.~Kontaxakis, A.~Parmigiani, D.~Lagrava, F.~Brogi,
  M.~B. Belgacem, Y.~Thorimbert, S.~Leclaire, S.~Li, F.~Marson, J.~Lemus,
  C.~Kotsalos, R.~Conradin, C.~Coreixas, R.~Petkantchin, F.~Raynaud, J.~Beny,
  and B.~Chopard.
\newblock Palabos: Parallel lattice {Boltzmann} solver.
\newblock \emph{Comput. Math. Appl.}, 81:\penalty0 334--350, 2021.
\newblock \doi{10.1016/j.camwa.2020.03.022}.

\bibitem[Yepez(2001{\natexlab{a}})]{yepez2001quantum}
J.~Yepez.
\newblock Quantum lattice-gas model for computational fluid dynamics.
\newblock \emph{Phys. Rev. E}, 63:\penalty0 046702, 2001{\natexlab{a}}.
\newblock \doi{10.1103/PhysRevE.63.046702}.

\bibitem[Berman et~al.(2002)Berman, Ezhov, Kamenev, and Yepez]{Berman2002}
G.~P. Berman, A.~A. Ezhov, D.~I. Kamenev, and J.~Yepez.
\newblock Simulation of the diffusion equation on a type-{II} quantum computer.
\newblock \emph{Phys. Rev. A}, 66:\penalty0 012310, 2002.
\newblock \doi{10.1103/PhysRevA.66.012310}.

\bibitem[Yepez(2001{\natexlab{b}})]{yepez2001type}
J.~Yepez.
\newblock Type-{II} quantum computers.
\newblock \emph{Int. J. Mod. Phys. C}, 12:\penalty0 1273--1284,
  2001{\natexlab{b}}.
\newblock \doi{10.1142/S0129183101002668}.

\bibitem[Love and Boghosian(2006)]{love2006type}
P.~J. Love and B.~M. Boghosian.
\newblock Type {II} quantum algorithms.
\newblock \emph{Phys. A: Stat. Mech. Appl.}, 362:\penalty0 210--214, 2006.
\newblock \doi{10.1016/j.physa.2005.09.017}.

\bibitem[Kocherla et~al.(2024)Kocherla, Song, Chrit, Gard, Dumitrescu, Alexeev,
  and Bryngelson]{Kocherla2024}
S.~Kocherla, Z.~Song, F.~E. Chrit, B.~Gard, E.~F. Dumitrescu, A.~Alexeev, and
  S.~H. Bryngelson.
\newblock Fully quantum algorithm for mesoscale fluid simulations with
  application to partial differential equations.
\newblock \emph{AVS Quantum Sci.}, 6:\penalty0 033806, 2024.
\newblock \doi{10.1116/5.0217675}.

\bibitem[Succi et~al.(2015)Succi, Fillion-Gourdeau, and
  Palpacelli]{succi2015quantum}
S.~Succi, F.~Fillion-Gourdeau, and S.~Palpacelli.
\newblock Quantum lattice {Boltzmann} is a quantum walk.
\newblock \emph{EPJ Quantum Technol.}, 2:\penalty0 12, 2015.
\newblock \doi{10.1140/epjqt/s40507-015-0025-1}.

\bibitem[Todorova and Steijl(2020)]{Todorova2020}
B.~N. Todorova and R.~Steijl.
\newblock Quantum algorithm for the collisionless {Boltzmann} equation.
\newblock \emph{J. Comput. Phys.}, 409:\penalty0 109347, 2020.
\newblock \doi{10.1016/j.jcp.2020.109347}.

\bibitem[Somma et~al.(2002)Somma, Ortiz, Gubernatis, Knill, and
  Laflamme]{somma2002simulating}
R.~Somma, G.~Ortiz, J.~E. Gubernatis, E.~Knill, and R.~Laflamme.
\newblock Simulating physical phenomena by quantum networks.
\newblock \emph{Phys. Rev. A}, 65:\penalty0 042323, 2002.
\newblock \doi{10.1103/PhysRevA.65.042323}.

\bibitem[Itani and Succi(2022)]{itani2022analysis}
W.~Itani and S.~Succi.
\newblock Analysis of {Carleman} linearization of lattice {Boltzmann}.
\newblock \emph{Fluids}, 7:\penalty0 24, 2022.
\newblock \doi{10.3390/fluids7010024}.

\bibitem[Itani et~al.(2024)Itani, Sreenivasan, and Succi]{Itani2024}
W.~Itani, K.~B. Sreenivasan, and S.~Succi.
\newblock Quantum algorithm for lattice {Boltzmann} ({QALB}) simulation of
  incompressible fluids with a nonlinear collision term.
\newblock \emph{Phys. Fluids}, 36:\penalty0 017112, 2024.
\newblock \doi{10.1063/5.0176569}.

\bibitem[Budinski(2021)]{Budinski2021}
L.~Budinski.
\newblock Quantum algorithm for the advection-diffusion equation simulated with
  the lattice {Boltzmann} method.
\newblock \emph{Quantum Inf. Process.}, 20:\penalty0 57, 2021.
\newblock \doi{10.1007/s11128-021-02996-3}.

\bibitem[Budinski(2022)]{ljubomir2022quantum}
L.~Budinski.
\newblock Quantum algorithm for the {Navier-Stokes} equations by using the
  streamfunction-vorticity formulation and the lattice {Boltzmann} method.
\newblock \emph{Int. J. Quantum Inf.}, 20:\penalty0 2150039, 2022.
\newblock \doi{10.1142/S0219749921500398}.

\bibitem[Schalkers and M{\"o}ller(2024)]{Schalkers2024}
M.~A. Schalkers and M.~M{\"o}ller.
\newblock On the importance of data encoding in quantum {Boltzmann} methods.
\newblock \emph{Quantum Inf. Process.}, 23:\penalty0 20, 2024.
\newblock \doi{10.1007/s11128-023-04216-6}.

\bibitem[Ferrari et~al.(2009)Ferrari, Dumbser, Toro, and Armanini]{Ferrari2009}
A.~Ferrari, M.~Dumbser, E.~F. Toro, and A.~Armanini.
\newblock A new {3D} parallel {SPH} scheme for free surface flows.
\newblock \emph{Comput. Fluids}, 38:\penalty0 1203--1217, 2009.
\newblock \doi{10.1016/j.compfluid.2008.11.012}.

\bibitem[Dom{\'i}nguez et~al.(2013)Dom{\'i}nguez, Crespo, and
  G{\'o}mez-Gesteira]{Dominguez2013}
J.~M. Dom{\'i}nguez, A.~J.~C. Crespo, and M.~G{\'o}mez-Gesteira.
\newblock Optimization strategies for {CPU} and {GPU} implementations of a
  smoothed particle hydrodynamics method.
\newblock \emph{Comput. Phys. Commun.}, 184:\penalty0 617--627, 2013.
\newblock \doi{10.1016/j.cpc.2012.10.015}.

\bibitem[Oger et~al.(2016)Oger, Touz{\'e}, Guibert, de~Leffe, Biddiscombe,
  Soumagne, and Piccinali]{Oger2016}
G.~Oger, D.~Le Touz{\'e}, D.~Guibert, M.~de~Leffe, J.~Biddiscombe, J.~Soumagne,
  and J.~G. Piccinali.
\newblock On distributed memory {MPI}-based parallelization of {SPH} codes in
  massive {HPC} context.
\newblock \emph{Comput. Phys. Commun.}, 200:\penalty0 1--14, 2016.
\newblock \doi{10.1016/j.cpc.2015.08.021}.

\bibitem[Dom{\'i}nguez et~al.(2022)Dom{\'i}nguez, Fourtakas, Altomare, Canelas,
  Tafuni, Garc{\'i}a-Feal, Mart{\'i}nez-Est{\'e}vez, Mokos, Vacondio, Crespo,
  Rogers, Stansby, and G{\'o}mez-Gesteira]{Dominguez2022}
J.~M. Dom{\'i}nguez, G.~Fourtakas, C.~Altomare, R.~B. Canelas, A.~Tafuni,
  O.~Garc{\'i}a-Feal, I.~Mart{\'i}nez-Est{\'e}vez, A.~Mokos, R.~Vacondio,
  A.~J.~C. Crespo, B.~D. Rogers, P.~K. Stansby, and M.~G{\'o}mez-Gesteira.
\newblock {DualSPHysics}: from fluid dynamics to multiphysics problems.
\newblock \emph{Comput. Part. Mech.}, 9:\penalty0 867--895, 2022.
\newblock \doi{10.1007/s40571-021-00404-2}.

\bibitem[Schaller et~al.(2024)Schaller, Borrow, Draper, Ivkovic, McAlpine,
  Vandenbroucke, Bah{\'e}, Chaikin, Chalk, Chan, Correa, van Daalen, Elbers,
  Gonnet, Hausammann, Helly, Hu{\v s}ko, Kegerreis, Nobels, Ploeckinger, Revaz,
  Roper, Ruiz-Bonilla, Sandnes, Uyttenhove, Willis, and Xiang]{Schaller2024}
M.~Schaller, J.~Borrow, P.~W. Draper, M.~Ivkovic, S.~McAlpine,
  B.~Vandenbroucke, Y.~Bah{\'e}, E.~Chaikin, A.~B.~G. Chalk, T.~K. Chan,
  C.~Correa, M.~van Daalen, W.~Elbers, P.~Gonnet, L.~Hausammann, J.~Helly,
  F.~Hu{\v s}ko, J.~A. Kegerreis, F.~S.~J. Nobels, S.~Ploeckinger, Y.~Revaz,
  W.~J. Roper, S.~Ruiz-Bonilla, T.~D. Sandnes, Y.~Uyttenhove, J.~S. Willis, and
  Z.~Xiang.
\newblock Swift: A modern highly-parallel gravity and smoothed particle
  hydrodynamics solver for astrophysical and cosmological applications.
\newblock \emph{Mon. Not. R. Astron. Soc.}, 530:\penalty0 2378--2419, 2024.
\newblock \doi{10.1093/mnras/stae922}.

\bibitem[Springel(2005)]{springel2005cosmological}
V.~Springel.
\newblock The cosmological simulation code {GADGET-2}.
\newblock \emph{Mon. Not. R. Astron. Soc.}, 364:\penalty0 1105--1134, 2005.
\newblock \doi{10.1111/j.1365-2966.2005.09655.x}.

\bibitem[Lind et~al.(2020)Lind, Rogers, and Stansby]{Lind2020}
S.~J. Lind, B.~D. Rogers, and P.~K. Stansby.
\newblock Review of smoothed particle hydrodynamics: towards converged
  {Lagrangian} flow modelling.
\newblock \emph{Proc. R. Soc. A}, 476:\penalty0 20190801, 2020.
\newblock \doi{10.1098/rspa.2019.0801}.

\bibitem[Vacondio et~al.(2021)Vacondio, Altomare, Leffe, Hu, Touz{\'e}, Lind,
  Marongiu, Marrone, Rogers, and Souto-Iglesias]{Vacondio2021}
R.~Vacondio, C.~Altomare, M.~De Leffe, X.~Hu, D.~Le Touz{\'e}, S.~Lind, J.-C.
  Marongiu, S.~Marrone, B.~D. Rogers, and A.~Souto-Iglesias.
\newblock Grand challenges for smoothed particle hydrodynamics numerical
  schemes.
\newblock \emph{Comp. Part. Mech.}, 8:\penalty0 575--588, 2021.
\newblock \doi{10.1007/s40571-020-00354-1}.

\bibitem[Au-Yeung et~al.(2024)Au-Yeung, Williams, Kendon, and
  Lind]{AuYeung2024}
R.~Au-Yeung, A.~J. Williams, V.~M. Kendon, and S.~J. Lind.
\newblock Quantum algorithm for smoothed particle hydrodynamics.
\newblock \emph{Comput. Phys. Commun.}, 294:\penalty0 108909, 2024.
\newblock \doi{10.1016/j.cpc.2023.108909}.

\bibitem[Buhrman et~al.(2001)Buhrman, Cleve, Watrous, and de~Wolf]{Buhrman2001}
H.~Buhrman, R.~Cleve, J.~Watrous, and R.~de~Wolf.
\newblock Quantum fingerprinting.
\newblock \emph{Phys. Rev. Lett.}, 87:\penalty0 167902, 2001.
\newblock \doi{10.1103/PhysRevLett.87.167902}.

\bibitem[Fanizza et~al.(2020)Fanizza, Rosati, Skotiniotis, Calsamiglia, and
  Giovannetti]{Fanizza2020}
M.~Fanizza, M.~Rosati, M.~Skotiniotis, J.~Calsamiglia, and V.~Giovannetti.
\newblock Beyond the swap test: Optimal estimation of quantum state overlap.
\newblock \emph{Phys. Rev. Lett.}, 124:\penalty0 060503, 2020.
\newblock \doi{10.1103/PhysRevLett.124.060503}.

\bibitem[Cincio et~al.(2018)Cincio, Suba{\k{s}}\i, Sornborger, and
  Coles]{Cincio2018}
L.~Cincio, Y.~Suba{\k{s}}\i, A.~T. Sornborger, and P.~J. Coles.
\newblock Learning the quantum algorithm for state overlap.
\newblock \emph{New J. Phys.}, 20:\penalty0 113022, 2018.
\newblock \doi{10.1088/1367-2630/aae94a}.

\bibitem[Aaronson and Rall(2019)]{Aaronson2019}
S.~Aaronson and P.~Rall.
\newblock Quantum approximate counting, simplified.
\newblock In \emph{{SOSA} 2020}, pages 24--32. SIAM, Philadelphia, PA, 2019.
\newblock \doi{10.1137/1.9781611976014.5}.

\bibitem[Dom{\'i}nguez et~al.(2011)Dom{\'i}nguez, Crespo, G{\'o}mez-Gesteira,
  and Marongiu]{Dominguez2011}
J.~M. Dom{\'i}nguez, A.~J.~C. Crespo, M.~G{\'o}mez-Gesteira, and J.~C.
  Marongiu.
\newblock Neighbour lists in smoothed particle hydrodynamics.
\newblock \emph{Int. J. Numer. Meth. Fluids}, 67:\penalty0 2026--2042, 2011.
\newblock \doi{10.1002/fld.2481}.

\bibitem[Courant et~al.(1967)Courant, Friedrichs, and Lewy]{Courant1967}
R.~Courant, K.~Friedrichs, and H.~Lewy.
\newblock On the partial difference equations of mathematical physics.
\newblock \emph{IBM J. Res. Dev.}, 11:\penalty0 215--234, 1967.
\newblock \doi{10.1147/rd.112.0215}.

\bibitem[Violeau and Leroy(2014)]{Violeau2014}
D.~Violeau and A.~Leroy.
\newblock On the maximum time step in weakly compressible {SPH}.
\newblock \emph{J. Comput. Phys.}, 256:\penalty0 388--415, 2014.
\newblock \doi{10.1016/j.jcp.2013.09.001}.

\bibitem[Violeau and Leroy(2015)]{Violeau2015}
D.~Violeau and A.~Leroy.
\newblock Optimal time step for incompressible {SPH}.
\newblock \emph{J. Comput. Phys.}, 288:\penalty0 119--130, 2015.
\newblock \doi{10.1016/j.jcp.2015.02.015}.

\bibitem[He et~al.(2021)He, Liu, Gan, Seaid, and Niu]{He2021}
L.~He, S.~Liu, Y.~Gan, M.~Seaid, and C.~Niu.
\newblock Development of time-space adaptive smoothed particle hydrodynamics
  method with {Runge-Kutta Chebyshev} scheme.
\newblock \emph{Eng. Anal. Bound. Elem.}, 126:\penalty0 55--67, 2021.
\newblock \doi{10.1016/j.enganabound.2021.02.004}.

\bibitem[Childs et~al.(2021)Childs, Liu, and Ostrander]{Childs2021p}
A.~M. Childs, J.-P. Liu, and A.~Ostrander.
\newblock High-precision quantum algorithms for partial differential equations.
\newblock \emph{Quantum}, 5:\penalty0 574, 2021.
\newblock \doi{10.22331/q-2021-11-10-574}.

\bibitem[Linden et~al.(2022)Linden, Montanaro, and Shao]{Linden2022}
N.~Linden, A.~Montanaro, and C.~Shao.
\newblock Quantum vs. classical algorithms for solving the heat equation.
\newblock \emph{Commun. Math. Phys.}, 395:\penalty0 601--641, 2022.
\newblock \doi{10.1007/s00220-022-04442-6}.

\bibitem[Liu et~al.(2021{\natexlab{a}})Liu, Wu, Wan, Pan, Qin, Gao, and
  Wen]{Liu2021v}
H.-L. Liu, Y.-S. Wu, L.-C. Wan, S.-J. Pan, S.-J. Qin, F.~Gao, and Q.-Y. Wen.
\newblock Variational quantum algorithm for the {Poisson} equation.
\newblock \emph{Phys. Rev. A}, 104:\penalty0 022418, 2021{\natexlab{a}}.
\newblock \doi{10.1103/PhysRevA.104.022418}.

\bibitem[Sato et~al.(2021)Sato, Kondo, Koide, Takamatsu, and Imoto]{Sato2021}
Y.~Sato, R.~Kondo, S.~Koide, H.~Takamatsu, and N.~Imoto.
\newblock Variational quantum algorithm based on the minimum potential energy
  for solving the {Poisson} equation.
\newblock \emph{Phys. Rev. A}, 104:\penalty0 052409, 2021.
\newblock \doi{10.1103/PhysRevA.104.052409}.

\bibitem[Robson et~al.(2022)Robson, Saha, Howington, Suh, and
  Nabrzyski]{Robson2022}
W.~Robson, K.~K. Saha, C.~Howington, I.~S. Suh, and J.~Nabrzyski.
\newblock Advanced quantum {Poisson} solver in the {NISQ} era.
\newblock In \emph{2022 QCE}, pages 741--744, USA, 2022. IEEE.
\newblock \doi{10.1109/QCE53715.2022.00103}.

\bibitem[Saha et~al.(2024)Saha, Robson, Howington, Suh, Wang, and
  Nabrzyski]{Saha2024}
K.~K. Saha, W.~Robson, C.~Howington, I.-S. Suh, Z.~Wang, and J.~Nabrzyski.
\newblock Enhancing scalability and accuracy of quantum poisson solver.
\newblock \emph{Quantum Inf. Process.}, 23:\penalty0 209, 2024.
\newblock \doi{10.1007/s11128-024-04420-y}.

\bibitem[Balducci et~al.(2022)Balducci, Chen, M{\"o}ller, Gerritsma, and
  Breuker]{Balducci2022}
G.~T. Balducci, B.~Chen, M.~M{\"o}ller, M.~Gerritsma, and R.~De Breuker.
\newblock Review and perspectives in quantum computing for partial differential
  equations in structural mechanics.
\newblock \emph{Front. Mech. Eng.}, 8:\penalty0 914241, 2022.
\newblock \doi{10.3389/fmech.2022.914241}.

\bibitem[Leyton and Osborne(2008)]{Leyton2008}
S.~K. Leyton and T.~J. Osborne.
\newblock A quantum algorithm to solve nonlinear differential equations, 2008.
\newblock \href{https://arxiv.org/abs/0812.4423}{arXiv:0812.4423}.

\bibitem[Liu et~al.(2021{\natexlab{b}})Liu, Kolden, Krovi, Loureiro, Trivisa,
  and Childs]{Liu2021p}
J.-P. Liu, H.~{\O}. Kolden, H.~K. Krovi, N.~F. Loureiro, K.~Trivisa, and A.~M.
  Childs.
\newblock Efficient quantum algorithm for dissipative nonlinear differential
  equations.
\newblock \emph{Proc. Natl. Acad. Sci.}, 118:\penalty0 e2026805118,
  2021{\natexlab{b}}.
\newblock \doi{10.1073/pnas.2026805118}.

\bibitem[Gaitan(2020)]{Gaitan2020}
F.~Gaitan.
\newblock Finding flows of a {Navier-Stokes} fluid through quantum computing.
\newblock \emph{npj Quantum Inf.}, 6:\penalty0 61, 2020.
\newblock \doi{10.1038/s41534-020-00291-0}.

\bibitem[Gaitan(2021)]{Gaitan2021}
F.~Gaitan.
\newblock Finding solutions of the {Navier-Stokes} equations through quantum
  computing--recent progress, a generalization, and next steps forward.
\newblock \emph{Adv. Quantum Technol.}, 4:\penalty0 2100055, 2021.
\newblock \doi{10.1002/qute.202100055}.

\bibitem[Kacewicz(1987)]{kacewicz1987optimal}
B.~Z. Kacewicz.
\newblock Optimal solution of ordinary differential equations.
\newblock \emph{J. Complex.}, 3:\penalty0 451--465, 1987.
\newblock \doi{10.1016/0885-064X(87)90011-2}.

\bibitem[Ingelman et~al.(2024)Ingelman, Bharadwaj, Pfeffer, Sreenivasan, and
  Schumacher]{ingelmann2024two}
J.~Ingelman, S.~S. Bharadwaj, P.~Pfeffer, K.~R. Sreenivasan, and J.~Schumacher.
\newblock Two quantum algorithms for solving the one-dimensional
  advection-diffusion equation.
\newblock \emph{Comput. Fluids}, 281:\penalty0 106369, 2024.
\newblock \doi{10.1016/j.compfluid.2024.106369}.

\bibitem[Weiss et~al.(2016)Weiss, Khoshgoftaar, and Wang]{weiss2016survey}
K.~Weiss, T.~M. Khoshgoftaar, and D.~Wang.
\newblock A survey of transfer learning.
\newblock \emph{J. Big Data}, 3:\penalty0 9, 2016.
\newblock \doi{10.1186/s40537-016-0043-6}.

\bibitem[Kordzanganeh et~al.(2023)Kordzanganeh, Sekatski, Fedichkin, and
  Melnikov]{kordzanganeh2023exponentially}
M.~Kordzanganeh, P.~Sekatski, L.~Fedichkin, and A.~Melnikov.
\newblock An exponentially-growing family of universal quantum circuits.
\newblock \emph{Mach. Learn.: Sci. Technol.}, 4:\penalty0 035036, 2023.
\newblock \doi{10.1088/2632-2153/ace757}.

\bibitem[Sedykh et~al.(2024)Sedykh, Podapaka, Sagingalieva, Pinto, Pflitsch,
  and Melnikov]{sedykh2024hybrid}
A.~Sedykh, M.~Podapaka, A.~Sagingalieva, K.~Pinto, M.~Pflitsch, and
  A.~Melnikov.
\newblock Hybrid quantum physics-informed neural networks for simulating
  computational fluid dynamics in complex shapes.
\newblock \emph{Mach. Learn.: Sci. Technol.}, 5:\penalty0 025045, 2024.
\newblock \doi{10.1088/2632-2153/ad43b2}.

\bibitem[Jaksch et~al.(2023)Jaksch, Givi, Daley, and Rung]{Jaksch2023}
D.~Jaksch, P.~Givi, A.~J. Daley, and T.~Rung.
\newblock Variational quantum algorithms for computational fluid dynamics.
\newblock \emph{AIAA J.}, 61:\penalty0 1885--1894, 2023.
\newblock \doi{10.2514/1.J062426}.

\bibitem[Mocz and Szasz(2021)]{Mocz2021}
P.~Mocz and A.~Szasz.
\newblock Toward cosmological simulations of dark matter on quantum computers.
\newblock \emph{Astrophys. J.}, 910:\penalty0 29, 2021.
\newblock \doi{10.3847/1538-4357/abe6ac}.

\bibitem[Leong et~al.(2024)Leong, Ewe, and Koh]{Leong2022}
F.~Y. Leong, W.-B. Ewe, and D.~E. Koh.
\newblock Variational quantum evolution equation solver.
\newblock \emph{Sci. Rep.}, 12:\penalty0 10817, 2024.
\newblock \doi{10.1038/s41598-022-14906-3}.

\bibitem[Sarma et~al.(2024)Sarma, Watts, Moosa, Liu, and McMahon]{Sarma2024}
A.~Sarma, T.~W. Watts, M.~Moosa, Y.~Liu, and P.~L. McMahon.
\newblock Quantum variational solving of nonlinear and multidimensional partial
  differential equations.
\newblock \emph{Phys. Rev. A}, 109:\penalty0 062616, 2024.
\newblock \doi{10.1103/PhysRevA.109.062616}.

\bibitem[Steijl and Barakos(2018)]{Steijl2018}
R.~Steijl and G.~N. Barakos.
\newblock Parallel evaluation of quantum algorithms for computational fluid
  dynamics.
\newblock \emph{Comput. Fluids}, 173:\penalty0 22--28, 2018.
\newblock \doi{10.1016/j.compfluid.2018.03.080}.

\bibitem[Gourianov et~al.(2022)Gourianov, Lubasch, Dolgov, van~den Berg,
  Babaee, Givi, Kiffner, and Jaksch]{Gourianov2022}
N.~Gourianov, M.~Lubasch, S.~Dolgov, Q.~Y. van~den Berg, H.~Babaee, P.~Givi,
  M.~Kiffner, and D.~Jaksch.
\newblock A quantum-inspired approach to exploit turbulence structures.
\newblock \emph{Nat. Comput. Sci.}, 2:\penalty0 30--37, 2022.
\newblock \doi{10.1038/s43588-021-00181-1}.

\bibitem[{HPCQS}(2023)]{HPCQS2023}
{HPCQS}.
\newblock {GENCI/CEA}, {FZJ}, and {PASQAL} announce significant milestone in
  hybrid computing {[Press release, 9 Nov 2023].}
\newblock
  \url{https://www.hpcqs.eu/news/genci-cea-fzj-and-pasqal-announce-significant-milestone-in-hybrid-computing}.,
  2023.
\newblock Accessed: 2023-12-01.

\bibitem[{University of Innsbruck}(2024)]{Innsbruck2024}
{University of Innsbruck}.
\newblock Supercomputer and quantum computer in harmony {[Press release, 23 Jul
  2024]}.
\newblock
  \url{https://www.uibk.ac.at/en/newsroom/2024/supercomputer-and-quantum-computer-in-harmony},
  2024.
\newblock Accessed: 2024-08-01.

\bibitem[{Google Quantum AI}(2023)]{Google2023}
{Google Quantum AI}.
\newblock Suppressing quantum errors by scaling a surface code logical qubit.
\newblock \emph{Nature}, 614:\penalty0 676--681, 2023.
\newblock \doi{10.1038/s41586-022-05434-1}.

\bibitem[Madsen et~al.(2022)Madsen, Laudenbach, Askarani, Rortais, Vincent,
  Bulmer, Miatto, Neuhaus, Helt, Collins, Lita, Gerrits, Nam, Vaidya, Menotti,
  Dhand, Vernon, Quesada, and Lavoie]{Madsen2022}
L.~S. Madsen, F.~Laudenbach, M.~F. Askarani, F.~Rortais, T.~Vincent, J.~F.~F.
  Bulmer, F.~M. Miatto, L.~Neuhaus, L.~G. Helt, M.~J. Collins, A.~E. Lita,
  T.~Gerrits, S.~W. Nam, V.~D. Vaidya, M.~Menotti, I.~Dhand, Z.~Vernon,
  N.~Quesada, and J.~Lavoie.
\newblock Quantum computational advantage with a programmable photonic
  processor.
\newblock \emph{Nature}, 606:\penalty0 75--81, 2022.
\newblock \doi{10.1038/s41586-022-04725-x}.

\bibitem[Blumoff et~al.(2022)Blumoff, Pan, Keating, Andrews, Barnes, Brecht,
  Croke, Euliss, Fast, Jackson, Jones, Kerckhoff, Lanza, Raach, Thomas,
  Velunta, Weinstein, Ladd, Eng, Borselli, Hunter, and Rakher]{Blumoff2022}
J.~Z. Blumoff, A.~S. Pan, T.~E. Keating, R.~W. Andrews, D.~W. Barnes, T.~L.
  Brecht, E.~T. Croke, L.~E. Euliss, J.~A. Fast, C.~A.~C. Jackson, A.~M. Jones,
  J.~Kerckhoff, R.~K. Lanza, K.~Raach, B.~J. Thomas, R.~Velunta, A.~J.
  Weinstein, T.~D. Ladd, K.~Eng, M.~G. Borselli, A.~T. Hunter, and M.~T.
  Rakher.
\newblock Fast and high-fidelity state preparation and measurement in
  triple-quantum-dot spin qubits.
\newblock \emph{PRX Quantum}, 3:\penalty0 010352, 2022.
\newblock \doi{10.1103/PRXQuantum.3.010352}.

\bibitem[Harty et~al.(2014)Harty, Allcock, Ballance, Guidoni, Janacek, Linke,
  Stacey, and Lucas]{Harty2014}
T.~P. Harty, D.~T.~C. Allcock, C.~J. Ballance, L.~Guidoni, H.~A. Janacek, N.~M.
  Linke, D.~N. Stacey, and D.~M. Lucas.
\newblock High-fidelity preparation, gates, memory, and readout of a
  trapped-ion quantum bit.
\newblock \emph{Phys. Rev. Lett.}, 113:\penalty0 220501, 2014.
\newblock \doi{10.1103/PhysRevLett.113.220501}.

\bibitem[Todaro et~al.(2021)Todaro, Verma, McCormick, Allcock, Mirin, Wineland,
  Nam, Wilson, Leibfried, and Slichter]{Todaro2021}
S.~L. Todaro, V.~B. Verma, K.~C. McCormick, D.~T.~C. Allcock, R.~P. Mirin,
  D.~J. Wineland, S.~W. Nam, A.~C. Wilson, D.~Leibfried, and D.~H. Slichter.
\newblock State readout of a trapped ion qubit using a trap-integrated
  superconducting photon detector.
\newblock \emph{Phys. Rev. Lett.}, 010501:\penalty0 126, 2021.
\newblock \doi{10.1103/PhysRevLett.126.010501}.

\bibitem[Kwon et~al.(2017)Kwon, Ebert, Walker, and Saffman]{Kwon2017}
M.~Kwon, M.~F. Ebert, T.~G. Walker, and M.~Saffman.
\newblock Parallel low-loss measurement of multiple atomic qubits.
\newblock \emph{Phys. Rev. Lett.}, 119:\penalty0 180504, 2017.
\newblock \doi{10.1103/PhysRevLett.119.180504}.

\bibitem[Istrati et~al.(2020)Istrati, Pilnyak, Loredo, Ant{\'o}n, Somaschi,
  Hilaire, Ollivier, Esmann, Cohen, Vidro, Millet, Lema{\^i}tre, Sagnes,
  Harouri, Lanco, Senellart, and Eisenberg]{Istrati2020}
D.~Istrati, Y.~Pilnyak, J.~C. Loredo, C.~Ant{\'o}n, N.~Somaschi, P.~Hilaire,
  H.~Ollivier, M.~Esmann, L.~Cohen, L.~Vidro, C.~Millet, A.~Lema{\^i}tre,
  I.~Sagnes, A.~Harouri, L.~Lanco, P.~Senellart, and H.~S. Eisenberg.
\newblock Sequential generation of linear cluster states from a single photon
  emitter.
\newblock \emph{Nat. Commun.}, 11:\penalty0 5501, 2020.
\newblock \doi{10.1038/s41467-020-19341-4}.

\bibitem[Cogan et~al.(2023)Cogan, Su, Kenneth, and Gershoni]{Cogan2023}
D.~Cogan, Z.-E. Su, O.~Kenneth, and D.~Gershoni.
\newblock Deterministic generation of indistinguishable photons in a cluster
  state.
\newblock \emph{Nat. Photon.}, 17:\penalty0 324--329, 2023.
\newblock \doi{10.1038/s41566-022-01152-2}.

\bibitem[Nickerson et~al.(2014)Nickerson, Fitzsimons, and
  Benjamin]{Nickerson2014}
N.~H. Nickerson, J.~F. Fitzsimons, and S.~C. Benjamin.
\newblock Freely scalable quantum technologies using cells of 5-to-50 qubits
  with very lossy and noisy photonic links.
\newblock \emph{Phys. Rev. X}, 4:\penalty0 041041, 2014.
\newblock \doi{10.1103/PhysRevX.4.041041}.

\bibitem[Zhong et~al.(2021)Zhong, Deng, Qin, Wang, Chen, Peng, Luo, Wu, Gong,
  Su, Hu, Hu, Yang, Zhang, Li, Li, Jiang, Gan, Yang, You, Wang, Li, Liu,
  Renema, Lu, and Pan]{Zhong2021}
H.-S. Zhong, Y.-H. Deng, J.~Qin, H.~Wang, M.-C. Chen, L.-C. Peng, Y.-H. Luo,
  D.~Wu, S.-Q. Gong, H.~Su, Y.~Hu, P.~Hu, X.-Y. Yang, W.-J. Zhang, H.~Li,
  Y.~Li, X.~Jiang, L.~Gan, G.~Yang, L.~You, Z.~Wang, L.~Li, N.-L. Liu, J.~J.
  Renema, C.-Y. Lu, and J.-W. Pan.
\newblock Phase-programmable {Gaussian} boson sampling using stimulated
  squeezed light.
\newblock \emph{Phys. Rev. Lett.}, 127:\penalty0 180502, 2021.
\newblock \doi{10.1103/PhysRevLett.127.180502}.

\bibitem[Long and Sun(2001)]{Long2001}
G.-L. Long and Y.~Sun.
\newblock Efficient scheme for initializing a quantum register with an
  arbitrary superposed state.
\newblock \emph{Phys. Rev. A}, 64:\penalty0 014303, 2001.
\newblock \doi{10.1103/PhysRevA.64.014303}.

\bibitem[Plesch and Brukner(2011)]{Plesch2011}
M.~Plesch and {\v C}.~Brukner.
\newblock Quantum-state preparation with universal gate decompositions.
\newblock \emph{Phys. Rev. A}, 83:\penalty0 032302, 2011.
\newblock \doi{10.1103/PhysRevA.83.032302}.

\bibitem[Sanders et~al.(2019)Sanders, Low, Scherer, and Berry]{Sanders2019}
Y.~R. Sanders, G.~H. Low, A.~Scherer, and D.~W. Berry.
\newblock Black-box quantum state preparation without arithmetic.
\newblock \emph{Phys. Rev. Lett.}, 122:\penalty0 020502, 2019.
\newblock \doi{10.1103/PhysRevLett.122.020502}.

\bibitem[Nakaji et~al.(2022)Nakaji, Uno, Suzuki, Raymond, Onodera, Tanaka,
  Tezuka, Mitsuda, and Yamamoto]{Nakaji2022}
K.~Nakaji, S.~Uno, Y.~Suzuki, R.~Raymond, T.~Onodera, T.~Tanaka, H.~Tezuka,
  N.~Mitsuda, and N.~Yamamoto.
\newblock Approximate amplitude encoding in shallow parameterized quantum
  circuits and its application to financial market indicators.
\newblock \emph{Phys. Rev. Res.}, 4:\penalty0 023136, 2022.
\newblock \doi{10.1103/PhysRevResearch.4.023136}.

\bibitem[Mitsuda et~al.(2024)Mitsuda, Ichimura, Nakaji, Suzuki, Tanaka,
  Raymond, Tezuka, Onodera, and Yamamoto]{Mitsuda2024}
N.~Mitsuda, T.~Ichimura, K.~Nakaji, Y.~Suzuki, T.~Tanaka, R.~Raymond,
  H.~Tezuka, T.~Onodera, and N.~Yamamoto.
\newblock Approximate complex amplitude encoding algorithm and its application
  to data classification problems.
\newblock \emph{Phys. Rev. A}, 109:\penalty0 052423, 2024.
\newblock \doi{10.1103/PhysRevA.109.052423}.

\bibitem[Ollitrault et~al.(2020{\natexlab{b}})Ollitrault, Kandala, Chen,
  Barkoutsos, Mezzacapo, Pistoia, Sheldon, Woerner, Gambetta, and
  Tavernelli]{Ollitrault2020e}
P.~J. Ollitrault, A.~Kandala, C.-F. Chen, P.~K. Barkoutsos, A.~Mezzacapo,
  M.~Pistoia, S.~Sheldon, S.~Woerner, J.~M. Gambetta, and I.~Tavernelli.
\newblock Quantum equation of motion for computing molecular excitation
  energies on a noisy quantum processor.
\newblock \emph{Phys. Rev. Res.}, 2:\penalty0 043140, 2020{\natexlab{b}}.
\newblock \doi{10.1103/PhysRevResearch.2.043140}.

\bibitem[Chan et~al.(2023)Chan, Meister, Jones, Tew, and Benjamin]{Chan2023}
H.~H.~S. Chan, R.~Meister, T.~Jones, D.~P. Tew, and S.~C. Benjamin.
\newblock Grid-based methods for chemistry simulations on a quantum computer.
\newblock \emph{Sci. Adv.}, 9:\penalty0 eabo7484, 2023.
\newblock \doi{10.1126/sciadv.abo7484}.

\bibitem[Jaques and Rattew(2023)]{Jaques2023}
S.~Jaques and A.~G. Rattew.
\newblock {QRAM}: A survey and critique, 2023.
\newblock \href{https://doi.org/10.48550/arxiv.2305.10310}{arXiv:2305.10310}.

\bibitem[Hann et~al.(2021)Hann, Lee, Girvin, and Jiang]{Hann2021}
C.~T. Hann, G.~Lee, S.~M. Girvin, and L.~Jiang.
\newblock Resilience of quantum random access memory to generic noise.
\newblock \emph{PRX Quantum}, 2:\penalty0 020311, 2021.
\newblock \doi{10.1103/PRXQuantum.2.020311}.

\bibitem[Giovannetti et~al.(2008{\natexlab{a}})Giovannetti, Lloyd, and
  Maccone]{Giovannetti2008}
V.~Giovannetti, S.~Lloyd, and L.~Maccone.
\newblock Quantum random access memory.
\newblock \emph{Phys. Rev. Lett.}, 100:\penalty0 160501, 2008{\natexlab{a}}.
\newblock \doi{10.1103/PhysRevLett.100.160501}.

\bibitem[Giovannetti et~al.(2008{\natexlab{b}})Giovannetti, Lloyd, and
  Maccone]{Giovannetti2008a}
V.~Giovannetti, S.~Lloyd, and L.~Maccone.
\newblock Architectures for a quantum random access memory.
\newblock \emph{Phys. Rev. A}, 78:\penalty0 052310, 2008{\natexlab{b}}.
\newblock \doi{10.1103/PhysRevA.78.052310}.

\bibitem[Ciliberto et~al.(2018)Ciliberto, Herbster, Ialongo, Pontil, Rocchetto,
  Severini, and Wossnig]{Ciliberto2018}
C.~Ciliberto, M.~Herbster, A.~D. Ialongo, M.~Pontil, A.~Rocchetto, S.~Severini,
  and L.~Wossnig.
\newblock Quantum machine learning: a classical perspective.
\newblock \emph{Proc. R. Soc. A}, 474:\penalty0 20170551, 2018.
\newblock \doi{10.1098/rspa.2017.0551}.

\bibitem[Arunachalam et~al.(2015)Arunachalam, Gheorghiu, Jochym-O'Connor,
  Mosca, and Srinivasan]{Arunachalam2015}
S.~Arunachalam, V.~Gheorghiu, T.~Jochym-O'Connor, M.~Mosca, and P.~V.
  Srinivasan.
\newblock On the robustness of bucket brigade quantum {RAM}.
\newblock \emph{New J. Phys.}, 17:\penalty0 123010, 2015.
\newblock \doi{10.1088/1367-2630/17/12/123010}.

\bibitem[Matteo et~al.(2020)Matteo, Gheorghiu, and Mosca]{DiMatteo2020}
O.~Di Matteo, V.~Gheorghiu, and M.~Mosca.
\newblock Fault-tolerant resource estimation of quantum random-access memories.
\newblock \emph{IEEE Trans. Quantum Eng.}, 1:\penalty0 1--13, 2020.
\newblock \doi{10.1109/TQE.2020.2965803}.

\bibitem[IEEE(2019)]{IEEE754}
IEEE.
\newblock {IEEE} standard for floating-point arithmetic.
\newblock \emph{{IEEE Std} 754-2019 (Revision of {IEEE} 754-2008)},
  2019:\penalty0 1--84, 2019.
\newblock \doi{10.1109/IEEESTD.2019.8766229}.

\bibitem[{IEEE Standards Association}(2019)]{IEEE7130}
{IEEE Standards Association}.
\newblock {P7130: Standard} for quantum technologies definitions.
\newblock \url{https://standards.ieee.org/ieee/7130/10680/}, 2019.
\newblock Accessed: 2023-12-01.

\bibitem[Seidel et~al.(2022)Seidel, Tcholtchev, Bock, Becker, and
  Hauswirth]{Seidel2022}
R.~Seidel, N.~Tcholtchev, S.~Bock, C.~K.-U. Becker, and M.~Hauswirth.
\newblock Efficient floating point arithmetic for quantum computers.
\newblock \emph{IEEE Access}, 10:\penalty0 72400--72415, 2022.
\newblock \doi{10.1109/ACCESS.2022.3188251}.

\bibitem[Gily{\'e}n et~al.(2019)Gily{\'e}n, Su, Low, and Wiebe]{Gilyen2019}
A.~Gily{\'e}n, Y.~Su, G.~H. Low, and N.~Wiebe.
\newblock Quantum singular value transformation and beyond: exponential
  improvements for quantum matrix arithmetics.
\newblock In \emph{Proc. 51st ACM SIGACT STOC}, pages 193--204. ACM, New York,
  NY, 2019.
\newblock \doi{10.1145/3313276.3316366}.

\bibitem[Camps and Beeumen(2020)]{Camps2020b}
D.~Camps and R.~Van Beeumen.
\newblock Approximate quantum circuit synthesis using block encodings.
\newblock \emph{Phys. Rev. A}, 102:\penalty0 052411, 2020.
\newblock \doi{10.1103/PhysRevA.102.052411}.

\bibitem[S{\"u}nderhauf et~al.(2024)S{\"u}nderhauf, Campbell, and
  Camps]{Sunderhauf2024}
C.~S{\"u}nderhauf, E.~Campbell, and J.~Camps.
\newblock Block-encoding structured matrices for data input in quantum
  computing.
\newblock \emph{Quantum}, 8:\penalty0 1226, 2024.
\newblock \doi{10.22331/q-2024-01-11-1226}.

\bibitem[Chakraborty et~al.(2019)Chakraborty, Gily{\'e}n, and
  Jeffery]{Chakraborty2019}
S.~Chakraborty, A.~Gily{\'e}n, and S.~Jeffery.
\newblock The power of block-encoded matrix powers: Improved regression
  techniques via faster {Hamiltonian} simulation.
\newblock In C.~Baier, I.~Chatzigiannakis, P.~Flocchini, and S.~Leonardi,
  editors, \emph{46th ICALP 2019}, volume 132 of \emph{LIPIcs}, pages
  33:1--33:14. Schloss Dagstuhl--Leibniz-Zentrum fuer Informatik, Dagstuhl,
  Germany, 2019.
\newblock \doi{10.4230/LIPIcs.ICALP.2019.33}.

\bibitem[Martyn et~al.(2021)Martyn, Rossi, Tan, and Chuang]{Martyn2021}
J.~M. Martyn, Z.~M. Rossi, A.~K. Tan, and I.~L. Chuang.
\newblock Grand unification of quantum algorithms.
\newblock \emph{PRX Quantum}, 2:\penalty0 040203, 2021.
\newblock \doi{10.1103/PRXQuantum.2.040203}.

\bibitem[Chancellor(2019)]{Chancellor2019}
N.~Chancellor.
\newblock Domain wall encoding of discrete variables for quantum annealing and
  {QAOA}.
\newblock \emph{Quantum Sci. Technol.}, 4:\penalty0 045004, 2019.
\newblock \doi{10.1088/2058-9565/ab33c2}.

\bibitem[Berwald et~al.(2023)Berwald, Chancellor, and Dridi]{Berwald2023}
J.~Berwald, N.~Chancellor, and R.~Dridi.
\newblock Understanding domain-wall encoding theoretically and experimentally.
\newblock \emph{Philos. Trans. R. Soc. A}, 381:\penalty0 20210410, 2023.
\newblock \doi{10.1098/rsta.2021.0410}.

\bibitem[Plewa et~al.(2021)Plewa, Sie{\'n}ko, and Rycerz]{Plewa2021}
J.~Plewa, J.~Sie{\'n}ko, and K.~Rycerz.
\newblock Variational algorithms for workflow scheduling problem in gate-based
  quantum devices.
\newblock \emph{Comput. Inform.}, 40:\penalty0 897--929, 2021.
\newblock \doi{10.31577/cai_2021_4_897}.

\bibitem[Barnes et~al.(2023)Barnes, Buyskikh, Chen, Gallardo, Ghibaudi,
  Ruszala, Underwood, Agarwal, Lall, Rungger, and Schoinas]{Barnes2023}
K.~M. Barnes, A.~Buyskikh, N.~Y. Chen, G.~Gallardo, M.~Ghibaudi, M.~J.~A.
  Ruszala, D.~S. Underwood, A.~Agarwal, D.~Lall, I.~Rungger, and N.~Schoinas.
\newblock Optimising the quantum/classical interface for efficiency and
  portability with a multi-level hardware abstraction layer for quantum
  computers.
\newblock \emph{EPJ Quantum Technol.}, 10:\penalty0 36, 2023.
\newblock \doi{10.1140/epjqt/s40507-023-00192-z}.

\bibitem[Statista(2019)]{Statista}
Statista.
\newblock Number of qubits achieved in quantum computers by
  company/organization from 1998 to 2019.
\newblock
  \url{https://www.statista.com/statistics/993634/quantum-computers-by-number-of-qubits/},
  2019.
\newblock Accessed: 2023-12-01.

\bibitem[Mandelbaum(2021)]{Mandelbaum2021}
R.~Mandelbaum.
\newblock Five years ago today, we put the first quantum computer on the cloud.
  {Here's} how we did it. {[Press release, 4 May 2021]}.
\newblock \url{https://research.ibm.com/blog/quantum-five-years}, 2021.
\newblock Accessed: 2023-11-01.

\bibitem[Castelvecchi(2023{\natexlab{b}})]{Castelvecchi2023}
D.~Castelvecchi.
\newblock {IBM} releases first-ever 1,000-qubit quantum chip {[Press release, 4
  Dec 2023]}.
\newblock \emph{Nature}, 624:\penalty0 238, 2023{\natexlab{b}}.
\newblock \doi{10.1038/d41586-023-03854-1}.

\bibitem[Leu et~al.(2023)Leu, Gely, Weber, Smith, Nadlinger, and
  Lucas]{Leu2023}
A.~D. Leu, M.~F. Gely, M.~A. Weber, M.~C. Smith, D.~P. Nadlinger, and D.~M.
  Lucas.
\newblock Fast, high-fidelity addressed single-qubit gates using efficient
  composite pulse sequences.
\newblock \emph{Phys. Rev. Lett.}, 131:\penalty0 120601, 2023.
\newblock \doi{10.1103/PhysRevLett.131.120601}.

\bibitem[Cong et~al.(2022)Cong, Levine, Keesling, Bluvstein, Wang, and
  Lukin]{Cong2022}
I.~Cong, H.~Levine, A.~Keesling, D.~Bluvstein, S.-T. Wang, and M.~D. Lukin.
\newblock Hardware-efficient, fault-tolerant quantum computation with {Rydberg}
  atoms.
\newblock \emph{Phys. Rev. X}, 12:\penalty0 021049, 2022.
\newblock \doi{10.1103/PhysRevX.12.021049}.

\bibitem[Wintersperger et~al.(2023)Wintersperger, Dommert, Ehmer, Hoursanov,
  Klepsch, Mauerer, Reuber, Strohm, Yin, and Luber]{Wintersperger2023}
K.~Wintersperger, F.~Dommert, T.~Ehmer, A.~Hoursanov, J.~Klepsch, W.~Mauerer,
  G.~Reuber, T.~Strohm, M.~Yin, and S.~Luber.
\newblock Neutral atom quantum computing hardware: performance and end-user
  perspective.
\newblock \emph{EPJ Quantum Technol.}, 10:\penalty0 32, 2023.
\newblock \doi{10.1140/epjqt/s40507-023-00190-1}.

\bibitem[Dhar(2020)]{Dhar2020}
P.~Dhar.
\newblock The carbon impact of artificial intelligence.
\newblock \emph{Nat. Mach. Intell.}, 2:\penalty0 423--425, 2020.
\newblock \doi{10.1038/s42256-020-0219-9}.

\bibitem[Debus et~al.(2023)Debus, Piraud, Streit, Theis, and
  G{\"o}tz]{Debus2023}
C.~Debus, M.~Piraud, A.~Streit, F.~Theis, and M.~G{\"o}tz.
\newblock Reporting electricity consumption is essential for sustainable {AI}.
\newblock \emph{Nat. Mach. Intell.}, 5:\penalty0 1176--1178, 2023.
\newblock \doi{10.1038/s42256-023-00750-1}.

\bibitem[Strubell et~al.(2019)Strubell, Ganesh, and McCallum]{Strubell2019}
E.~Strubell, A.~Ganesh, and A.~McCallum.
\newblock Energy and policy considerations for deep learning in {NLP}.
\newblock In \emph{Proc. 57th ACL}, pages 3645--3650, Stroudsburg, PA, 2019.
  ACL.
\newblock \doi{10.18653/v1/P19-1355}.

\bibitem[{Royal Academy of Engineering}(2024)]{RAEng2024}
{Royal Academy of Engineering}.
\newblock Quantum infrastructure review: An independent review of the {UK}'s
  quantum sector's infrastructure requirements for the next decade, 2024.
\newblock
  \url{https://raeng.org.uk/policy-and-resources/research-and-innovation/how-can-we-improve-quantum-infrastructure-in-the-uk}.

\bibitem[Chen(2023)]{Chen2023}
S.~Chen.
\newblock The future is quantum: universities look to train engineers for an
  emerging industry.
\newblock \emph{Nature}, 623:\penalty0 653--655, 2023.
\newblock \doi{10.1038/d41586-023-03511-7}.

\bibitem[Goorney et~al.(2024)Goorney, Sarantinou, and Sherson]{Goorney2024}
S.~Goorney, M.~Sarantinou, and J.~Sherson.
\newblock The quantum technology open master: widening access to the quantum
  industry.
\newblock \emph{EPJ Quantum Technol.}, 11:\penalty0 7, 2024.
\newblock \doi{10.1140/epjqt/s40507-024-00217-1}.

\bibitem[Heidt(2024)]{Heidt2024}
A.~Heidt.
\newblock Quantum computing aims for diversity, one qubit at a time.
\newblock \emph{Nature}, 632:\penalty0 464--465, 2024.
\newblock \doi{10.1038/d41586-024-02541-z}.

\bibitem[Meyer et~al.(2024)Meyer, Passante, and Wilcox]{Meyer2024}
J.~C. Meyer, G.~Passante, and B.~Wilcox.
\newblock Disparities in access to {U.S.} quantum information education.
\newblock \emph{Phys. Rev. Phys. Educ. Res.}, 20:\penalty0 010131, 2024.
\newblock \doi{10.1103/PhysRevPhysEducRes.20.010131}.

\bibitem[Rosenberg et~al.(2024)Rosenberg, Holincheck, and
  Colandene]{Rosenberg2024}
J.~L. Rosenberg, N.~Holincheck, and M.~Colandene.
\newblock Science, technology, engineering, and mathematics undergraduates'
  knowledge and interest in quantum careers: Barriers and opportunities to
  building a diverse quantum workforce.
\newblock \emph{Phys. Rev. Phys. Educ. Res.}, 20:\penalty0 010138, 2024.
\newblock \doi{10.1103/PhysRevPhysEducRes.20.010138}.

\bibitem[Holter et~al.(2023)Holter, Inglesant, and Jirotka]{TenHolter2023}
C.~Ten Holter, P.~Inglesant, and M.~Jirotka.
\newblock Reading the road: challenges and opportunities on the path to
  responsible innovation in quantum computing.
\newblock \emph{Technol. Anal. Strat. Manag.}, 35:\penalty0 844--856, 2023.
\newblock \doi{10.1080/09537325.2021.1988070}.

\bibitem[{World Economic Forum} et~al.(2024){World Economic Forum}, {IBM}, and
  {SandboxAQ}]{WEF2024}
{World Economic Forum}, {IBM}, and {SandboxAQ}.
\newblock Quantum economy blueprint.
\newblock Insight report, World Economic Forum, 2024.
\newblock
  \url{https://www.weforum.org/publications/quantum-economy-blueprint/}.

\bibitem[Neumann et~al.(2021)Neumann, van Heesch, Phillipson, and
  Smallegange]{Neumann2021}
N.~M.~P. Neumann, M.~P.~P. van Heesch, F.~Phillipson, and A.~A.~P. Smallegange.
\newblock Quantum computing for military applications.
\newblock In \emph{2021 ICMCIS}, pages 1--8. IEEE, 2021.
\newblock \doi{10.1109/ICMCIS52405.2021.9486419}.

\end{thebibliography}


\end{document}